\documentclass[a4paper,12pt]{book}
\usepackage{amsfonts}
\usepackage{amsmath}
\usepackage[dvips]{graphicx}
\usepackage{epsfig}
\usepackage[section]{placeins}

\def\frac#1#2{{#1\over#2}}

\def\half{\ifinner {\scriptstyle {1 \over 2}}%
          \else {\textstyle {1 \over 2}}\fi}

\def\simge{
    \mathrel{\rlap{\raise 0.511ex
        \hbox{$>$}}{\lower 0.511ex \hbox{$\sim$}}}}

\def\simle{
    \mathrel{\rlap{\raise 0.511ex
        \hbox{$<$}}{\lower 0.511ex \hbox{$\sim$}}}}

\def\therefore{
   \setbox0=\hbox{$.\kern.2em.$}\dimen0=\wd0    %
   \mathrel{\rlap{\raise.25ex\hbox to\dimen0{\hfil$\cdotp$\hfil}}%
   \copy0}}

\def\|{\ifmmode\Vert\else \char`\|\fi}

\def\tr{\mathop{\rm tr}\nolimits}       

\def\sc#1{\setbox0=\hbox{$#1$}           
   \dimen0=\wd0                                 
   \setbox1=\hbox{/} \dimen1=\wd1               
   \ifdim\dimen0>\dimen1                        
      \rlap{\hbox to \dimen0{\hfil/\hfil}}      
      #1                                        
   \else                                        
      \rlap{\hbox to \dimen1{\hfil$#1$\hfil}}   
      /                                         
   \fi}                                         %

\def\subrightarrow#1{
  \setbox0=\hbox{
    $\displaystyle\mathop{}
    \limits_{#1}$}
  \dimen0=\wd0
  \advance \dimen0 by .5em
  \mathrel{
    \mathop{\hbox to \dimen0{\rightarrowfill}}
       \limits_{#1}}}                           

\def\vbig#1#2{{\vbigd@men=#2\divide\vbigd@men by 2%
   \hbox{$\left#1\vbox to \vbigd@men{}\right.\n@space$}}}

\usepackage{fancyhdr}

\makeatletter
\def\cleardoublepage{\clearpage\if@twoside \ifodd\c@page\else
        \hbox{}
        \thispagestyle{empty}
        \newpage
        \if@twocolumn\hbox{}\newpage\fi\fi\fi}
\makeatother

\def\bea{\begin{eqnarray}}
\def\eea{\end{eqnarray}}
\def\beq{\begin{equation}}
\def\eeq{\end{equation}}

\def\as{\alpha_s}
\def\az{\alpha_0}
\relax
\def\abs#1{\left|#1\right|}

\def    \hepph  #1 {{\tt hep-ph/#1}}
\def    \hepex  #1 {{\tt hep-ex/#1}}

\newsavebox\tmpfig

\begin{document}

\thispagestyle{empty}
\begin{center}

{\bf
UNIVERSIT\`A DEGLI STUDI DI MILANO\\
\vskip 0.2truecm
Facolt\`a di Scienze Matematiche Fisiche e Naturali \\
\vskip 0.2truecm
Dipartimento di Fisica\\
\vskip 0.2truecm
Corso di Dottorato di Ricerca in Fisica, Astrofisica e Fisica Applicata \\
\vskip 0.2truecm
Ciclo XIX\vspace{3.cm} } \\
{\bf TESI DI DOTTORATO DI RICERCA:\\
 \vskip 0.5truecm
\large{Sudakov resummation in QCD}}\\
\vskip 0.5truecm
Settore scientifico: FIS/02
\end{center}

\vspace{2cm}

\begin{flushleft}
{\bf
Tutor:\hspace{2.23cm}Prof. Stefano FORTE\\
\vskip 0.2truecm
Referee:\hspace{1.87cm}Prof. Giovanni RIDOLFI\\
\vskip 0.2truecm
Coordinatore:\hspace{0.7cm}Prof. Gianpaolo BELLINI\\
\ }
\end{flushleft}

\vspace{2cm}

\begin{flushleft}
{\bf
\hspace{8.5cm}Tesi di dottorato di:\\
\vskip 0.1truecm
\hspace{8.5cm}Paolo BOLZONI \\
\vskip 0.1truecm
\hspace{8.5cm}Matr. R05430}
\end{flushleft}
\vspace{1.5cm}

\begin{center}
{\bf Anno Accademico 2005-2006}
\end{center}

\thispagestyle{empty}

\begin{flushright}

\end{flushright}

\begin{center}
\vspace*{0.5cm}
{\Large \bf Sudakov resummation in QCD} \\
\vspace*{1.5cm}
Paolo Bolzoni\\
\vspace{0.6cm}  
\vspace*{1.5cm}

{\bf Abstract}
\end{center}

\noindent

In this PhD thesis, we analyze and generalize the renormalization group approach 
to the resummation of large logarithms in  the perturbative expansion due 
to soft and collinear multiparton emissions. In particular, we present a 
generalization of this approach to prompt photon production.  It is interesting to see that also with the more intricate 
two-scale kinematics that characterizes prompt photon production in the 
soft limit, it remains true that
resummation simply follows from general kinematic properties of the 
phase space. Also, this approach does not require a separate treatment of 
individual colour structures when more than one
colour structure contributes to fixed order results. 
However, the resummation formulae obtained here turn out to be less predictive 
than previous results: this depends on the fact that here neither specific 
factorization properties of the cross section in the soft limit is assumed, 
nor that soft emission satisfies eikonal-like relations. We also derive  
resumation formulae to all logarithmic accuracy and valid for all values of
rapidity for the prompt photon production and 
the Drell-Yan rapidity distributions. We show that for the fixed-target
experiment E866/NuSea, the NLL resummation corrections are comparable to NLO
fixed-order corrections and are crucial to obtain agreement with the data. 
Finally we outline also possible future applications 
of the renormalization group approach. \vspace*{1cm}

\vfill
\noindent


\frontmatter

\title{Sudakov resummation in QCD}
\author{Paolo Bolzoni}
\date{}

\maketitle \tableofcontents
\chapter{Introduction}

The current theory that describes the strong subnuclear
interactions is the non-abelian gauge theory of the SU(3) local
symmetry group. This theory, called Quantum Chromo-Dynamics (QCD),
has been tested to great accuracy in many experiments. Its
perturbative regime will be crucial at the imminent high-energy
proton-proton collider LHC. The main target of LHC, is to find
signals of the Standard Model Higgs boson and of new physics, i.e.
supersymmetry, new interactions predicted by great unified
theories, extra-dimensions, new gauge bosons and mini-black holes.
In order to accomplish all this, an excellent understanding of QCD
is necessary, both because LHC is a proton collider and because
QCD backgrounds must be described accurately.

In perturbative QCD (pQCD), it is well known that, when one
approaches to the boundary of the phase space, the cross section
receives logarithmical-enhanced contributions of soft and
collinear origin at all orders. These large tems at order
$\emph{O}(\alpha_{s}^{n})$ of the strong coupling constant in pQCD
have, in general, the form
\begin{equation}
\alpha_{s}^{n}\left[\frac{\log^{m}(1-z)}{1-z}\right]_{+},\quad
m\leq 2n-1\nonumber,
\end{equation}
where $z$ is a parameter that becomes close to 1 near to the phase
space boundary and ``$+$'' denotes the plus distribution. These
terms become important in the limit $z\rightarrow 1$ spoiling the
validity of the perturbative fixed-order QCD expansion. Hence,
they should be resummed to all-orders of pQCD in order to get
accurate predictions of the cross sections of QCD processes. An
example of the importance that can have resummation is given by
the Higgs boson production at LHC near the threshold of its
production. In this case $z$ is given by the ratio of the searched
Higgs mass over the center-of-mass energy of the partons in the
colliding hadrons and the large logarithms arise from soft-gluon
emissions.

These large terms have been resummed a long time ago for the
classes of inclusive hadronic processes of the type of
deep-inelastic scattering (DIS) and Drell-Yan (DY)
\cite{Sterman:1986aj,Catani:1989ne,Catani:1990rp}. Threshold
resummation of inclusive processes can affect significantly cross
sections and the extraction of parton densities
\cite{Corcella:2005us,Corcella:2005ig}. For the case of small
transverse momentum distributions in Drell-Yan processes, it has
been shown that resummation is necessary to reproduce the correct
behavior of the cross section \cite{Collins:1984kg}.

These results has been obtained using the eikonal approaximation
of Ref.\cite{Catani:1989ne} which generalizes to QCD the Sudakov
exponentiation of soft photons emissions in electrodynamics
\cite{Sudakov:1954sw} or assuming suitable factorization
properties of the QCD cross section \cite{Sterman:1986aj}. More
recently two other approaches to resummation have been proposed.
The first is the renormalization group approach of
Ref.\cite{Forte:2002ni} and the other is the effective field
theoretic (EFT) approach of
Refs.\cite{Manohar:2003vb,Idilbi:2005ky}. There the resummation of
the large logarithms for full inclusive deep-inelastic and
Drell-Yan processes is obtained.

In this thesis we will concentrate mostly on the study of the
renormalization group approach of Ref.\cite{Forte:2002ni} and on
its applications and generalizations. This approach has the
advantage of being valid to all logarithmic orders, and
self-contained,  in that it does not require any factorization
beyond the standard factorization of collinear singularities. It
relies on an essentially kinematical analysis of the phase space
for the given process in the soft limit, which is used to
establish the result that the dependence on the resummation
variable only appears through a given fixed dimensionful
combination. This provides a second dimensionful variable, along
with the hard scale of the process, which can be resummed using
standard renormalization group techniques. Beyond the leading log
level, the resummed result found within this approach turns out to
be somewhat less predictive than the result obtained with the
other methods. In the other approaches references resummed results
at a certain logarithmic accuracy is fully determined by a ceratin
fixed order computation, whereas a higher fixed order computation
is needed to determine all coefficients in the resummed formula of
Ref.\cite{Forte:2002ni}. The more predictive result is recovered
within this approach if the dependence of the perturbative
coefficients on the two dimensionful variables factorizes, i.e. if
the two-scale factorization mentioned above holds.

We shall show the generalization of the renormalization group
approach to the resummation of  the inclusive transverse momentum
spectrum of prompt photons produced in hadronic collisions in the
region where the transverse momentum is close to its maximal
value. Prompt photon production is a less inclusive process than
Drell-Yan or deep-inelastic scattering, and it is especially
interesting from the point of view of the renormalization group
approach, because the large logs which must be resummed turn out
to depend on two independent dimensionful variables, on top of the
hard scale of the process: hence, prompt photon production is
characterized by three scales.  The possibility that the general
factorization Ref.\cite{Collins:1989gx} might extend to prompt
photon production was discussed in Ref.\cite{Laenen:1998qw}, based
on previous generalizations \cite{Kidonakis:1998nf} of
factorization, and used to derive the corresponding resummed
results. Resummation formulae for prompt photon production in the
approach of Ref.\cite{Catani:1989ne} were also proposed in
Ref.\cite{Catani:1998tm}, and some arguments which might support
such resummation were presented in Ref.\cite{Bonciani:2003nt}. Our
treatment will provide a full proof of resummation to all
logarithmic orders. Our resummation formula does not require the
factorization proposed in Refs.\cite{Laenen:1998qw,Catani:1998tm},
and it is accordingly less predictive. Because of the presence of
two scales, it is also weaker than the result of
Ref.\cite{Forte:2002ni} for DIS and Drell-Yan production.
Increasingly more predictive results are recovered if increasingly
restrictive forms of factorization hold.

Moreover, we shall prove an all-logarithmic orders resummation
formula for differential rapidity of Drell-Yan and prompt photon
production processes. The differential rapidity Drell-Yan cross
section is used for the extraction of the ratio $\bar{d}/\bar{u}$
of parton densities. The accurate knowledge of these functions is
needed to study Higgs boson production and the asymmetry
$W^{\pm}$. The resummation of Drell-Yan rapidity distributions was
first considered in 1992 \cite{Laenen:1992ey}. At that time, it
was suggested a resummation formula for the case of zero rapidity.
Very recently, thanks to the analysis of the full NLO calculation
of the Drell-Yan rapidity distribution, it has been shown
\cite{Mukherjee:2006uu}, that the result given in
\cite{Laenen:1992ey} is valid at next-to-leading logarithmic
accuracy (NLL) for all rapidities. In this thesis, we shall give a
simple proof of an all-order resummation formula valid  for all
values of rapidity. To do this, we will use the technique of the
double Fourier-Mellin moments developed in \cite{Sterman:2000pt}.
In particular, we will show that the resummation can be reduced to
that of the rapidity-integrated process, which is given in terms
of a dimensionless universal function for both DY and $W^{\pm}$
and $Z^{0}$ production. Then, we implement numerically the
resummation formula and give predictions of the full
rapidity-dependent NLL Drell-Yan cross section for the case of the
fixed-target E866/NuSea experiment. We find that resummation at
the NLL level is necessary and that its agreement with the
experimental data is better than the NNLO calculation of
Ref.\cite{Anastasiou:2003yy}. In this case, we find also that the
NLL resummation reduced the cross section instead of enhancing it
for the parameter choices of this
experiment. Threshold corrections to Higgs, $Z^{0}$ and
$W^{\pm}$ production rapidity distributions at high energy hadron colliders
have also been studied in \cite{Ravindran:2006bu,Ravindran:2007av}.

Finally, we shall discuss the application of the renormalization
group approach to the resummation another class of large
logarithms that arise in Drell-Yan processes for small tranverse
momentum distribution of the produces lepton pairs. These
logarithmic-enhanced terms at order $\emph{O}(\alpha_{s}^{n})$
have, in general, the form
\begin{equation}
\alpha_{s}^{n}\left[\frac{\log^{m}(q_{\perp}^{2})}{q_{\perp}^{2}}\right]_{+},\quad
m\leq 2n-1\nonumber,
\end{equation}
where $q_{\perp}$ is the transverse momentum of the produced
Drell-Yan pair. Also in this case the resummed results using the
renormalization group approach are less predictive than results
obtained with the approach of Ref.\cite{Sterman:1986aj}, as it is
shown in Ref.\cite{Collins:1984kg}. Furthermore the conditions
that reduce our results to those of Ref.\cite{Collins:1984kg} in
terms of factorization properties is still an interesting open
question.

This thesis is organized as follows. In Chapter \ref{QCD}, we
review the basics concepts of pQCD. In particular the construction
of the QCD Lagrangian, the asymptotic freedom of strong
interactions, the structure of the cross section when initial
state hadrons are present and the evolution equation of the parton
densities. In Chapter \ref{RI}, we discuss the importance of
resummation at hadron colliders. Then, we describe how in the
various approaches the large logarithms are exponentiated and
resummed. They are the renormalization group approach, the eikonal
approximation approach, the approach of non standard factorization
properties and the effective field theoretic approach. In Chapter
\ref{DISDY2}, we show in the detail the renormalization group
approach to the resummation of all inclusive deep-inelastic and
Drell-Yan processes and its generalization to the prompt photon
process at large transverse photon momentum is shown in Chapter
\ref{DP}. In Chapter \ref{DYDPR}, we prove the all-logarithmic
orders resummation formula for the Drell-Yan and prompt photon
production processes. We show also the impact of the NLL
resummation for the DY process at the E866/NuSea experiment and
discuss the numerical results. Then in Chapter \ref{QT}, we turn
to discuss the generalization of the renormalization group
approach to resummation in the case of the small transverse
momentum differential cross section for the Drell-Yan process.
Finally, in the last Chapter, we summarize and determine the
predictive power of the resummation formulae obtained with the
different approaches, namely, the fixed-order computation needed
to determine completely a resummation formula for an arbitrary
logarithmic accuracy.

\mainmatter
\chapter{General aspects of perturbative QCD}\label{QCD}

\section{Quarks, Gluons and QCD}

The quarks (the constituents of hadrons) are Dirac fermions. In
the Standard Model (SM), as far as the electroweak interactions
are concerned, the properties of quarks and leptons are similar.
Indeed, as for the leptons, the six quark flavors are grouped into
three $SU_{L}(2)$ left-handed doublets
\begin{equation}
\left(\begin{array}{c}u\\d\end{array}\right)_{L},\quad
\left(\begin{array}{c}c\\s\end{array}\right)_{L},\quad\left(\begin{array}{c}t\\b\end{array}\right)_{L}
\end{equation}
and six $SU(2)_{L}$ singlets, which are the right-handed parts of
each flavor. Both, quarks and leptons, interact in a similar way
with the electroweak gauge bosons ($\gamma$, $W^{\pm}$ and
$Z^{0}$) of the group $SU_{L}(2)\times U_{Y}(1)$. The main
difference is that each quark flavor eigenstate is a unitary
mixing of the quark mass eigenstate, while, according to the SM,
this is not the case for charged leptons and massless neutrinos.
However, in this decade it has been proven that neutrinos have
mass and that there is also mixing in the neutrino sector thanks
to observations of their oscillations.

The peculiarity of quarks is that they have a specific property,
the color charge, which is absent for leptons. Indeed, a quark of
a given flavor has three different color states with equal masses
and electroweak charges. The interaction of the quarks is mediated
by the eight gauge bosons (gluons $g$) of the color group
$SU_{C}(3)$. So, the quarks belong to the fundamental
representation of $SU_{C}(3)$ and the gluons to the adjoint one.
The gauge theory of this non-abelian group is called Quantum
Chromodynamics (QCD) and is the current theory of strong
interactions.

Specifically, a  gauged $SU_{C}(3)$ transformation of a quark
($q_{a}(x)$ with $a=1,2,3$) is given by
\begin{eqnarray}\label{gaugetrans}
q_{a}(x)&\rightarrow& q_{a}'(x)=U_{ab}(x)q_{b}(x)\\
\bar{q}_{a}(x)&\rightarrow&\bar{q}_{a}'(x)=\bar{q}_{b}U^{\dag}_{ba}(x),
\end{eqnarray}
where the $3\times 3$ matrix $U_{ik}(x)$ is the fundamental
representation of the $SU_{C}(3)$ group that acts on an internal
space defined at each space-time coordinate $x$. It satisfies
\begin{equation}
UU^{\dag}=U^{\dag}U=1,\qquad \det(U)=1.\label{unitarity}
\end{equation}
In this section the sum over all the repeated indices is implicit
and the sum over spinor indices is omitted for brevity. The usual
exponential representation of the gauge transformation matrix in
terms of the basis of matrices (the generators of $SU_{C}(3)$) of
the corresponding algebra $su_{C}(3)$ is:
\begin{equation}\label{gaugetrans2}
U(x)=e^{-\frac{i}{2}\vec{\alpha}(x)\cdot\vec{\lambda}}=e^{-\vec{\alpha}(x)\cdot\vec{t}},
\end{equation}
where $\vec{\alpha}(x)=(\alpha_{1}(x),\dots,\alpha_{8}(x))$ are
the eight arbitrary parameters of the gauge transformation,
$\vec{\lambda}$ are the eight elements of the basis of the algebra
$su_{C}(2)$ (or equivalently the eight generators of the group)
and $\vec{t}$ are the eight color operators (in analogy to the
spin operators of the group $SU(2)$). It is clear that, in order
to respect Eq.(\ref{unitarity}), the generators of the group must
be hermitian and traceless. The normalization of the color
operators depends on the representation $r$
\begin{equation}\label{norm}
\tr(t^{A}_{r}t^{B}_{r})=T_{r}\delta^{AB}.
\end{equation}
The form chosen by Gell-Mann for the $su_{C}(3)$ basis in the
fundamental representation:
\begin{eqnarray}
\lambda^{1}=\left(\begin{array}{ccc}0&1&0\\1&0&0\\0&0&0\end{array}\right),\quad
\lambda^{2}=\left(\begin{array}{ccc}0&-i&0\\i&0&0\\0&0&0\end{array}\right),\quad
\lambda^{3}=\left(\begin{array}{ccc}1&0&0\\0&-1&0\\0&0&0\end{array}\right),\nonumber\\
\lambda^{4}=\left(\begin{array}{ccc}0&0&1\\0&0&0\\1&0&0\end{array}\right),\quad
\lambda^{5}=\left(\begin{array}{ccc}0&0&-i\\0&0&0\\i&0&0\end{array}\right),\quad
\lambda^{6}=\left(\begin{array}{ccc}0&0&0\\0&0&1\\0&1&0\end{array}\right),\nonumber\\
\lambda^{7}=\left(\begin{array}{ccc}0&0&0\\0&0&-i\\0&i&0\end{array}\right),\quad
\lambda^{8}=\frac{1}{\sqrt{3}}\left(\begin{array}{ccc}1&0&0\\0&1&0\\0&0&-2\end{array}\right).
\end{eqnarray}
With these definitions, the color matrices satisfy the following
relations
\begin{eqnarray}
[t^{A},t^{B}]&=&if^{ABC}t^{C}\\
\tr(t^{A}t^{B})&=&T_{F}\delta^{AB},\quad
T_{F}=\frac{1}{2}\label{normf}
\end{eqnarray}
where $T_{F}$ is the normalization of the color matrices in the
fundamental representation and $f^{ABC}$ are called the structure
constants of the algebra $su_{C}(3)$ which are totally
antisymmetric in $\{A,B,C\}$. The independent non-vanishing
structure constants are given by:
\begin{eqnarray}
f^{123}=1,\quad f^{147}=\frac{1}{2},\quad f^{156}=-\frac{1}{2},\quad f^{246}=\frac{1}{2},\quad f^{257}=\frac{1}{2}\\
f^{345}=\frac{1}{2},\quad f^{367}=-\frac{1}{2},\quad
f^{458}=\frac{\sqrt{3}}{2},\quad f^{678}=\frac{\sqrt{3}}{2}.
\end{eqnarray}
Furthermore, the structure constants provide the adjoint
representation of the $su_{C}(3)$ algebra (the one which has the
same dimension of the algebra). Indeed, if we define the adjoint
representation as $T^{A}_{BC}=-if^{ABC}$, we can verify explicitly
that this is the adjoint representatio because
\begin{eqnarray}
[T^{A},T^{B}]&=&if^{ABC}T^{C}\\
\tr(T^{A}T^{B})&=&T_{A}\delta^{AB},\quad T_{A}=3
\end{eqnarray}
The Casimir operator $C_{r}$ (the one which commutes with all
elements of the algebra), for a certain representation $r$, is
constructed as
\begin{equation}
t^{A}_{r}t^{A}_{r}=C_{r}1_{d_{r}\times d_{r}}.
\end{equation}
Now, since the contraction of this last equation is equal to the
contraction of Eq.(\ref{norm}), we get
\begin{equation}
d_{r}C_{r}=8T_{r}.
\end{equation}
In particular, we find the Casimir operators are given by
\begin{equation}
C_{F}=\frac{4}{3},\qquad C_{A}=3
\end{equation}
for the fundamental and the adjoint representation respectively.

We shall now show that the  symmetry with respect to the gauge
transformations Eqs.(\ref{gaugetrans},\ref{gaugetrans2}) (together
with the Lorentz invariance), can be used as guiding principle to
construct the QCD Lagrangian. We start from the usual Dirac free
Lagrangian for each quark mass and color eigenstate
\begin{equation}\label{diracfreelag}
\mathcal{L}_{D}(x)=\bar{\psi}_{fa}(x)\left(i\sc{\partial}-m_{f}\right)\psi_{fa}(x),
\end{equation}
where $f$ is the flavor index and $a$ is the color quark index.
This term is not gauge invariant. In fact under the gauge
transformations Eqs.(\ref{gaugetrans},\ref{gaugetrans2}), the
Dirac free lagrangian Eq.(\ref{diracfreelag}) transform as
\begin{equation}
\mathcal{L}_{D}(x)\rightarrow
\mathcal{L}_{D}(x)+\bar{\psi}_{fb}\left[iU^{\dag}_{ba}(x)\partial_{\mu}U_{ac}(x)\right]\gamma^{\mu}
\psi_{fc}(x).\label{symmbreak}
\end{equation}
To restore gauge invariance, we introduce a gauge field matrix
$A_{\mu\,ab}(x)$ made up of eight gauge fields $A_{\mu}^{A}(x)$ in
this way:
\begin{equation}
A_{\mu\, ab}(x)=t^{A}_{ab}A_{\mu}^{A}.
\end{equation}
We assign to this matrix field the following interaction
Lagrangian
\begin{equation}
\mathcal{L}_{I}(x)=g_{s}\bar{\psi}_{fa}A_{\mu\,ab}\gamma^{\mu}\psi_{fb}.
\end{equation}
Here $g_{s}$ is the gauge dimensionless coupling analogous to the
electric charge in QED. In order to cancel the symmetry breaking
term of Eq.(\ref{symmbreak}), the gauge transformation of the
field matrix $A_{\mu\,ab}(x)$ has to be
\begin{equation}
A_{\mu\,ab}(x)\rightarrow
U_{ac}(x)A_{\mu\,cd}(x)U^{\dag}_{db}(x)-\frac{i}{g_{s}}\partial_{\mu}U_{ac}(x)U^{\dag}_{cb}(x).
\end{equation}
Hence, with the introduction of the field matrix $A_{\mu\,ab}$,
the sum $\mathcal{L}_{D}+\mathcal{L}_{I}$ is now gauge invariant.
To complete the Lagrangian one has to add the pure gauge invariant
term, which is
\begin{equation}
\mathcal{L}_{G}(x)=-\frac{1}{2}\tr\left(G_{\mu\nu}(x)G^{\mu\nu}(x)\right),
\end{equation}
where the gluon field-strength tensor is given by
\begin{equation}
G_{\mu\nu\,ab}(x)=\partial_{\mu}A_{\nu\,ab}-\partial_{\nu}A_{\mu\,ab}-ig_{s}\left[A_{\mu},A_{\nu}\right]_{ab}.
\end{equation}
Note that the third term of the gluon field-strength gives rise to
the self interactions of gluons and that its origin stands in the
fact that the gauge group $SU_{C}(3)$ is non-abelian. The final
form of the QCD classical Lagrangian is obtained by adding the
three pieces introduced above and using Eq.(\ref{normf}):
\begin{equation}\label{classicqcd}
\mathcal{L}_{QCD}^{\rm
cl}(x)=-\frac{1}{4}G_{\mu\nu}^{A}(x)G^{A\,\mu\nu}(x)+\bar{\psi}_{fa}(x)(iD_{\mu\,ab}\gamma^{\mu}-m_{f}\delta_{ab})\psi_{fb}(x),
\end{equation}
where
\begin{equation}
G_{\mu\nu}^{A}=\partial_{\mu}A_{\nu}^{A}-\partial_{\nu}A_{\mu}^{A}+g_{s}f^{ABC}A_{\mu}^{B}A_{\nu}^{C}
\end{equation}
and
\begin{equation}
D_{\mu ab}=\delta_{ab}\partial_{\mu}-ig_{s}A_{\mu\,ab}
\end{equation}
is the covariant derivative in the sense that
$D_{\mu\,ab}\psi_{fb}(x)$ transforms as $\psi_{fa}(x)$ under the
gauge group.

\begin{figure}
\begin{center}
\includegraphics[scale=0.8, angle=180]{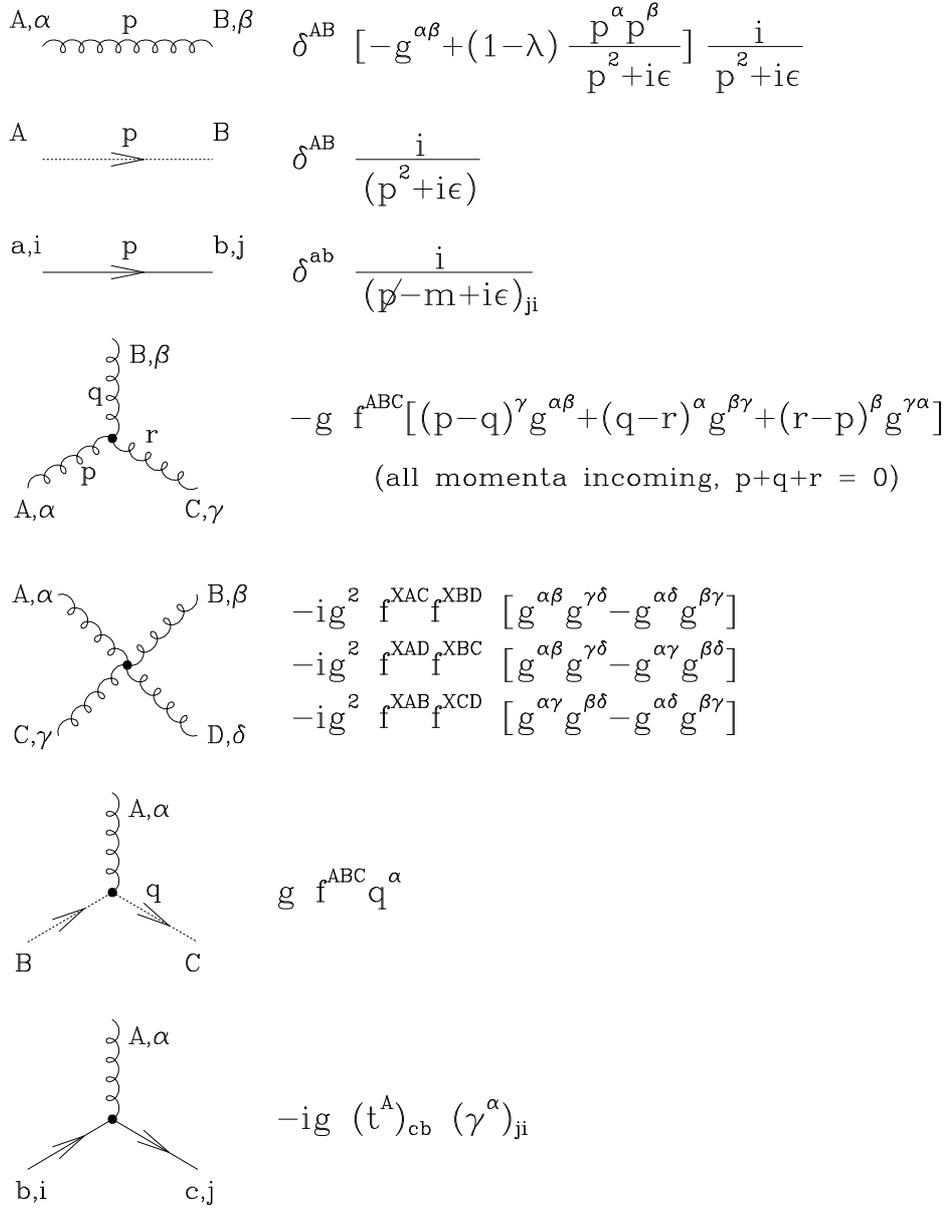}
\caption{\footnotesize{Feynman rules for QCD in a covariant gauge
for gluons (curly lines), quarks (solid lines) and ghosts (dotted
 lines). Here $A,B,C,D$ are the color indexes in the adjoint
representation, $a,b,c$ in the fundamental one,
$\alpha,\beta,\gamma,\delta$ are the gluon polarization indexes
and $i,j$ are the spinorial indexes.}\label{qcdfnmrules}}
\end{center}
\end{figure}

The quantization of the classical gauge invariant theory of QCD
Eq.(\ref{classicqcd}), can be done with the Fadeev-Popov
procedure. This procedure takes care of the fact that the equation
of motion of the gluon field $A_{\mu}^{A}$ can not be inverted and
this prohibits to find the propagator. However, this is a
consequence of gauge invariance which implies that the physical
massless gluon has only two polarizations/spin states whereas the
field $A_{\mu}^{A}$ has four components. To make things work, an
additional constraint on the gluon field is introduced, the so
called gauge fixing condition which uses gauge invariance to
define properly the gluon propagator. In QCD, however, this
constraint is not linear and one should add specially designed
fictitious particles (the so called Fadeev-Popov ghosts) which are
Lorentz scalar anti-commuting fields and appear only in the loops.
After this procedure (usually performed in the functional
formalism), we have that the QCD Lagrangian from which we can
calculate directly the Feynman rules is given by (see e.g.
\cite{Michael1995} page 514):
\begin{equation}
\mathcal{L}_{QCD}^{FP}(x)=\mathcal{L}_{QCD}^{cl}(x)-
\frac{1}{2\lambda}(\partial^{\mu}\partial^{\nu}A_{\mu}^{A}(x)A_{\nu}^{A}(x))
-\bar{c}^{A}\partial^{\mu}D_{\mu}^{AC}c^{C},
\end{equation}
where $\lambda$ is the gauge fixing parameter, $c^{A}$ is the
complex colored scalar ghost field and $D_{\mu}^{AB}$ is the
covariant derivative in the adjoint representation:
\begin{equation}
D_{\mu}^{AB}=\delta^{AB}\partial_{\mu}-g_{s}f^{ABC}A_{\mu}^{C}.
\end{equation}
 The quark, gluon and
ghost propagators and vertices for QCD are collected in Figure
\ref{qcdfnmrules}.

\section{Asymptotic freedom and perturbative QCD}\label{appA}

Feynman diagrams in QCD are obtained employing the vertices and
propagators as building blocks. However, the use of diagrams makes
sense only if the perturbative expansion in $g_{s}$ is meaningful.
To respect this condition, the coupling
\begin{equation}
\alpha_{s}=\frac{g_{s}}{4\pi},
\end{equation}
the QCD analog of the electromagnetic coupling
$\alpha=e^{2}/4\pi$, has to be sufficiently small. We shall now
show that the method of perturbation theory in QCD are useful at
high energy. Indeed, the coupling constant is large at low energy
and becomes smaller at high energy (asymptotic freedom).

The simplest way to introduce the running coupling, is to consider
a dimensionless physical observable $R$ which depends on a single
energy scale $\sqrt{Q^{2}}$. This is the case, for example of the
ratio of the annihilation cross section of electron-positron into
hadron with the annihilation into muons where $Q^{2}=S$ the
center-of-mass energy. We assume that this scale is much bigger
than the quark masses that can be therefore neglected. Now,
dimensional analysis should implies that a dimensionless
observable is independent of $Q^{2}$. However, higher order
corrections produce divergences and so the perturbation series
requires renormalization to remove ultraviolet divergences that in
$d=4-2\epsilon$ dimensions are regularized as $1/\epsilon$ poles.
This poles can be removed defining a renormalized coupling
constant at a certain renormalization scale $\mu^{2}_{r}$.
Consequently, in the finite $\epsilon \rightarrow 0$ limit, $R$
depends in general on the ratio $Q^{2}/\mu^{2}_{r}$ and the
renormalized coupling $\alpha_{s}$ depends on $\mu^{2}_{r}$; we
call the renormalized coupling at the scale $\mu^{2}_{r}$,
$\alpha_{\mu_{r}^{2}}$. Since the renormalization scale is an
arbitrary parameter introduced only to define the theory at the
quantum level, we conclude that $R$ has to be
$\mu^{2}_{r}$-independent. Formally this independence is expressed
as follows:
\begin{equation}\label{rengroupeq}
\mu^{2}_{r}\frac{d}{d\mu^{2}_{r}}R\left(\frac{Q^{2}}{\mu^{2}_{r}},\alpha_{\mu^{2}_{r}}\right)
=\left[\mu^{2}_{r}\frac{\partial}{\partial\mu^{2}_{r}}+\beta(\alpha_{\mu^{2}_{r}})\frac{\partial}{\partial\alpha_{\mu^{2}_{r}}}\right]R=0,
\end{equation}
where
\begin{equation}
\beta(\alpha_{\mu^{2}_{r}})=\mu^{2}_{r}
\frac{\partial\alpha_{\mu^{2}_{r}}}{\partial\mu^{2}_{r}}
\end{equation}
Eq.(\ref{rengroupeq}) is a first order differential equation with
the initial condition (at $Q^{2}=\mu^{2}_{r}$)
$R(1,\alpha_{\mu^{2}_{r}})$. This means that if we find a solution
of Eq.(\ref{rengroupeq}), it is the only possible solution. This
solution is easily found defining a function $\alpha_{s}(Q^{2})$
such that
\begin{equation}
\alpha_{s}(\mu^{2}_{r})=\alpha_{\mu^{2}_{r}}
\end{equation}
and that
\begin{equation}
\ln\left(\frac{Q^{2}}{\mu^{2}_{r}}\right)=\int_{\alpha_{\mu^{2}_{r}}}^{\alpha_{s}(Q^{2})}
\frac{dx}{\beta(x)}.
\end{equation}
In fact differentiating this equation, we find that
\begin{eqnarray}\label{run}
Q^{2}\frac{\partial\alpha_{s}(Q^{2})}{\partial
Q^{2}}&=&\beta(\alpha_{s}(Q^{2}))\\
\frac{\partial\alpha_{s}(Q^{2})}{\partial\alpha_{\mu^{2}_{r}}}&=&\frac{\beta(\alpha_{s}(Q^{2}))}{\beta(\alpha_{\mu^{2}_{r}})}
\end{eqnarray}
and that $R(1,\alpha_{s}(Q^{2}))$ is the desired solution of
Eq.(\ref{rengroupeq}), because
\begin{equation}
\frac{\partial}{\partial\alpha_{\mu^{2}_{r}}}=\frac{\partial\alpha_{s}(Q^{2})}{\partial\alpha_{\mu^{2}_{r}}}
\frac{\partial}{\partial\alpha_{s}(Q^{2})},\quad
\mu^{2}_{r}\frac{\partial}{\partial\mu^{2}_{r}}=-Q^{2}\frac{\partial}{\partial
Q^{2}}=-Q^{2}\frac{\partial\alpha_{s}(Q^{2})}{\partial
Q^{2}}\frac{\partial}{\partial\alpha_{s}(Q^{2})}.
\end{equation}
This shows that all of the scale dependence in $R$ enters through
the running of the coupling constant $\alpha_{s}(Q^{2})$. To find
explicitly this function, we need to know the $\beta$-function so
that we can solve Eq.(\ref{run}). The $\beta$-function can be
calculate perturbatively from the counterterms of the
renormalization precedure and a knowledge to order
$\alpha_{s}^{n+1}$ requires a $n$-loop computation. The
perturbative expansion of the $\beta$-function is given by:
\begin{equation}\label{betafunc2}
\beta(\alpha_{s})=-\alpha_{s}\sum_{n=0}^{\infty}\beta_{n}\left(\frac{\alpha_{s}}{4\pi}\right)^{n+1}.
\end{equation}
At the moment, the QCD $\beta$-function is known to order
$\alpha_{s}^{5}$ \cite{vanRitbergen:1997va}: where in the
$\overline{MS}$ scheme,
\begin{eqnarray}\label{betapar}
\beta_{0}&=&11-\frac{2}{3}N_{f},\quad\beta_{1}=102-\frac{38}{3}N_{f},\\
\beta_{2}&=&\frac{2857}{2}-\frac{5033}{18}N_{f}+\frac{325}{54}N_{f}^{2}\\
\beta_{3}&=&\left(\frac{149753}{6}+3564\xi_{3}\right)-\left(\frac{1078361}{162}+\frac{6508}{27}\xi_{3}\right)N_{f}\nonumber\\
&&+\left(\frac{50065}{162}+\frac{6472}{81}\xi_{3}\right)N_{f}^{2}+\frac{1093}{729}N_{f}^{3},
\end{eqnarray}
with $N_{f}$ the numbers of flavors and $\xi$ is the Riemann
zeta-function ($\xi_{3}=1.202056903\dots$). The two loop solution
of Eq.(\ref{run}) is given by:
\begin{eqnarray}
\alpha_{s}(Q^{2})&=&\frac{\alpha_{s}(\mu^{2}_{r})}{1+(\beta_{0}/4\pi)
\alpha_{s}(\mu^{2}_{r})\log\frac{Q^{2}}{\mu^{2}_{r}}}\bigg[1-\frac{\beta_{1}}{4\pi\beta_{0}}
\frac{\alpha_{s}(\mu^{2}_{r})\log(1+(\beta_{0}/4\pi)
\alpha_{s}(\mu^{2}_{r})\log\frac{Q^{2}}{\mu^{2}_{r}})}{1+(\beta_{0}/4\pi)\alpha_{s}(\mu^{2}_{r})\log\frac{Q^{2}}{\mu^{2}_{r}}}\bigg]\nonumber\\
&+&\emph{O}(\alpha_{s}^{k+3}\log^{k}
\frac{Q^{2}}{\mu^{2}_{r}})\label{tls}.
\end{eqnarray}

\begin{figure}
\begin{center}
\includegraphics[scale=0.6]{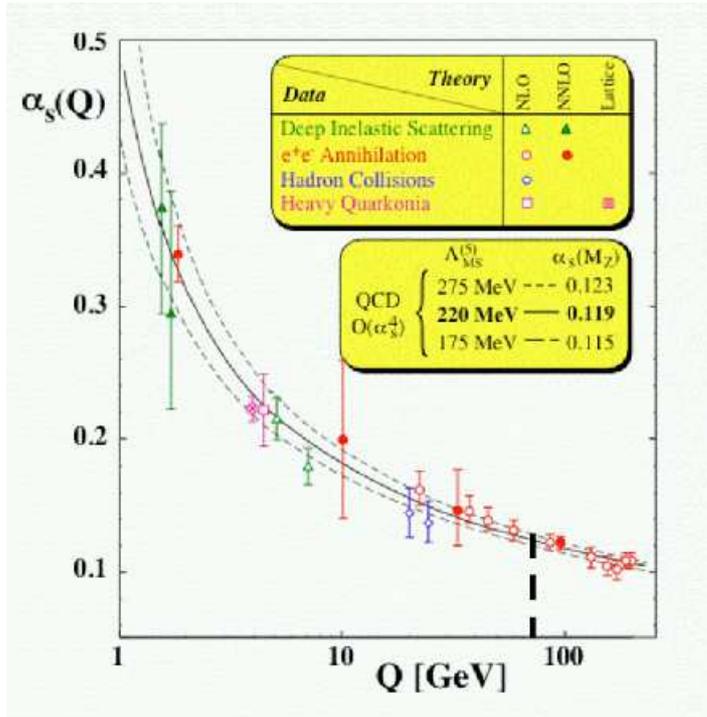}
\caption{\footnotesize{The running of the strong coupling
constant. The asymptotic freedom is confirmed by the experiments
}\label{af}}
\end{center}
\end{figure}

For simplicity, in many cases, we will use another parametrization
of the coefficients $\beta_{n}$, which is obtained with the
substitution:
\begin{equation}
\beta_{n}=b_{n}(4\pi)^{n+1}.
\end{equation}
With this parametrization, the perturbative expansion of the
$\beta$-function Eq.(\ref{betafunc2}) becomes
\begin{equation}\label{betafunc}
\beta(\alpha_{s})=-\sum_{n=0}^{\infty}b_{n}\alpha_{s}^{n+2}.
\end{equation}
From Eq.(\ref{tls}), we see that $\alpha_{s}(Q^{2})$ is a
monotonically decreasing function of $Q^{2}$, because the
coefficients $\beta_{0}$ and $\beta_{1}$ are positive (with
$N_{f}\leq 6$). The running of $\alpha_{s}(Q^{2})$ has been
measured with great accuracy (see Figure \ref{af}). The fact that
at high energy, the running coupling becomes small is a
peculiarity of non-abelian gauge theories and is called asymptotic
freedom.

Hence, perturbative QCD can be applied when the relevant scale of
a certain process is high enough such that the running coupling
becomes small. A typical example is given by the annihilation of
two high-energy electrons into hadrons. Perturbative QCD can also
be applied to processes in which hadrons are present also in the
initial state thanks to the factorization theorem. According to
this theorem \cite{Collins:1989gx}, the cross section for the
production of some final state with high invariant mass $Q^{2}$
(the scale at which the running coupling constant is small) with
two incoming hadrons is given by:
\begin{equation}\label{factheo1}
\sigma(P_{1},P_{2},Q^{2})=\sum_{a,b}\int_{0}^{1}dx_{1}dx_{2}F^{H_{1}}_{a}(x_{1},\mu^{2})
F^{H_{2}}_{b}(x_{2},\mu^{2})\hat{\sigma}_{ab}(x_{1}P_{1},x_{2}P_{2},\alpha_{s}(\mu^{2}),Q^{2},\mu^{2}).
\end{equation}
For processes with a single incoming hadron the factorization
theorem is simpler. For example for the deep inelastic scattering
(DIS) of a lepton that exchanges a high square momentum $Q^{2}$
with the hadron, the cross section takes the form:
\begin{equation}\label{factheo2}
\sigma(P,Q^{2})=\sum_{a}\int_{0}^{1}dxF_{a}^{H}(x,\mu^{2})\hat{\sigma}_{a}(xP,
\alpha_{s}(\mu^{2}),Q^{2},\mu^{2}).
\end{equation}
In Eqs.(\ref{factheo1},\ref{factheo2}), $P_{i}$ is the momentum of
the incoming hadron $H_{i}$. A beam of hadrons of type $H_{i}$ is
equivalent to a beam of constituents (or partons) which are quarks
or gluons. These constituents carry a longitudinal momentum
$x_{i}P_{i}$ characterized by the parton densities
$F_{a}^{H_{i}}(x_{i},\mu^{2})$. That is, given a hadron $H_{i}$
with momentum $P_{i}$, the probability density to find in $H_{i}$
the parton $a$ with momentum $x_{i}P_{i}$ is given by
$F_{a}^{H_{i}}(x_{i},\mu^{2})$. Furthermore, these functions are
universal in the sense that they are process independent. The
parton densities depends also on the so called factorization scale
$\mu^{2}$. This scale is introduced to separate off the
non-perturbative part of the cross section (the parton densities)
from the perturbative one $\hat{\sigma}_{a(b)}$. This is exactly
the cross section where the incoming particles are the partons $a$
(and $b$) and can be calculated as a perturbative expansion in
$\alpha_{s}(\mu^{2}_{r})$. The parton densities have a mild
dependence on the scale $\mu^{2}$ determined by the
Dokshitzer-Gribov-Lipatov-Altarelli-Parisi (DGLAP) equations (see
section \ref{dglapeq}). Here, we have chosen the renormalization
scale $\mu^{2}_{r}$ equal to the factorization scale $\mu^{2}$ for
simplicity. Anyway, in order to reintroduce the scale
$\mu^{2}_{r}$, we have only to rewrite $\alpha_{s}(\mu^{2})$ in
terms of $\mu^{2}_{r}$ (see Eq.(\ref{tls})) and expand it
consistently with the order of the calculation. The $\mu^{2}$
dependence in the parton densities is compensated by the $\mu^{2}$
dependence in the partonic cross section $\hat{\sigma}$. However,
with a fixed-order computaction of the partonic coefficient
function at order $\alpha_{s}^{k}$ the hadronic cross section will
still depend on $\mu^{2}$ with a dependence which should be of
order $\alpha_{s}^{k+1}$. Hence, this dependence can be used to
estimate the theoretical error of a fixed-order computation. A
simple discussion about the dependence of the hadronic cross
section on the factorization and renormalization scale is given in
Ref.\cite{Ridolfi:1999vr}.

\section{NLO DY and DIS cross sections}\label{disdycs}

\begin{figure}
\begin{center}
\includegraphics[scale=0.4,angle=-90]{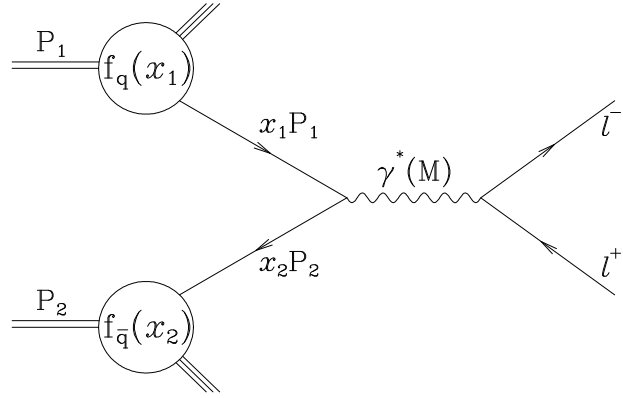}
\caption{\footnotesize{Drell-Yan pair production. Here
$Q=M$.}\label{dy}}
\end{center}
\end{figure}

We consider for simplicity the classical Drell-Yan (DY) hadronic
process for the production of a dimuon pair through a virtual
photon $\gamma^{*}$ (see Figure \ref{dy}): \bea\label{hpdy}
H_{1}(P_{1})+H_{2}(P_{2})=\gamma^{*}(Q)+\emph{X}(K), \eea where
$H_{1}$ and $H_{2}$ are the colliding hadrons with momementum
$P_{1}$ and $P_{2}$ respectively, $Q$ is the momentum of the
virtual photon and $\emph{X}$ is any number of additional hadrons
with total momentum $K$. For the process of Eq.(\ref{hpdy}), we
define \bea\label{defx} x\equiv\frac{Q^{2}}{S}, \eea where
$S=(P_{1}+P_{2})^{2}$ is the usual Mandelstam invariant, which can
be viewed as the hadronic center-of-mass energy. It is clear that
Eq.(\ref{defx}) represents the fraction of energy that the hadrons
transfer to the photon and, hence, $0\leq x \leq 1$.

According to the factorized expression of the QCD cross section
Eq.(\ref{factheo1}), the LO $Q^{2}$ differential cross section is
given by:
\begin{eqnarray}
\frac{d\sigma}{dQ^{2}}(x,
Q^{2})&=&\sum_{i}\int_{0}^{1}dx_{1}dx_{2}\left[q_{i}(x_{1},\mu^{2})\bar{q}_{i}(x_{2},\mu^{2})+
\bar{q}_{i}(x_{1})q_{i}(x_{2},\mu^{2}))\right]\frac{d\hat{\sigma}_{i}}{dQ^{2}},\label{LOdy}\\
\frac{d\hat{\sigma}_{i}}{dQ^{2}}&=&\sigma_{0}^{DY}(Q^{2},x)Q_{q_{i}}^{2}\delta(x_{1}x_{2}-x)\label{LOdyp},\quad
\sigma_{0}^{DY}(Q^{2},x)=\frac{4\pi\alpha^{2}}{9Q^{4}}x
\end{eqnarray}
where the functions
$q_{i}(x_{j},\mu^{2})$($\bar{q}_{i}(x_{j},\mu^{2})$)are the parton
densities of a quark (or an anti-quark) of flavor $i$ in the
hadron $j=1,2$ at the scale $\mu^{2}$, $\alpha$ is the
fine-structure constant and $Q_{q_{i}}$ is the fraction of
electronic charge of the quark $q_{i}$. Now, if we define the
dimensionaless cross section $\sigma(x,Q^{2})$ as,
\begin{equation} \sigma^{DY}(x,Q^{2})\equiv
\frac{1}{\sigma_{0}^{DY}}\frac{d\sigma}{dQ^{2}}(x, Q^{2}),
\end{equation}
and use the identity,
\begin{equation}
\delta(x_{1}x_{2}-x)=\int_{0}^{1}dz\delta(1-z)\delta(x_{1}x_{2}z-x),
\end{equation}
then Eqs.(\ref{LOdy},\ref{LOdyp}) become:
\begin{eqnarray}
\sigma^{DY}(x,Q^{2})=\qquad\qquad\qquad\qquad\qquad\qquad\qquad\qquad\qquad\qquad\quad
\qquad\nonumber\\
=\sum_{i}\int_{0}^{1}dx_{1}dx_{2}dz\left[q_{i}(x_{1})\bar{q}_{i}(x_{2})+
\bar{q}_{i}(x_{1})q_{i}(x_{2})\right]Q_{q_{i}}^{2}C_{qq}(z)\delta(x_{1}x_{2}z-x)\label{LOaddy}\\
=\sum_{i}\int_{x}^{1}\frac{dx_{1}}{x_{1}}\int_{x/x_{1}}^{1}\frac{dx_{2}}{x_{2}}
\left[q_{i}(x_{1})\bar{q}_{i}(x_{2})+\bar{q}_{i}(x_{1})q_{i}(x_{2}))\right]
Q_{q_{i}}^{2}C_{qq}\left(\frac{x}{x_{1}x_{2}}\right),\\
C_{qq}(z)=\delta(1-z)\label{LOcf},\qquad\qquad\qquad\qquad\qquad\qquad\qquad\qquad\qquad\qquad\qquad
\end{eqnarray}
where $C_{qq}(z)$ is the LO Drell-Yan coefficient function. From
Eq.(\ref{LOaddy}), we see that the new variable $z$ that we have
introduced is in general given by
\begin{equation}
z=\frac{x}{x_{1}x_{2}}.
\end{equation}
This means that at the partonic level, $z$ can be viewed as the
fraction of energy that the colliding partons transfer to the
virtual photon. At LO it is clear that $z=1$ as can be explicitely
seen from Eq.(\ref{LOcf}), because there is no emission but the
virtual photon.

Beyond the LO the extra radiated partons in the final state can
carry away some energy
 (so $z<1$) and  the gluon channel contributes. The NLO coefficient functions $C_{ab}(z)$ ($a,b=q,g$)
 recieves contributions that have infrared and ultraviolet.
Infrared singularities cancel out (see e.g. \cite{Weinberg1996}).
The ultraviolet ones are reabsobed by renormalization of the bare
parameters of the QCD Lagrangian, thus defining a renormalized
strong coupling constant $\alpha_{\mu^{2}_{r}}$ at an arbitraty
renormalization scale $\mu_{r}^{2}$ (see section \ref{appA}).
Collinear divergences are cut off by infrared physics. They can be
absorbed multiplicatively in redefinition of the parton densities
\cite{Curci:1980uw}, thus reabsorbing all dependence on soft
physics in the parton distributions. The parton densities at a
certain scale are determined by a reference process and their
scale dependence is determined by the DGLAP equations (see section
\ref{dglapeq}). However, there is an ambiguity on how to define
the reference process, related to the fact that collinear
divergences can always be factorized together with finite terms.
The choice of these finite terms defines a factorization scheme.
The most common factorization scheme is the $\overline{MS}$ scheme
in which the collinear divergence (which is in $d=4-2\epsilon$
dimensions a single pole $1/\epsilon$) is factorized together the
finite terms $-\gamma_{E}+\log 4\pi$, where $\gamma_{E}=0.5772...$
is the Euler gamma.  Now, in order to avoid the perturbative
expansion to receive large contributions, the factorization and
the renormalization scales are expected to be chosen of the same
order of the scale of the process $Q^{2}$. Here, for simplicity,
we choose the factorization scale $\mu^{2}$ equal to the
renormalization scale $\mu_{r}^{2}$.
 We report the NLO Drell-Yan cross section
(see e.g. \cite{Ellis1989,Field1989}):
\begin{eqnarray}
&&\sigma^{DY}(x,Q^{2})=\sum_{i}Q_{q_{i}}^{2}\int_{x}^{1}\frac{dx_{1}}{x_{1}}\int_{x/x_{1}}^{1}\frac{dx_{2}}{x_{2}}\qquad\nonumber\\
&&\times\bigg\{\left[q_{i}(x_{1},\mu^{2})\bar{q}_{i}(x_{2},\mu^{2})+(1\leftrightarrow
2)\right]
C_{qq}\left(z,\frac{Q^{2}}{\mu^{2}},\alpha_{s}(\mu^{2})\right)\qquad\nonumber\\
&&+\left[g(x_{1},\mu^{2})\left(q_{i}(x_{2},\mu^{2})+\bar{q}_{i}(x_{2},\mu^{2})\right)+(1\leftrightarrow
2)\right]C_{qg}\left(z,\frac{Q^{2}}{\mu^{2}},\alpha_{s}(\mu^{2})\right)\bigg\},\label{cqqnlo0}\qquad
\end{eqnarray}
where, in the $\overline{MS}$ scheme,
\begin{eqnarray}
C_{qq}\left(z,\frac{Q^{2}}{\mu^{2}},\alpha_{s}(\mu^{2})\right)&=&\delta(1-z)+\frac{\alpha_{s}(\mu^{2})}{2\pi}
\bigg\{\frac{4}{3}\bigg[\left(\frac{2\pi^{2}}{3}-8\right)\delta(1-z)\nonumber\\
&&+4(1+z^{2})\left[\frac{\log(1-z)}{1-z}\right]_{+}-2\frac{1+z^{2}}{1-z}\log
z\bigg]\nonumber\\
&&+\frac{8}{3}\left[\frac{1+z^{2}}{[1-z]_{+}}+\frac{3}{2}\delta(1-z)\right]\log\left(\frac{Q^{2}}{\mu^{2}}\right)\bigg\},\label{cqqnlo}
\end{eqnarray}
and
\begin{eqnarray}
C_{qg}\left(z,\frac{Q^{2}}{\mu^{2}},\alpha_{s}(\mu^{2})\right)&=&\frac{\alpha_{s}(\mu^{2})}{2\pi}\bigg\{\frac{1}{2}
\bigg[(z^{2}+(1-z)^{2})\log\frac{(1-z)^{2}}{z}
+\frac{1}{2}+3z\nonumber\\
&&-\frac{7}{2}z^{2}\bigg]+\frac{1}{2}[z^{2}+(1-z)^{2}]\log\left(\frac{Q^{2}}{\mu^{2}}\right)\bigg\},\label{cqgnlo}
\end{eqnarray}
where the ``$+$'' distribution is defined as follows:
\begin{equation}
\int_{0}^{1}dzf(z)[g(z)]_{+}\equiv\int_{0}^{1}dz[f(z)-f(1)]g(z).
\end{equation}

\begin{figure}
\begin{center}
\includegraphics[scale=0.4,angle=-90]{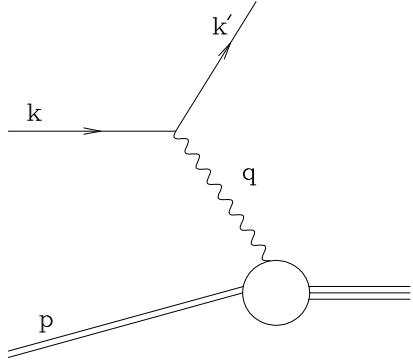}
\caption{\footnotesize{Deep inelastic electron-proton
scattering}\label{dis}}
\end{center}
\end{figure}

Also for the case of the deep-inelastic scattering (DIS), we
consider the simplest process in which a high energy electron
scatters from a hadron exchanging with it a virtual photon
$\gamma^{*}$ (see Figure \ref{dis}):
\begin{equation}
H(P)+e(k)\rightarrow e(k')+X(K),
\end{equation}
where $H$ is typically a proton with momentum $P$, $e$ is the
scattered electron and $X$ is any collection of hadrons. The
standard parametrization of DIS is done in terms of three relevant
parameters:
\begin{eqnarray}
Q^{2}&\equiv& -q^{2}\equiv -(k-k')^{2}\\
y&\equiv&\frac{P\cdot q}{P\cdot k};\quad 0\leq y\leq 1\\
x&\equiv& x_{Bj}=\frac{Q^{2}}{2P\cdot
q}=\frac{Q^{2}}{(P+q)^{2}+Q^{2}};\quad 0\leq x\leq 1,
\end{eqnarray}
where in the last line we have neglected the proton mass. $Q^{2}$
is the virtuality of the photon exchanged between the electron and
the proton and $y$ is the fraction of energy that the incoming
electron transfer to the proton. The Bjorken scaling variable $x$
has a simple physical interpretation: it is the fraction of
longitudinal momentum of the LO incoming quark of the partonic
subprocess.

Indeed, the most general parametrization of the  $Q^{2}$
differential cross section is given by:
\begin{equation}
\frac{d\sigma}{dQ^{2}}(x,Q^{2},y)=\frac{4\pi\alpha^{2}}{Q^{4}}
\left[[1+(1-y)^{2}]F_{1}(x,Q^{2})+\frac{(1-y)}{x}(F_{2}(x,Q^{2})-2xF_{1}(x,Q^{2}))\right],
\end{equation}
 The
functions $F_{1(2)}$ are called structure functions and contains
the information about the structure of the proton. In fact they
are determined by the photon-proton sub-process in this way:
\begin{eqnarray}
F_{1}(x,Q^{2})&=&\frac{Q^{2}}{16\pi^{2}\alpha
x}\left[\sigma_{\Sigma}(\gamma^{*}P)+\sigma_{L}(\gamma^{*}P)\right]\\
F_{2}(x,Q^{2})&=&2xF_{1}(x,Q^{2})+F_{L}(x,Q^{2})\\
F_{L}(x,Q^{2})&=&\frac{Q^{2}}{4\pi^{2}\alpha}\sigma_{L}(\gamma^{*}P),
\end{eqnarray}
where $\sigma_{\Sigma}(\gamma^{*}P)$ and
$\sigma_{\Sigma}(\gamma^{*}P)$ are the cross sections of the
photon-proton process determined summing over all the virtual
photon polarization and over only the longitudinal one
respectively. At LO
\begin{eqnarray}
F_{1}(x,Q^{2})&=&\frac{1}{2}\sum_{i}Q_{q_{i}}^{2}[q_{i}(x,Q^{2})+\bar{q}_{i}(x,Q^{2})]\\
F_{2}(x,Q^{2})&=&2xF_{1}(x,\mu^{2}),
\end{eqnarray}
where $q_{i}$ and $\bar{q}_{i}$ are the parton densities. In
general the structure functions should depend on both $x$ and
$Q^{2}$, because these are the relevant kinematic variable of the
photon-proton sub-process. Now we want, as we have done for the
Drell-Yan case, rewrite the structure functions in terms of parton
densities and of a coefficient function that can be computed in
perturbative QCD. If we use the identity
\begin{equation}
\delta(y-x)=\int_{0}^{1}dz\,\delta(1-z)\delta(yz-x),
\end{equation}
we have for the LO structure functions $F_{2}$ and $F_{L}$:
\begin{eqnarray}
F_{2}(x,Q^{2})&=&x\sum_{i}\int_{0}^{1}dydz\,\left[q_{i}(y,\mu^{2})+\bar{q}_{i}(y,\mu^{2})\right]Q_{q_{i}}^{2}C_{q}(z)\delta(yz-x)\label{zdis}\\
&=&x\sum_{i}\int_{x}^{1}\frac{dy}{y}\,\left[q_{i}(y,\mu^{2})+\bar{q}_{i}(y,\mu^{2})\right]Q_{q_{i}}^{2}C_{q}\left(\frac{x}{y}\right)\\
C_{q}(z)&=&\delta(1-z)\label{cdislo}\\
F_{L}(x,Q^{2})&=&0,
\end{eqnarray}
where $C_{q}(z)$ is the LO DIS coefficient function for $F_{2}$.
From Eq.(\ref{zdis}), we see that the variable $z$ is in general
given by
\begin{equation}
z=\frac{x}{y}.
\end{equation}
This means that at the partonic level, $z$ can be viewed as the
longitudinal momentum of the incoming parton before it scatters
with the virtual photon. At LO it is clear that $z=1$ as can be
explicitely seen from
Eq.(\ref{cdislo}), because there is no extra
emissions.

Beyond the LO the extra radiated partons in the final state can
carry some energy
 (so $z<1$) and also the gluon channel contributes. At NLO we have
 ultraviolet, infrared and collinear singularities. They
must be regularized and treated as in the Drell-Yan case. We
report the NLO structure functions (see e.g.
\cite{Ellis1989,Field1989}) with the renormalization scale equal
to the factorization scale:
\begin{eqnarray}
F_{2}(x,Q^{2})&=&x\sum_{i}Q_{q_{i}}^{2}\int_{x}^{1}\frac{dy}{y}\,\bigg\{\left[q_{i}(y,\mu^{2})+\bar{q}_{i}(y,\mu^{2})\right]
C_{q}\left(z,\frac{Q^{2}}{\mu^{2}},\alpha_{s}(\mu^{2})\right)\bigg\}\nonumber\\
&&+x\sum_{i}Q_{q_{i}}^{2}\int_{x}^{1}\frac{dy}{y}\,g(y,\mu^{2})C_{g}\left(z,\frac{Q^{2}}{\mu^{2}},\alpha_{s}(\mu^{2})\right)\bigg\}\label{f2nlo}
\end{eqnarray}
where, in the $\overline{MS}$ scheme,
\begin{eqnarray}
C_{q}\left(z,\frac{Q^{2}}{\mu^{2}},\alpha_{s}(\mu^{2})\right)&=&\delta(1-z)+
\frac{\alpha_{s}(\mu^{2})}{2\pi}\bigg\{\frac{4}{3}\bigg[2\left[\frac{\ln(1-z)}{1-z}\right]_{+}-
\frac{3}{2}\left[\frac{1}{1-z}\right]_{+}\nonumber\\
&&-(1+z)\ln(1-z)
-\frac{1+z^{2}}{1-z}\ln\,z-\left(\frac{\pi^{2}}{3}+\frac{9}{2}\right)\delta(1-z)\nonumber\\
&&+3+2z\bigg]
+\frac{4}{3}\left[\frac{1+z^{2}}{[1-z]_{+}}+\frac{3}{2}\delta(1-z)\right]\log\left(\frac{Q^{2}}{\mu^{2}}\right)\bigg\},\label{cqnlo}
\end{eqnarray}
\begin{eqnarray}
C_{g}\left(z,\frac{Q^{2}}{\mu^{2}},\alpha_{s}(\mu^{2}\right)&=&\frac{\alpha_{s}(\mu^{2})}{2\pi}\bigg\{\frac{1}{2}\bigg[\left((1-z)^{2}
+z^{2}\right)\ln\left(\frac{1-z}{z}\right)-8z^{2}\nonumber\\
&&+8z-1+[z^{2}+(1-z)^{2}]\log\left(\frac{Q^{2}}{\mu^{2}}\right)\bigg]\bigg\}\label{cgnlo}
\end{eqnarray}
and
\begin{eqnarray}
F_{L}(x,Q^{2})&=&x\sum_{i}Q_{q_{i}}^{2}\int_{x}^{1}\frac{dy}{y}\,\left[q_{i}(y,\mu^{2})+\bar{q}_{i}(y,\mu^{2})\right]
\frac{\alpha_{s}(\mu^{2})}{2\pi}\bigg\{\frac{8}{3}z\nonumber\\
&&+\frac{4}{3}\left[\frac{1+z^{2}}{[1-z]_{+}}+\frac{3}{2}\delta(1-z)\right]\log\left(\frac{Q^{2}}{\mu^{2}}\right)\bigg\}\nonumber\\
&&+x\sum_{i}Q_{q_{i}}^{2}\int_{x}^{1}\frac{dy}{y}\,g(y,\mu^{2})\frac{\alpha_{s}(\mu^{2})}{2\pi}\bigg\{2z(1-z)\nonumber\\
&&+\frac{1}{2}[z^{2}+(1-z)^{2}]\log\left(\frac{Q^{2}}{\mu^{2}}\right)\bigg\},
\end{eqnarray}
is factorization scheme independent at the lowest non-vanishing
order.

\section{The DGLAP equations}\label{dglapeq}

The coefficient function and the parton densities depend on the
factorization scale in such a way that the resulting hadronic
cross section is $\mu^{2}$-independent. The equations that fix the
$\mu^{2}$-dependence of parton densities (the so called
Dokshitzer-Gribov-Lipatov-Altarelli-Parisi (DGLAP) equations) can
be found imposing the $\mu^{2}$-independence of the DY cross
section or of the DIS structure functions. For example, imposing
this condition to the explicit expression for the NLO $F_{2}$ (see
Eqs.(\ref{f2nlo},\ref{cqnlo},\ref{cgnlo}), we find the LO DGLAP
evolution equations for the quark parton densities:
\begin{eqnarray}
\mu^{2}\frac{\partial
q_{i}(x,\mu^{2})}{\partial\mu^{2}}&=&\frac{\alpha_{s}(\mu^{2})}{4\pi}\bigg\{\sum_{j}
\int_{x}^{1}\frac{dy}{y}\,\left[P_{q_{i}q_{j}}^{(0)}\left(\frac{x}{y}\right)q_{j}(y,\mu^{2})+
P_{q_{i}\bar{q}_{j}}^{(0)}\left(\frac{x}{y}\right)\bar{q}_{j}(y,\mu^{2})\right]\nonumber\\
&&+\int_{x}^{1}\frac{dy}{y}\,P_{q_{i}g}^{(0)}\left(\frac{x}{y}\right)g(y,\mu^{2})\bigg\}+\emph{O}(\alpha_{s}^{2}),
\end{eqnarray}
where
\begin{eqnarray}
P_{q_{i}q_{j}}^{(0)}(z)&=&\delta_{ij}P_{qq}^{(0)}(z),\\
P_{q_{i}\bar{q}_{j}}^{(0)}(z)&=&0\\
P_{qq}^{(0)}(z)&=&\frac{8}{3}\left[\frac{1+z^{2}}{[1-z]_{+}}+\frac{3}{2}\delta(1-z)\right],\label{p0qq}
\end{eqnarray}
and
\begin{equation}\label{p0qg}
P_{q_{i}g}^{(0)}(z)=\frac{1}{N_{f}}P_{qg}^{(0)}(z)=z^{2}+(1-z)^{2},
\end{equation}
with $N_{f}$ the number of active flavors. The functions
$P_{pp'}^{(0)}(z)$ are called LO splitting functions. They can be
viewed as the probability per unit of $\ln(\mu^{2}/Q^{2})$ to find
a parton $p$ in a parton $p'$. The LO evolution equation for the
gluon can be calculated from the LO splitting diagrams for a quark
into another quark and a gluon and for a gluon into two gluons.
Furthermore, we simplify the notation introducing the convolution
product $\otimes$, defined in this way:
\begin{equation}\label{defconv}
(f_{1}\otimes f_{2}\otimes\cdots\otimes
f_{n})(x)=\int_{0}^{1}dx_{1}dx_{2}\cdots
dx_{n}f_{1}(x_{1})f_{2}(x_{2})\cdots
f_{n}(x_{n})\delta(x_{1}x_{2}\cdots x_{n}-x).
\end{equation}
We report here the full result for the DGLAP evolution equations:
\begin{equation}\label{fulldglapeq}
\mu^{2}\frac{\partial}{\partial\mu^{2}}\left(\begin{array}{c}
q_{i}(z,\mu^{2})\\g(z,\mu^{2})\end{array}\right)=
\sum_{q_{j},\bar{q}_{j}}\left(\begin{array}{cc}
P_{q_{i}q_{j}}(z,\mu^{2})&P_{q_{i}g}(z,\mu^{2})\\
P_{gq_{j}}(z,\mu^{2})&P_{gg}(z,\mu^{2})
\end{array}\right)\otimes \left(\begin{array}{c}q_{j}(z,\mu^{2})\\
g(z,\mu^{2})\end{array}\right),
\end{equation}
where $q_{i}$ can be also a quark or anti-quark and where the
splitting functions $P_{pp'}$ have the following perturbative
expansion:
\begin{eqnarray}
P_{q_{i}q_{j}}(z,\mu^{2})&=&P_{\bar{q}_{i}\bar{q}_{j}}(z,\mu^{2})=\frac{\alpha_{s}(\mu^{2})}{4\pi}\delta_{ij}P_{qq}^{V\,(0)}(z)\nonumber\\
&&+\sum_{k=1}^{\infty}\left(\frac{\alpha_{s}(\mu^{2})}{4\pi}\right)^{k+1}\left(\delta_{ij}P^{V\,(k)}_{qq}(z)+P^{S\,(k)}_{qq}(z)\right),\label{prima}\\
P_{q_{i}\bar{q}_{j}}(z,\mu^{2})&=&P_{\bar{q_{i}}q_{j}}(z,\mu^{2})=\left(\frac{\alpha_{s}(\mu^{2})}{4\pi^{2}}\right)^{2}\left(\delta_{ij}
P_{q\bar{q}}^{V\,(1)}+P_{qq}^{S\,(1)}\right)\nonumber\\
&&+\sum_{k=1}^{\infty}\left(\frac{\alpha_{s}(\mu^{2})}{4\pi}\right)^{k+2}\left(\delta_{ij}P^{V\,(k+1)}_{q\bar{q}}(z)+P^{S\,(k+1)}_{q\bar{q}}(z)\right),
\end{eqnarray}
\begin{eqnarray}
P_{q_{i}g}(z,\mu^{2})&=&P_{\bar{q}_{i}g}(z,\mu^{2})=\frac{1}{N_{f}}
\sum_{k=1}^{\infty}\left(\frac{\alpha_{s}(\mu^{2})}{4\pi}\right)^{k}P_{qg}^{(k-1)},\\
P_{gq_{i}}(z,\mu^{2})&=&P_{g\bar{q_{i}}}=\sum_{k=1}^{\infty}\left(\frac{\alpha_{s}(\mu^{2})}{4\pi}\right)^{k}P_{gq}^{(k-1)},\\
P_{gg}(z,\mu^{2})&=&\sum_{k=1}^{\infty}\left(\frac{\alpha_{s}(\mu^{2})}{4\pi}\right)^{k}P_{gg}^{(k-1)},\label{ultima}
\end{eqnarray}
where $N_{f}$ is the number of active flavors.
Eq.(\ref{fulldglapeq}) represents a system of $2N_{f}+1$
integro-differential equations equations. The solution to this
system however can be calculated analytically for a certain
fixed-order. In fact it can be translated into a system of
ordinary differential equation performing a Mellin transform:
\begin{eqnarray}
F_{p}(N,\mu^{2})&=&\int_{0}^{1}dz\,z^{N-1}F_{p}(z,\mu^{2}),\\
\mu^{2}\frac{\partial
F_{p}(N,\mu^{2})}{\partial\mu^{2}}&=&\sum_{p'}\gamma^{AP}_{pp'}(N,\mu^{2})F_{p'}(N,\mu^{2}),\label{fulldglapeq2}
\end{eqnarray}
where $p,p'=q_{i},\bar{q}_{j},g$ and
\begin{equation}
\gamma^{AP}_{pp'}(N,\mu^{2})=\int_{0}^{1}dzP_{pp'}(z,\mu^{2}).
\end{equation}
After that, these equations can be decoupled searching linear
combinations of parton densities that depends on the independent
splitting functions of Eqs.(\ref{prima}-\ref{ultima}) and that
diagonalize the system. For example, at LO, there are $4$
independent splitting functions which are
$P_{qq}^{V\,(0)}=P_{qq}^{(0)}$, $P_{qg}^{(0)}$ (given in
Eqs.(\ref{p0qq},\ref{p0qg}) respectively) and
\begin{eqnarray}
P_{gq}^{(0)}(z)&=&\frac{8}{3}\left[\frac{1+(1-z)^{2}}{z}\right]\\
P_{gg}^{(0)}(z)&=&12\left[\frac{z}{[1-z]_{+}}+\frac{1-z}{z}+z(1-z)\right]\nonumber\\
&&+\left(\frac{44}{9}-\frac{2}{3}N_{f}\right)\delta(1-z).
\end{eqnarray}
At NLO there are $6$ independent splitting functions which are for
example $P^{V}_{qq}(z)$, $P^{S}_{qq}(z)$, $P^{V}_{q\bar{q}}(z)$,
$P_{qg}(z)$, $P_{gq}(z)$ and $P_{gg}(z)$. They are given in
Ref.\cite{Ellis1989} pages $111$ and $112$. The LO and the NLO
solution to the DGLAP equations (\ref{fulldglapeq2}) in Mellin
space is computed in the next Section.

\section{NLO solution of the DGLAP evolution equations}\label{appB}

In this Section, we want to solve the NLO DGLAP equations
(Eq.(\ref{fulldglapeq2}) of section \ref{dglapeq}):
\begin{equation}\label{fulldglap}
\mu^{2}\frac{\partial
F_{p}(N,\mu^{2})}{\partial\mu^{2}}=\sum_{p'}\gamma_{pp'}(N,\mu^{2})F_{p'}(N,\mu^{2}),
\end{equation}
where all the splitting functions defined here (and in the
following) have the same perturbative expansion:
\begin{equation}\label{pertsplitt}
\gamma_{pp'}(N,\mu^{2})=\frac{\alpha_{s}(\mu^{2})}{4\pi}\gamma^{(0)}_{pp'}(N)
+\left(\frac{\alpha_{s}(\mu^{2})}{4\pi}\right)^{2}\gamma^{(1)}_{pp'}(N)+\emph{O}(\alpha_{s}^{3}).
\end{equation}
This is a system of $2N_{f}+1$ coupled equations with $N_{f}$ the
number of active flavors. At NLO, there are $6$ independent
splitting functions defined through the following equations:
\begin{eqnarray}
\gamma_{gq_{i}}&=&\gamma_{g\bar{q}_{i}}\equiv\gamma_{gq}\label{primagamma}\\
\gamma_{q_{i}g}&=&\gamma_{\bar{q}_{i}g}\equiv 1/N_{f}\gamma_{qg}\\
\gamma_{q_{i}q_{k}}&=&\gamma_{\bar{q}_{i}\bar{q}_{k}}\equiv\delta_{ik}\gamma^{V}_{qq}+\gamma^{S}_{qq}\\
\gamma_{q_{i}\bar{q}_{k}}&=&\gamma_{\bar{q}_{i}q_{k}}\equiv\delta_{ik}\gamma^{V}_{q\bar{q}}+\gamma^{S}_{qq}\label{bnlo}\\
\gamma_{gg}&\equiv&\gamma_{gg},\label{ultimagamma}
\end{eqnarray}
where $i,k$ are a flavor index. We omit the dependence on $N$ and
$\mu^{2}$ for brevity of notation. Note that beyond the NLO, there
is one more independent splitting function. In fact in
Eq.(\ref{bnlo}) we should substitute $\gamma^{S}_{qq}$ with
$\gamma^{S}_{q\bar{q}}$ which are different beyond the NLO
\cite{Vogt:2004mw,Moch:2004pa}. We note also that at LO
$\gamma^{S}_{qq}=\gamma^{S}_{q\bar{q}}=\gamma^{V}_{q\bar{q}}=0$
and hence at LO there are only $4$ independent splitting
functions.

Now, we define the $2N_{f}-1$ so called  non-singlet (NS)
combinations
\begin{eqnarray}
q^{\pm}_{(NS)k}&=&\sum_{i=1}^{k}(q_{i}\pm \bar{q}_{i})-k(q_{k}\pm \bar{q}_{k});\quad k=2,\dots,N_{f}\\
q^{V}_{(NS)}&=&\sum_{i=1}^{N_{f}}(q_{i}-\bar{q}_{i})
\end{eqnarray}
and the $2$ so called singlet (S) combinations: $g$ and
\begin{eqnarray}
q_{(S)}&=&\sum_{i=1}^{N_{f}}(q_{i}+\bar{q}_{i}).
\end{eqnarray}
With this definitions, from Eq.(\ref{fulldglap}) and
Eqs.(\ref{primagamma}-\ref{ultimagamma}), we find that for the
non-singlet combinations
\begin{eqnarray}
\mu^{2}\frac{\partial
q^{\pm}_{(NS)k}}{\partial\mu^{2}}&=&\gamma^{\pm}q^{\pm}_{(NS)k}\\
\mu^{2}\frac{\partial
q^{V}_{(NS)}}{\partial\mu^{2}}&=&\gamma^{-}q^{V}_{(NS)},
\end{eqnarray}
where
\begin{equation}
\gamma^{\pm}=\gamma^{V}_{qq}\pm\gamma^{V}_{q\bar{q}}.
\end{equation}
For the $2$ remaining singlet combinations, we find in the same
way that
\begin{equation}\label{fulldglapeq3}
\mu^{2}\frac{\partial}{\partial\mu^{2}}\left(\begin{array}{c}
q_{(S)}\\g\end{array}\right)= \left(\begin{array}{cc}
\gamma_{qq}&\gamma_{qg}\\
\gamma_{gq}&\gamma_{gg}
\end{array}\right)\left(\begin{array}{c}q_{(S)}\\
g\end{array}\right),
\end{equation}
where
\begin{equation}
\gamma_{qq}=\gamma^{+}+\gamma_{PS},\qquad \gamma_{PS}\equiv
2N_{f}\gamma^{S}_{q\bar{q}}.
\end{equation}

The NLO Mellin splitting functions can be found in
Ref.\cite{Floratos:1981hs} written in terms of harmonic sums. In
many cases, however, their analytic continuation to all the
complex plane is useful (see e.g.
\cite{Gluck:1989ze,Gonzalez-Arroyo:1979df}) For the NNLO solution
of the DGLAP equations and the NNLO splitting functions we refer
to \cite{Vogt:2004mw,Moch:2004pa}. The techniques  for the
analytic continuations of the NNLO splitting functions can be
found in Ref.\cite{Blumlein:1998if}.

For the NS combinations, the solution is easy to obtain. Indeed,
making the change of variable
\begin{equation}\label{changevarapp}
\frac{d\mu^{2}}{\mu^{2}}=\frac{d\alpha_{s}(\mu^{2})}{\beta(\alpha_{s}(\mu^{2}))},
\end{equation}
where $\beta(\alpha_{s})$ is the $\beta$ function defined in
section \ref{appA}, we get:
\begin{equation}\label{nlosolns}
\frac{q^{\pm}_{(NS)k}(\mu^{2})}{q^{\pm}_{(NS)k}(\mu^{2}_{0})}=
\left(\frac{\alpha_{s}(\mu^{2})}{\alpha_{s}(\mu^{2}_{0})}\right)^{-\gamma^{
(0)\pm}/\beta_{0}} \left[1+\left(\frac{\gamma^{
(1)\pm}}{\beta_{0}}-\frac{\beta_{1}\gamma^{
(0)\pm}}{\beta_{0}^{2}}\right)\left(\frac{\alpha_{s}(\mu^{2}_{0})}{4\pi}-\frac{\alpha_{s}(\mu^{2})}{4\pi}\right)\right]
\end{equation}
and
\begin{equation}\label{nlosolnsv}
\frac{q^{V}_{(NS)}(\mu^{2})}{q^{V}_{(NS)}(\mu^{2}_{0})}=
\left(\frac{\alpha_{s}(\mu^{2})}{\alpha_{s}(\mu^{2}_{0})}\right)^{-\gamma^{
(0)-}/\beta_{0}} \left[1+\left(\frac{\gamma^{
(1)-}}{\beta_{0}}-\frac{\beta_{1}\gamma^{
(0)-}}{\beta_{0}^{2}}\right)\left(\frac{\alpha_{s}(\mu^{2}_{0})}{4\pi}-\frac{\alpha_{s}(\mu^{2})}{4\pi}\right)\right],
\end{equation}
where we have omitted the $N$-dependence of the splitting
functions for brevity. For the S combinations, some linear algebra
is needed. We, first, define the singlet vector and the splitting
matrix:
\begin{equation}
\vec{q}_{S}\equiv\left(\begin{array}{c}
q_{(S)}\\g\end{array}\right),\qquad
\tilde{\gamma}_{S}\equiv\left(\begin{array}{cc}
\gamma_{qq}&\gamma_{qg}\\
\gamma_{gq}&\gamma_{gg}
\end{array}\right).
\end{equation}
Using the NLO splitting matrix and the change of variable
Eq.(\ref{changevarapp}), we find immeditely the formal solution,
which is
\begin{equation}\label{formalsol}
\vec{q}_{S}(\mu^{2})=\exp\left\{-R_{0}\ln\frac{\alpha_{s}(\mu^{2})}{\alpha_{s}(\mu^{2}_{0})}+
R_{1}\left(\frac{\alpha_{s}(\mu^{2}_{0})}{4\pi}-\frac{\alpha_{s}(\mu^{2})}{4\pi}\right)\right\}\vec{q}_{S}(\mu^{2}_{0}),
\end{equation}
where
\begin{equation}
R_{0}=\frac{\tilde{\gamma}_{S}^{(0)}}{\beta_{0}},\qquad
R_{1}=\frac{\tilde{\gamma}_{S}^{(1)}}{\beta_{0}}-\frac{\beta_{1}\tilde{\gamma}_{S}^{(0)}}{\beta_{0}^{2}}.
\end{equation}
The two matrices $R_{0}$ and $R_{1}$ in Eq.(\ref{formalsol})
cannot be diagonalized simultaneously, as they do not commute.
Hence, in order to extract the NLO solution from
Eq.(\ref{formalsol}), we use the following Ansatz:
\begin{equation}\label{ansatz}
\vec{q}_{S}(\mu^{2})=U(\alpha_{s}(\mu^{2}))\left(\frac{\alpha_{s}(\mu^{2})}{\alpha_{s}(\mu^{2}_{0})}\right)
^{-R_{0}}U^{-1}(\alpha_{s}(\mu^{2}_{0}))\vec{q}_{S}(\mu^{2}_{0}),
\end{equation}
where the matrix $U$ has the perturbative expansion:
\begin{equation}
U(\alpha_{s}(\mu^{2}))=1+\frac{\alpha_{s}(\mu^{2})}{4\pi}U_{1}+\emph{O}(\alpha_{s}^{2}).
\end{equation}
The condition that the matrix $U_{1}$ should satisfy can be easily
obtained imposing that the derivative with respect to
$\alpha_{s}(\mu^{2})$ of Eq.(\ref{formalsol}) and of
Eq.(\ref{ansatz}) are equal at NLO. Thus, we get
\begin{equation}\label{u1cond}
[U_{1},R_{0}]=U_{1}+R_{1}.
\end{equation}
We write $R_{0}$ in terms of its $2$ eigenvalues
\begin{equation}
R^{\pm}=\frac{1}{2\beta_{0}}\left[(\gamma_{qq}^{(0)}+\gamma_{gg}^{(0)})
\pm\sqrt{(\gamma_{qq}^{(0)}-\gamma_{gg}^{(0)})^{2}+4\gamma_{qg}^{(0)}\gamma_{gq}^{(0)}}\right]
\end{equation}
and of the $2$ corresponding eigenspaces projectors
$\emph{P}_{\pm}$:
\begin{equation}\label{diagr0}
R_{0}=R^{+}\emph{P}_{+}+R^{-}\emph{P}_{-}.
\end{equation}
The explicit expression for the projectors can be obtained using
the completeness relation $\emph{P}_{+}+\emph{P}_{-}=1$. We find
\begin{equation}
\emph{P}_{\pm}=\frac{1}{R^{\pm}-R^{\mp}}[R_{0}-R^{\mp}].
\end{equation}
Now, writing the matrices $U_{1}$ and $R_{1}$ in terms of these
projectors
\begin{eqnarray}
U_{1}&=&\emph{P}_{-}U_{1}\emph{P}_{-}+\emph{P}_{-}U_{1}\emph{P}_{+}+\emph{P}_{+}U_{1}\emph{P}_{-}
+\emph{P}_{+}U_{1}\emph{P}_{+}\\
R_{1}&=&\emph{P}_{-}R_{1}\emph{P}_{-}+\emph{P}_{-}R_{1}\emph{P}_{+}+\emph{P}_{+}R_{1}\emph{P}_{-}
+\emph{P}_{+}R_{1}\emph{P}_{+},
\end{eqnarray}
substituting them and Eq.(\ref{diagr0}) into Eq.(\ref{u1cond}) and
comparing each matrix element, we find
\begin{equation}\label{u1sol}
U_{1}=-(\emph{P}_{-}R_{1}\emph{P}_{-}+\emph{P}_{+}R_{1}\emph{P}_{+})
+\frac{\emph{P}_{+}R_{1}\emph{P}_{-}}{R^{-}-R^{+}-1}+\frac{\emph{P}_{-}R_{1}\emph{P}_{+}}{R^{+}-R^{-}-1}.
\end{equation}
Thanks to this result, we can now write the NLO solution of the
singlet doublet in a form which is useful for practical
calculations. Indeed, if we substitute Eq.(\ref{u1sol}) in
Eq.(\ref{ansatz}), we get (at NLO)
\begin{eqnarray}
\vec{q}_{S}(\mu^{2})&=&\bigg\{\left(\frac{\alpha_{s}(\mu^{2})}{\alpha_{s}(\mu^{2}_{0})}\right)^{-R_{-}}
\bigg[\emph{P}_{-}+\left(\frac{\alpha_{s}(\mu^{2}_{0})}{4\pi}-\frac{\alpha_{s}(\mu^{2})}{4\pi}\right)
\emph{P}_{-}R_{1}\emph{P}_{-}\nonumber\\
&&-\left(\frac{\alpha_{s}(\mu^{2}_{0})}{4\pi}-\frac{\alpha_{s}(\mu^{2})}{4\pi}
\left(\frac{\alpha_{s}(\mu^{2})}{\alpha_{s}(\mu_{0}^{2})}\right)^{R^{-}-R^{+}}\right)
\frac{\emph{P}_{-}R_{1}\emph{P}_{+}}{R^{+}-R^{-}-1}\bigg]\nonumber\\
&&+(+\leftrightarrow
-)\bigg\}\vec{q}_{S}(\mu^{2}_{0}).\label{nlosols}
\end{eqnarray}
After the evolution of the NS and S combinations has been
performed from a certain scale $\mu^{2}_{0}$ to the scale
$\mu^{2}$, we need to return to the parton distributions for all
the quarks but the gluon. These are obtained straightforwardly
with the following relations
\begin{eqnarray}
q_{k}+\bar{q}_{k}&=&\frac{1}{N_{f}}q_{S}-\frac{1}{k}q^{+}_{(NS)k}+\sum_{i=k+1}^{N_{f}}\frac{1}{i(i-1)}q^{+}_{(NS)i},\qquad
k=1,\dots,N_{f}\\
q_{k}-\bar{q}_{k}&=&\frac{1}{N_{f}}q_{(NS)}^{V}-\frac{1}{k}q^{-}_{(NS)k}+\sum_{i=k+1}^{N_{f}}\frac{1}{i(i-1)}q^{-}_{(NS)i},\qquad
k=1,\dots,N_{f}.
\end{eqnarray}
However, Eqs.(\ref{nlosolns},\ref{nlosolnsv},\ref{nlosols})
represent the NLO solution of the DGLAP equations
Eq.(\ref{fulldglap}) in the case when the number of active flavors
$N_{f}$ has been kept fixed. This is the so called fixed flavor
scheme solution. If we want to take into account the thresholds of
the heavy quark flavors, we can evolve up the NS a S combinations
from the scale $\mu^{2}_{0}$ (with a certain number $N_{f}$ of
active flavors) to the scale of production of a new flavor. Then,
we can take the result of this evolution as the starting point of
a second evolution (with $N_{f}+1$ active flavors this time) above
the production scale of the new flavor, assuming that the new
flavor vanishes at threshold. This is the most simple way to
generate dynamically a new flavor.

Finally, we note that the procedure outlined in this appendix can
be easily generalized beyond the NLO order \cite{Vogt:2004ns}.
Furthermore, in many cases, it is interesting to study the
dependence on the renormalization scale, in order to estimate the
theoretical error of the evolution. Here the renormalization scale
$\mu^{2}_{r}$ has been chosen equal to the factorization one
$\mu^{2}$ for simplicity. To restore the implicit
$\mu^{2}_{r}$-dependence in parton densities, we need only to
rewrite the running coupling constant $\alpha_{s}(\mu^{2})$ in
terms of $\mu^{2}_{r}$ (see Eq.(\ref{tls}) in section \ref{appA})
in the splitting functions. Making this substitution, we have that
the perturbative expansion of a generic splitting function
Eq.(\ref{pertsplitt}) becomes
\begin{equation}\label{pertsplitt}
\gamma_{pp'}(N,\mu^{2},k'
)=\frac{\alpha_{s}(k'\mu^{2})}{4\pi}\gamma^{(0)}_{pp'}(N)
+\left(\frac{\alpha_{s}(k'\mu^{2})}{4\pi}\right)^{2}(\gamma^{(1)}_{pp'}(N)+
\beta_{0}\gamma^{(0)}_{pp'}\ln k')+\emph{O}(\alpha_{s}^{3}),
\end{equation}
where $k'=\mu^{2}_{r}/\mu^{2}$. Hence, in
Eqs.(\ref{nlosolns},\ref{nlosolnsv},\ref{nlosols}), we should
perform the following substitutions:
\begin{equation}
\gamma^{(1)}\rightarrow \gamma^{(1)}+\beta_{0}\gamma^{(0)}\ln
k',\quad
\alpha_{s}(\mu^{2})\rightarrow\alpha_{s}(k'\mu^{2}),\quad\alpha_{s}(\mu^{2}_{0})\rightarrow\alpha_{s}(k'\mu^{2}_{0})
\end{equation}
and use $k'\mu^{2}$ as reference scale for new flavors production.

\chapter{High order QCD and resummation}\label{RI}

\section{When is NLO not enough?}\label{nlonotenough}

\begin{figure}
\begin{center}
\includegraphics[scale=0.5]{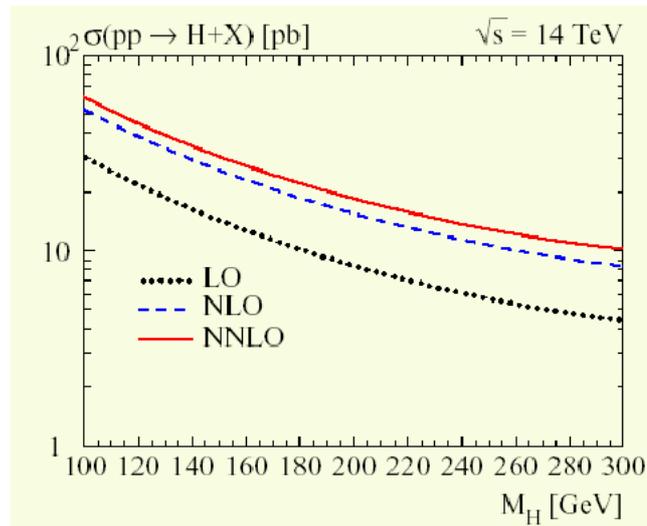}
\caption{\footnotesize{Total cross section for the Higgs boson
production at LHC at (from bottom to top) at LO, NLO, NNLO in the
gluon fusion channel \cite{Harlander:2002wh}.}} \label{fig1}
\end{center}
\end{figure}

In section \ref{disdycs}, we have discussed briefly the analytic
NLO calculation of the full inclusive DIS and DY cross sections.
However, in many cases, the NLO pQCD computation turns out not to
be enough. This is, for example, often the case at LHC where the
Higgs boson production has to be distinguished from the
background. A computation beyond the NLO is needed also when the
NLO corrections are large and higher-order calculation permit us
to test the convergence of the perturbative expansion. In figure
\ref{fig1} the total cross section of the production of the Higgs
boson at LHC \cite{Harlander:2002wh} is plotted and we note
convergence in going from LO to NLO and to NNLO.

This can also happen when a new parton level subprocess first
appear at NLO. This is the case for example for the rapidity DY
distributions at Tevatron (shown in figure \ref{fig2}) and at the
fixed-target experiment E866/NuSea (shown in figure \ref{fig3}).
The agreement with the data of figure \ref{fig2} has represented
an important test of the NNLO splitting functions
\cite{Vogt:2004mw,Moch:2004pa}. We note also that going from the
LO to the NNLO the factorization scale dependence is significantly
reduced.

\begin{figure}
\begin{center}
\includegraphics[scale=0.5]{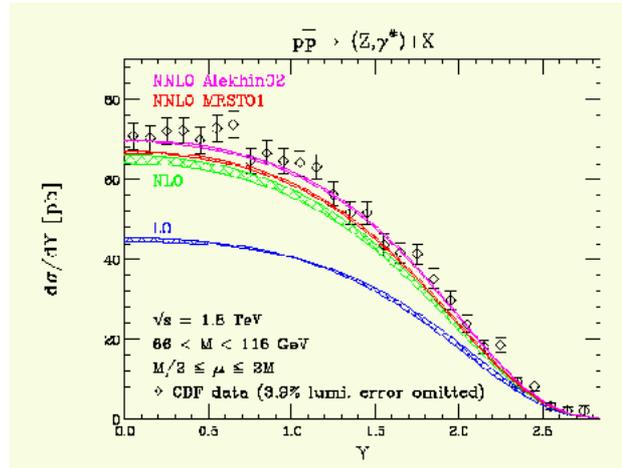}
\caption{\footnotesize{DY rapidity distribution for proton
anti-proton collisions at Tevatron at (from bottom to top) LO,
NLO, NNLO, together with the CDF data
\cite{Anastasiou:2003ds}.}\label{fig2}}
\end{center}
\end{figure}

\begin{figure}
\begin{center}
\includegraphics[scale=0.5]{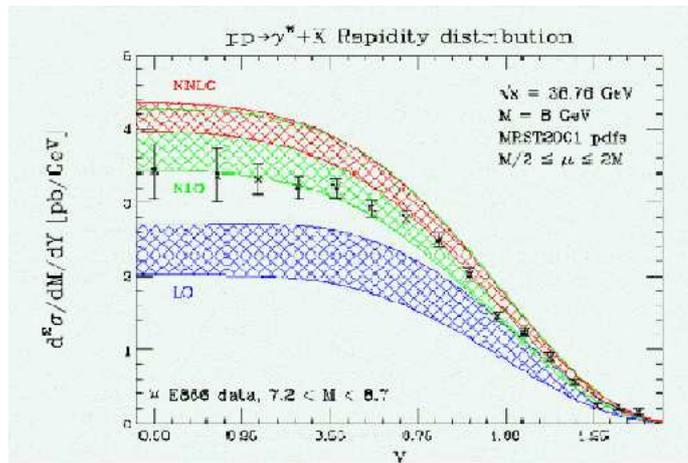}
\caption{\footnotesize{DY rapidity distribution for proton proton
collisions at fixed-target experiment E866/NuSea at (from bottom
to top) LO, NLO, NNLO, together with the data
\cite{Anastasiou:2003yy,Webb:2003ps}.}\label{fig3}}
\end{center}
\end{figure}

Calculations beyond the NLO can be important also in processes
which involve large logaritms when different significant scales
appear. In these cases, these large logarithms should be resummed
and this is the topic of this thesis. A first example of these
large logarithms has appeared in section \ref{disdycs}. In fact,
from Eqs.(\ref{cqqnlo},\ref{cqnlo})of section \ref{disdycs}, we
see that there are contributions that become large when
$z\rightarrow 1$ from the quark-antiquark channel in the DY case
and from the quark channel for the structure function $F_{2}$ in
the DIS case. These are the terms proportional to
\begin{equation}\label{nloenhanced}
\alpha_{s}\left[\frac{\log(1-z)}{1-z}\right]_{+},\quad
\alpha_{s}\left[\frac{1}{1-z}\right]_{+}.
\end{equation}
The terms of the type of Eq.(\ref{nloenhanced}) arise from the
infrared cancellation between virtual and real emissions. It can
be shown that enhanced contributions of the same type arise at all
orders. In fact, at order $\emph{O}(\alpha_{s}^{n})$ there are
contributions proportional  to
\cite{Sterman:1986aj,Catani:1989ne,Catani:1990rp}:
\begin{equation}\label{logterm}
\alpha_{s}^{n}\left[\frac{\log^{m}(1-z)}{1-z}\right]_{+},\quad
m\leq 2n-1. \end{equation}
These terms become important in the
limit $z\rightarrow 1$ spoiling the validity of the perturbative
fixed-order QCD expansion and, thus, should be resummed to
all-orders of pQCD.

The limit $z\rightarrow 1$ corresponds in general to the kinematic
boundary where emitted partons are all soft (as it happens in the
DY case) or collinear (as in the DIS and the prompt photon case as
we shall see in section \ref{kindpp}).

In fact, in the DY case, if we consider a contribution to the
coefficient function with $n$ radiated extra partons with momenta
$k_{1},\dots ,k_{n}$, the squaring of four-momentum conservation
($p_{1}+p_{2}=Q+k_{1}+\cdots +k_{n}$) implies
\begin{eqnarray}
x_{1}x_{2}S(1-z)&=&\sum_{i,j=1}^{n}k_{i}\cdot k_{j}+2\sum_{i=1}^{n}Q\cdot k_{i}\\
&=&\sum_{i,j=1}^{n}k_{i}^{0}k_{j}^{0}(1-\cos\theta_{ij})+2\sum_{i=1}^{n}k_{i}^{0}
(\sqrt{Q^{2}+|\vec{Q}|^{2}}-|\vec{Q}|\cos\theta_{i})\label{kinlim},
\end{eqnarray}
where $\theta_{ij}$ is the angle between $\vec{k}_{i}$ and
$\vec{k}_{j}$ and $\theta_{i}$ is the angle between $\vec{k}_{i}$
and $\vec{Q}$. Since all the terms in the first sum of
Eq.(\ref{kinlim}) are positive semi-definite and the terms
$(\sqrt{Q^{2}+|\vec{Q}|^{2}}-|\vec{Q}|\cos\theta_{i})$ in the
second sum are positive for all possible values of $\theta_{i}$,
we have that the limit $z=1$ is achieve only for $k_{i}^{0}=0$ for
all $i$. This means that when $z$ approaches $1$ all the emitted
partons in the Drell-Yan processes are soft and that we have
reached the threshold for the production of a virtual photon or a
real vector boson.

In the DIS case, at the partonic level, we have
\begin{equation}
p+q=k_{1}+\dots+k_{n}+k_{n+1},
\end{equation}
where $k_{n+1}$ is the LO outgoing parton. If we square this last
equation, we get
\begin{equation}\label{zto1dis}
\frac{Q^{2}(1-z)}{z}=\sum_{i,j=1}^{n+1}k_{i}^{0}k_{j}^{0}(1-\cos\theta_{ij}),
\end{equation}
where $\theta_{ij}$ is the angle between $\vec{k}_{i}$ and
$\vec{k}_{j}$. Eq.(\ref{zto1dis}) tells us that in the
$z\rightarrow 1$ limit, there can be not only soft radiated
partons in the final state, but there can be also a set of partons
collinear to each other. However, in Section \ref{kindy} we will
show with a more detailed analysis of the DIS kinematics and phase
space that the collinear partons are also soft in the
$z\rightarrow$ limit for the deep-inelastic process.

\begin{figure}
\begin{center}
\includegraphics[scale=0.42]{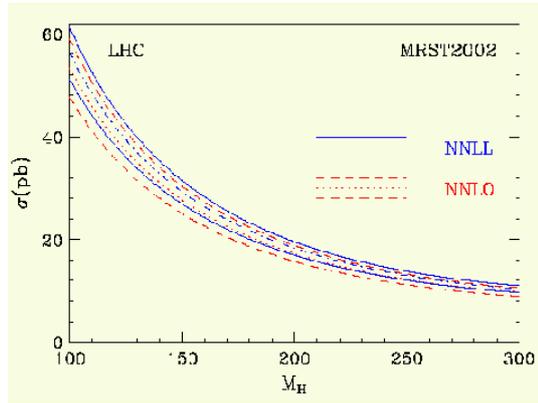}
\caption{\footnotesize{Total cross section for the Higgs boson
production at LHC at (from bottom to top) at NNLO and NNLO
improved with NNLL resummation in the gluon fusion channel
\cite{Catani:2003zt}.}\label{fig4}}
\end{center}
\end{figure}

An example of the impact of resummation can be seen in figure
\ref{fig4}. There, the total cross section for the Higgs
production at LHC is plotted at NNLO with its NNLL resummation
improvement improvement \cite{Catani:2003zt}. The scale
uncertainty reduced to about $15\%$ at NNLO is further reduced to
$10\%$ by the NNLL resummation. The resummed large logaritms in
this case are of the class of the DY-like soft emissions.

\begin{figure}
\begin{center}
\includegraphics[scale=0.42]{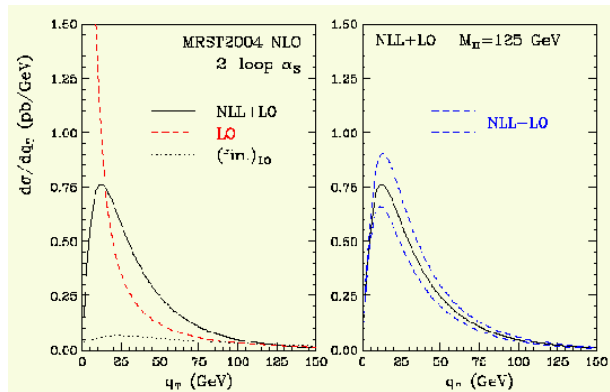}
\caption{\footnotesize{Total cross section for the transverse
momentum Higgs boson production at LHC at LO and LO improved with
NLL resummation in the gluon fusion channel
\cite{Bozzi:2005wk}.}\label{fig5}}
\end{center}
\end{figure}

Resummation of another class of large logarithms plays a crucial
role in transverse momentum distributions. Indeed, in figure
\ref{fig5}, we observe that resummation changes substantially the
behavior of the cross section for the production of the Higgs
boson at small transverse momentum. In these case the large
logarithms of $q_{\perp}^{2}/M_{H}^{2}$ with $M_{H}$ the Higgs
mass are resummed.

\begin{figure}
\begin{center}
\includegraphics[scale=0.6]{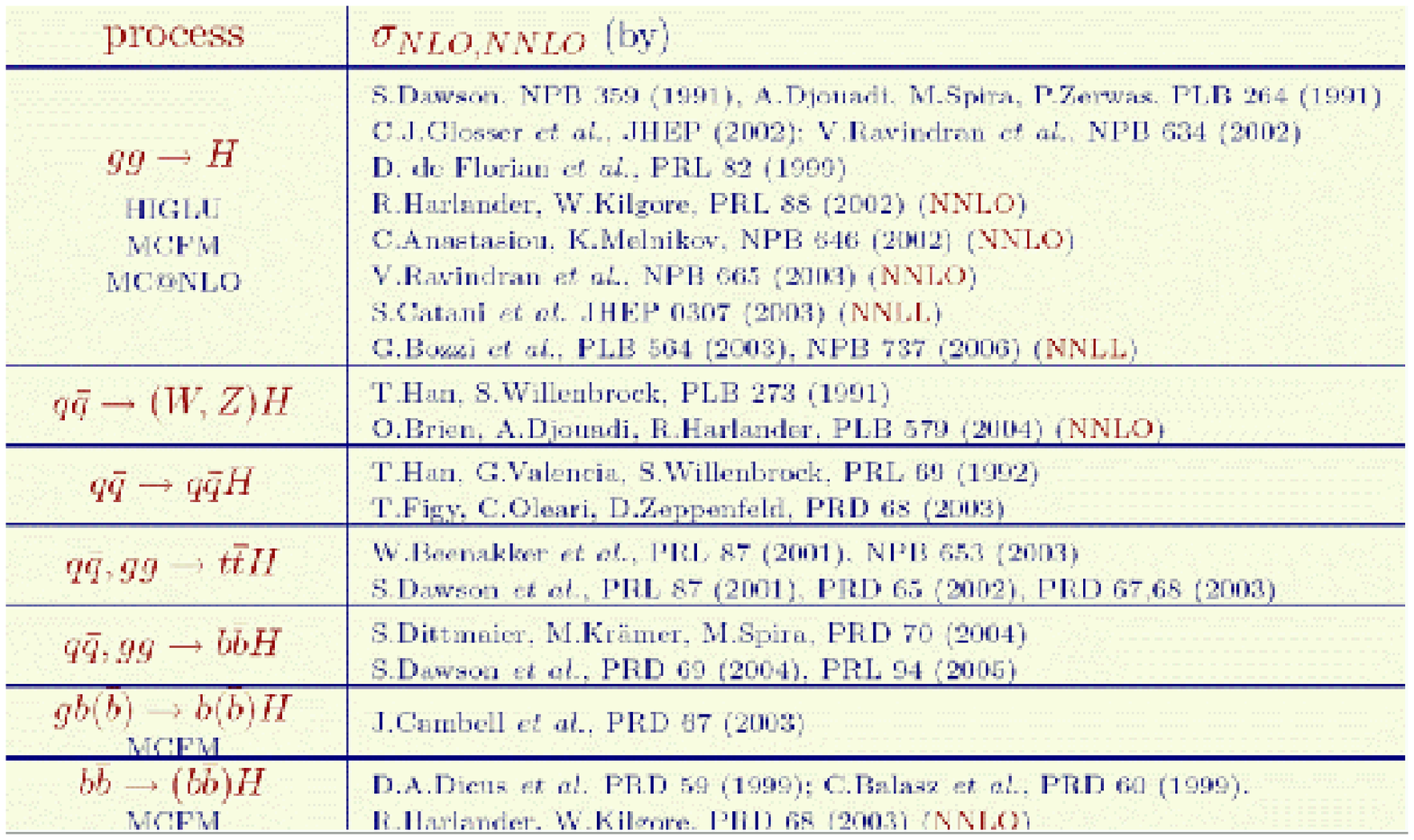}
\caption{\footnotesize{State of art of QCD predictions for Higgs
boson production at hadron colliders.}\label{stateofart1}}
\end{center}
\end{figure}

The state of art of QCD predictions for Higgs boson production at
LHC is reported in figure \ref{stateofart1} as it was summarized
by Laura Reina at the CTEQ summer shool 2006 on QCD analysis and
phenomenology, where also the Monte Carlo event generators are
indicated.

Furthermore, at LHC, multi-particles/jet production will be the
inescapable background to Higgs searches and searches for new
physics. This means that we should have a precise knowledge of the
QCD background. As seen previously, we know many QCD processes up
to the NNLO. However, we have at the moment limited NLO knowledge
of some important final states that will constitute background.
They are
\begin{eqnarray}
&\rightarrow&W/Z+\textrm{jets}\quad (2j)\nonumber \\
&\rightarrow&WW/ZZ/WZ+\textrm{jets}\quad (0j)\nonumber\\
&\rightarrow&WWW/ZZZ/WZZ+\textrm{jets}\quad (0j)\nonumber\\
&\rightarrow&Q\bar{Q}+\textrm{jets}\quad (0j)\nonumber\\
&\rightarrow&\gamma+\textrm{jets}\quad (1j)\nonumber\\
&\rightarrow&\gamma\gamma+\textrm{jets}\nonumber\\
&\rightarrow&Z\gamma\gamma+\textrm{jets},\nonumber
\end{eqnarray}
where in parenthesis is indicated the NLO knowledge.

Finally, we also note that in higher order contributions to the
splitting functions ($P_{gg}^{1}$,$P_{gq}^{1}$ for example), it
can be shown that there can appear also terms proportional to
\begin{equation}
\alpha_{s}^{n}\ln^{m}\frac{1}{z};\qquad m\leq n.
\end{equation}
These contributions spoil the convergence when $z\rightarrow 0$
and, in order to have reliable predictions, must be resummed. The
inclusion of the terms with $m=n$ defines a $LL_{z}$ resummation,
the inclusion of also the terms with $m=n-1$ defines a $NLL_{z}$
resummation. This resummation is realized by the
Balitsky-Fadin-Kuraev-Lipatov (BFKL) equation. Anyway, in this
thesis, we will not concentrate on this resummation. We will give
a briefly description of the various techniques to resum the large
soft logs giving attention to the renormalization group approach
and studying in detail its applications.

\section{The renormalization group approach to
resummation}\label{introrga}

The aim of resummation is to include all the logarithmic enhanced
terms of the form of Eq.(\ref{logterm}) of Section
\ref{nlonotenough} with a certain hierarchy of logarithms that we
shall define in the current section.

From Eqs.(\ref{cqqnlo0},\ref{f2nlo}) of Section \ref{disdycs} we
see that the QCD cross section (up to dimensional overall factors)
can in general be written as a convolution of the parto densities
$F_{a}^{H_{i}}(x_{i},\mu^{2})$ and of the dimensionless partonic
cross section, i.e. the coefficient function
$C(z,Q^{2}/\mu^{2},\alpha_{s}(\mu^{2}))$:
\begin{equation}\label{convdy}
\sigma_{\rm
DY}(x,Q^{2})=\sum_{a,b}\left[F_{a}^{H_{1}}(\mu^{2})\otimes
F_{b}^{H_{2}}(\mu^{2})\otimes
C_{ab}(Q^{2}/\mu^{2},\alpha_{s}(\mu^{2}))\right](x),
\end{equation}
for the DY case; and
\begin{equation}\label{convdis}
\sigma_{\rm DIS}(x,Q^{2})=\sum_{a}\left[F_{a}^{H}(\mu^{2})\otimes
C_{a}(Q^{2}/\mu^{2},\alpha_{s}(\mu^{2}))\right](x),
\end{equation}
for the DIS case. The convolution product $\otimes$ has been
defined in Eq.(\ref{defconv}) of Section \ref{dglapeq}. Performing
the Mellin transformation
\begin{equation}
\sigma(N,Q^{2})=\int_{0}^{1}dx\,x^{N-1}\sigma(x,Q^{2})
\end{equation}
we turn the convolution products of
Eqs.(\ref{convdy},\ref{convdis}) into ordinary products:
\begin{eqnarray}
\sigma_{\rm
DY}(N,Q^{2})&\equiv&\sum_{a,b}\sigma_{ab}(N,Q^{2})\nonumber\\
&=&\sum_{a,b}F_{a}^{H_{1}}(N,\mu^{2}) F_{b}^{H_{2}}(N,\mu^{2})
C_{ab}(N,Q^{2}/\mu^{2},\alpha_{s}(\mu^{2})),\label{maddi}\\
 \sigma_{\rm
DIS}(N,Q^{2})&\equiv&\sum_{a}\sigma_{a}(N,Q^{2})=\sum_{a}F_{a}^{H}(N,\mu^{2})
C_{a}(N,Q^{2}/\mu^{2},\alpha_{s}(\mu^{2}))\label{maddi2},
\end{eqnarray}
where
\begin{eqnarray}
C_{a(b)}\left(N,\frac{Q^{2}}{\mu^{2}},\alpha_{s}(\mu^{2})\right)&=&\int_{0}^{1}
dzz^{N-1}C_{a(b)}\left(z,\frac{Q^{2}}{\mu^{2}},\alpha_{s}(\mu^{2})\right),\\
F^{H_{j}}_{a}(N,\mu^{2})&=&\int_{0}^{1}dxx^{N-1}F^{H_{j}}_{a}(x,\mu^{2}),
\end{eqnarray}
and where the second index in brackets $(b)$ is involved only when
there are two hadrons in the initial state as the the DY case.

The large logs of $1-z$ of Eq.(\ref{logterm}) in Section
\ref{nlonotenough} are mapped to the large logs of $N$ by the
Mellin transform. This fact and the relations between the large
logs of $1-z$ and the large logs of $N$ are shown in detail in
Appendix \ref{lognlogx}.

Whereas the cross section $\sigma(N,Q^2)$ is clearly
$\mu^2$-independent, this is not the case for each contribution
$\sigma_{a(b)}(N,Q^{2})$. However, the $\mu^2$ dependence of each
contribution to the sum over $a,(b)$ in
Eqs.(\ref{maddi},\ref{maddi2}) is proportional to the off-diagonal
anomalous dimensions $\gamma_{qg}$ and $\gamma_{gq}$. In the large
$N$ limit, these are suppressed by a power of $\frac{1}{N}$ in
comparison to $\gamma_{gg}$ and $\gamma_{qq}$, or, equivalently,
the corresponding splitting functions are suppressed by a factor
of $1-z$ in the large $z$ limit (see for example
Eqs.(\ref{p0qq},\ref{p0qg}) in Section \ref{dglapeq}). Hence, in
the large $N$ limit each parton subprocess can be treated
independently, specifically, each $C_{a(b)}$ is separately
renormalization-group invariant. Because we are interested in the
behaviour of $C_{a(b)}(N,Q^2/\mu^2,\as(\mu^2))$ in the limit $N\to
\infty$ we can treat each subprocess independently.

Because resummation takes the form of an exponentiation, we define
a so-called physical anomalous dimension defined implicitly
through the equation
\begin{equation} \label{physad2}
Q^2\frac{\partial \sigma_{a(b)}(N,Q^2)}{\partial Q^2}=
\gamma_{a(b)}(N,\as(Q^2))\,\sigma_{a(b)}(N,Q^2).
\end{equation}
The physical anomalous dimensions $\gamma_{a(b)}$
Eq.(\ref{physad2}) is independent of factorization scale, and it
is related to the diagonal standard anomalous dimension
$\gamma^{AP}_{cc}$, defined by
\begin{equation} \label{evolF}
\mu^2\frac{\partial F_{c}(N,\mu^2)}{\partial \mu^2}= \gamma^{\rm
AP}_{cc}(N,\as(\mu^2)) F_{c}(N,\mu^2),
\end{equation}
according to
\begin{eqnarray}
\gamma_{a(b)}(N,\as(Q^2)) &=&\frac{\partial\ln
C_{a(b)}(N,Q^2/\mu^2,\as(\mu^2))}{\partial\ln Q^2} =\gamma^{\rm
AP}_{aa}(N,\as(Q^2))\label{16}\\
&&+\gamma^{\rm AP}_{(bb)}(N,\as(Q^2)) +\frac{\partial\ln
C_{a(b)}(N,1,\as(Q^2))}{\partial\ln Q^2}.\label{generic}
\end{eqnarray}
We recall that both the standard anomalous dimensions
(Altarelli-Parisi splitting functions) and the coefficient
function are computable in perturbation theory. Hence, the
physical anomalous dimensions differs from the standard anomalous
dimensions only beyond the LO in $\alpha_{s}$ as can be seen
directly from Eq.(\ref{generic}). In terms of the physical
anomalous dimensions, the cross section can be written as
\begin{eqnarray} \label{pertic}
\sigma(N,Q^2)&=&\sum_{a,(b)}K_{a(b)}(N;Q_0^2,Q^2)\,\sigma_{a(b)}(N,Q_0^2)\\
&&=\sum_{a,(b)}\exp\left[E_{a(b)}(N;Q_0^2,Q^2)\right]\sigma_{a(b)}(N,Q_0^2),
\end{eqnarray}
where
\begin{eqnarray}
\label{esplit} E_{a(b)}(N;Q_0^2,Q^2)
&=&\int_{Q_0^2}^{Q^2}\frac{dk^2}{k^2}\gamma_{a(b)}(N,\as(k^2))\label{edefic}\\
&=&\int_{Q_0^2}^{Q^2} \frac{dk^2}{k^2}[\gamma^{\rm AP}_{aa}(N,\as(k^2))+\gamma^{\rm AP}_{(bb)}(N,\as(k^2))]\nonumber\\
&&+\ln C_{a(b)}(N,1,\as(Q^2))-\ln
C_{a(b)}(N,1,\as(Q_0^2))\label{edefic2}.
\end{eqnarray}

We now concentrate on the single subprocess with incoming partons
$a,(b)$.  Resummation of the large logs of $N$ in the cross
section is obtained performing their resummation in the physical
anomalous dimension:
\begin{equation}
\sigma^{res}(N,Q^{2})=\exp\left\{\int_{Q_{0}^{2}}^{Q^{2}}\frac{dk^{2}}{k^{2}}\gamma^{res}(N,\alpha_{s}(k^{2}))\right\}\sigma^{res}(N,Q_{0}^{2}).
\end{equation}
 This shows how in general the large logs of $N$ can be
 exponentiated. For the DY case only the quark-anti-quark channel
 should be resummed and in the DIS case only the quark one. This
 is a consequence of the fact that the off-diagonal splitting
 functions are suppressed in the large $N$ limit as discussed before.

 The accuracy of resummation here depends on the
 accuracy at which the physical anomalous dimension $\gamma$ is
 computed. We say that the physical anomalous dimension is
 resummed at the next$^{k-1}$-to-leading-logarithmic accuracy
 ($N^{k-1}LL$) when all the contributions of the form
 \begin{equation}
 \alpha_{s}^{n+m}(Q^{2})\ln^{m} N;\qquad n=0,\dots,k-1
 \end{equation}
 are included in its determination. The goal of resummation is to
 determine the resummed physical anomalous dimension from at most
 a finite fixed-order computation of it.
 Clearly, once the resummed physical anomalous dimension is
 determined, it can then predict the leading,
 next-to-leading...logarithmic contributions to the cross section
 at all orders.

Here, in this Section, we shall only outline the key ideas of the
renormalization group approach to resummation. However, throughout
this thesis we shall show in detail how this method works and how
the resummed physical anomalous dimensions obtained with such an
approach can be fully determined expanding it to a certain finite
fixed-order and comparing this expansion with Eq.(\ref{generic})
obtained from explicit computations.

The renormalization group approach to resummation is essentially
divided in two steps. The first is to analyze the generic phase
space measure in $d=4-2\epsilon$ dimensions thus finding where the
large logs are originated in the coefficient function and in the
physical anomalous dimension. The second consists in resumming
them imposing the renormalization group invariance of the physical
anomalous dimension.

So, let's consider a generic phase space measure $d\phi_{n}$ for
the emission of $n$ massless partons with momenta
$p_{1},\dots,p_{n}$. In Appendix \ref{phasespacedec}, we show that
this phase space can be decomposed in terms of two-body phase
space. Roughly speaking, the phase spaces measure for the emission
of $n$ partons can be viewed as the emission of two partons (one
with momentum $p_{n}$ and the other with momentum
$P_{n}=p_{1}+\dots+_{n-1}$ and invariant mass $P_{n}^{2}$ times
the phase space measure where the momentum $P_{n}$ is incoming and
the momenta $p_{1},\dots,p_{n}$ are outgoing. The price to pay for
this is the introduction of an integration over the invariant mass
$P_{n}^{2}$. Then using recursively this procedure, we obtain that
the $n$-body phase space measure is decomposed in $n-1$ two-body
phase space measures. This means that we have reduced the study of
the soft emission of $n$-body phase space measure to the study of
the soft emission of two-body phase spaces.

The two-body phase space with an incoming momentum $P$ and two
outgoing momenta $Q$ and $p$ in $d=4-2\epsilon$ dimensions
(derived explicitly in Eq.(\ref{twobody}) of Section
\ref{phasespacedec}) is given by
\begin{equation}
d\phi_{2}(P;Q,p)=N(\epsilon)(P^{2})^{-\epsilon}\left(1-\frac{Q^{2}}{P^{2}}\right)^{1-2\epsilon}d\Omega_{d-1},\quad
N(\epsilon)=\frac{1}{2(4\pi)^{2-2\epsilon}},\label{twobody2}
\end{equation}
where $d\Omega_{d-1}$ is the solid angle in $d-1$ dimensions. For
definiteness, we can think that this is the phase space measure
for a single DIS-like or DY-like soft emission with with momentum
$p$. Thus, we have
\begin{eqnarray}
\textrm{DIS-like emission:}&& P^{2}\propto (1-z_{_{DIS}});\quad Q^{2}=0\\
\textrm{DY-like
emission:}&&\left(1-\frac{Q^{2}}{P^{2}}\right)\propto
(1-z_{_{DY}});\quad P^{2}=s_{_{DY}},
\end{eqnarray}
where $z_{_{DIS}}$, $z_{_{DY}}$ are close to one for a soft
emission. Hence, we have that the two-body phase space measure for
a single soft emission contributes with a factor
$(1-z)^{-a\epsilon}$ with $a=1$ for a DIS-like emission and $a=2$
for DY-like emission. In the case of the prompt-photon process, we
will see in Chapter \ref{DP} that there are both types of
emission. The large logs of $1-z$ are originated by the
interference with the infrared poles in $\epsilon=0$ in the square
modulus amplitude in the $\epsilon\rightarrow 0$ limit. For
example
\begin{equation}\label{lognint}
\frac{1}{\epsilon}\,(1-z)^{-a\epsilon}=\frac{1}{\epsilon}-\ln(1-z)^{a}+\emph{O}(\epsilon),
\end{equation}
for the case of interference with a pole of order $1$.

Now, since each factor of $(1-z)^{-a\epsilon}$ that comes from the
phase space measure is associated to a single real emission then
it will appear in the coefficient function together with a power
of the bare strong coupling constant $\alpha_{0}$. In
$d$-dimensions, the coupling constant is dimensionful, and thus on
dimensional grounds each emission is accompanied by a factor
\begin{equation}\label{chiaras}
\alpha_{0}\,\left[Q^{2}(1-z)^{a}\right]^{-\epsilon},
\end{equation}
where $Q^{2}$ is now the typical perturbative scale of a certain
process. Upon Mellin transformation, this becomes
\begin{equation}
\alpha_{0}\,\left[\frac{Q^{2}}{N^{a}}\right]^{-\epsilon}.
\end{equation}
Furthermore an analysis of the structure of diagrams shows that in
the soft (large $N$) limit, all dependence on $N$ appears through
the variable $Q^{2}/N^{a}$ also in the amplitude. Finally, a
renormalization group argument shows that all this dependence can
be reabsorbed in the running of the strong coupling. Indeed, the
first order renormalization of the bare coupling constant at the
renormalization scale $\mu$
\begin{equation}
\alpha_{0}=\mu^{2\epsilon}\alpha_{s}(\mu^{2})+\emph{O}(\alpha_{s}^{2})
\end{equation}
and the renormalization group invariance of the physical anomalous
dimension imply that
\begin{eqnarray}
\alpha_{0}\,\left[\frac{Q^{2}}{N^{a}}\right]^{-\epsilon}&=&
\alpha_{s}(\mu^{2})\,\left[\frac{Q^{2}}{\mu^{2}N^{a}}\right]^{-\epsilon}
+\emph{O}(\alpha_{s}^{2})\nonumber\\
&=&\alpha_{s}(Q^{2}/N^{a})+\emph{O}(\alpha_{s}^{2}),
\end{eqnarray}
where $\alpha_{0}$ is the bare coupling, $\alpha_{s}(\mu^{2})$ the
renormalized coupling and the higher order terms contain
divergences which cancel those in the cross section. Following
this line of argument one may show that the finite expression of
the renormalized cross section in terms of the renormalized
coupling is a function of $\alpha_{s}(Q^{2})$ and
$\alpha_{s}(Q^{2}/N^{a})$ with numerical coefficients, up to
$\emph{O}(1/N)$ corrections. We shall see this in detail in
Chapter \ref{DISDY2}.

\section{Alternative approaches}\label{alternative}

The exponentiation of the large soft logs and their resummation
has been demonstrated in QCD with the eikonal approximation
\cite{Catani:1989ne} or thanks to strong non-standard
factorization properties of the cross section in the soft limit
\cite{Sterman:1986aj}. Recently, also the effective field
theoretic (EFT) approach has been applied to QCD resummation in
Refs.\cite{Manohar:2003vb,Idilbi:2005ky} for DIS and DY and in
Ref.\cite{Bauer:2000ew} for the $B$ meson decay $B\rightarrow
X_{s}\,\gamma$. In this Section, we shall only give a brief
description of these alternative approaches to the resummation of
the large perturbative logarithms.

\subsection{Eikonal approach}\label{eiconalapp}

We first consider the simpler case of QED. In QED the
exponentiation of the large soft logs has been proved thanks to
the eikonal approximation in Ref.\cite{Sudakov:1954sw}. We report
the basic steps of the proof for the QED case and a brief
description of the generalization to the QCD case.

Consider a final fermion line with momentum $p'$ of a generic QED
Feynman diagram. We attach $n$ soft photons to this fermion line
with momenta $k_{1},\dots,k_{n}$. For the moment we do not care
whether these are external photons, virtual photons connected to
each other, or virtual photons connected to vertices on other
fermion lines. The amplitude for such a diagram has the following
structure in the soft limit:
\begin{eqnarray}\label{stasescuola}
\bar{u}(p')(-ie\gamma^{\mu_{1}})\frac{i\sc{p}'}{2p'\cdot
k_{1}}(-ie\gamma^{\mu_{2}})\frac{i\sc{p}'}{2p'\cdot(k_{1}+k_{2})}\cdots
(-ie\gamma{\mu_{n}})\frac{i\sc{p}'}{2p'\cdot (k_{1}+\dots
+k_{n})}i\mathcal {M}_{h},
\end{eqnarray}
where $e=-|e|$ is the electron charge and $i\mathcal{M}_{h}$ is
the amplitude of the hard part of the process without the final
fermion line we are considering. We note that here we have
neglected the electron mass. Then we can push the factors of
$\sc{p}'$ to the left and use the Dirac equation
$\bar{u}(p')\sc{p}'=0$:
\begin{equation}
\bar{u}(p')\gamma^{\mu_{1}}\sc{p}'\gamma^{\mu_{2}}\sc{p}'\cdots
\gamma^{\mu_{n}}\sc{p}'=\bar{u}(p')2p'^{\mu_{1}}\gamma^{\mu_{2}}\sc{p}'\cdots
\gamma^{\mu_{n}}\sc{p}'=\bar{u}(p')2p'^{\mu_{1}}2p'^{\mu_{2}}\cdots
2p'^{\mu_{n}}.
\end{equation}
Thus Eq.(\ref{stasescuola}) becomes
\begin{equation}
e^{n}\bar{u}(p')\left(\frac{p'^{\mu_{1}}}{p'\cdot
k_{1}}\right)\left(\frac{p'^{\mu_{2}}}{p'\cdot
(k_{1}+k_{2})}\right)\cdots \left(\frac{p'^{\mu_{n}}}{p'\cdot
(k_{1}+\dots+k_{n})}\right)i\mathcal{M}_{h}.\label{ledzeppelin}
\end{equation}
Still working with only a final fermion line, we must now sum over
all possible orderings of momenta $k_{1},\dots,k_{n}$. There are
$n!$ different diagrams to sum, corresponding to the $n!$
permutations of the $n$ photon momenta. Let $P$ denote one such
permutation, so that $P(i)$ is the number between $1$ and $n$ that
$i$ is taken to. Now, using the identity
\begin{equation}
\sum_{P}\frac{1}{p\cdot k_{P(1)}}\, \frac{1}{p\cdot
(k_{P(1)}+k_{P(2)})}\cdots \frac{1}{p\cdot
(k_{P(1)})+\dots+k_{P(n)})}=\frac{1}{p\cdot k_{1}}\cdots
\frac{1}{p\cdot k_{n}},
\end{equation}
the sum over all the permutations of the photons of
Eq.(\ref{ledzeppelin}) is:
\begin{equation}
e^{n}\bar{u}(p')\left(\frac{p'^{\mu_{1}}}{p'\cdot
k_{1}}\right)\left(\frac{p'^{\mu_{2}}}{p'\cdot k_{2}}\right)\cdots
\left(\frac{p'^{\mu_{n}}}{p'\cdot
k_{n}}\right)i\mathcal{M}_{h}.\label{ledzeppelin2}
\end{equation}

At this point, we consider an initial fermion line with momentum
$p$. In this case the photon momenta in the denominators of the
fermion propagators have an opposite sign. Therefore, if we sum
over all the diagrams containing a total of $n$ soft photons,
connected in any possible order to an arbitrary number of initial
and final fermion lines, Eq.(\ref{ledzeppelin2}) becomes:
\begin{equation}
e^{n}i\mathcal{M}_{0}\prod_{r=1}^{n}\sum_{i}\frac{\eta_{i}p^{\mu_{i}}}{p_{i}\cdot
k_{r}},
\end{equation}
where $i\mathcal{M}_{0}$ is the full amplitude of the hard part of
the process and where the index $r$ runs over the radiated photons
and the index $j$ runs over the initial and final fermion lines
with
\begin{equation}
\eta_{i}=\left\{\begin{array}{ll}1&\textrm{for a final fermion line}\\
-1&\textrm{for an initial fermion line}\end{array}\right.
\end{equation}

If only a real soft photon is radiated, we must multiply by its
polarization vector, sum over polarizations, and integrate the
squared matrix element over the photon phase space. In the Feynman
gauge this gives a factor
\begin{equation}
Y=\int\frac{d^{3}k}{(2\pi)^{3}2k^{0}}e^{2}\left(\sum_{i}\frac{\eta_{i}p_{i}}{p_{i}\cdot
k}\right)^{2}\label{factoreic}
\end{equation}
in the final cross section. If $n$ real photons are emitted, we
get $n$ such $Y$ factors Eq.(\ref{factoreic}), and also a symmetry
factor $1/n!$ since there are $n$ identical bosons in the final
state. The cross section resummed for the emission of any number
of soft photons is therefore
\begin{equation}
\sigma^{res}(i\rightarrow f)=\sigma_{0}(i\rightarrow
f)\sum_{n=0}^{\infty}\frac{Y^{n}}{n!}=\sigma_{0}(i\rightarrow
f)e^{Y},
\end{equation}
where $\sigma_{0}(i\rightarrow f)$ is the cross section for the
hard process without extra soft emissions. This result shows that
all the possible soft real emissions exponentiate and that only
the single emission contributes to the exponent. However, this is
not the end of the story, because the exponent $Y$
Eq.(\ref{factoreic}) is infrared divergent. Indeed, to obtain a
reliable finite result, we must include also loop corrections to
all orders. For a detailed analysis about the inclusion of loops
see for example Ref.\cite{Yennie:1961ad}. Here, we just give the
final result which reads:
\begin{equation}
\sigma^{res}(i\rightarrow f)=\sigma_{0}(i\rightarrow
f)e^{\sigma^{(1)}},\label{resformqed}
\end{equation}
where $\sigma^{(1)}$ is the cross section relative to the single
soft emission from the hard process. Clearly, the accuracy of this
resummation formula for soft photon emission Eq.(\ref{resformqed})
depends on the accuracy at which the exponent for the single
emission is computed.

In Ref.\cite{Catani:1989ne} the exponentiation of the soft
emissions, here outlined for QED, is generalized to the QCD case.
Differently from QED, QCD is a non-abelian gauge theory and this
implies that this generalization is highly non-trivial. Indeed,
the gluons can interact with each other. This fact makes the
exponentiation mechanism much more difficult since the three gluon
vertex color factor is different from that of the quark-gluon
vertex. In order to exponentiate the single emission cross section
(as it happens in QED), one should prove that these gluon
correlations cancels out order by order in perturbation theory.
This is shown for example in Ref.\cite{Amati:1980ch}. According to
this result, it has been shown in Ref.\cite{Catani:1989ne} how the
the exponentiation of soft emission works in QCD resummation. We
report here the result for the NLL resummed coefficient function
in Mellin space for inclusive DIS and DY processes in the
$\overline{MS}$ scheme in a compact form:
\begin{eqnarray}
C_{\rm
NLL}(N,Q^{2}/\mu^{2},\alpha_{s}(\mu^{2}))&=&\exp\bigg\{a\int_{0}^{1}dx\,
\frac{x^{N-1}-1}{1-x}\bigg[\int_{\mu^{2}}^{Q^{2}(1-x)^{a}}\frac{dk^{2}}{k^{2}}A(\alpha_{s}(k^{2}))\nonumber\\
&&+B^{(a)}(\alpha_{s}(Q^{2}(1-x)^{a}))\bigg]\bigg\}\label{fanta},
\end{eqnarray}
where
\begin{eqnarray}
A(\alpha_{s})&=&A_{1}\alpha_{s}+A_{2}\alpha_{s}^{2}+\dots\\
B^{(a)}(\alpha_{s})&=&B_{1}^{(a)}\alpha_{s}+\dots
\end{eqnarray}
with
\begin{equation}\label{bah}
A_{1}=\frac{C_{F}}{\pi},\qquad
A_{2}=\frac{C_{F}}{2\pi^{2}}\left[C_{A}\left(\frac{67}{18}-\frac{\pi^{2}}{6}\right)-\frac{5}{9}N_{f}\right],\qquad
B_{1}^{(a)}=-\frac{(2-a)3C_{F}}{4\pi}.
\end{equation}
Here $a=1$ for the DIS structure functio $F_{2}$ and $a=2$ for the
DY case. How this result is strictly connected to the resummed
results that can be obtained with the renormalization group
approach will be discussed in Section \ref{nllresummation}.

\subsection{Resummation from strong factorization
properties}\label{vivamaddy2}

This is the approach of Ref.\cite{Sterman:1986aj}. Also in this
approach the results given in Eq.(\ref{fanta}) of Section
\ref{eiconalapp} are recovered. Here we give only a rough
description of this method based on strong factorization
properties of the QCD cross section.

It is essentially assumed that at the boundary of the phase space,
the cross section is factorized in a hard and in a soft part and
eventually in an other factor associated to final collinear jets
as in the DIS case where there is an outgoing emitting quark. The
final result is then obtained exponentiating the soft and
collinear factors. This is done solving their evolution equations.

In Ref.\cite{Sterman:1986aj} it is shown that the semi-inclusive
cross section can be factorized in three factors relative to the
three different regions in the momentum space of the process: the
off-shell partons that participate to the partonic hard process,
the collinear and soft on-shell radiated partons. The cross
section is given by
\begin{eqnarray}
\sigma(w)&=&H\left(\frac{p_{1}}{\mu},\frac{p_{2}}{\mu},\zeta_{i}\right)\int
\frac{dw_{1}}{w_{1}}\frac{dw_{2}}{w_{2}}\frac{dw_{3}}{w_{3}}J_{1}
\left(\frac{p_{1}\cdot\zeta_{1}}{\mu},w_{1}\left(\frac{Q}{\mu}\right)^{a}\right)
\nonumber\\
&&J_{2}\left(\frac{p_{2}\cdot\zeta_{2}}{\mu},w_{2}\left(\frac{Q}{\mu}\right)^{a}\right)
S\left(w_{s}\frac{Q}{\mu},\zeta_{i}\right)\delta(w-w_{1}-w_{2}-w_{3}),\label{checazzoe}
\end{eqnarray}
where $a$ is the number of hadrons in the initial state, $\mu$ is
the factorization scale and $\zeta_{i}$ are gauge-fixing
parameters; the integration variables $w_{1}$, $w_{2}$ and $w_{3}$
are referred to the two collinear jets and to soft radiation
respectively. Each factor of Eq.(\ref{checazzoe}) is evaluated at
the typical scale of the momentum space region which is associated
to. The delta function imposes that
\begin{equation}
w=w_{1}+w_{2}+w_{3}=\left\{\begin{array}{ll}1-x_{Bj},&\textrm{for
the DIS case}\\1-Q^{2}/S&\textrm{for the DY
case}\end{array}\right.
\end{equation}
The convolution of Eq.(\ref{checazzoe}) is turned into an ordinary
product performing the Mellin transform:
\begin{eqnarray}
\sigma(N)&=&\int_{0}^{\infty}dw\,e^{-Nw}\sigma(w)=H\left(\frac{p_{1}}{\mu},\frac{p_{2}}{\mu},\zeta_{i}\right)
S\left(\frac{Q}{\mu N},\zeta_{i}\right)\nonumber\\
&&\times J_{1}\left(\frac{p_{1}\cdot\zeta_{1}}{\mu},\frac{Q}{\mu
N^{1/a}}\right)J_{2}\left(\frac{p_{2}\cdot\zeta_{2}}{\mu},\frac{Q}{\mu
N^{1/a}}\right).
\end{eqnarray}
Each factor $H$, $J_{i}$, $S$ satisfy the following evolution
equations
\begin{eqnarray}
\mu^{2}\frac{\partial}{\partial\mu^{2}}\ln
H&=&-\gamma_{H}(\alpha_{s}(\mu^{2})),\label{chiara}\\
\mu^{2}\frac{\partial}{\partial\mu^{2}}\ln
S&=&-\gamma_{S}(\alpha_{s}(\mu^{2})),\\
\mu^{2}\frac{\partial}{\partial\mu^{2}}\ln
J_{i}&=&-\gamma_{J_{i}}(\alpha_{s}(\mu^{2})),\label{chiara2}
\end{eqnarray}
where the physical anomalous dimensions $\gamma_{H}(\alpha_{s})$,
$\gamma_{J_{i}}(\alpha_{s})$ and $\gamma_{S}(\alpha_{s})$ are
calculable in perturbation theory and must satisfy, according to
the renormalization group invariance of the cross section the
relation
\begin{equation}\label{chiara3}
\gamma_{H}(\alpha_{s})+\gamma_{S}(\alpha_{S})+\sum_{i=1}^{2}\gamma_{J_{i}}(\alpha_{s})=0.
\end{equation}
Solving Eqs.(\ref{chiara}-\ref{chiara2}) and imposing
renormalization group invariance Eq.(\ref{chiara3}), the resummed
section can be written in the form
\begin{eqnarray}
\sigma(N)&=&\exp\bigg\{D_{1}(\alpha_{s}(Q^{2}))+D_{2}\left(\alpha_{s}\left(\frac{Q^{2}}{N^{2}}\right)\right)
-\frac{2}{a-1}\int_{\frac{Q}{N}}^{\frac{Q}{N^{1/a}}}\frac{d\xi}{\xi}\ln\left(\frac{\xi
N}{Q}\right)A(\alpha_{s}\xi^{2})\nonumber\\
&&-2\int_{\frac{Q}{N^{1/a}}}^{Q^{2}}\frac{d\xi}{\xi}\left[\ln\left(\frac{Q}{\xi}\right)A(\alpha_{s}(\xi^{2}))
-B(\alpha_{s}(\xi^{2}))\right],
\end{eqnarray}
where the functions $A(\alpha_{s})$, $B(\alpha_{s})$,
$D_{i}(\alpha_{s})$ are determined in terms of the anomalous
dimensions and the beta function. Finally, it can be shown that
this result can be casted in the form of Eq.(\ref{fanta}) in
Section \ref{eiconalapp} for the resummed coefficient function.

\subsection{Resummation from Effective Field Theory}

This is the approach of Refs.\cite{Manohar:2003vb,Idilbi:2005ky}.
This EFT methodology to resum threshold logarithms is made
concrete due to the recently developed ``soft collinear effective
theory" (SCET)
\cite{Bauer:2000yr,Bauer:2001yt,Chay:2002vy,Bauer:2002nz}. The
SCET describes interactions between soft and collinear partons.

The starting point (considering the DY case as an example) is the
collinearly factorized inclusive cross section in Mellin space:
\begin{equation}
\sigma(N,Q^{2})=\sigma_{0}\,C(N,Q^{2}/\mu^{2},\alpha_{s}(\mu^{2}))F_{1}(N,\mu^{2})F_{2}(N,\mu^{2}),
\end{equation}
where $\sigma_{0}$ is the born level cross section and
$F_{i}(N,\mu^{2})$ are the parton densities at the factorization
scale $\mu$. Here, the basic idea is to write the coefficient
function $C(N,Q^{2}/\mu^{2},\alpha_{s}(\mu^{2}))$ in terms of an
intermediate scale $\mu_{I}$:
\begin{equation}\label{021}
C(N,Q^{2}/\mu^{2},\alpha_{s}(\mu^{2}))=C(N,Q^{2}/\mu^{2}_{I},\alpha_{s}(\mu_{I}^{2}))
\exp\left[-2\int_{\mu_{I}^{2}}^{\mu^{2}}\frac{dk^{2}}{k^{2}}\gamma^{\rm
AP}_{qq}(N,\alpha_{s}(k^{2}))\right]
\end{equation}
and then to compute
$C(N,Q^{2}/\mu^{2}_{I},\alpha_{s}(\mu_{I}^{2}))$ with the ``soft
collinear effective theory" with the intermediate scale
$\mu_{I}^{2}$ equal to the typical scale of the soft-collinear
emission, i.e. $\mu_{I}=Q^{2}/N^{2}$ in the DY case and
$\mu_{I}^{2}=Q^{2}/N$ in the DIS case. In SCET,
$C(N,Q^{2}/\mu^{2}_{I},\alpha_{s}(\mu_{I}^{2}))$ has the following
general structure:
\begin{equation}\label{coefffuncscet}
C\left(N,\frac{Q^{2}}{\mu^{2}_{I}},\alpha_{s}(\mu_{I}^{2})\right)=
\left|\tilde{C}\left(\frac{Q^{2}}{\mu_{I}^{2}},\alpha_{s}(\mu_{I}^{2})\right)\right|^{2}
\mathcal{M}(N,\alpha_{s}(\mu_{I}^{2})),
\end{equation}
Here $\tilde{C}(Q^{2}/\mu_{I}^{2},\alpha_{s}(\mu_{I}^{2}))$ is the
effective coupling that matches the full QCD theory currents
\begin{equation}
J_{QCD}=\tilde{C}(Q^{2}/\mu_{I}^{2},\alpha_{s}(\mu_{I}^{2}))J_{eff}(\mu_{I}^{2}).
\end{equation}
We note that the effective coupling $\tilde{C}$ contains the
perturbative contribution between the scale $Q^{2}$ and
$\mu_{I}^{2}$ and $J_{eff}(\mu_{I}^{2})$ contains the soft and
collinear contributions below the scale $\mu_{I}^{2}$. Then
$\mathcal{M}(N,\alpha_{s}(\mu_{I}^{2})$ is the matching
coefficient that guarantees that the EFT used generates the full
QCD results in the appropriate kinematical limit. In SCET the
matching coefficient $\mathcal{M}$ can be computed perturbativly
and is free of any logarithms.

The effective coupling $\tilde{C}$ satisfy to a certain evolution
equation
\begin{equation}\label{guido}
\mu^{2}\frac{\partial}{\partial\mu^{2}}\ln\tilde{C}(Q^{2}/\mu^{2},\alpha_{s}(\mu^{2}))=-\frac{1}{2}\gamma_{1}(\alpha_{s}(\mu^{2}))
\end{equation}
where the physical anomalous dimension $\gamma_{1}$ is computable
perturbativly in SCET. Finally, solving the evolution equation
Eq.(\ref{guido}) one finds that Eq.(\ref{coefffuncscet}) becomes
\begin{eqnarray}
C\left(N,\frac{Q^{2}}{\mu^{2}_{I}},\alpha_{s}(\mu_{I}^{2})\right)&=&\left|\tilde{C}(1,\alpha_{s}(Q^{2}))\right|^{2}
e^{E_{1}(Q^{2}/\mu_{I}^{2},alpha_{s}(Q^{2}))}\times\nonumber\\
&&\times\mathcal{M}(N,\alpha_{s}(Q^{2}))e^{E_{2}(Q^{2}/\mu_{I}^{2},\alpha_{s}(Q^{2}))},\label{022}
\end{eqnarray}
where
\begin{eqnarray}
E_{1}\left(\frac{Q^{2}}{\mu_{I}^{2}},\alpha_{s}(Q^{2})\right)&=&-\frac{1}{4}\int_{\mu_{I}^{2}}^{Q^{2}}
\frac{dk^{2}}{k^{2}}\gamma_{1}(\alpha_{s}(k^{2})),\\
E_{2}\left(\frac{Q^{2}}{\mu_{I}^{2}},\alpha_{s}(Q^{2})\right)&=&
\int_{\mu_{I}^{2}}^{Q^{2}}\frac{dk^{2}}{k^{2}}\beta(\alpha_{s}(k^{2}))\frac{d\ln\mathcal{M}(N,\alpha_{s}(k^{2}))}{d\ln\alpha_{s}(k^{2})},
\end{eqnarray}
with $\beta(\alpha_{s})$ the beta function of Eq.(\ref{betafunc})
of Section \ref{appA}. $\tilde{C}(1,\alpha_{s}(Q^{2}))$ contains
the non-logarithmic contribution of the purely virtual diagrams
and the exponent $E_{1}$ contains all the logarithms originating
from the same type of diagrams. $E_{2}$ encodes all the
contributions due to the running of the coupling constant between
the scale $\mu_{I}^{2}$ and $Q^{2}$. All the logarithms appear
only in the exponents (see Eqs.(\ref{021},\ref{022})) and the term
$|\tilde{C}(1,\alpha_{s}(Q^{2}))|^{2}\mathcal{M}(N,\alpha_{s}(Q^{2}))$
is free of any large logarithms.

The various approaches can be related one to the other according
to factorization properties of the QCD cross section in the soft
limit. In this way, it is possible to show that all the approaches
are equivalent except for the renormalization group approach that
produces correct but less restrictive results.

\chapter{Renormalization group resummation of DIS and DY}\label{DISDY2}

In this chapter, we analyze in detail how the renormalization
group approach to resummation works in the case of the
all-inclusive Drell-Yan (DY) and deeep inelastic scattering (DIS).
We recall that this is done only for the quark-anti-quark channel
in the DY case and only for the quark channel in the DIS case as
discussed in Section \ref{introrga}. First, we determine the N
dependence of the regularized coefficient function in the
large-$N$ limit. Then we show that, given this form of the
$N$-dependence of the regularized cross section, renormalization
group invariance fixes the all-order dependence of the physical
anomalous dimension in such a way that the infinite class of
leading, next-to-leading etc. resummations can be found in terms
of fixed order computations. This approach will lead us to
resummation formulae valid to all logrithmic orders.

\section{Kinematics of inclusive DIS in the soft limit}\label{kindis}

In the case of deep-inelastic scattering, the relevant parton
subprocesses are:
\begin{eqnarray}
\gamma^{*}(q)+\mathcal{Q}(p)&\rightarrow& \mathcal{Q}(p')+\mathcal{X}(K)\label{proc1}\\
\gamma^{*}(q)+\mathcal{G}(p)&\rightarrow&
\mathcal{Q}(p')+\mathcal{X}(K),\label{proc2}
\end{eqnarray}
where $\mathcal{Q}$ is a quark or an anti-quark, $\mathcal{G}$ is
a gluon and $\mathcal{X}$ is any collection of quarks and gluons.
We are interested in the most singular parts in the limit
$z\rightarrow 1$ of the phase space and of the amplitude for the
generic processes Eqs.(\ref{proc1},\ref{proc2}). We treat first
the tree level processes and then we will introduce the loops.

Using Eq.(\ref{gendecps}) in Appendix \ref{phasespacedec} with
$m=1$ recursively, we can express the phase space for a generic
process in terms of two-body phase space integrals. For the DIS
processes Eqs.(\ref{proc1},\ref{proc2}) with $n$ extra emissions
($K=k_{1}+\cdots +k_{n}$) we have
\begin{eqnarray}
&&d\phi_{n+1}(p+q;k_{1},\dots,k_{n},p')\nonumber \\
&=&\int_{0}^{s}\frac{dM_{n}^{2}}{2\pi}d\phi_{2}(p+q;k_{n},P_{n})d\phi_{n}(P_{n};k_{1},\dots,k_{n-1},p')\nonumber
\end{eqnarray}
\begin{eqnarray}
&=&\int_{0}^{s}\frac{dM_{n}^{2}}{2\pi}d\phi_{2}(p+q;k_{n},P_{n})\nonumber\\
&&\times\int_{0}^{M_{n}^{2}}\frac{dM_{n-1}^{2}}{2\pi}d\phi_{2}(P_{n};k_{n-1},P_{n-1})d\phi_{n-1}
(P_{n-1};k_{1},\dots,k_{n-2},p')\nonumber\\
&=&\int_{0}^{s}\frac{dM_{n}^{2}}{2\pi}d\phi_{2}(P_{n+1};k_{n},P_{n})\int_{0}^{M_{n}^{2}}\frac{dM_{n-1}}{2\pi}d\phi_{2}(P_{n};k_{n-1},P_{n-1})
\nonumber\\
&&\times\cdots\times
\int_{0}^{M_{3}^{2}}\frac{dM_{2}^{2}}{2\pi}d\phi_{2}(P_{3};k_{2},P_{2})d\phi_{2}(P_{2};k_{1},P_{1}),\label{disps}
\end{eqnarray}
where we have defined $M_{n}^{2}\equiv P_{n}^{2}$, $P_{1}\equiv
p'$, $P_{n+1}\equiv p+q$ and
\begin{equation}\label{sxdep}
s\equiv M_{n+1}^{2}=(p+q)^{2}=Q^{2}\frac{(1-z)}{z};\qquad
z=\frac{Q^{2}}{2p\cdot q}.
\end{equation}
Now, according to Eq.(\ref{twobody}) in Appendix
\ref{phasespacedec} we have for each two-body phase space
\begin{equation}
d\phi_{2}(P_{i+1};k_{i},P_{i})=N(\epsilon)(M_{i+1})^{-2\epsilon}\left(1-\frac{M_{i}^{2}}{M_{i+1}^{2}}\right)^{1-2\epsilon}d\Omega_{i};\quad
i=1,\dots n,
\end{equation}
where
\begin{equation}
N(\epsilon)=\frac{1}{2(4\pi)^{2-2\epsilon}}
\end{equation}
and $\Omega_{i}$ is the solid angle in the center-of-mass frame of
$P_{i+1}$. We perform the change of variables
\begin{equation}\label{changedis}
z_{i}=\frac{M_{i}^{2}}{M_{i+1}};\quad M_{i}^{2}=sz_{n}\dots
z_{i};\quad i=2,\dots,n.
\end{equation}
From the fact that $M_{i+1}^{2}\geq M_{i}^{2}$ (we have one more
real particle in $P_{i+1}$ than in $P_{i}$), it follows that
\begin{equation}
0\leq z_{i}\leq 1.
\end{equation}
From Eq.(\ref{changedis}) we get
\begin{equation}\label{sost1}
dM_{n}^{2}dM_{n-1}^{2}\cdots dM_{2}^{2}=\det\left(\frac{\partial
M_{i}^{2}}{\partial z_{j}}\right)dz_{n}dz_{n-1}\cdots dz_{2},
\end{equation}
where
\begin{equation}\label{sost2}
\det\left(\frac{\partial M_{i}^{2}}{\partial
z_{j}}\right)=\frac{\partial M_{n}^{2}}{\partial
z_{n}}\cdots\frac{\partial M_{2}^{2}}{\partial z_{2}}
=s^{n-1}z_{n}^{n-2}z_{n-1}^{n-3}\cdots z_{3}.
\end{equation}
Furthermore, in these new variables the two-body phase space
becomes
\begin{equation}\label{sost3}
d\phi_{2}(P_{i+1};k_{i},P_{i})=N(\epsilon)s^{-\epsilon}(z_{n}z_{n-1}\cdots
z_{i+1})^{-\epsilon}(1-z_{i})^{1-2\epsilon}d\Omega_{i}.
\end{equation}
Substituting Eqs.(\ref{sost1},\ref{sost2},\ref{sost3}) into the
generic phase space Eq.(\ref{disps}), we finally get
\begin{eqnarray}
d\phi_{n+1}(p+q;k_{1},\dots,k_{n},p')=2\pi\left[\frac{N(\epsilon)}{2\pi}\right]^{n}s^{n-1-n\epsilon}d\Omega_{n}\cdots d\Omega_{1}&&\nonumber\\
\times\int_{0}^{1}dz_{n}z_{n}^{(n-2)-(n-1)\epsilon}(1-z_{n})^{1-2\epsilon}\cdots\int_{0}^{1}dz_{2}z_{2}^{-\epsilon}(1-z_{2})^{1-2\epsilon}.&&
\label{disps2}
\end{eqnarray}
The dependence of the phase space on $1-z$ comes entirely from the
prefactor of $s^{n-1-n\epsilon}$ according to Eq.(\ref{sxdep}).
Indeed the dependence on $z$ and $Q^{2}$ has been entirely removed
from the integration range thanks to the change of variables of
Eq.(\ref{changedis}).

Now, the amplitude whose square modulus is integrated with the
phase space Eq.(\ref{disps2}) is in general a function:
\begin{equation}
A_{n+1}=A_{n+1}(Q^{2},s,z_{2},\dots,z_{n},\Omega_{1},\dots,\Omega_{n}).
\end{equation}
The number of independent variables for a process with 2 incoming
particle (one on-shell and the other virtual) and $n+1$ outgoing
real particles is given by the number of parameters minus the
on-shell conditions and the ten parameters of the Poincare' group:
\begin{equation}
4(n+3)-(n+2)-10=3n.
\end{equation}
These $3n$ variable correspond in this case to
\begin{equation}
Q^{2},s,z_{2},\dots,z_{n},\Omega_{1},\dots,\Omega_{n},
\end{equation}
where an azimutal angle is arbitrary. In the $z\rightarrow 1$
limit, $s\rightarrow 0$ and the dominant contribution is given by
terms which are most singular as $s$ vanishes. Because of
cancellation of infrared singularities
\cite{Kinoshita:1962ur,Lee:1964is}, $|A_{n+1}|^{2}\sim
s^{-n+\emph{O}(\epsilon)}$ when $s\rightarrow 0$. Indeed, a
stronger singularity would lead to powerlike infrared divergences
and a weaker singularity would lead to suppressed terms in the
$z\rightarrow 1$ limit.

Hence only terms in the square amplitude which behave as
$s^{-n+\emph{O}(\epsilon)}$ contribute in the $z\rightarrow 1$
limit. In $d$ dimensions, these pick up an
$s^{-1-n\epsilon+\emph{O}(\epsilon)}$ prefactor from the phase
space Eq.(\ref{disps2}). Let's consider the simplest case, that is
the tree level case where we have only purely real soft emission.
In this case $\emph{O}(\epsilon)=0$ and thus we get that the
contribution to the coefficient function from the tree level
diagrams with $n$ extra radiated partons behaves as:
\begin{eqnarray}
|A_{n+1}|^{2}d\phi_{n+1}\sim
s^{-1-n\epsilon}\int_{0}^{1}dz_{n}\cdots
dz_{2}z_{n}^{(n-2)-(n-1)\epsilon}(1-z_{n})^{1-2\epsilon}\cdots
z_{2}^{-\epsilon}(1-z_{2})^{1-2\epsilon}\nonumber\\
\times \int d\Omega_{1}\cdots
d\Omega_{n}.\label{realem}\qquad\qquad\qquad\qquad\qquad\qquad\qquad\qquad\qquad\qquad
\end{eqnarray}
We note that each $z$ integration can produce at most a
$1/\epsilon$ pole from the soft region and that each angular
integration can produce at most an additional $1/\epsilon$ pole
from the collinear region (see Eq.(\ref{domega}) in Appendix
\ref{phasespacedec}). Therefore, from the contribution of tree
level diagrams with $n$ extra radiated partons there come at most
\begin{equation}
\frac{1}{\epsilon^{n-1}}\,\frac{1}{\epsilon^{n}}=\frac{1}{\epsilon^{2n-1}}
\end{equation}
poles in $\epsilon=0$. All this means that we can write the
$\emph{O}(\alpha_{s}^{n})$ to the bare coefficient function in $d$
dimensions in the following form:
\begin{equation}
C_{n}^{'(0)}(z,Q^{2},\epsilon)=(Q^{2})^{-n\epsilon}\frac{C_{nn}^{(0)}(\epsilon)}{\Gamma(-n\epsilon)}\,(1-z)^{-1-n\epsilon},
\end{equation}
where the factor $(Q^{2})^{-n\epsilon}$ is due to elementary
dimensional analysis, $C_{nn}^{(0)}(\epsilon)$ are coefficients
with poles in $\epsilon=0$ of order at most of $2n$ and the
$\Gamma$ function factor $\Gamma^{-1}(-n\epsilon)$ has been
introduced for future convenience. For the LO tree level case (see
Eq.(\ref{cdislo}) of Chapter \ref{QCD}) we have that
\begin{equation}
C_{0}^{'(0)}(z,Q^{2},\epsilon)=\delta(1-z).
\end{equation}
Hence, the tree level coefficient function $C^{(0)}_{\rm tree}$ in
the $z\rightarrow 1$ limit has the form:
\begin{eqnarray}
C^{(0)}_{\rm tree}(z,Q^{2},\alpha_{0},\epsilon)&=&\sum_{n=0}^{\infty}\alpha_{0}^{n}C_{n}^{'(0)}(z,Q^{2},\epsilon)\\
&=&\delta(1-z)+\sum_{n=1}^{\infty}\alpha_{0}^{n}(Q^{2})^{-n\epsilon}\frac{C_{nn}^{'(0)}(\epsilon)}{\Gamma(-n\epsilon)}\,(1-z)^{-1-n\epsilon}\nonumber\\
&&+\emph{O}((1-z)^{0}),\label{coefftree}
\end{eqnarray}
where $\alpha_{0}$ is the bare i.e. the unrenormalized strong
coupling constant.

We will now study how the result of Eq.(\ref{coefftree}) is
modified by the inclusion of loops. To this purpose, we notice
that a generic amplitude with loops can be viewed as a tree-level
amplitude formed with proper vertices. Contributions to the
dimensionless coefficient function with powers of $s^{\epsilon}$
can only arise from loop integrations in the proper vertices. We
thus consider only purely scalar loop integrals, since numerators
of fermion or vector propagators and vertex factors cannot induce
any dependence on $s^{\epsilon}$. Let us therefore consider an
arbitrary proper diagram $G$ in a massless scalar theory with $E$
external lines, $I$ internal lines and $V$ vertices. It can be
shown (see e.g. section 6.2.3 of \cite{C.1980} and references
therein) that, denoting with $P$ the set of $E$ external momenta
and $P_{E}$ the set of independent invariants, the corresponding
amplitude $\tilde{A}_{G}(P_{E})$ has the form
\begin{eqnarray}
\tilde{A}_{G}(P_{E})&=&K(2\pi)^{d}\delta_{d}\left(P\right)A_{G}(P_{E}),\nonumber\\
A_{G}(P_{E})&=&\frac{i^{I-L(d-1)}}{(4\pi)^{dL/2}}\Gamma(I-dL/2)\nonumber\\
&&\times
\prod_{l=1}^{I}\left[\int_{0}^{1}d\beta_{l}\right]\frac{\delta\left(1-\sum_{l=1}^{I}\beta_{l}\right)}{[P_{G}(\beta)]
^{d(L+1)/2-I}[D_{G}(\beta,P_{E})]^{I-dL/2}}.\label{amplitude}
\end{eqnarray}
Here, $\beta_{l}$ are the usual Feynman parameters, $P_{G}(\beta)$
is a homogeneous polynomial of degree $L$ in the $\beta_{l}$,
$D_{G}(\beta,P_{E})$ is a homogeneous polynomial of degree $L+1$
in the $\beta_{l}$ with coefficients which are linear functions of
the scalar products of the set $P_{E}$, i.e. with dimensions of
$(\textrm{mass})^{2}$, and $K$ collects all overall factors, such
as couplings and symmetry factors.

The amplitude $\tilde{A}_{G}(P_{E})$ Eq.(\ref{amplitude}) depends
on $s$ only through $D_{G}(\beta,P_{E})$, which, in turn is linear
in $s$. We can determine in general the dependence of
$A_{G}(P_{E})$ by considering two possible cases. The first
possibility is that $D_{G}(\beta,P_{E})$ is independent of all
invariants except $s$, i.e. $D_{G}(\beta,P_{E})=sd_{G}(\beta)$. In
such case, $A_{G}(P_{E})$ depends on $s$ as
\begin{equation}\label{loopscontr}
A_{G}(P_{E})=\left(\frac{1}{s}\right)^{I-dL/2}a_{G}=\left(\frac{1}{s}\right)^{I-2L+L\epsilon}a_{G},
\end{equation}
where $a_{G}$ is a numerical constant, obtained performing the
Feynman parameters integrals. The second possibility is that
$D_{G}(\beta,P_{E})$ depends on some of the other invariants. In
such case, $A_{G}(P_{E})$ is manifestly an analytic function of
$s$ at $s=0$, and thus it can be expanded in Taylor series around
$s=0$, with coefficients which depend on the other invariants. In
the former case, Eq.(\ref{loopscontr}) implies that the $s$
dependence induced by loops integration in the square amplitude is
given by integer powers of $s^{-\epsilon}$. In the latter case,
the $s$ dependence induced by loops integrations in the square
amplitude is given by integer positive powers of $s$.

Therefore, we conclude that each loop integration can carry at
most a factor of $s^{-\epsilon}$ and that Eq.(\ref{coefftree}),
after the inclusion of loops, becomes:
\begin{eqnarray}
C^{(0)}(z,Q^{2},\alpha_{0},\epsilon)&=&\sum_{n=0}^{\infty}\alpha_{0}^{n}C_{n}^{(0)}(z,Q^{2},\epsilon),\label{coeffx1}\\
C_{n}^{(0)}(z,Q^{2},\epsilon)&=&(Q^{2})^{-n\epsilon}\left[C_{n0}^{(0)}(\epsilon)\delta(1-z)+
\sum_{k=1}^{n}\frac{C_{nk}^{(0)}(\epsilon)}{\Gamma(-k\epsilon)}(1-z)^{-1-k\epsilon}\right]\nonumber\\
&&+\emph{O}((1-z)^{0}),\label{coeffx2}
\end{eqnarray}
where again for future convenience we have defined the
coefficients of $(1-z)^{-1-k\epsilon}$ in terms of
$\Gamma^{-1}(-k\epsilon)$ and where $\emph{O}((1-z)^{0})$ denotes
terms which are not divergent as $z\rightarrow 1$ in the
$\epsilon\rightarrow 0$ limit.

Using the identity
\begin{equation}\label{fattorigamma3}
\int_{0}^{1}dz\,z^{N-1}(1-z)^{-1-k\epsilon}=\frac{\Gamma(N)\Gamma(-k\epsilon)}{\Gamma(N-k\epsilon)}
\end{equation}
and the Stirling expansion Eq.(\ref{stirling}) of Appendix
\ref{lognlogx} we get that the Mellin transform of
Eqs.(\ref{coeffx1},\ref{coeffx2}) in the large-$N$ limit is given
by
\begin{eqnarray}
C^{(0)}(N,Q^{2},\alpha_{0},\epsilon)=\sum_{n=0}^{\infty}\alpha_{0}^{n}C_{n}^{(0)}(N,Q^{2},\epsilon),\\
C_{n}^{(0)}(N,Q^{2},\epsilon)=\sum_{k=0}^{n}C_{nk}^{(0)}(\epsilon)(Q^{2})^{-(n-k)\epsilon}\left(\frac{Q^{2}}{N}\right)^{-k\epsilon}
+\emph{O}\left(\frac{1}{N}\right)\label{ncoeff}.
\end{eqnarray}
The content of this result is that, in the large-$N$ limit, the
dependence of the regularized cross section on $N$ only goes
through integer powers of the dimensionful variable
$(Q^{2}/N)^{-\epsilon}$.

\section{Kinematics of inclusive DY in the soft limit}\label{kindy}

In the Drell-Yan case the argument follows in an analogous way
with minor modification which account for the different
kinematics. In this case the relevant parton subprocesses are:
\begin{eqnarray}
\mathcal{Q}(p)+\mathcal{Q}(p')&\rightarrow&
\gamma^{*}(Q)+\mathcal{X}\\
\mathcal{Q}(p)+\mathcal{G}(p')&\rightarrow&
\gamma^{*}(Q)+\mathcal{X}\\
\mathcal{G}(p)+\mathcal{G}(p')&\rightarrow&
\gamma^{*}(Q)+\mathcal{X}.
\end{eqnarray}
The recursive application of Eq.(\ref{gendecps}) in Appendix
\ref{phasespacedec} with $m=1$, in this case gives:
\begin{eqnarray}
&&d\phi_{n+1}(p+p';Q,k_{1},\dots,k_{n})\nonumber \\
&=&\int_{0}^{s}\frac{dM_{n}^{2}}{2\pi}d\phi_{2}(p+p';k_{n},P_{n})d\phi_{n}(P_{n};Q,k_{1},\dots,k_{n-1})\nonumber
\end{eqnarray}
\begin{eqnarray}
&=&\int_{0}^{s}\frac{dM_{n}^{2}}{2\pi}d\phi_{2}(p+p';k_{n},P_{n})\nonumber\\
&&\times\int_{0}^{M_{n}^{2}}\frac{dM_{n-1}^{2}}{2\pi}d\phi_{2}(P_{n};k_{n-1},P_{n-1})d\phi_{n-1}
(P_{n-1};Q,k_{1},\dots,k_{n-2})\nonumber\\
&=&\int_{0}^{s}\frac{dM_{n}^{2}}{2\pi}d\phi_{2}(P_{n+1};k_{n},P_{n})\int_{0}^{M_{n}^{2}}\frac{dM_{n-1}}{2\pi}d\phi_{2}(P_{n};k_{n-1},P_{n-1})
\nonumber\\
&&\times\cdots\times
\int_{0}^{M_{3}^{2}}\frac{dM_{2}^{2}}{2\pi}d\phi_{2}(P_{3};k_{2},P_{2})d\phi_{2}(P_{2};k_{1},P_{1}),\label{dyps}
\end{eqnarray}
where now we have defined $P_{n+1}\equiv p+p'$, so $M_{n+1}^{2}=s$
and $P_{1}\equiv Q$. The change of variables which separates off
the dependence on $(1-z)$, where now $z=Q^{2}/s$, is
\begin{eqnarray}
z_{i}&=&\frac{M_{i}^{2}-Q^{2}}{M_{i+1}^{2}-Q^{2}};\qquad
i=2,\dots,n\label{changedy}\\
M_{i}^{2}-Q^{2}&=&(s-Q^{2})z_{n}\cdots z_{i}\label{changedy2}.
\end{eqnarray}
Also here all $z_{i}$ range between $0$ and $1$, because
$M_{i}^{2}\leq M_{i+1}^{2}\leq Q^{2}$. From Eq.(\ref{changedy}),
we get:
\begin{equation}\label{sost4}
dM_{n}^{2}dM_{n-1}^{2}\cdots dM_{2}^{2}=\det\left(\frac{\partial
(M_{i}^{2}-Q^{2})}{\partial z_{j}}\right)dz_{n}dz_{n-1}\cdots
dz_{2},
\end{equation}
where
\begin{equation}\label{sost5}
\det\left(\frac{\partial (M_{i}^{2}-Q^{2})}{\partial
z_{j}}\right)=\frac{\partial (M_{n}^{2}-Q^{2})}{\partial
z_{n}}\cdots\frac{\partial (M_{2}^{2}-Q^{2})}{\partial z_{2}}
=(s-Q^{2})^{n-1}z_{n}^{n-2}z_{n-1}^{n-3}\cdots z_{3}.
\end{equation}
In this case in the new variables Eq.(\ref{changedy}) the two-body
phase space becomes
\begin{eqnarray}
&&d\phi_{2}(P_{i+1};k_{i},P_{i})\nonumber\\
&&\quad=N(\epsilon)(M_{i+1}^{2})^{-1+\epsilon}\left[(M_{i+1}^{2}-Q^{2})-(M_{i}^{2}-Q^{2})\right]^{1-2\epsilon}d\Omega_{i}\nonumber\\
&&\quad=N(\epsilon)(Q^{2})^{-1+\epsilon}(s-Q^{2})^{1-2\epsilon}(z_{n}\cdots
z_{i+1})^{1-2\epsilon}(1-z_{i})^{1-2\epsilon}d\Omega_{i},\label{sost6}
\end{eqnarray}
where in the last step we have replaced
$(M_{i+1}^{2})^{-1+\epsilon}$ by $(Q^{2})^{-1+\epsilon}$ in the
$z\rightarrow 1$ limit as can be seen comparing with
Eq.(\ref{changedy2}). Now, substituting
Eqs.(\ref{sost4},\ref{sost5},\ref{sost6}) into Eq.(\ref{dyps}), we
finally get
\begin{eqnarray}
d\phi_{n+1}(p+p';Q,k_{1},\dots,k_{n})=2\pi\left[\frac{N(\epsilon)}{2\pi}\right]^{n}(Q^{2})^{-n(1-\epsilon)}(s-Q^{2})^{2n-1-2n\epsilon}
d\Omega_{n}\cdots d\Omega_{1}&&\nonumber\\
\times\int_{0}^{1}dz_{n}z_{n}^{(n-2)-(n-1)\epsilon}(1-z_{n})^{1-2\epsilon}\cdots\int_{0}^{1}dz_{2}z_{2}^{-\epsilon}(1-z_{2})^{1-2\epsilon}.&&
\label{dyps2}
\end{eqnarray}
The dependence on $1-z$ is now entirely contained in the phase
space prefactor
\begin{equation}
(Q^{2})^{-n(1-\epsilon)}(s-Q^{2})^{2n-1-2n\epsilon}=\frac{z^{1-2n+2n\epsilon}}{Q^{2}(1-z)}\left[Q^{2}(1-z)^{2}\right]^{n-n\epsilon}.
\end{equation}
As before, this proves that the coefficient function for real
emission at tree level is given by
\begin{eqnarray}
C^{(0)}_{\rm tree}(z,Q^{2},\alpha_{0},\epsilon)&=&\sum_{n=0}^{\infty}\alpha_{0}^{n}C_{n}^{'(0)}(z,Q^{2},\epsilon)\\
&=&\delta(1-z)+\sum_{n=1}^{\infty}\alpha_{0}^{n}(Q^{2})^{-n\epsilon}\frac{C_{nn}^{'(0)}(\epsilon)}{\Gamma(-2n\epsilon)}\,
(1-z)^{-1-2n\epsilon}\nonumber\\
&&+\emph{O}((1-z)^{0})\label{coefftree2},
\end{eqnarray}

In this case the introduction of loops requires more care than for
the deep-inelastic-scattering case. We shall now show the main
difference between the DIS case and the DY case as far as the
introduction of loops is concerned. The two-body kinematics (see
Eq.(\ref{pezzo2}) in Appendix \ref{phasespacedec} together with
Eqs.(\ref{disps},\ref{changedis}) states that the radiated partons
in the DIS case are all soft:
\begin{eqnarray}
k_{i}^{0}&=&\frac{M_{i+1}}{2}\left(1-\frac{M_{i}^{2}}{M_{i+1}^{2}}\right)=\frac{\sqrt{s}}{2}
(z_{n}\cdots z_{i+1})^{1/2}(1-z_{i});\quad 1\leq i\leq n-1\label{softdis}\\
k_{n}^{0}&=&\frac{\sqrt{s}}{2}\left(1-\frac{M_{n}^{2}}{s}\right)=\frac{\sqrt{s}}{2}(1-z_{n});\quad
s=Q^{2}\frac{1-z}{z}.\label{softdis2}
\end{eqnarray}
This confirms the validity of the argument of Section \ref{kindis}
for the introduction of loops, because all the invariants that can
appear in the function $D_{G}(\beta,P_{E})$ in
Eq.(\ref{amplitude}) can be expresses in terms of the following
ones
\begin{eqnarray}
q^{2}=-Q^{2}\\
p^{2}=p^{'2}=k_{i}^{2}=0\\
p\cdot p'\sim p\cdot k_{i}\sim Q^{2}\\
k_{i}\cdot k_{j}\sim Q^{2}(1-z)\label{esplicita1},
\end{eqnarray}
which are either constant or proportional to $1-z$, i.e. to s.
Here we have used Eqs.(\ref{softdis},\ref{softdis2}) of this
Section, Eq.(\ref{zto1dis}) in Chapter \ref{RI} and the fact that
the definition of $z$ in the DIS case Eq.(\ref{sxdep}) implies
that
\begin{equation}
(p^{0})^{2}=\frac{Q^{2}}{4z(1-z)}.
\end{equation}
In the DY case things are quite different, because in this case
two-body kinematics (see Eq.(\ref{pezzo2}) in Appendix
\ref{phasespacedec} together with Eqs.(\ref{dyps},\ref{changedy})
gives
\begin{eqnarray}
k_{i}^{0}&=&\frac{\sqrt{s}}{2}(1-z)z_{n}\cdots
z_{i+1}(1-z_{i})+\emph{O}((1-z)^{2});\quad 1\leq i\leq n-1\\
k_{n}^{0}&=&\frac{\sqrt{s}}{2}(1-z)(1-z_{n});\quad
s=\frac{Q^{2}}{z}
\end{eqnarray}
and this implies that all the invariants that can appear in the
function $D_{G}(\beta,P_{E})$ in Eq.(\ref{amplitude}) can be
expresses in terms of the following ones
\begin{eqnarray}
q^{2}=Q^{2}\\
p\cdot p'=s/2\\
p\cdot k_{i}\sim p'\cdot k_{i}\sim s(1-z)\label{esplicita2}\\
k_{i}\cdot k_{j}\sim s(1-z)^{2}\label{esplicita3}.
\end{eqnarray}
Hence, we see that, in general, both odd and even powers of
$(1-z)^{-\epsilon}$ may arise adding the loops contribution to the
tree level coefficient function Eq.(\ref{coefftree2}). Here, in
this thesis, we will assume that odd powers of $(1-z)^{-\epsilon}$
do not arise, because it can be shown by explicit computations
that it is the case up to order $\emph{O}(\alpha_{s}^{2})$.
However, there are possible indications that at higher orders this
assumption could not be true. Anyway the investigations of these
aspects is beyond the aim of this thesis.

Thus, after the inclusions of loops and with our assumptions,
Eq.(\ref{coefftree2}) becomes
\begin{eqnarray}
C^{(0)}(z,Q^{2},\alpha_{0},\epsilon)&=&\sum_{n=0}^{\infty}\alpha_{0}^{n}C_{n}^{(0)}(z,Q^{2},\epsilon),\label{coeffx1}\\
C_{n}^{(0)}(z,Q^{2},\epsilon)&=&(Q^{2})^{-n\epsilon}\left[C_{n0}^{(0)}(\epsilon)\delta(1-z)+
\sum_{k=1}^{n}\frac{C_{nk}^{(0)}(\epsilon)}{\Gamma(-2k\epsilon)}(1-z)^{-1-2k\epsilon}\right]\nonumber\\
&&+\emph{O}((1-z)^{0}).\label{coeffxx}
\end{eqnarray}
Its Mellin transform can be written in a compact way together with
that of the DIS case Eq.(\ref{ncoeff}):
\begin{eqnarray}\label{coeffndisdy}
C^{(0)}(N,Q^{2},\alpha_{0},\epsilon)=\sum_{n=0}^{\infty}\sum_{k=0}^{n}C_{nk}^{(0)}(\epsilon)
[\alpha_{0}(Q^{2})^{-\epsilon}]^{(n-k)}\left[\alpha_{0}\left(\frac{Q^{2}}{N^{a}}\right)^{-\epsilon}\right]^{k}
+\emph{O}\left(\frac{1}{N}\right),
\end{eqnarray}
where $a=1$ for the DIS case and $a=2$ for the DY case and where
the coefficients are those that could be obtained from the
parton-level cross sections for the partonic subprocesses that
contribute to the given process.

\section{Resummation from renormalization group
improvement}\label{DISDY3}

In this section, we want to impose the restrictions that
renormalization group invariance imposes on the cross section. Our
only assumption is that the coefficient function can be
multiplicatively renormalized. This means that all divergences can
be removed from the bare coefficient function
$C^{(0)}(N,Q^{2},\alpha_{0},\epsilon)$ Eq.(\ref{coeffndisdy}) by
defining a renormalized running coupling $\alpha_{s}(\mu^{2})$
according to the implicit equation
\begin{equation}\label{alfaren}
\alpha_{0}(\mu^{2},\alpha_{s}(\mu^{2}),\epsilon)=\mu^{2\epsilon}\alpha_{s}(\mu^{2})Z^{(\alpha_{s})}(\alpha_{s}(\mu^{2}),\epsilon)
\end{equation}
and a renormalized coefficient function
\begin{equation}
C\left(N,\frac{Q^{2}}{\mu^{2}},\alpha_{s}(\mu^{2},\epsilon)\right)=Z^{(
C)}(N,\alpha_{s}(\mu^{2}),\epsilon)C^{(0)}(N,Q^{2},\alpha_{0},\epsilon),
\end{equation}
where $\mu$ is the renormalization scale (here chosen equal to the
factorization one) and
$Z^{(\alpha_{s})}(\alpha_{s}(\mu^{2}),\epsilon)$ and $Z^{\rm
C}(N,\alpha_{s}(\mu^{2}),\epsilon)$ are computable in perturbation
theory and have multiple poles in $\epsilon=0$. The renormalized
coefficient function
$C(N,Q^{2}/\mu^{2},\alpha_{s}(\mu^{2},\epsilon))$ is finite in
$\epsilon=0$ and it can only depend on $Q^{2}$ through
$Q^{2}/\mu^{2}$, because $\alpha_{s}(\mu^{2})$ is dimensionless.

The physical anomalous dimension is given by
\begin{eqnarray}
\gamma(N,\alpha_{s}(Q^{2}),\epsilon)&=&Q^{2}\frac{\partial}{\partial
Q^{2}}\ln
C\left(N,\frac{Q^{2}}{\mu^{2}},\alpha_{s}(\mu^{2}),\epsilon\right)=
Q^{2}\frac{\partial}{\partial Q^{2}}\ln
C^{(0)}\left(N,Q^{2},\alpha_{0},\epsilon\right)\nonumber\\
&=&-\epsilon
(\alpha_{0}Q^{-2\epsilon})\frac{\partial}{\partial(\alpha_{0}Q^{-2\epsilon})}
\ln
C^{(0)}\left(N,Q^{2},\alpha_{0},\epsilon\right),\label{defdimanom}
\end{eqnarray}
where we have exploited the fact that $C^{(0)}$
Eq.(\ref{coeffndisdy}) depends on $Q^{2}$ through the combination
$\alpha_{0}Q^{-2\epsilon}$. This implies that the physical
anomalous dimension $\gamma$ has the following perturbative
expression:
\begin{equation}\label{gammabare}
\gamma(N,\alpha_{s}(Q^{2}),\epsilon)=
\sum_{i=0}^{\infty}\sum_{j=0}^{n}\gamma_{ij}(\epsilon)
[\alpha_{0}(Q^{2})^{-\epsilon}]^{(i-j)}\left[\alpha_{0}\left(\frac{Q^{2}}{N^{a}}\right)^{-\epsilon}\right]^{j}
+\emph{O}\left(\frac{1}{N}\right).
\end{equation}

The renormalized expression of the physical anomalous dimension is
found expressing in this equation the bare coupling constant in
terms of the renormalized one by means of Eq.(\ref{alfaren}). Now,
the functions
\begin{eqnarray}
(Q^{2})^{-\epsilon}\alpha_{0}&=&\left(\frac{Q^{2}}{\mu^{2}}\right)^{-\epsilon}\alpha_{s}(\mu^{2})
Z^{(\alpha_{s})}(\alpha_{s}(\mu^{2}),\epsilon)\\
(Q^{2}/N^{a})^{-\epsilon}\alpha_{0}&=&\left(\frac{Q^{2}/N^{a}}{\mu^{2}}\right)^{-\epsilon}\alpha_{s}(\mu^{2})
Z^{(\alpha_{s})}(\alpha_{s}(\mu^{2}),\epsilon)
\end{eqnarray}
are manifestly renormalization group invariant, i.e.
$\mu^{2}$-independent. Thus, it follows that
\begin{eqnarray}
(Q^{2})^{-\epsilon}\alpha_{0}&=&\alpha_{s}(Q^{2})
Z^{(\alpha_{s})}(\alpha_{s}(Q^{2}),\epsilon)\label{paura}\\
(Q^{2}/N^{a})^{-\epsilon}\alpha_{0}&=&\alpha_{s}(Q^{2}/N^{a})
Z^{(\alpha_{s})}(\alpha_{s}(Q^{2}/N^{a}),\epsilon).\label{paura1}
\end{eqnarray}
The renormalized physical anomalous dimension is then found by
substituting Eqs.(\ref{paura},\ref{paura1}) into
Eq.(\ref{gammabare}) and re-expanding $Z^{(\alpha_{s})}$ in powers
of the renormalized coupling. We obtain:
\begin{equation}\label{paura2}
\gamma(N,\alpha_{s}(Q^{2}),\epsilon)=\sum_{m=1}^{\infty}\sum_{n=0}^{m}\gamma^{\rm
R}_{mn}(\epsilon)\alpha_{s}^{m-n}(Q^{2})\alpha_{s}^{n}(Q^{2}/N^{a})+\emph{O}\left(\frac{1}{N}\right).
\end{equation}

At this point, we cannot yet conclude that the four-dimensional
physical anomalous dimension admits an expression of the form of
Eq.(\ref{paura2}), because the coefficients $\gamma^{\rm
R}_{mn}(\epsilon)$ are not necessarily finite as
$\epsilon\rightarrow 0$. In order to understand this, it is
convenient to separate off the $N$-independent terms in the
renormalized physical anomalous dimension, i.e. the terms with
$n=0$ in the internal sum in Eq.(\ref{paura2}). Namely, we write
\begin{equation}\label{paura3}
\gamma(N,\alpha_{s}(Q^{2}),\epsilon)=\hat{\gamma}^{(l)}(\alpha_{s}(Q^{2}),\alpha_{s}(Q^{2}/N^{a}),\epsilon)+
\hat{\gamma}^{(c)}(\alpha_{s}(Q^{2}),\epsilon)+\emph{O}\left(\frac{1}{N}\right),
\end{equation}
where we have defined
\begin{eqnarray}
\hat{\gamma}^{(l)}(\alpha_{s}(Q^{2}),\alpha_{s}(Q^{2}/N^{a}),\epsilon)&=&\sum_{m=0}^{\infty}\sum_{n=1}^{\infty}\gamma^{\rm
R}_{m+nn}(\epsilon)\alpha_{s}^{m}(Q^{2})\alpha_{s}^{n}(Q^{2}/N^{a})\label{prima}\\
\hat{\gamma}^{(c)}(\alpha_{s}(Q^{2}),\epsilon)&=&\sum_{m=1}^{\infty}\gamma^{\rm
R}_{m0}(\epsilon)\alpha_{s}^{m}(Q^{2}).
\end{eqnarray}
Whereas $\gamma(N,\alpha_{s}(Q^{2}),\epsilon)$ is finite in the
limit $\epsilon\rightarrow 0$, where it coincides with the
four-dimensional physical anomalous dimension,
$\hat{\gamma}^{(l)}$ and $\hat{\gamma}^{(c)}$ are not necessarily
finite as $\epsilon\rightarrow 0$. However, Eq.(\ref{paura3})
implies that $\hat{\gamma}^{(l)}$ and $\hat{\gamma}^{(c)}$ can be
made finite by adding and subtracting the counterterm
\begin{equation}
Z^{(\gamma)}(\alpha_{s}(Q^{2}),\epsilon)=\hat{\gamma}^{(l)}(\alpha_{s}(Q^{2}),\alpha_{s}(Q^{2}),\epsilon).
\end{equation}
In this way the physical anomalous dimension Eq.(\ref{paura3})
becomes
\begin{equation}\label{paura4}
\gamma(N,\alpha_{s}(Q^{2}),\epsilon)=\gamma^{(l)}(\alpha_{s}(Q^{2}),\alpha_{s}(Q^{2}/N^{a}),\epsilon)+
\gamma^{(c)}(\alpha_{s}(Q^{2}),\epsilon)+\emph{O}\left(\frac{1}{N}\right),
\end{equation}
where
\begin{eqnarray}
\gamma^{(l)}(\alpha_{s}(Q^{2}),\alpha_{s}(Q^{2}/N^{a}),\epsilon)&=&
\hat{\gamma}^{(l)}(\alpha_{s}(Q^{2}),\alpha_{s}(Q^{2}/N^{a}),\epsilon)+\nonumber\\
&&-\hat{\gamma}^{(l)}(\alpha_{s}(Q^{2}),\alpha_{s}(Q^{2}),\epsilon),\\
\gamma^{(c)}(\alpha_{s}(Q^{2}),\epsilon)&=&\hat{\gamma}^{(c)}(\alpha_{s}(Q^{2}),\epsilon)+
\hat{\gamma}^{(l)}(\alpha_{s}(Q^{2}),\alpha_{s}(Q^{2}),\epsilon).\label{456}
\end{eqnarray}
Now, $\gamma^{(c)}$ is clearly finite in $\epsilon=0$, because at
$N=1$ $\gamma^{(l)}$ vanishes and it is $N$-independent. This also
implies that $\gamma^{(l)}$ is finite for all $N$, because
$\gamma$ shoul be finite for all $N$. Therefore, $\gamma^{(l)}$
provides an expression of the resummed physical anomalous
dimension in the large $N$ limit, up to non-logarithmic terms:
\begin{equation}
\gamma(N,\alpha_{s}(Q^{2}),\epsilon)=\gamma^{(l)}(\alpha_{s}(Q^{2}),\alpha_{s}(Q^{2}/N^{a}),\epsilon)+\emph{O}(N^{0}).
\end{equation}
It is apparent from Eq.(\ref{456}) that $\gamma^{(c)}$ is a power
series in $\alpha_{s}(Q^{2})$ with finite coefficients in the
$\epsilon\rightarrow 0$ limit. In order to understand the
perturbative structure of $\gamma^{(l)}$ as well, define
implicitly the function
$g(\alpha_{s}(Q^{2}),\alpha_{s}(Q^{2}/n),\epsilon)$ as
\begin{equation}
\gamma^{(l)}(\alpha_{s}(Q^{2}),\alpha_{s}(Q^{2}/N^{a}),\epsilon)=\int_{1}^{N^{a}}
\frac{dn}{n}\,g(\alpha_{s}(Q^{2}),\alpha_{s}(Q^{2}/n),\epsilon),
\end{equation}
where
\begin{eqnarray}
g(\alpha_{s}(Q^{2}),\alpha_{s}(\mu^{2}),\epsilon)&\equiv&
-\mu^{2}\frac{\partial}{\partial\mu^{2}}\hat{\gamma}^{(l)}(\alpha_{s}(Q^{2}),\alpha_{s}(\mu^{2}),\epsilon)\\
&=&-\beta^{(d)}(\alpha_{s}(\mu^{2}),\epsilon)\frac{\partial}{\partial\alpha_{s}(\mu^{2})}
\hat{\gamma}^{(l)}(\alpha_{s}(Q^{2}),\alpha_{s}(\mu^{2}),\epsilon),\label{ultima}
\end{eqnarray}
with $\beta^{(d)}(\alpha_{s})$ is the $d$-dimensional beta
function
\begin{equation}
\beta^{(d)}(\alpha_{s}(\mu^{2}),\epsilon)-\epsilon\alpha_{s}(\mu^{2})+\beta(\alpha_{s}(\mu^{2}))
\end{equation}
and where we have performed the change of variable
$n=Q^{2}/\mu^{2}$. It immediately follows from
Eqs.(\ref{prima}-\ref{ultima}) that $g$ is a power series in
$\alpha_{s}(Q^{2})$ and $\alpha_{s}(\mu^{2})$ with finite
coefficients in the limit $\epsilon \rightarrow 0$:
\begin{eqnarray}
\lim_{\epsilon\rightarrow
0}g(\alpha_{s}(Q^{2}),\alpha_{s}(\mu^{2}),\epsilon)&\equiv&
g(\alpha_{s}(Q^{2}),\alpha_{s}(\mu^{2}))\nonumber\\
&=&\sum_{m=0}^{\infty}\sum_{n=1}^{\infty}g_{mn}\alpha_{s}^{m}(Q^{2})\alpha_{s}^{n}(\mu^{2}).
\end{eqnarray}
 Hence, our final result for the four-dimensional all-rder resummed
 physical anomalous dimension is given by
 \begin{equation}\label{disdyres}
 \gamma^{\rm
 res}(N,\alpha_{s}(Q^{2}))=\int_{1}^{N^{a}}\frac{dn}{n}\,g(\alpha_{s}(Q^{2}),\alpha_{s}(Q^{2}/n))+\emph{O}(N^{0}).
 \end{equation}

This result can be compared to the all-order resummation formula
derived in Ref.\cite{Contopanagos:1996nh}. This resummation has
the form of Eq.(\ref{disdyres}), but with $g$ a function of
$\alpha_{s}(\mu^{2})$ only, i.e. with all $g_{mn}=0$ when $m>0$
and our result is thus less predictive in the sense that it
requires a higher fixed-order computation of the physical
anomalous dimension in order to extract the resummation
coefficients $g_{mn}$. The predictive power of the resummation
formulae is analyzed in detail in Chapter \ref{predictivity}.
 According to the more
restrictive result of Ref.\cite{Contopanagos:1996nh},
Eq.(\ref{disdyres}) becomes
\begin{equation}\label{disdyresfac}
 \gamma^{\rm
 res}(N,\alpha_{s}(Q^{2}))=\int_{1}^{N^{a}}\frac{dn}{n}\,g(\alpha_{s}(Q^{2}/n))+\emph{O}(N^{0}),
\end{equation}
where
\begin{equation}
g(\alpha_{s}(\mu^{2}))=\sum_{n=1}^{\infty}g_{0n}\alpha_{s}^{n}(\mu^{2}).
\end{equation}

The conditions under which the more restrictive result of
Ref.\cite{Contopanagos:1996nh} holds can be understood by
comparing to our approach the derivation of that result. The
approach of Ref.\cite{Contopanagos:1996nh} is based on assuming
the validity of the factorization formula Eq.(\ref{checazzoe}) of
Sec.\ref{vivamaddy2} which is more restrictive than the standard
collinear factorization. This factorization was proven for a wide
class of processes in Ref.\cite{Collins:1989gx}, and implies that
the perturbative coefficient function Eq.(\ref{coeffndisdy}) in
the large $N$ limit can be factored as:
\begin{equation}\label{fuc}
C^{(0)}(N,Q^{2},\alpha_{0},\epsilon)=C^{(0,l)}(Q^{2}/N^{a},\alpha_{0},\epsilon)C^{(0,c)}(Q^{2},\alpha_{0},\epsilon).
\end{equation}
We notice that this can happen if and only if the coefficients
$C^{(0)}_{nk}(\epsilon)$ in Eq.(\ref{coeffndisdy}) can be written
in the form
\begin{equation}
C_{nk}^{(0)}(\epsilon)=F_{k}(\epsilon)G_{n-k}(\epsilon).
\end{equation}
The validity of factorization Eq.(\ref{checazzoe}) of
Sec.\ref{vivamaddy2} to all orders and for various processes is
based on assumptions whose reliability will not be discussed here.
Anyway, Eq.(\ref{fuc}) implies that the physical anomalous
dimension Eq.(\ref{defdimanom}) becomes
\begin{equation}
\gamma(N,\alpha_{s}(Q^{2}),\epsilon)=\gamma^{(l)}(\alpha_{s}(Q^{2}/N^{a}),\epsilon)
+\gamma^{(c)}(\alpha_{s}(Q^{2}),\epsilon).
\end{equation}
Then, proceeding as before, one then ends up with the resummation
formula Eq.(\ref{disdyresfac}).

\section{NLL resummation}\label{nllresummation}

In this Section, we shall give explicit expressions of the
reummation formulae at NLL for the deep-inelastic structure
function $F_{2}$ and for the Drell-Yan cross section. These
explicit expressions are useful for practical computations and we
shall use them in Chapter \ref{DYDPR}.

The expression of the resummed physical anomalous dimension in
Eq.(\ref{disdyres}) can be used to compute the resummed evolution
factor $K^{res}(N;Q_{0}^{2},Q^{2})$ Eq.(\ref{pertic}) in Section
\ref{introrga}. At NLL, we get
\begin{equation}\label{resevfact}
K_{\rm NLL}(N;Q_{0}^{2},Q^{2})=\exp\left[E_{\rm
NLL}(N;Q_{0}^{2},Q^{2})\right]=\exp
\int_{Q_{0}^{2}}^{Q^{2}}\frac{dk^{2}}{k^{2}}\gamma^{\rm res}_{\rm
NLL}(N,\alpha_{s}(k^{2})),
\end{equation}
where
\begin{equation}\label{nll}
\gamma^{\rm res}_{\rm
NLL}(N,\alpha_{s}(k^{2}))=\int_{1}^{N^{a}}\frac{dn}{n}\,\left[g_{01}\alpha_{s}(k^{2}/n)
+g_{02}\alpha_{s}^{2}(k^{2}/n)\right];\qquad g_{11}=0.
\end{equation}
The fact that the resummation coefficient $g_{11}=0$ for both the
deep-inelastic and the Drell-Yan case can be shown by explicit
computations of the fixed-order anomalous dimension (see
Ref.\cite{Forte:2002ni} in Section 4.3). This shows that at NLL
level Eq.(\ref{disdyresfac}) holds. However this does not mean
that this is the case at all logarithmic orders.

Many times in literature, the resummed results are given in terms
of the Mellin transform of a resummed physical anomalous dimension
in $z$ space. To rewrite Eq.(\ref{nll}) as the Mellin transform of
a function of $x$, we can use the all-orders relations between the
logs of $N$ and the logs of $1-z$ that are given in Appendix
\ref{lognlogx}. In particular, we can use
Eqs.(\ref{fiorentina},\ref{fiorentina2}) in Appendix
\ref{lognlogx} to rewrite Eq.(\ref{nll}) at NLL in the following
form:
\begin{equation}
\gamma^{\rm res}_{\rm
NLL}(N,\alpha_{s}(k^{2}))=a\int_{0}^{1}dx\,\frac{x^{N-1}-1}{1-x}\left[\hat{g}_{01}
\alpha_{s}(k^{2}(1-x)^{a})+\hat{g}_{02}
\alpha_{s}^{2}(k^{2}(1-x)^{a})\right],
\end{equation}
where
\begin{equation}
\hat{g}_{01}=-g_{01},\qquad\quad
\hat{g}_{02}=-(g_{02}+a\gamma_{E}b_{0}g_{01}).
\end{equation}
and where we have used the definition of the beta function
Eq.(\ref{betafunc}) in Section \ref{appA}. As a consequence, the
NLL resummed exponent in Eq.(\ref{resevfact}) can be rewritten in
the following form
\begin{equation}\label{stanco}
E_{\rm
NLL}(N;Q_{0}^{2},Q^{2})=a\int_{0}^{1}dx\,\frac{x^{N-1}-1}{1-x}\int_{Q_{0}^{2}(1-x)^{a}}^{Q^{2}(1-x)^{a}}
\frac{dk^{2}}{k^{2}}\hat{g}(\alpha_{s}(k^{2})),
\end{equation}
where
\begin{equation}
\hat{g}(\alpha_{s})=\hat{g}_{01}\alpha_{s}(k^{2})+\hat{g}_{02}\alpha_{s}^{2}(k^{2})
\end{equation}
Beyond leading order the standard anomalous dimension differs from
the physical one, so $\hat{g}_{02}$ receives a contribution both
from the standard anomalous dimension and from the coefficient
function. It is thus natural to rewrite the resummation formula
Eq.(\ref{stanco}) separating off the contribution the contribution
which originates from the anomalous dimension $\gamma^{\rm AP}$
Eq.(\ref{generic}) of Section \ref{introrga}. This is done
defining two functions of $\alpha_{s}$, $A(\alpha_{s})$ and
$B^{a}(\alpha_{s})$ in such a way that
\begin{equation}
\hat{g}(\alpha_{s})=A(\alpha_{s})+\frac{\partial
B^{(a)}(\alpha_{s}(k^{2}))}{\partial\ln k^{2}},\quad
A(\alpha_{s})=A_{1}\alpha_{s}+A_{2}\alpha_{s}^{2},\quad
B^{(a)}(\alpha_{s})=B_{1}^{(a)}\alpha_{s}.
\end{equation}
It is clear that the constant $A_{i}$ are obtained directly form
the coefficients of the $1/[1-x]_{+}$ terms of the $i$-loop
quark-quark splitting functions and that the coefficients
$B_{i}^{a}$ depends on the particular process ($a=1$ for DIS and
$a=2$ for DY).  To the NLL order, we find that
\begin{equation}
\hat{g}_{01}=A_{1},\qquad\quad \hat{g}_{02}=A_{2}-b_{0}B_{1}^{(a)}
\end{equation}
and that
\begin{eqnarray}
E_{\rm
NLL}(N;Q_{0}^{2},Q^{2})=a\int_{0}^{1}dx\,\frac{x^{N-1}-1}{1-x}\bigg[\int_{Q_{0}^{2}(1-x)^{a}}^{Q^{2}(1-x)^{a}}\frac{dk^{2}}{k^{2}}
A(\alpha_{s}(k^{2}))\nonumber\\
+B^{(a)}(\alpha_{s}(Q^{2}(1-x)^{a}))-B^{(a)}(\alpha_{s}(Q_{0}^{2}(1-x)^{a}))\bigg].\label{214}
\end{eqnarray}
We can then rewrite the resummed cross section
\begin{equation}
\sigma_{\rm NLL}(N,Q^{2})=\exp\left[E_{\rm
NLL}(N;Q_{0}^{2},Q^{2})\right]\sigma_{\rm NLL}(N,Q_{0}^{2})
\end{equation}
in a factorized form according to Eqs.(\ref{maddi},\ref{maddi2})
in Section \ref{introrga} by collecting all $Q^{2}$-dependent
contributions to the resummation Eq.(\ref{214}) into a resummed
perturbative coefficient function $C_{\rm NLL}$:
\begin{equation}
\sigma_{\rm NLL}(N,Q^{2})=C_{\rm
NLL}(N,Q^{2}/\mu^{2},\alpha_{s}(\mu^{2}))F(N,\mu^{2}),
\end{equation}
where
\begin{eqnarray}
C_{\rm
NLL}(N,Q^{2}/\mu^{2},\alpha_{s}(\mu^{2}))&=&\exp\bigg\{a\int_{0}^{1}dx\,
\frac{x^{N-1}-1}{1-x}\bigg[\int_{\mu^{2}}^{Q^{2}(1-x)^{a}}\frac{dk^{2}}{k^{2}}A(\alpha_{s}(k^{2}))\nonumber\\
&&+B^{(a)}(\alpha_{s}(Q^{2}(1-x)^{a}))\bigg]\bigg\}\label{vogliofinire},
\end{eqnarray}
which ha the same form of the resummed results discussed in
Section \ref{eiconalapp}. The precise definition of the parton
distribution $F$ and the factorization scale $\mu^{2}$ will depend
on the choice of factorization scheme: according to the choice of
scheme, the resummed terms will be either part of the hard
coefficient function $C_{\rm NLL}$, or of the evolution of the
parton distribution $F$. In the $\overline{MS}$ scheme the NLL
coefficients $A_{1}$, $A_{2}$ and $B_{1}^{(a)}$ are given in
Eq.(\ref{bah}) of Section \ref{eiconalapp} and the NNLL ones are
given for example in Ref.\cite{Vogt:2000ci}. These coefficients
can also be obtained comparing a fixed order computation of the
physical anomalous dimension with a fixed order expansion of
Eq.(\ref{disdyres}) as is shown explicitly in Section 4.3 of
Ref.\cite{Forte:2002ni}.

To compute explicitly Eq.(\ref{vogliofinire}), we first exploit
Eq.(\ref{calcespl}) in Appendix \ref{lognlogx} at NLL level, thus
finding
\begin{equation}
C_{\rm
NLL}(N,Q^{2}/\mu^{2},\alpha_{s}(\mu^{2}))=\exp\left\{-\int_{1}^{N^{a}}
\frac{dn}{n}\left[\int_{n\mu^{2}}^{Q^{2}}\frac{dk^{2}}{k^{2}}\,A(\alpha_{s}(k^{2}/n))
+\tilde{B}^{(a)}(\alpha_{s}(Q^{2}/n))\right]\right\},\label{vogliofinire2}
\end{equation}
where
$\tilde{B}^{(a)}(\alpha_{s})=B^{(a)}(\alpha_{s})-a\gamma_{E}A_{1}\alpha_{s}$.
The explicit expression of Eq.(\ref{vogliofinire}) is then
obtained performing the changes of variables,
 \begin{equation}\label{cov}
\frac{dk^{2}}{k^{2}}=\frac{d\alpha_{s}(k^{2}/n)}{\beta(\alpha_{s}(k^{2}/n))},\quad\frac{dn}{n}=-\frac{d\alpha_{s}(Q^{2}/n)}{\beta(\alpha_{s}(Q^{2}/n))},
\end{equation}
to evaluate the integrals in Eq.(\ref{vogliofinire2}) and using
the two loop solution of the renormalitazion-group equation for
the running of $\alpha_{s}$ given in Eq.(\ref{tls}) of Section
\ref{appA}, Now, after some algebra we find for the integral in
Eq.(\ref{vogliofinire2}):
\begin{eqnarray}
-\int_{1}^{N^{a}}\frac{dn}{n}
\left[\int_{n\mu^{2}}^{Q^{2}}\frac{dk^{2}}{k^{2}}\left(A_{1}\alpha_{s}(\frac{k^{2}}{n})+A_{2}\alpha_{s}^{2}(\frac{k^{2}}{n})\right)+
\tilde{B}_{1}^{(a)}\alpha_{s}(\frac{Q^{2}}{n})\right]\nonumber\\
=\log N g_{1}(\lambda,a)+g_{2},(\lambda,a)\qquad\qquad\qquad\qquad
\end{eqnarray}
where $\lambda=b_{0}\alpha_{s}(\mu^{2}_{r})\log N$ and
\begin{eqnarray}
g_{1}(\lambda,a)&=&\frac{A_{1}}{b_{0}\lambda}[a\lambda+(1-a\lambda)\log(1-a\lambda)]\label{g1}\\
g_{2}(\lambda,a)&=&-\frac{A_{1}a\gamma_{E}-B_{1}^{(a)}}{b_{0}}\log(1-a\lambda)+\frac{A_{1}b_{1}}{b_{0}^{3}}
[a\lambda+\log(1-a\lambda)+\frac{1}{2}\log^{2}(1-a\lambda)]\nonumber\\
&&-\frac{A_{2}}{b_{0}^{2}}[a\lambda+\log(1-a\lambda)]+\log\left(\frac{Q^{2}}{\mu^{2}_{r}}\right)\frac{A_{1}}{b_{0}}\log(1-a\lambda)\nonumber\\
&&+\log\left(\frac{\mu^{2}}{\mu^{2}_{r}}\right)
\frac{A_{1}}{b_{0}}a\lambda\label{g2},
\end{eqnarray}
where $a=1$ for the DIS case and $a=2$ for the DY case. Evidently,
in Eqs.(\ref{g1},\ref{g2}) there is a dependence on the
renormalization scale.  To obtain the desired result, we simply
have to keep the renormalization scale equal to the factorization
scale $\mu^{2}_{r}=\mu^{2}$. Thus, for the explicit analytic
expression of Eq.(\ref{vogliofinire2}), we get
\begin{equation}\label{explform2}
C_{\rm
NLL}\left(N,\frac{Q^{2}}{\mu^{2}},\alpha_{s}(\mu^{2})\right)=\left[\exp\{\log
N
g_{1}(\lambda,a)+g_{2}(\lambda,a)\}\right]_{\mu^{2}_{r}=\mu^{2}},
\end{equation}
with the resummation coefficients given in Eq.(\ref{bah}) in
Section \ref{eiconalapp} for the $\overline{MS}$ factorization
scheme choice. However, a general analysis of the factorization
scheme choices and changes for the resummation formulae is given
for example in Section 6 of Ref.\cite{Forte:2002ni}.

\chapter{Renormalization group resummation of prompt photon
production}\label{DP}

In this chapter, we prove the all-order exponentiation of soft
logarithmic corrections to prompt photon production in hadronic
collisions, by generalizing the renormalization group approach of
chapter \ref{DISDY2}. Here, we will show that all large logs in
the soft limit can be expressed in terms of two dimensionful
variables. Then, we use the renormalization group to resum them.
The resummation formulae that we obtain are more general though
less predictive than those that can be obtained with other
approaches discussed in chapter \ref{RI}.

\section{Kinematics and notation}\label{kinp}

We consider the process
\begin{equation} \label{HHg}
H_1(P_1)+H_2(P_2)\rightarrow \gamma(p_\gamma)+X,
\end{equation}
of two colliding hadrons $H_{1}$ and $H_{2}$ with momentum $P_{1}$ and
$P_{2}$ respectively into a real photon with momentum $p_{\gamma}$ and any
collection of hadrons $X$.
More specifically, we are interested in
the differential cross section
$p_\perp^3\frac{d\sigma}{dp_\perp}(x_{\perp},p_{\perp}^{2})$,
where $p_\perp$ is the transverse momentum of the photon with
respect to the direction of the colliding hadrons $H_1$ and $H_2$,
and
\begin{equation} \label{x1}
x_{\perp}=\frac{ 4p^2_\perp}{S};\qquad
S=(P_1+P_2)^2.
\end{equation}
The scaling variable $x$ can be viewed as the squared fraction of
transverse energy that the hadrons transfer to the outgoing particles
(hence $0\leq x_{\perp}\leq 1$) and $S$ is the hadronic center-of-mass
energy.
We parametrize the momentum of the photon in terms of its partonic
center-of-mass pseudorapidity and its transverse momentum
$\vec{p}_{\perp}$. The pseudorapidiry of a massless particle is
defined in terms its scattering angle $\theta$ in the
center-of-mass frame as follows
\begin{equation}
\eta=-\ln(\tan(\theta/2)).
\end{equation}
So, in the partonic center-of-mass frame, we can write:
\begin{equation}\label{parfot}
p_{\gamma}=(p_{\perp}\cosh\hat{\eta}_{\gamma},\vec{p}_{\perp},p_{\perp}\sinh\hat{\eta}_{\gamma}).
\end{equation}
In the same frame, the incoming partons' momenta can be written as
\begin{equation}\label{parin2}
p_{1}=x_{1}P_{1}=\frac{\sqrt{s}}{2}(1,\vec{0}_{\perp},1),\qquad
p_{2}=x_{2}P_{2}=\frac{\sqrt{s}}{2}(1,\vec{0}_{\perp},-1),
\end{equation}
where $x_{1(2)}$ are the longitudinal fraction of momentum of the
parton $1(2)$ in the hadron $H_{1(2)}$ and
$s=(p_{1}+p_{2})^{2}=x_{1}x_{2}S$ is the center-of-mass energy of
the partonic process. The relation between the hadronic
center-of-mass pseudorapidity and the partonic one is obtained
performing a boost along the collision axis:
\begin{equation}\label{etaetahat}
\hat{\eta}_{\gamma}=\eta_{\gamma}-\frac{1}{2}\ln\frac{x_{1}}{x_{2}}.
\end{equation}

The factorized expression for this cross section in perturbative
QCD is
\begin{eqnarray} \label{crosssec}
p_\perp^3\frac{d\sigma}{dp_\perp}(x_{\perp},p_\perp^2)
&=&\sum_{a,b}\int_0^1dx_1\,dx_2\,dz\,x_1F_{a}^{H_1}(x_1,\mu^2)\,
x_2F_{b}^{H_2}(x_2,\mu^2)\nonumber\\
&&\times C_{ab}\left(z,\frac{Q^2}{\mu^2},\as(\mu^2)\right)\delta
\left(x_{\perp}-z x_1x_2\right),
\end{eqnarray}
where $F_a^{H_1}(x_1,\mu^2)$, $F_b^{H_2}(x_2,\mu^2)$ are the
distribution functions of partons $a,b$ in the colliding hadrons.
Here we have defined the perturbative scale $Q^{2}$ and the partonic
scaling variable $z$ as the squared fraction of transverse energy that
the parton $a,b$ transfer to the outgoing partons ($ 0\leq z\leq 1$):
\begin{eqnarray} \label{qdef}
Q^2&=&4p_\perp^2,   \\
z&=&\frac{Q^2}{s}=\frac{Q^{2}}{x_{1}x_{2}S}.
\label{xdef}
\end{eqnarray}
 The coefficient function\phantom{a}
$C_{ab}(z,\frac{Q^2}{\mu^2},\as(\mu^2))$ is defined in terms of
the partonic cross section for the process where partons $a$, $b$
are incoming as \beq
C_{ab}\left(z,\frac{Q^2}{\mu^2},\as(\mu^2)\right)
=p_\perp^3\frac{d\hat\sigma_{ab}}{dp_\perp}. \eeq

\section{Leading order calculation}\label{loc}

\begin{figure}
\begin{center}
\includegraphics[scale=0.28]{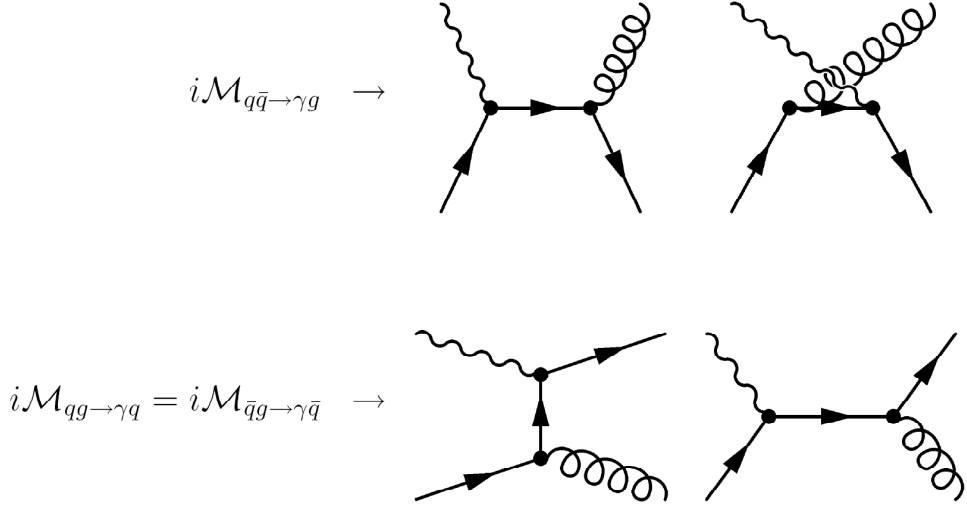}
\caption{Feynman graphs that contribute to the amplitude of the
partonic process at the leading order}
\label{diagrammi}\end{center}\end{figure}

The hard-scattering subprocesses that contribute to the amplitude of
the prompt-photon prodution at the leading order are:
\begin{eqnarray}
q(p_{1})+\bar{q}(p_{2})&\rightarrow& g(p') +\gamma (p_{\gamma}) \\
q(p_{1})+g(p_{2}) &\rightarrow& q(p') + \gamma (p_{\gamma}) \\
\bar{q}(p_{1})+g(p_{2}) &\rightarrow & \bar{q}(p')+\gamma
(p_{\gamma})
\end{eqnarray}
The corresponding Feynman graphs are shown in figure
\ref{diagrammi}.

We now want to obtain the coefficient function for these two
elementary subprocesses at the leading order. Using the QCD
Feynman rules to evaluate the first amplitude in figure \ref{diagrammi}, we have:
\begin{equation}\label{amp3}
i\emph{M}_{q\bar{q}\rightarrow\gamma g}= -i\bar{v}(p_{2})Q_{q}eg
\biggr[\frac{\gamma^{\rho}\sc{p}_{\gamma}\gamma^{\mu}-2\gamma^{\rho}p^{\mu}_{1}}{2p_{1}\cdot
p_{\gamma}}+\frac{\gamma^{\mu}\sc{p'}\gamma^{\rho}-2\gamma^{\mu}p_{1}^{\rho}}{2p_{1}\cdot
p'}\biggl]\epsilon^{*}_{\mu}(p_{\gamma})\epsilon^{*}_{\rho}(p')t^{a}_{ji}u(p_{1}),
\end{equation}
where $e=|e|$ is the electrical charge, $g$ is the strong charge
and $Q_{q}$ is the electrical charge of the quark $q$ in units of
$e$. Taking the square modulus of equation (\ref{amp3}) and
averaging over the two quarks polarizations and colours we obtain:
\begin{equation}\label{mod1}
\frac{1}{4}\frac{1}{N_{C}^{2}}\sum_{\text{pol,col}}|\mathcal{M}_{q\bar{q}\rightarrow\gamma
g }|^{2}=2\frac{C_{F}}{N_{C}}Q_{q}^{2}e^{2}g^{2}\biggl[
\frac{p_{1}\cdot p_{\gamma}}{p_{1}\cdot p'}+\frac{p_{1}\cdot
p'}{p_{1}\cdot p_{\gamma}}\biggr],
\end{equation}
where $N_{C}$ is the number of quark colours and $C_{F}\equiv
(N_{C}^{2}-1)/2N_{C}$ is the Casimir operator with respect to the
colour matrix $t^{a}_{ji}$.

Proceeding in the same way for the Feynman graphs contributing to the second amplitude
in figure \ref{diagrammi} we
could obtain the averaged square modulus for this other
subprocess, but, we note that it can be immediately obtained using
crossing simmetry. More precisely, we only have to substitute $p'$
with $p_{2}$ and take into account the fact that we must now
average not over two quarks' colours but over the colours of a
quark and a gluon. Therefore we arrive at the following expression
for the averaged square modulus of the second amplitude of
figure \ref{diagrammi}:

\begin{eqnarray}\label{mod2}
\frac{1}{4}\frac{1}{2N_{C}C_{F}}\frac{1}{N_{C}}\sum_{\text{pol,col}}|\mathcal{M}_{q(\bar{q})g\rightarrow\gamma
q(\bar{q})}|^{2}&=&\frac{1}{4}\frac{1}{2N_{C}C_{F}}\frac{1}{N_{C}}\sum_{\text{pol,col}}|\mathcal{M}_{q\bar{q}\rightarrow\gamma
g}|^{2}\bigg|_{p'\rightarrow p_{2}}=\nonumber \\
&=&\frac{1}{N_{C}}Q_{q}^{2}e^{2}g^{2}\biggl[\frac{p_{1}\cdot
p_{\gamma}}{p_{1}\cdot p_{2}}+\frac{p_{1}\cdot p_{2}}{p_{1}\cdot
p_{\gamma}}\biggr].
\end{eqnarray}
We will now rewrite equations (\ref{mod1}) and (\ref{mod2}) in
terms of the kinematic parameters defined in section \ref{kinp}.
For this purpose we first note that momentum conservation and  the
parametrizations defined in Eqs.(\ref{parfot},\ref{parin2}) imply (in the centre of mass of the incident
partons):
\begin{eqnarray}
p'&=&(p_{\perp}\cosh \label{p'}
\hat{\eta}_{\gamma},-\vec{p}_{\perp},-p_{\perp}\sinh
\hat{\eta}_{\gamma})\\
s&=&(p_{1}+p_{2})^{2}=(p_{\gamma}+p')^{2}=
4p_{\perp}^{2}\cosh^{2}\hat{\eta}_{\gamma}\\
z&=&\frac{4p_{\perp}^{2}}{s}=\frac{1}{\cosh^{2}\hat{\eta}_{\gamma}}.\label{x}
\end{eqnarray}
From the last equation we see that in the limit $z\rightarrow 1$,
$\hat{\eta}_{\gamma}\rightarrow 0$. Physically this is because in
this limit all the centre of mass energy is transverse and so the
photon cannot have a non zero pseudorapidity. Now, using again
Eqs.(\ref{parfot},\ref{parin2}) and the equation for $p'$
(\ref{p'}) we have:
\begin{eqnarray}
p_{1}\cdot p_{2}&=&\frac{s}{2}\nonumber\\
p_{1}\cdot
p'&=&\frac{\sqrt{s}}{2}p_{\perp}e^{\hat{\eta}_{\gamma}} \label{scal}\\
p_{1}\cdot
p_{\gamma}&=&\frac{\sqrt{s}}{2}p_{\perp}e^{-\hat{\eta}_{\gamma}}
\nonumber\end{eqnarray} A combination of equations (\ref{x}) and
(\ref{scal}) yields our final result for the averaged square
modulus of the amplitudes:
\begin{eqnarray}
\frac{1}{4}\frac{1}{N_{C}^{2}}\sum_{\text{pol,col}}|\mathcal{M}_{q\bar{q}\rightarrow\gamma
g
}|^{2}&=&4\frac{C_{F}}{N_{C}}Q_{q}^{2}e^{2}g^{2}\frac{(2-z)}{z}\label{mod3}\\
\frac{1}{4}\frac{1}{2N_{C}C_{F}}\frac{1}{N_{C}}\sum_{\text{pol,col}}|\mathcal{M}_{q(\bar{q})g\rightarrow\gamma
q(\bar{q})}|^{2}&=&\frac{1}{2N_{C}}Q_{q}^{2}e^{2}g^{2}
 \biggl[1\pm\sqrt{1-z}\nonumber\\
 &&+\frac{4}{1\pm\sqrt{1-z}}\biggr],
\label{mod4}\end{eqnarray} where the plus sign has to be chosen
for positive values of the pseudorapidity $\hat{\eta}_{\gamma}$
and the minus sign for negative values. The two-body phase space
is:
\begin{eqnarray}
d\phi
(p_{1}+p_{2};p_{\gamma},p')&=&\frac{d^{3}p_{\gamma}}{(2\pi)^{3}2E_{p_{\gamma}}}\frac{d^{3}p'}{(2\pi)^{3}2E_{p'}}(2\pi)^{4}\delta_{(4)}
(p'+p_{\gamma}-p_{1}-p_{2})\nonumber\\
&=&\frac{d^{3}p_{\gamma}}{4E_{p_{\gamma}}E_{p'}}\frac{1}{(2\pi)^{2}}\delta_{(1)}(E_{p'}+E_{p_{\gamma}}-E_{p_{1}}-E_{p_{2}}).
\end{eqnarray}
Imposing the conservation of spatial momentum
($E_{p_{\gamma}}=E_{p'}=|\vec{p_{\gamma}}|$) the two-body phase
space becomes:
\begin{eqnarray}\label{fase1}
d\phi
(p_{1}+p_{2};p_{\gamma},p')&=&\frac{d^{3}p_{\gamma}}{4|\vec{p_{\gamma}}|^{2}(2\pi)^{2}}\delta_{(1)}(2|\vec{p_{\gamma}}|-\sqrt{\hat{s}}) \nonumber\\
&=&\frac{1}{16\pi}d\cos\theta
d|\vec{p_{\gamma}}|\delta_{(1)}(|\vec{p_{\gamma}}|-\sqrt{\hat{s}}/2)\nonumber
\\
&=&\frac{1}{16\pi}\delta_{(1)}(p_{\perp}\cosh\hat{\eta}_{\gamma}-\sqrt{\hat{s}}/2)d\cos\theta
d|\vec{p_{\gamma}}|,
\end{eqnarray}
where $\theta$ is the scattering angle of the photon with respect
to the collision axis. Because of the fact that we want to
integrate over the pseudorapidity of the photon $\eta_{\gamma}$ at
fixed $p_{\perp}$, we must perform a change of variables. In
particular we must rewrite the two-body phace space, expressed in
terms of $\cos\theta$ and $|\vec{p_{\gamma}}|$, in terms of the
new variables $\cosh\hat{\eta}_{\gamma}$ and $p_{\perp}$. This
change of variables is given by the equations:
\begin{displaymath}\left\{
\begin{array}{ll}
\cos\theta =& \tanh\hat{\eta}_{\gamma} \\
|\vec{p_{\gamma}}|=& p_{\perp}\cosh\hat{\eta}_{\gamma}.
\end{array}
\right.
\end{displaymath}
The determinant of the Jacobian matrix is easily obtained and is
given by:
\begin{equation}\label{determinant}
|J|=\left\|\frac{\partial(\cos\theta,|\vec{p_{\gamma}}|)}{\partial
(\cosh\hat{\eta}_{\gamma},p_{\perp})
}\right\|=\frac{1}{\cosh\hat{\eta}_{\gamma}\sqrt{\cosh^{2}\hat{\eta}_{\gamma}-1}}.
\end{equation}
Thanks to equation (\ref{x}), we obtain this determinant in term of
$z$:
\begin{equation}
|J|=\frac{z}{\sqrt{1-z}}
\end{equation}
Using this last result, equation (\ref{fase1}) becomes:
\begin{eqnarray}\label{fase3}
d\phi
(p_{1}+p_{2};p_{\gamma},p')&=&\frac{1}{16\pi}\frac{z}{\sqrt{1-z}}\delta_{(1)}(\cosh\hat{\eta}_{\gamma}-\sqrt{s}/2p_{\perp})
\frac{dp_{\perp}}{p_{\perp}}d\cosh\hat{\eta}_{\gamma}\nonumber\\
&=&\frac{1}{16\pi}\frac{z}{\sqrt{1-z}}[\delta_{(1)}(\hat{\eta}_{\gamma}-\hat{\eta}_{+})+\delta_{(1)}(\hat{\eta}_{\gamma}-\hat{\eta}_{-})]\nonumber\\
&&\times
\frac{dp_{\perp}}{p_{\perp}}d\hat{\eta}_{\gamma},\label{ps6}
\end{eqnarray}
where $\hat{\eta}_{+}$ and $\hat{\eta}_{-}$ are the two solutions
(one positive and one negative) of the equation imposed by the
delta function $p_{\perp}\cosh\hat{\eta}_{\gamma}=\sqrt{s}/2$
which are:
\begin{equation}
\hat{\eta}_{\pm}=\ln\left(\frac{\sqrt{s}}{2p_{\perp}}\pm\sqrt{\frac{s}{4p_{\perp}^{2}}-1}\right).
\end{equation}
The flux factor $\Phi$ is immediately obtained from equations
(\ref{parin2}) and (\ref{x}):
\begin{equation}\label{flusso}
\Phi\equiv 4(p_{1}\cdot p_{2})=2s=\frac{8p_{\perp}^{2}}{z}.
\end{equation}
Remembering the definition of the QED and QCD coupling constants,
\begin{equation*}
\alpha\equiv\frac{e^{2}}{4\pi}\qquad ;\qquad
\alpha_{s}\equiv\frac{g^{2}}{4\pi},
\end{equation*}
and putting together expressions (\ref{mod3}), (\ref{mod4}),
(\ref{fase3}), (\ref{flusso}) and performing the integration over
$\hat{\eta}_{\gamma}$, we obtain our final result for the
coefficient function at the leading order for the two subprocesses
in figure \ref{diagrammi}:

\begin{eqnarray}
C^{(LO)}_{q\bar{q}\rightarrow\gamma
g}(z,\alpha_{s})&=&\alpha\alpha_{s}Q_{q}^{2}\pi\frac{C_{F}}{N_{C}}\frac{z}{\sqrt{1-z}}(2-z)\\
C^{(LO)}_{q(\bar{q})g\rightarrow\gamma
q(\bar{q})}(z,\alpha_{s})&=&\alpha\alpha_{s}Q_{q}^{2}\pi\frac{1}{2N_{C}}\frac{z}{\sqrt{1-z}}\left(1+\frac{z}{4}\right).
\end{eqnarray}

\section{The soft limit}\label{kindpp}

We will study the cross section Eq.~(\ref{crosssec}) in the
threshold transverse limit, when the transverse transverse energy of
outgoing particles is
close to its maximal value ($x_{\perp}\rightarrow 1$ at the
hadronic level or, equivalently, $z\rightarrow 1$ at the partonic level).
The convolution in
Eq.~(\ref{crosssec}) is turned into an ordinary product by Mellin
transformation:
\begin{eqnarray}
\sigma(N,Q^2)
&=&\sum_{a,b}\sigma_{ab}(N,Q^{2})\\
&=&\sum_{a,b}F_{a}^{H_1}(N+1,\mu^2)\,F_{b}^{H_2}(N+1,\mu^2)\nonumber\\
&&\times C_{ab}\left(N,\frac{Q^{2}}{\mu^{2}},\as(\mu^2)\right).\label{cdef}
\end{eqnarray}

As discussed in Section \ref{introrga}, in the large $N$ limit
each parton subprocess can be treated independently, specifically,
each $C_{ab}$ is separately renormalization-group invariant.

At this point, it is interesting to discuss the differences in the large
$N$ behavior of the partonic subprocesses. The cross sections for the partonic channels
with two quarks of different flavors
($ab=$ $q\bar{q}'$, $\bar{q}q'$, $qq$, $qq'$, $\bar{q}\bar{q}$, $\bar{q}\bar{q}'$)
vanish at LO and are hence suppressed by a factor of $\alpha_{s}$ with respect to the
subprocesses with $ab=$ $q\bar{q}$, $qg$, $\bar{q}g$. Moreover, in the large
$N$ limit this relative suppression is further enhanced by a factor of $\emph{O}(1/N)$
because the processes with two different quark flavors involve the off-diagonal Alatarelli-Parisi
splitting functions. Therefore, these partonic channels
do not contribute in the large $N$ limit. The partonic channel $ab=gg$ has a different large $N$
behavior. It begins to contribute at NLO via the partonic process $g+g\rightarrow \gamma+q+\bar{q}$,
which again leads to a suppression effect of $\emph{O}(1/N)$ with respect to the LO suprocesses. However,
owing to the photon-gluon coupling through a fermion box, the partonic subprocess
$g+g\rightarrow \gamma+g$ is also permitted. This subprocess is logarithmically-enhanced
by multiple soft-gluon radiation in the final state, but it starts to contribute
only at NNLO in perturbation theory. It follows that the partonic channel
$ab=$ $gg$ is suppressed by a factor $\alpha_{s}^{2}$ with respect to the LO partonic channels
$ab=$ $q\bar{q}$, $qg$, $\bar{q}g$ and it will enter the resummed cross secion only at NNLL
order.
In conclusion, the partonic channels
that should be resummed are $ab=$ $q\bar{q}$, $qg$, $\bar{q}g$, $gg$,
where the last channel is that coupled to the gluon via a fermion box and eneters
resummation only at NNLL.

Furthermore, on
top of Eqs.~(\ref{crosssec},~\ref{cdef}) the physical process
Eq.~(\ref{HHg}) receives another factorized contribution, in which
the final-state photon is produced by fragmentation of a primary
parton produced in the partonic sub-process. However, the cross
section for this process is also suppressed by a factor of
$\frac{1}{N}$ in the large $N$ limit. This is due to the fact that
the fragmentation function carries this suppression, for the same
reason why the anomalous dimensions $\gamma_{qg}$ and
$\gamma_{gq}$ are suppressed. Therefore, we will disregard the
fragmentation contribution.

According to Eqs.(\ref{pertic}-\ref{edefic2}) of Section \ref{introrga}, the cross section
can be written in terms of the physical anomalous dimensions:
\begin{eqnarray} \label{pert2}
\sigma(N,Q^2)&=&\sum_{a,b}K_{ab}(N;Q_0^2,Q^2)\,\sigma_{ab}(N,Q_0^2)\\
&&=\sum_{a,b}\exp\left[E_{ab}(N;Q_0^2,Q^2)\right]\sigma_{ab}(N,Q_0^2),
\end{eqnarray}
where
\begin{eqnarray}
\label{esplit} E_{ab}(N;Q_0^2,Q^2)
&=&\int_{Q_0^2}^{Q^2}\frac{dk^2}{k^2}\gamma_{ab}(N,\as(k^2))\label{edef}\\
&=&\int_{Q_0^2}^{Q^2} \frac{dk^2}{k^2}[\gamma^{\rm AP}_{aa}(N,\as(k^2))+\gamma^{\rm AP}_{bb}(N,\as(k^2))]\nonumber\\
&&+\ln C_{ab}(N,1,\as(Q^2))-\ln C_{ab}(N,1,\as(Q_0^2)).
\end{eqnarray}

In the large-$x_{\perp}$ limit, the order-$n$ coefficient of the
perturbative expansion of the hadronic cross section is dominated by terms
proportional to
$\left[\frac{\ln^k(1-x_{\perp})}{1-x_{\perp}}\right]_+$, with $k\leq 2n-1$, that
must be resummed to all orders. Upon Mellin transformation, these
lead to contributions proportional to powers of $\ln\frac{1}{N}$.
In the sequel, we will consider the resummation of these
contributions to all logarithmic orders, and disregard all
contributions to the cross section which are suppressed by powers
of $(1-x_{\perp})$, i.e., upon Mellin transformation, by powers of
$\frac{1}{N}$.

The resummation is performed in two steps as in chapter
\ref{DISDY2}. First, we show that the origin of the large logs
is essentially kinematical: we identify the configurations which
contribute in the soft limit, we show by explicit computation that
large Sudakov logs are produced by the phase-space for real
emission with the required kinematics as logs of two dimensionful
variables, and we show that this conclusion is unaffected by
virtual corrections. Second, we resum the logs of these variables
using the renormalization group.

The $l$-th order correction to the leading $O(\alpha_s)$ partonic
process receives contribution from the emission of up to $l+1$
massless partons with momenta $k_1,\ldots,k_{l+1}$. Four-momentum
conservation implies:
\begin{equation}
\label{cons} p_1+p_2=p_\gamma+k_1+\ldots k_{l+1}.
\end{equation}
In the
partonic center-of-mass frame, according to Eqs.(\ref{parfot},\ref{parin2}),
 we have
\begin{equation} \label{cons2}
(p_1+p_2-p_\gamma)^2=\frac{Q^2}{z}(1-\sqrt{z}\cosh\hat{\eta}_\gamma)
=\sum_{i,j=1}^{l+1} k_i^0k_j^0(1-\cos\theta_{ij})\geq 0,
\end{equation}
where $\theta_{ij}$ is the angle between $\vec{k}_i$ and
$\vec{k}_j$. Hence,
\begin{equation}
 1\leq \cosh\hat{\eta}_\gamma\leq\frac{1}{\sqrt
z}. \label{etagammalim}
\end{equation}
 Therefore,
 \begin{equation}
\label{soft}\sum_{i,j=1}^{l+1} k_i^0k_j^0(1-\cos\theta_{ij})
=\frac{Q^2}{2}(1-z)+O\left[(1-z)^2\right].
\end{equation}

Equation~(\ref{soft}) implies that in the soft limit the sum of
scalar products of momenta $k_i$ of emitted partons
Eq.~(\ref{cons2}) must vanish. However, contrary to the case of
deep-inelastic scattering or Drell-Yan, not all momenta $k_i$ of
the emitted partons can be soft as $z\rightarrow 1$, because the
three-momentum of the photon must be balanced. Assume thus that
momenta $k_i,i=1,\ldots,n; n<l+1$ are soft in the $z\rightarrow 1$ limit,
while momenta $k_i,i>n$ are non-soft. For the sake of simplicity,
we relabel non-soft momenta as
\begin{equation} k'_j =k_{n+j};\qquad 1\leq
j\leq m+1;\qquad m=l-n.
\end{equation}
The generic kinematic configuration
in the $z=1$ limit is then
\begin{eqnarray} &&k_i=0\qquad 1\leq i\leq n
\nonumber\\
&&\theta_{ij}=0;\qquad \sum_{j=1}^{m+1} k^{'0}_j=p_\perp\qquad
\qquad 1\leq i,j\leq m+1.
\end{eqnarray}
for all $n$ between $1$ and $l$,
namely, the configuration where at least one momentum is not soft,
and the remaining momenta are either collinear to it, or soft.

With this labelling of the momenta, the phase space can be written,
using twice the phase space decomposition of Eq.(\ref{gendecps}) in Appendix \ref{phasespacedec},
as
\begin{eqnarray}
&&d\phi_{n+m+2}(p_1+p_2;p_\gamma,k_1,\ldots,k_n,k'_1,\ldots,k'_{m+1})\nonumber
\\
&=&\int_0^s \frac{dq^2}{2\pi}d\phi_{n+1}(p_1+p_2;q,k_1,\ldots,k_n)
\nonumber\\
&&\times \int_0^{q^2}\frac{d{k'}^2}{2\pi}\,
d\phi_{m+1}(k';k'_1,\dots,k'_{m+1})\,d\phi_2(q;p_\gamma,k').
\label{genid}
\end{eqnarray}
We shall now compute the phase space in the $z\rightarrow 1$
limit in $d=4-2\epsilon$ dimensions.
Consider first the two-body phase space $d\phi_2$ in
Eq.~(\ref{genid}). In the rest frame of $q$ we have
\begin{eqnarray}
d\phi_2(q;p_\gamma,k') &=&\frac{d^{d-1}k'}{(2\pi)^{d-1}2k^{'0}}\,
\frac{d^{d-1}p_\gamma}{(2\pi)^{d-1}2p_\gamma^0}\,
(2\pi)^d\delta^{(d)}(q-k'-p_\gamma)
\nonumber\\
&=&\frac{(4\pi)^\epsilon}{8\pi\Gamma(1-\epsilon)}
\frac{P^{1-2\epsilon}}{\sqrt{q^2}}\,\sin^{-2\epsilon}\theta_\gamma\,
d\abs{\vec p_\gamma}\,d\cos\theta_\gamma\,
\delta(|\vec{p}_\gamma|-P),\nonumber\\
\label{psfd}
\end{eqnarray}
where
\begin{equation}
P=\frac{\sqrt{q^2}}{2}\left(1-\frac{{k'}^2}{q^2}\right).
\label{pgammacm}
\end{equation}
Because momenta $k_i$, $i\leq n$ are soft,
up to terms suppressed by powers of $1-z$, the rest frame of $q$
is the same as the center-of-mass frame of the incoming partons,
in which
\begin{eqnarray}
&&|\vec{p}_\gamma|=p_\perp\,\cosh\hat{\eta}_\gamma
\\
&&\cos\theta_\gamma=\tanh\hat{\eta}_\gamma.
\end{eqnarray}
Hence,
\begin{equation}
d\phi_2(q;p_\gamma,k')
=\frac{(4\pi)^\epsilon}{8\pi\Gamma(1-\epsilon)}
\frac{(Q^2/4)^{-\epsilon}}{\sqrt{q^2}} \,dp_\perp\,d\hat{\eta}_\gamma\,
\delta\left(\cosh\hat{\eta}_\gamma-\frac{2P}{\sqrt{Q^2}}\right).
\label{ps3}
\end{equation}
The conditions
\begin{equation}
\cosh\hat{\eta}_\gamma=\frac{2P}{\sqrt{Q^2}}\geq 1; \qquad {k'}^2\geq 0,
\label{M2softlim}
\end{equation}
together with Eq.(\ref{pgammacm}), restrict the integration range to
\begin{eqnarray}
Q^2\leq &q^2&\leq s\label{regint3}\\
0\leq &{k'}^2&\leq q^2-\sqrt{Q^2q^2}\label{regint4}.
\end{eqnarray}
It is
now convenient to define new variables $u,v$
\begin{eqnarray}
q^2=Q^2+u(s-Q^2)&=&Q^2\left[1+u(1-z)\right]+O((1-z)^2)
\label{Q2soft}
\\
{k'}^2=v(q^2-\sqrt{Q^2q^2})&=&Q^2 \frac{1}{2}uv(1-z)+O((1-z)^2)
\label{M2soft}
\\
0\leq u\leq 1&;&0\leq v\leq 1,
\end{eqnarray}
in terms of which
\begin{equation}
P=\frac{\sqrt{Q^2}}{2}\left[1+\frac{1}{2}u(1-v)(1-z)\right]+O\left[(1-z)^2\right].
\end{equation}
Thus, the two-body phase space Eq.~(\ref{ps3})  up to
subleading terms is given by
\begin{equation}
d\phi_2(q;p_\gamma,k')
=\frac{(4\pi)^\epsilon}{8\pi\Gamma(1-\epsilon)}
\frac{(Q^2/4)^{-\epsilon}}{\sqrt{Q^2}} \,dp_\perp\,d\hat{\eta}_\gamma\,
\frac{\delta(\hat{\eta}_{\gamma}-\hat{\eta}_{+})+\delta(\hat{\eta}_{\gamma}-\hat{\eta}_{-})}
{\sqrt{u(1-v)(1-z)}}, \label{ps4}
\end{equation}
where
\begin{equation}\label{etapm}
\hat{\eta}_{\pm}=\ln\left(\frac{2P}{\sqrt{Q^2}}\pm
\sqrt{\frac{4P^2}{Q^2}-1}\right)=\pm\sqrt{u(1-v)(1-z)}.
\end{equation}

We now note  that the phase-space element
$d\phi_{n+1}(p_1+p_2;q,k_1,\ldots,k_n)$ contains in the final
state a system with large invariant mass $q^2\geq Q^2$, plus a
collection of $n$ soft partons: this same configuration is
encountered in the case of Drell-Yan pair production in the limit
$z_{DY}=q^2/s\to 1$, discussed in Section \ref{kindy} . Likewise, the
phase space for the set of collinear partons
$d\phi_{m+1}(k';k'_1,\ldots,k'_{m+1})$ is the same as the phase
space for deep-inelastic scattering (discussed in Section \ref{kindis}), where the invariant mass of
the initial state ${k'}^2$ vanishes as $1-z$ (see
Eq.~(\ref{M2soft})). We may therefore use the results obtained in
chapter \ref{DISDY2}, where, in the case of deep-inelastic scattering,
one of the outgoing parton momenta ($k'_{m+1}$,
say) was identified with the momentum of the leading-order outgoing
quark $p'$. Hence Eq.~(\ref{discase}) is obtained from the
corresponding result in chapter \ref{DISDY2} for deep-inelastic scattering
by the replacement $p'\to
k'_{m+1}$:
\begin{eqnarray}
d\phi_{n+1}(p_1+p_2;q,k_1,\dots,k_n)&=&
2\pi\left[\frac{N(\epsilon)}{2\pi}\right]^{n}(q^2)^{-n(1-\epsilon)}
(s-q^2)^{2n-1-2n\epsilon}\nonumber\\
&&\times d\Omega^{(n)}(\epsilon) \label{dycase}
\\
d\phi_{m+1}(k';k'_1,\ldots,k'_{m+1})&=&
2\pi\left[\frac{N(\epsilon)}{2\pi}\right]^m({k'}^2)^{m-1-m\epsilon}
d\Omega^{'(m)}(\epsilon), \label{discase}
\end{eqnarray}
where
$N(\epsilon)=1/(2(4\pi)^{2-2\epsilon})$ and
\begin{eqnarray}
d\Omega^{(n)}(\epsilon)&=&d\Omega_1\dots
d\Omega_n
\int_0^1dz_{n}z_{n}^{(n-2)+(n-1)(1-2\epsilon)}
(1-z_{n})^{1-2\epsilon}\dots
\nonumber\\
&&\times\int_0^1dz_2z_2^{1-2\epsilon}(1-z_2)^{1-2\epsilon}
\\
d\Omega^{'(m)}(\epsilon)&=&d\Omega'_1\dots
d\Omega'_m
\int_0^1dz'_mz_m^{'(m-2)-(m-1)\epsilon}
(1-z'_m)^{1-2\epsilon}\dots
\nonumber\\
&&\times\int_0^1dz'_2z_2^{'-\epsilon}(1-z'_2)^{1-2\epsilon}.
\end{eqnarray}
Here, the variables $z_i,z'_i$ are defined as in chapter \ref{DISDY2}
for the Drell-Yan and deep-inelastic scattering respectively.

Equations~(\ref{Q2soft},\ref{M2soft}) imply that the phase space
depends on $(1-z)^{-\epsilon}$ through the two variables
\begin{eqnarray}
&&{k'}^2\propto Q^2 (1-z)\\
&&\frac{(s-q^2)^2}{q^2}\propto Q^2 (1-z)^2,
\end{eqnarray}
where the
coefficients of proportionality are dimensionless and
$z$-independent. By explicitly combining the two-body phase space
Eq.~(\ref{ps4}) and the phase spaces for soft radiation
Eq.~(\ref{dycase}) and for collinear radiation Eq.~(\ref{discase})
we get
\begin{eqnarray}
&&d\phi_{n+m+2}(p_1+p_2;p_\gamma,k_1,\ldots,k_n,k'_1,\ldots,k'_{m+1})\nonumber\\
&&=(Q^2)^{n+m-(n+m+1)\epsilon} \frac{dp_\perp}{p_\perp}\,
\frac{(1-z)^{2n+m-(2n+m)\epsilon}}{\sqrt{1-z}}
\nonumber\\
&&\times 2^{-m+m\epsilon}\,
\frac{(16\pi)^{-1+\epsilon}}{\Gamma(1-\epsilon)}
\left[\frac{N(\epsilon)}{2\pi}\right]^{n+m}\,d\hat{\eta}_{\gamma}\,
d\Omega^{(n)}(\epsilon)\,d\Omega^{'(m)}(\epsilon)
\nonumber\\
&&\times \int_0^1du\,
\frac{u^{m-m\epsilon}(1-u)^{2n-1-2n\epsilon}}{\sqrt{u}}
\int_0^1dv\, \frac{v^{m-1-m\epsilon}}{\sqrt{1-v}}\,
[\delta(\hat{\eta}_{\gamma}-\hat{\eta}_{+})+\delta(\hat{\eta}_{\gamma}-\hat{\eta}_{-})].\nonumber\\
\label{dphi}
\end{eqnarray}
In the limiting cases $n=0$ and $m=0$ we have
\begin{eqnarray}
d\phi_1(p_1+p_2;q)&=&2\pi\delta(s-q^2)=\frac{2\pi}{Q^2(1-z)}\,\delta(1-u)
\label{dycase0}
\\
d\phi_1(k';p')&=&2\pi\delta({k'}^2)=\frac{4\pi}{Q^2u(1-z)}\,\delta(v);
\label{discase0}
\end{eqnarray}
the corresponding expressions for the phase
space are therefore obtained by simply replacing
\begin{equation}
(1-u)^{-1}\,d\Omega^{(n)}(\epsilon)\to \delta(1-u); \qquad
v^{-1}\,d\Omega^{'(m)}(\epsilon)\to \delta(v)
\end{equation}
in
Eq.~(\ref{dphi}) for $n=0$, $m=0$ respectively.

The logarithmic dependence of the four-dimensional cross section
on $1-z$ is due to interference between powers of
$(1-z)^{-\epsilon}$ and $\frac{1}{\epsilon}$ poles in the
$d$-dimensional cross section. Hence, we must classify the
dependence of the cross section on powers of $(1-z)^{-\epsilon}$.
We have established that in the phase space each real emission
produces a factor of $\left[Q^2(1-z)^2\right]^{1-\epsilon}$ if the
emission is soft and a factor of
$\left[Q^2(1-z)\right]^{1-\epsilon}$  if the emission is
collinear. The squared amplitude can only depend on
$(1-z)^{-\epsilon}$ because of loop integrations. This dependence
for a generic proper Feynman diagram $G$ will in general
appear, as discussed in chapter \ref{DISDY2}, through the coefficient
(see Eq.(\ref{amplitude}) of Section \ref{kindis})
\begin{equation}\label{loops}
\left[D_G(P_{E})\right]^{dL/2-I},
\end{equation}
where $L$ and $I$ are
respectively the number of loops and internal lines in $G$, and
$D_G(P_E)$ is a linear combination of all scalar products $P_E$ of
external momenta. In the soft limit all scalar
products which vanish as $z\to 1$ are either proportional to
$Q^2(1-z)$ or to $Q^2(1-z)^2$ as shown in Eqs.(\ref{esplicita1},\ref{esplicita2},\ref{esplicita3})
of Section \ref{kindy} . Equation~(\ref{loops}) then implies
that each loop integration can carry at most a factor of
$\left[Q^2(1-z)^2\right]^{-\epsilon}$ or
$\left[Q^2(1-z)\right]^{-\epsilon}$.

This then proves that the perturbative expansion of the bare
coefficient function, for each sub-process which involves partons $a,b$,
takes the form
\begin{eqnarray}\label{cofun}
C^{(0)}(z,Q^2,\az,\epsilon)&=& \alpha\az(Q^2)^{-\epsilon}
\sum_{l=0}^{\infty}\az^l\,C_l^{(0)}(z,Q^2,\epsilon)
\\
C_l^{(0)}(z,Q^2,\az,\epsilon)&=&
\frac{(Q^2)^{-l\epsilon}}{\Gamma(1/2)\sqrt{1-z}} \sum_{k=0}^l
\sum_{k'=0}^{l-k} C_{lkk'}^{(0)}(\epsilon)
(1-z)^{-2k\epsilon-k'\epsilon}\label{gencoefffunc},
\end{eqnarray}
where the factor $1/\Gamma (1/2)$ was introduced for
later convenience and terms $C_{lkk'}^{(0)}$ with $k+k'<l$ at order
$\as^l$ are present in general because of loops.
The coefficients $C_{lkk'}^{(0)}$ have poles in $\epsilon=0$ up to
order $2l$. To understand this, we have to count the independent
variables for the
prompt photon process. We have $2$ incoming particles
and $l+2$ outgoing partons (a leading-order parton, the photon and
$l$ extra emissions). Therefore, imposing the on-shell conditions and
the constraints due to Poincarè invariance, we get
\begin{equation}\label{varind}
4(l+4)-(l+4)-10=3l+2,
\end{equation}
independent variables. Now, we need to understand which are
these independent variables: the general expression of the phase
space in the threshold limit Eq.(\ref{dphi}) is written in terms
of $3l+3$ variables which are
\begin{equation}
s,4p_{\perp}^{2},u,v,z_{2},\dots,z_{n},z'_{2},\dots,z_{m}',\Omega_{1},\dots,\Omega_{n},\Omega_{1}',\dots,\Omega_{m}',\hat{\eta}_{\gamma},
\end{equation}
where $n+m=l$ and each solid angle depends on two parameters.
Clearly, one of them must be a function of some of the others
because of Eq. {\ref{varind}}. In fact, from Eq.(\ref{etapm}),
we know that $\hat{\eta}_{\gamma}$ depends on $u,v,4p_{\perp}^{2},s$.
Thus, the $3l+2$ independent variables on which depends the square modulus
amplitude can be chosen as
\begin{equation}
s,4p_{\perp}^{2},u,v,z_{2},\dots,z_{n},z'_{2},\dots,z_{m}',\Omega_{1},\dots,\Omega_{n},\Omega_{1}',\dots,\Omega_{m}'.
\end{equation}
Now, each of the $l$ integrations over a solid angle can produce a pole
$1/\epsilon$ from the collinear region. Furthermore, each of the $l$ integrations
over a dimensionless variable $u,v,z_{2},\dots,z_{n},z'_{2},\dots,z_{m}'$
can produce a pole $1/\epsilon$ from the soft region. This explains
why the coefficients $C_{lkk'}^{(0)}$ can have poles in $\epsilon=0$ up to
order $2l$.

\section{Resummation from renormalization group improvement}\label{DPresform}

The Mellin transform of Eq.~(\ref{cofun}) can be performed using
\begin{equation}
\int_0^1dzz^{N-1}(1-z)^{-1/2-2k\epsilon-k'\epsilon}
=\frac{\Gamma(1/2)}{\sqrt{N}}N^{2k\epsilon}N^{k'\epsilon}+O\left(\frac{1}{N}\right),
\end{equation}
with the result
\begin{eqnarray}
&&C^{(0)}(N,Q^2,\az,\epsilon)\nonumber\\
&&=\frac{\alpha\az(Q^2)^{-\epsilon}}{\sqrt{N}}\sum_{l=0}^\infty
\sum_{k=0}^l \sum_{k'=0}^{l-k} C_{lkk'}^{(0)}(\epsilon)
\left[\left(\frac{Q^2}{N^2}\right)^{-\epsilon}\az\right]^k
\left[\left(\frac{Q^2}{N}\right)^{-\epsilon}\az\right]^{k'}\nonumber\\
&&\times\left[(Q^2)^{-\epsilon}\az\right]^{l-k-k'}+O\left(\frac{1}{N}\right)\label{C0}.
\end{eqnarray}

Equation~(\ref{C0}) shows that indeed as $N\to\infty$, up to
$\frac{1}{N}$ corrections, the  coefficient function depends on
$N$ through the two dimensionful variables $\frac{Q^2}{N^2}$ and
$\frac{Q^2}{N}$.   The argument henceforth follows in the same way as
in chapter \ref{DISDY2}, in this more general situation. The
argument is based on the observation that, because of collinear
factorization, the physical anomalous dimension
\begin{equation}
\gamma(N,\as(Q^2))= Q^2\frac{\partial}{\partial Q^2 }\ln
C(N,Q^2/\mu^2,\as(\mu^2)) \label{paddef}
\end{equation}
is renormalization-group invariant and finite when expressed in terms
of the renormalized coupling $\as(\mu^2)$, related to $\alpha_0$
by
\begin{eqnarray}
\az(\mu^2,\as(\mu^2))=\mu^{2\epsilon}\as(\mu^2)\,
Z^{(\as)}(\as(\mu^2),\epsilon), \label{renas}
\end{eqnarray}
where
$Z^{(\as)}(\as(\mu^2),\epsilon)$ is a power series in
$\as(\mu^2)$. Because $\alpha_0$ is manifestly independent of
$\mu^2$, Eq.~(\ref{renas}) implies that the dimensionless
combination $(Q^2)^{-\epsilon}\az(\as(\mu^2),\mu^2)$ can depend on
$Q^2$ only through $\alpha_s(Q^2)$:
\begin{equation}
(Q^2)^{-\epsilon}\az(\as(\mu^2),\mu^2)
=\as(Q^2)\,Z^{(\as)}(\as(Q^2),\epsilon). \label{rencou}
\end{equation}

Using Eq.~(\ref{rencou}) in Eq.~(\ref{C0}), the coefficient
function and consequently the physical anomalous dimension are
seen to be given by a power series in $\as(Q^2)$, $\as(Q^2/N)$ and
$\as(Q^2/N^2)$:
\begin{equation}
\gamma(N,\as(Q^2),\epsilon)
=\sum_{m=0}^\infty\sum_{n=0}^\infty\sum_{p=0}^\infty
\gamma^R_{mnp}(\epsilon)\,
\as^m(Q^2)\,\as^n(Q^2/N^2)\,\as^p(Q^2/N). \label{gamma}
\end{equation}
Even
though the anomalous dimension is finite as $\epsilon\to0$ for
all $N$, the
individual terms in the expansion Eq.~(\ref{gamma}) are not
separately finite. However, if we separate $N$-dependent and
$N$-independent terms in Eq.~(\ref{gamma}):
\begin{equation}
\gamma(N,\as(Q^2),\epsilon)
=\hat\gamma^{(c)}(\as(Q^2),\epsilon)+\hat\gamma^{(l)}(N,\as(Q^2),\epsilon),
\label{gammahat}
\end{equation}
we note that the two functions
\begin{eqnarray}
\gamma^{(c)}(\as(Q^2),\epsilon)
&\equiv&\hat\gamma^{(c)}(\as(Q^2),\epsilon)+\hat\gamma^{(l)}(1,\as(Q^2),\epsilon)
\\
\gamma^{(l)}(N,\as(Q^2),\epsilon)
&\equiv&\hat\gamma^{(l)}(N,\as(Q^2),\epsilon)-\hat\gamma^{(l)}(1,\as(Q^2),\epsilon)
\end{eqnarray}
must be separately finite.
In fact,
\begin{equation}
\gamma(N,\as(Q^2),\epsilon)=\gamma^{(c)}(\as(Q^2),\epsilon)+\gamma^{(l)}(N,\as(Q^2),\epsilon),
\label{gamdec}
\end{equation}
is finite for all $N$ and $\gamma^{(l)}$
vanishes for $N=1$. This implies that $\gamma^{(c)}(\as(Q^2),\epsilon)$
is finite in $\epsilon=0$ and that $\gamma^{(l)}(N,\as(Q^2),\epsilon)$
is also finite in $\epsilon=0$ for all $N$ because of the $N$-independence
of $\gamma^{(c)}(\as(Q^2),\epsilon)$.

We can rewrite conveniently
\begin{equation}
\gamma^{(l)}(N,\as(Q^2),\epsilon)
=\int_1^N
\frac{dn}{n}\,g(\as(Q^2),\as(Q^2/n^2),\as(Q^2/n),\epsilon),
\end{equation}
where
\begin{equation}
g(\as(Q^2),\as(Q^2/n^2),\as(Q^2/n),\epsilon)=n\frac{\partial}{\partial
n} \hat\gamma^{(l)}(n,\as(Q^2),\epsilon). \label{gdef}
\end{equation}
is a
Taylor series in its arguments whose coefficients remain finite as
$\epsilon\to0$. In four dimension we have thus
\begin{eqnarray}
\gamma(N,\as(Q^2))&=&\gamma^{(l)}(N,\as(Q^2),0)
+\gamma^{(c)}(N,\as(Q^2),0)+O\left(\frac{1}{N}\right)
\nonumber\\
&=&\gamma^{(l)}(N,\as(Q^2),0)+O\left(N^0\right)\nonumber\\
&=&\int_1^N \frac{dn}{n}\,
g(\as(Q^2),\as(Q^2/n^2),\as(Q^2/n))+O\left(N^0\right),\nonumber\\
\label{resformula}
\end{eqnarray}
where
\begin{equation}
g(\as(Q^2),\as(Q^2/n^2),\as(Q^2/n))\equiv
\lim_{\epsilon\to0}g(\as(Q^2),\as(Q^2/n^2),\as(Q^2/n),\epsilon)
\end{equation}
is a generic Taylor series of its arguments.

Renormalization group invariance thus implies that the physical
anomalous dimension $\gamma$ Eq.~(\ref{paddef}) depends on its
three arguments $Q^2$, $Q^2/N$ and $Q^2/N^2$ only through
$\alpha_s$. Clearly, any function of $Q^2$ and $N$ can be
expressed as a function of $\as(Q^2)$ and $\as(Q^2/N)$  or
$\as(Q^2/N^2)$. The nontrivial statement, which endows
Eq.~(\ref{resformula}) with predictive power, is that the log
derivative of $\gamma$, $g(\as(Q^2),\as(Q^2/n^2),\as(Q^2/n))$
Eq.~(\ref{gdef}), is analytic in its three arguments. This
immediately implies that when $\gamma$ is computed at (fixed)
order $\alpha_s^k$, it is a polynomial in $\ln\frac{1}{N}$ of
$k$-th order at most.

In order to discuss the factorization properties of our result we
write the function $g$ as
\begin{eqnarray}
g(\as(Q^2),\as(Q^2/n^2),\as(Q^2/n))
&=&g_1(\as(Q^2),\as(Q^2/n))\nonumber\\
&&+\quad g_2(\as(Q^2),\as(Q^2/n^2))
\nonumber\\
&&+\quad g_3(\as(Q^2),\as(Q^2/n),\as(Q^2/n^2)),\nonumber\\
\end{eqnarray}
where
\begin{eqnarray}
g_1(\as(Q^2),\as(Q^2/n))&=& \sum_{m=0}^\infty\sum_{p=1}^\infty
g_{m0p}\,\as^m(Q^2)\,\as^p(Q^2/n)\label{ggdef}
\\
g_2(\as(Q^2),\as(Q^2/n^2))&=&\sum_{m=0}^\infty\sum_{n=1}^\infty
g_{mn0}\, \as^m(Q^2)\,\as^n(Q^2/n^2)\nonumber\\
\label{ggdef2}\\
g_3(\as(Q^2),\as(Q^2/n),\as(Q^2/n^2))
&=&\sum_{m=0}^\infty\sum_{n=1}^\infty\sum_{p=1}^\infty
g_{mnp}\,\as^m(Q^2)\nonumber\\
&&\times\as^n(Q^2/n^2)\,\as^p(Q^2/n).\label{ggdef3}
\end{eqnarray}
The dependence on the resummation variables $Q^{2}$, $Q^{2}/N$ and
$Q^{2}/N^{2}$ is fully factorized if
the bare coefficient functions has the factorized structure
\begin{equation}
\label{fstrong}
C^{(0)}(N,Q^2,\az,\epsilon)=C^{(0,c)}(Q^2,\az,\epsilon)\,
C_1^{(0,l)}(Q^2/N,\az,\epsilon)\,
C_2^{(0,l)}(Q^2/N^2,\az,\epsilon).
\end{equation}
This is argued to be the
case in the approach of Refs.\cite{Catani:1998tm,Laenen:1998qw}.
If this happens,
the resummed anomalous dimension is given by
Eq.~(\ref{resformula}) with all $g_{mnp}=0$ except
$g_{0n0},g_{00p}$:
\begin{equation} \label{resformulanason}
\gamma(N,\as(Q^2))
=\int_1^N\frac{dn}{n}\,g_1(0,\as(Q^2/n))
+\int_1^N\frac{dn}{n}\,g_2(0,\as(Q^2/n^2)).
\end{equation}
We recall that
the coefficient function depends on the parton subprocess
in which the incoming partons are $a,b$ (compare
Eq.~(\ref{crosssec})). So, the factorization Eq.~(\ref{fstrong})
applies to the coefficient function corresponding to each
subprocess, and the decomposition Eq.~(\ref{resformulanason}) to
the physical anomalous dimension computed from each of these
coefficient functions.

A weaker form of factorization  is obtained assuming that in the
soft limit the $N$-dependent and $N$-independent parts of the
coefficient function factorize:
\begin{equation} \label{fweak}
C^{(0)}(N,Q^2,\epsilon)=C^{(0,c)}(Q^2,\az,\epsilon)\,
C^{(0,l)}(Q^2/N^2,Q^2/N,\az,\epsilon).
\end{equation}
This condition turns
out to be satisfied~\cite{Forte:2002ni} in Drell-Yan and deep-inelastic
scattering to order $\alpha_s^2$. It holds in QED to all
orders~\cite{Weinberg1995} as a consequence of the fact that each
emission in the soft limit can be described by  universal
(eikonal) factors, independent of the underlying diagram. This
eikonal structure of Sudakov radiation has been argued in
Refs.~\cite{Catani:1989ne,Catani:1998tm} to apply also to QCD.
If the factorized form
Eq.~(\ref{fweak}) holds, the coefficients
 $g_{mnp}$ Eqs.~(\ref{ggdef},\ref{ggdef2},\ref{ggdef3})  vanish for all
$m\not= 0$, and  the physical anomalous dimension takes the form
\begin{eqnarray} \label{resformulaint} \gamma(N,\as(Q^2))
&=&\int_1^N\frac{dn}{n}\,g_1(0,\as(Q^2/n))
+\int_1^N\frac{dn}{n}\,g_2(0,\as(Q^2/n^2))
\nonumber\\
&&+\int_1^N\frac{dn}{n}\,g_3(0,\as(Q^2/n^2),\as(Q^2/n)).
\end{eqnarray}

It is interesting to observe that in the approach of
Refs.~\cite{Catani:1998tm,Laenen:1998qw} for processes where more than
one colour structure contributes to the cross-section, the factorization
Eq.~(\ref{fstrong}) of the coefficient function is argued to take
place separately for each colour structure. This means that in
such case the exponentiation takes place for each colour structure
independently, i.e. the resummed cross section for each parton
subprocess is in turn expressed as a sum of factorized terms of
the form of Eq.~(\ref{fstrong}). This happens for instance in the
case of heavy quark production~\cite{Catani:1998tm,Bonciani:1998vc}.
In prompt photon
production different colour structures appear for the gluon-gluon
subprocess which starts at next-to-next-to-leading order, hence
their separated exponentiation would be relevant for
next-to-next-to-leading log resummed results.

When several colour structures contribute to a given parton
subprocess, the coefficients of the perturbative expansion
Eq.~(\ref{C0}) for that process take the form
\begin{equation}
C_{lkk'}^{(0)}(\epsilon)=C_{lkk'}^{(0)\bf{1}}(\epsilon)
+C_{lkk'}^{(0)\bf{8}}(\epsilon), \label{colco}
\end{equation}
(assuming for
definiteness that a colour singlet and octet contribution are
present) so that the coefficient function can be written as a sum
$C^{(0)}=C^{(0)}_{\bf{1}}+C^{(0)}_{\bf{8}}$. The argument which
leads from Eq.~(\ref{C0}) to the resummed result
Eq.~(\ref{resformula}) then implies that exponentiation takes
place for each colour structure independently if and only if
\begin{equation}
\gamma_{\bf{1}}\equiv\partial\ln C^{(0)}_{\bf{1}}/\partial\ln
Q^2, \qquad\gamma_{\bf{8}}\equiv\partial\ln
C^{(0)}_{\bf{8}}/\partial\ln Q^2
\end{equation}
are separately finite.

This, however, is clearly a more restrictive assumption than that
under which we have derived the result Eq.~(\ref{resformula}),
namely that the full anomalous dimension $\gamma$ is finite. It
follows that exponentiation of each colour structure must be a
special case of our result. However, this can only be true if the
coefficients $g_{ijk}$ of the expansion Eq.~(\ref{ggdef}) of the
physical anomalous dimension satisfy suitable relations. In
particular, at the leading log level, it is easy to see that
exponentiation of each colour structure is compatible with
exponentiation of their sum only if the leading order coefficients
are the same for the given colour structures:
$g_{001}^{\bf{1}}=g_{001}^{\bf{8}}$ and
$g_{010}^{\bf{1}}=g_{010}^{\bf{8}}$. This is indeed the case for
heavy quark production (where $g_{001}=0$).

Note that, however, if the factorization holds for each colour
structure separately it will not apply to the sum of colour
structures. For instance, the weaker form of factorization
Eq.~(\ref{fweak}) requires that
\begin{equation}
C_{lkk'}^{(0)}(\epsilon)=F_{k+k'}(\epsilon)G_{l-k-k'}(\epsilon),
\end{equation}
but
\begin{equation}
F^{\bf{1}}_{k+k'}(\epsilon)G^{\bf{1}}_{l-k-k'}(\epsilon)+F^{\bf{8}}_{k+k'}(\epsilon)G^{\bf{8}}_{l-k-k'}(\epsilon)\neq
F_{k+k'}(\epsilon)G_{l-k-k'}(\epsilon).
\end{equation}
Hence, our result
Eq.~(\ref{resformula}) for the sum of colour structures is more
general than the separate exponentiation of individual colour
structures, but it leads to results which have weaker
factorization properties.

\section{Comparison with previous results}

In this section, we want to make a comparison with the resummation formula for prompt photon production previously released.
In order to do this, we need to rewrite the NLL result of  Ref.\cite{Catani:1998tm} in our formalism. The physical anomalous dimension
can be obtained performing the $Q^{2}$-logarithmic derivative of the NLL resummed exponent in the
$\overline{MS}$ scheme of Ref.\cite{Catani:1998tm}. We obtain for a particular partonic sub-process:
\begin{eqnarray}
\gamma(N,\alpha_{s}(Q^{2}))&=&\int_{0}^{1}dx\,\frac{x^{N-1}-1}{1-x}\big[\hat{g}_{2}
\alpha_{s}(Q^{2}(1-x)^{2})+\hat{g}'_{2}\alpha_{s}^{2}(Q^{2}(1-x)^{2})\nonumber\\
&&+\hat{g}_{1}\alpha_{s}(Q^{2}(1-x))+\hat{g}'_{1}\alpha_{s}^{2}(Q^{2}(1-x-))\big],\label{cmn}
\end{eqnarray}
where
\begin{eqnarray}
\hat{g}_{2}&=&\frac{A_{d}^{(1)}+A_{b}^{(1)}-A_{d}^{(1)}}{\pi}\label{iniz}\\
\hat{g}'_{2}&=&\frac{A_{a}^{(2)}+A_{b}^{2}-A_{d}^{(2)}}{\pi^{2}}-\frac{\beta_{0}}{4\pi}\left[\frac{A_{a}^{(1)}+A_{b}^{(1)}-A_{d}^{(1)}}{\pi}\right]\ln 2\\
\hat{g}_{1}&=&\frac{A_{d}^{(1)}}{\pi}\\
\hat{g}'_{1}&=&\frac{A_{d}^{(2)}}{\pi^{2}}-\frac{\beta_{0}}{4\pi}\,\frac{B_{d}^{(1)}}{2\pi}.\label{finito}
\end{eqnarray}
Here, $A_{a}^{i}$ is the coefficient of $\ln(1/N)$ in the Mellin transform of the $P_{aa}$ Altarelli-Parisi splitting function at order
$\alpha_{s}^{i}$, $\beta_{0}$ is the $\alpha_{s}^{2}$ coefficient of the $\beta$ function (Eq.(\ref{betapar}) in section \ref{appA})
and $B_{d}^{(1)}$ is a constant to be determined from the comparison with the fixed-order calculation.
In Eqs.(\ref{iniz}-\ref{finito}) $a,b$ are the incoming partons (on which the coefficient function
implicitly depends) and $d$ is the LO outgoing parton uniquely determined by the incoming ones.
For completeness, we list explicitly these coefficients:
\begin{eqnarray}
A_{a=q,\bar{q}}^{(1)}&=&\frac{4}{3},\qquad\qquad\,\, A_{a=g}^{1}=3\\
A_{a}^{(2)}&=&\frac{1}{2}A_{a}^{(1)}K,\qquad K=\frac{67}{6}-\frac{\pi^{2}}{6}-\frac{5}{9}N_{f}\\
B_{d=q,\bar{q}}^{(1)}&=&-2,\qquad\qquad\,\, B_{d=g}^{(1)}=-\frac{11}{2}+\frac{1}{3}N_{f},
\end{eqnarray}
where $N_{f}$ is the number of active flavors. Now, performing
 the change of
variable
\begin{equation}
n=\frac{1}{1-x}
\end{equation}
in the integral Eq.(\ref{cmn}), we obtain at NLL
\begin{eqnarray}
\gamma(N,\alpha_{s}(Q^{2}))&=&\int_{0}^{1}\frac{dn}{n}\big[g_{010}\alpha_{s}\left(\frac{Q^{2}}{n^{2}}\right)
+g_{001}\alpha_{s}\left(\frac{Q^{2}}{n}\right)\nonumber\\
&&+g_{020}\alpha_{s}^{2}\left(\frac{Q^{2}}{n^{2}}\right)
+g_{002}\alpha_{s}^{2}\left(\frac{Q^{2}}{n}\right)\big],\label{cmnb}
\end{eqnarray}
where
\begin{eqnarray}
g_{010}&=&-\hat{g}_{2},\qquad\quad g_{020}=-\left(\hat{g}'_{2}-\frac{\gamma_{E}\beta_{0}}{2\pi}\hat{g}_{2}\right)\label{ctb}\\
g_{001}&=&-\hat{g}_{1},\qquad\quad g_{002}=-\left(\hat{g}'_{1}-\frac{\gamma_{E}\beta_{0}}{4\pi}\hat{g}_{1}\right),\label{ctb2}
\end{eqnarray}
with $\gamma_{E}$ the Euler constant. Hence, according to Ref.\cite{Catani:1998tm}, we know exactly the value of the
coefficients $g_{010}$, $g_{020}$, $g_{001}$ and $g_{002}$. This enables us to compute predictions
of high-order logarithmic contributions to the physical anomalous dimension performing a fixed order expansion of $\gamma$.

We shall now show that the resummation formula of Ref.\cite{Catani:1998tm} predicts the coefficient of $\alpha_{s}^{3}
\ln^{2}(1/N)$ of the fixed order expansion of $\gamma$, while in our approach it is required in order to perform
a NLL resummation.
We need to expand Eq.(\ref{cmnb}) to order $\alpha_{s}^{3}$ and this is obtained using
the change of variable
\begin{equation}
\frac{dn}{n}=-\frac{d\alpha_{s}(Q^{2}/n^{a})}{a\beta{\alpha_{s}}},\qquad a=1,2
\end{equation}
to perform the integral and expanding the two loops running of $\alpha_{s}$
(see Eq.(\ref{betafunc}) in section \ref{appA}). We find
\begin{eqnarray}
\gamma&=&[-(g_{001}+g_{010})]\alpha_{s}(Q^{2})\ln\frac{1}{N}\nonumber\\
&&+[-(g_{002}+g_{020})-(b_{1}/b_{0})(g_{001}+g_{010})]\alpha_{s}^{2}(Q^{2})\ln\frac{1}{N}\nonumber\\
&&+\left[\frac{b_{0}}{2}(g_{001}+2g_{010})\right]\alpha_{s}^{2}(Q^{2})\ln^{2}\frac{1}{N}+[-(g_{003}+g_{030})]\alpha_{s}^{3}(Q^{2})\ln\frac{1}{N}\nonumber\\
&&+[(3b_{1}/2)(g_{001}+2g_{010})+b_{0}(g_{002}+2g_{020})]\alpha_{s}^{3}(Q^{2})\ln^{2}\frac{1}{N}\nonumber\\
&&+[-(b_{0}^{2}/3)(g_{001}+4g_{010})]\alpha_{s}^{3}\ln^{3}\frac{1}{N}+\emph{O}(\alpha_{s}^{4}).
\end{eqnarray}
In our approach, in order to determine the NLL resummation coefficients
$g_{010}$, $g_{020}$, $g_{001}$ and $g_{002}$, we must compare this expansion to a fixed order
computation of the physical anomalous dimension, which in the general has the form
\begin{equation}
\gamma_{\rm FO}(N,\as)=\sum_{i=1}^{k}\as^i\,\sum_{j=1}^i
\gamma^i_j\,\ln^j\frac{1}{N}+O(\as^{k+1})+O(N^0),
\end{equation}
where $k$ is the fixed-order at which it has been computed (see Chapter \ref{predictivity} for a
general discussion about the determination of the resummation coefficients). Hence, we determine the $4$ NLL
resummation coefficients through the following $4$ independent conditions:
\begin{eqnarray}
g_{001}+g_{010}&=&-\gamma^{1}_{1}\\
g_{001}+2g_{010}&=&\frac{2}{b_{0}}\gamma^{2}_{2}\\
g_{002}+g_{020}&=&-\gamma^{2}_{1}-\frac{b_{1}}{b_{0}}(g_{001}+g_{010})\\
g_{002}+2g_{020}&=&\frac{1}{b_{0}}\gamma^{3}_{2}-\frac{3b_{1}}{2b_{0}}(g_{001}+2g_{010}).\label{prediction}
\end{eqnarray}
Thus, according to our formalism, all the coefficients $\gamma^{1}_{1}$, $\gamma^{2}_{2}$, $\gamma^{2}_{1}$
and $\gamma^{3}_{2}$ should be known. The first three coefficients are all known thanks to the explicit
$\emph{O}(\alpha_{s}^{2})$ calculation of the prompt photon cross section
\cite{Aurenche:1983ws,Aurenche:1987fs,Gordon:1993qc}.
The last one ($\gamma^{3}_{2}$), is not yet known from explicit
$\emph{O}(\alpha_{s}^{3})$ calculation, but, according to the approach
of Ref.\cite{Catani:1998tm}, it is predicted using Eqs.(\ref{prediction},\ref{ctb},\ref{ctb2}):
\begin{eqnarray}
\gamma_2^3&=&-b_0\Bigg[\frac{2(A_a^{(2)}+A_b^{(2)})-A_d^{(2)}}{\pi^2}
-b_0(2\ln2+4\gamma_{E})\frac{A_a^{(1)}+A_b^{(1)}}{\pi}\\ &&\quad+b_0
(2\ln2+3\gamma_{E})\frac{A_d^{(1)}}{\pi} \nonumber
-\frac{b_0B_d^{(1)}}{2\pi}\Bigg]-b_{1}\frac{3}{2}
\left[\frac{2(A_a^{(1)}+A_b^{(1)})-A_d^{(1)}}{\pi}\right].
\end{eqnarray}
The correctness of this result could be tested by an order $\alpha_{s}^{3}$ calculation. If it
were to fail, the more general resummation formula with $g_{020}$ determined
by Eq.(\ref{prediction}) should be used, or one of the resummations which do not
assume the factorization Eq.(\ref{fstrong}).

\chapter{Resummation of rapidity distributions}\label{DYDPR}

In this chapter, we present a derivation of the threshold
resummation formula for the Drell-Yan  and the prompt photon
production rapidity distributions. Our arguments are valid for all
values of rapidity and to all orders in perturbative QCD. For the
case of the Drell-Yan process, resummation is realized in a
universal way, i.e. both for the production of a virtual photon
$\gamma^{*}$ and the production of a vector boson $W^{\pm}$,
$Z^{0}$. We will show that for the fixed-target proton-proton
Drell-Yan experiment E866/NuSea used in current parton fits, the
NLL resummation corrections are comparable to NLO fixed-order
corrections and are crucial to obtain agreement with the data.
This means that the NLL resummation of rapidity distributions is
necessary and turns out to give better results than
high-fixed-order calculations. We consider first the resummation
of the DY case and its phenomenology and then the resummation of
the prompt photon case.

\section{Threshold resummation of DY rapidity distributions}

\subsection{General kinematics of Drell-Yan rapidity distributions}

We consider the general Drell-Yan process in which the collisions
of two hadrons ($H_{1}$ and $H_{2}$) produce a virtual photon
$\gamma^{*}$ (or an on-shell vector boson $V$) and any collection
of hadrons (X): \bea H_{1}(P_{1})+H_{2}(P_{2})\rightarrow
\gamma^{*}(V)(Q) +X(K). \eea In particular, we are interested in
the differential cross section
$\frac{d\sigma}{dQ^{2}dY}(x,Q^{2},Y)$, where $Q^{2}$ is the
invariant mass of the photon or of the vector boson, $x$ is
defined as usual as the fraction of invariant mass that the
hadrons transfer to the photon (or to the vector boson) and $Y$ is
the rapidity of $\gamma^{*}$ ($V$) in the hadronic
centre-of-mass-frame: \bea\label{rap} x\equiv\frac{Q^{2}}{S},\quad
S=(P_{1}+P_{2})^{2}, \quad
Y\equiv\frac{1}{2}\ln\left(\frac{E+Q_{z}}{E-Q_{z}}\right), \eea
where $E$ and $p_{z}$ are the energy and the longitudinal momentum
of $\gamma^{*}(V)$ respectively. In this frame, the four-vector
$Q$ of $\gamma^{*}(V)$ can be written in terms of its rapidity and
its tranverse momentum \bea
Q=(Q^{0},\vec{Q})=(\sqrt{Q^{2}+Q_{\perp}^{2}}\cosh Y,
\vec{Q}_{\perp}, \sqrt{Q^{2}+Q_{\perp}^{2}}\sinh Y), \label{finalmente}\eea or in
terms of the scattering angle $\theta$ \bea
Q=(Q^{0},\vec{Q})=(\sqrt{Q^{2}+|\vec{Q}|^{2}},
\vec{Q}_{\perp},|\vec{Q}|\cos(\theta)).\label{qtheta}\eea For
completeness, we recall that when $Q^{2}=0$ (which is not our
case), the rapidity $Y$ defined in Eq.(\ref{rap}) reduces to the
pseudorapidity $\eta$ according to Eq.(\ref{qtheta}): \bea
\eta=-\ln(\tan(\theta/2)). \eea

At the partonic level, a parton 1(2) in the hadron $H_{1}$
($H_{2}$) carries a fraction of momentum $x_{1}$ ($x_{2}$): \bea
p_{1}=x_{1}P_{1}=x_{1}\frac{\sqrt{S}}{2}(1,\vec{0}_{\perp},1)),
\quad
p_{2}=x_{2}P_{2}=x_{2}\frac{\sqrt{S}}{2}(1,\vec{0}_{\perp},-1)).
\eea It is clear that the hadronic center-of-mass frame does not
coincide with the partonic one, because $x_{1}$ is, in general,
different from $x_{2}$. Furthermore, from Eq.(\ref{rap}), we see
that the rapidity is not an invariant. Hence, in order to define
the rapidity in the partonic center-of-mass frame ($y$), we have
to perform a boost of $Y$ which connects the two frames. This
provides us a relation between the rapidity in these two frames:
\bea\label{yY} y=Y-\frac{1}{2}\ln(\frac{x_{1}}{x_{2}}). \eea

In order to understand the kinematic configurations in terms of
rapidity, it is convenient to define a new variable $u$,
\bea\label{u} u\equiv\frac{Q\cdot p_{1}}{Q\cdot
p_{2}}=e^{-2y}=\frac{x_{1}}{x_{2}}e^{-2Y}. \eea With no partons
radiated as in the case of the LO, the rapidity is obviously zero.
Beyond the LO, one or more partons can be radiated. Now, if these
partons are radiated collinear with the incoming parton 2, then
the partonic rapidity reaches its maximum value and $u$ its
minimum one. Similarly the minimum value of $y$ and the maximum
value of $u$ is achieved when the radiated partons are collinear
with the incoming parton 1. To be more precise, suppose that in
the first case the radiated partons (collinear with the parton 2)
carry away a fraction of momentum equal to $(1-z)p_{2}$, so that
by momentum conservation $Q=p_{1}+zp_{2}$. In this case, we obtain
immediately the lower bound for $u$, which is $z$. In the second
case the collinear radiated partons have momentum $(1-z)p_{1}$,
hence $Q=zp_{1}+p_{2}$ and the upper bound of $u$ is $1/z$. So,
$z$ can be interpreted as the fraction of invariant mass that
incoming partons transfer to $\gamma^{*}(V)$. In fact:
\bea\label{z} z=\frac{Q^{2}}{2p_{1}\cdot
p_{2}}=\frac{Q^{2}}{(p_{1}+p_{2})^{2}}=\frac{x}{x_{1}x_{2}}, \eea
where we have neglected the quark masses. Therefore, we have
that the kinematic constraints of $u$ are: \bea\label{boundary}
z\leq u\leq \frac{1}{z}.\eea Then, since $x_{1}<1$ and $x_{2}<2$,
the lower and upper bounds of $z$ are: \bea x\leq z\leq 1. \eea
Thanks to Eqs.(\ref{yY},\ref{u}), the first relation can be
translated directly into a relation for the upper and lower limit
of the partonic center-of-mass rapidity: \bea\label{boundy}
\frac{1}{2}\ln z\leq y\leq \frac{1}{2}\ln\frac{1}{z}. \eea Now, we
need to obtain the boundaries of the hadronic center-of-mass
rapidity. Substituting Eqs.(\ref{u}, \ref{z}) into the two
conditions $u\geq z$ and $u\leq 1/z$, we obtain the lower
kinematical bound for $x_{1}$ and $x_{2}$: \bea x_{1}\geq
\sqrt{x}e^{Y}\equiv x_{1}^{0}, \quad x_{2}\geq
\sqrt{x}e^{-Y}\equiv x_{2}^{0} \eea and the obvious requirment
that $x_{1(2)}^{0}\leq 1$ implies that the hadronic rapidity has a
lower and an upper bound: \bea\label{boundY} \frac{1}{2}\ln x\leq
Y\leq \frac{1}{2}\ln\frac{1}{x}. \eea

\subsection{The universality of resummation in Drell-Yan processes}

According to standard factorization of collinear singularities of
perturbative QCD, the expression for the hadronic differential
cross section in rapidity has the form, \bea
\frac{d\sigma}{dQ^{2}dY}&=&\sum_{i,j}\int_{x_{1}^{0}}^{1}dx_{1}\int_{x_{2}^{0}}^{1}
dx_{2}F^{H_{1}}_{i}(x_{1},\mu^{2})F^{H_{2}}_{j}(x_{2},\mu^{2})
\nonumber\\
&\times&\frac{d\hat{\sigma}_{ij}}{dQ^{2}dy}\left(x_{1},x_{2},\frac{Q^{2}}{\mu^{2}},\alpha_{s}(\mu^{2}),y\right),\label{hadrcs}
\eea where $y$ depends on $Y$, $x_{1}$ and $x_{2}$ according to
Eq.(\ref{yY}). The sum runs over all possible partonic
subprocesses, $F^{(1)}_{i}$ ,$F^{(2)}_{j}$ are respectively the
parton densities of the hadron $H_{1}$ and $H_{2}$, $\mu$ is the
factorization scale (chosen equal to renormalization scale for
simplicity) and $d\hat{\sigma}_{ij}/(dQ^{2}dy)$ is the partonic
cross section.  Even if the cross section Eq.(\ref{hadrcs}) is
$\mu^{2}$-independent, this is not the case for each  parton
subprocess. However, the $\mu^{2}$-dependence of each contribution
is proportional to the off-diagonal anomalous dimensions (or
splitting functions), which in the threshold limit, ($z\rightarrow
1$) are suppressed by factors of $1-z$. Therefore, each partonic
subprocess can be treated independently and is separately
renormalization-group invariant. Furthermore, the suppression, in
the threshold limit, of the off-diagonal splitting functions
implies also that only the gluon-quark channels are suppressed.
So, in order to study resummation, we will consider only the quark-
anti-quark channel, which can be related to the same dimensionless
coefficient function $C(z,Q^{2}/\mu^{2},\alpha_{s}(\mu^{2}),y)$
for both, the production of a virtual photon and the production of
a on-shell vector boson. In fact, if for the production of a
virtual photon, we define
$C(z,Q^{2}/\mu^{2},\alpha_{s}(\mu^{2}),y)$ through the equation,
\bea
x_{1}x_{2}\frac{d\hat{\sigma}^{\gamma^{*}}_{q\bar{q}'}}{dQ^{2}dy}\left(x_{1},x_{2},\frac{Q^{2}}{\mu^{2}},
\alpha_{s}(\mu^{2}),y\right)=\frac{4\pi\alpha^{2}c_{q\bar{q}'}}{9Q^{2}S}
C\left(z,\frac{Q^{2}}{\mu^{2}},\alpha_{s}(\mu^{2}),y\right), \eea
where the prefactor $x_{1}x_{2}$ has been introduced for future
convenience, we find that for the case of the production of a real
vector boson, \bea\label{Vprod}
x_{1}x_{2}\frac{d\hat{\sigma}^{V}_{q\bar{q}'}}{dQ^{2}dy}\left(x_{1},x_{2},\frac{Q^{2}}{\mu^{2}},\alpha_{s}(\mu^{2}),y\right)&=&
\frac{\pi
G_{F}Q^{2}\sqrt{2}c_{q\bar{q}'}}{3S}\delta(Q^{2}-M^{2}_{V})\nonumber\\
&\times&C\left(z,\frac{Q^{2}}{\mu^{2}},\alpha_{s}(\mu^{2}),y\right),
\eea where $G_{F}$ is the Fermi constant, $M_{V}$ is the mass of
the produced vector boson. The coefficients $c_{q\bar{q}'}$, for
the different Drell-Yan processes, are given by:
\begin{eqnarray}
c_{q\bar{q}'}&=&Q_{q}^{2}\delta_{q\bar{q}}\quad\textrm{for $\gamma^{*}$},\label{gamma}\\
c_{q\bar{q}'}&=&|V_{qq'}|^{2}\quad \textrm{for  $W^{\pm}$},\label{W}\\
c_{q\bar{q}'}&=&4[(g_{v}^{q})^{2}+(g_{a}^{q})^{2}]\delta_{q\bar{q}}
\quad\textrm{for $Z^{0}$}.\label{Z}
\end{eqnarray}
Here, $Q_{q}^{2}$ is the square charge of the quark $q$, $V_{qq'}$
are the CKM mixing factors for the quark flavors $q,q'$ and
\begin{eqnarray}
g_{v}^{q}&=&\frac{1}{2}-\frac{4}{3}\sin^{2}\theta_{W},\quad g_{a}^{q}=\frac{1}{2}\quad\textrm{for an up-type quark},\\
g_{v}^{q}&=&-\frac{1}{2}+\frac{2}{3}\sin^{2}\theta_{W},\quad
g_{a}^{q}=-\frac{1}{2}\quad\textrm{for a down-type quark},
\end{eqnarray}
with $\theta_{W}$  the Weinberg weak mixing angle. As a
consequence of these facts, resummation has to be performed only
for the quark-anti-quark channels omitting  the overall dimensional
factors of $C(z,Q^{2}/\mu^{2},\alpha_{s}(\mu^{2}),y)$ in the
different Drell-Yan processes. Thus, we are left with the
following dimensionless cross section, which has the form: \bea
\sigma(x,Q^{2},Y)&\equiv&
\int_{x_{1}^{0}}^{1}\frac{dx_{1}}{x_{1}}\int_{x_{2}^{0}}^{1}\frac{dx_{2}}{x_{2}}F_{1}^{H_{1}}(x_{1},\mu^{2})
F_{2}^{H_{2}}(x_{2},\mu^{2})\nonumber\\
&\times&C\left(z,\frac{Q^{2}}{\mu^{2}},\alpha_{s}(\mu^{2}),y\right),\label{universal}
\eea where $F_{1}$ and $F_{2}$ are quark or anti-quark parton
densities in the hadron $H_{1}$ and $H_{2}$ respectively.  This
shows the universality of resummation in Drell-Yan processes in
the sense that only the renormalization-group invariant quantity
defined in Eq.(\ref{universal}) has to be resummed.

\subsection{Factorization properties and the Mellin-Fourier transform}

For the case of the rapidity-integrated cross section, resummation
is usually done in Mellin space transforming the variable $x$ into its
conjugate variable $N$, because the Mellin transformation turns
 convolution products into ordinary products. Furthermore, the
Mellin space is the natural space where to define resummation of
leading, next-to-leading and so on logarithmic contribution,
because in this space momentum conservation is respected as shown
in \cite{Catani:1996yz}. In the case of the rapidity distribution,
the Mellin transformation is not sufficient. In fact, rewriting
Eq.(\ref{universal}) in this form \bea
\sigma(x,Q^{2},Y)&=&\int_{0}^{1}dx_{1}dx_{2}dzF_{1}^{H_{1}}(x_{1},\mu^{2})
F_{2}^{H_{2}}(x_{2},\mu^{2})\nonumber\\
&\times&C\left(z,\frac{Q^{2}}{\mu^{2}},\alpha_{s}(\mu^{2}),y\right)\delta(x-x_{1}x_{2}z),\label{universal2}
\eea we see that the Mellin transform with respect to $x$,
\bea\label{mellintrans} \sigma(N,Q^{2},Y)\equiv
\int_{0}^{1}dxx^{N-1}\sigma(x,Q^{2},Y), \eea does not  diagonalize
the triple integral in Eq.(\ref{universal2}). This is due to
the fact that the partonic center-of-mass rapidity $y$ depends
on $x_{1}$ and $x_{2}$  through Eq.(\ref{yY}). The ordinary
product in Mellin space can be recovered performing the Mellin
transform with respect to $x$ of the Fourier transform of
Eq.(\ref{universal2}) with respect to $Y$. Calling the Fourier
moments $M$, using Eq.(\ref{yY}) the relations
(\ref{boundy},\ref{boundY}) and the identity \bea
C\left(z,\frac{Q^{2}}{\mu^{2}},\alpha_{s}(\mu^{2}),Y-\frac{1}{2}\ln\frac{x_{1}}{x_{2}}\right)&=&\int_{\ln\sqrt{z}}^{\ln1/\sqrt{z}}dy
C\left(z,\frac{Q^{2}}{\mu^{2}},\alpha_{s}(\mu^{2}),y\right)\nonumber\\
&\times&\delta\left(y-Y+\frac{1}{2}\ln\frac{x_{1}}{x_{2}}\right),
\eea
 we find that
\begin{eqnarray}
\sigma(N,Q^{2},M)&\equiv&\int_{0}^{1}dxx^{N-1}\int_{\ln\sqrt{x}}^{\ln1/\sqrt{x}}dYe^{iMY}\sigma(x,Q^{2},Y)\\
&=&F_{1}^{H_{1}}(N+iM/2,\mu^{2})F_{2}^{H_{2}}(N-iM/2,\mu^{2})\nonumber\\
&\times&C\left(N,\frac{Q^{2}}{\mu^{2}},\alpha_{s}(\mu^{2}),M\right),\label{fact}
\end{eqnarray}
where
\begin{eqnarray}
F_{i}^{H_{i}}(N\pm iM/2,\mu^{2})&=&\int_{0}^{1}dxx^{N-1\pm iM/2}F_{i}^{H_{i}}(x,\mu^{2}),\label{mfpd}\\
C\left(N,\frac{Q^{2}}{\mu^{2}},\alpha_{s}(\mu^{2}),M\right)&=&\int_{0}^{1}dzz^{N-1}\int_{\ln\sqrt{z}}^{\ln1/\sqrt{z}}dye^{iMy}\nonumber\\
&\times&
C\left(z,\frac{Q^{2}}{\mu^{2}},\alpha_{s}(\mu^{2}),y\right).\label{mfcoefffunc2}
\end{eqnarray}
Eq.(\ref{fact}) shows that performing the Mellin-Fourier moments
of the hadronic dimensionless cross section Eq.(\ref{universal}),
we recover an ordinary product of the Mellin-Fourier trnsform of
the coefficient function and the Mellin moments of the parton
densities translated outside the real axis by  $\pm iM/2$.
Because the coefficient function is symmetric
in $y$, we can rewrite Eq.(\ref{mfcoefffunc2}) in this way:
\bea
C\left(N,\frac{Q^{2}}{\mu^{2}},\alpha_{s}(\mu^{2}),M\right)&=&2\int_{0}^{1}dzz^{N-1}\int_{0}^{\ln1/\sqrt{z}}dy\cos(My)\nonumber\\
&\times&
C\left(z,\frac{Q^{2}}{\mu^{2}},\alpha_{s}(\mu^{2}),y\right)\label{cosfact}.
\eea
From this last equation and Eq.(\ref{mfpd}), we see that the dependence on $M$, the Fourier conjugate of the rapidity
$y$, originates from the parton densities, that depend on $N\mp iM/2$, and from the factor of $\cos(My)$
in the integrand of Eq.(\ref{cosfact}).

\subsection{The all-order resummation formula and its NLL implementation}\label{resproof}

In this section, we show that the resummed expression of Eq.(\ref{fact}) is obtained by simply replacing
the coefficient function $C\left(N,\frac{Q^{2}}{\mu^{2}},\alpha_{s}(\mu^{2}),M\right)$ with its integral over $y$, resummed to
the desired logarithmic accuracy. This is equivalent to saying that the factor of $\cos(My)$ in  Eq.(\ref{cosfact}) is irrelevant in
the large-N limit. Indeed, one can expand $\cos(My)$ in powers of $y$,
\bea
\cos(My)=1-\frac{M^{2}y^{2}}{2}+\emph{O}(M^{4}y^{4}).\label{expansion}
\eea
and observe that the first term of this expansion leads to a convergent integral (the rapidity-integrated cross section), while
the subsequent terms are suppressed by powers of $(1-z)$, since the upper integration bound in Eq.(\ref{cosfact}) is
\bea
\ln\frac{1}{\sqrt{z}}=\frac{1}{2}(1-z)+\emph{O}((1-z)^2).
\eea
Hence, up to terms suppressed
by factors $1/N$, Eq.(\ref{mfcoefffunc2}) is equal to the Mellin transform of the rapidity-integrated Drell-Yan coefficient function that we call
$C_{I}(N,Q^{2}/\mu^{2},\alpha_{s}(\mu^{2}))$.
This completes our proof. We get
\bea\label{rescrosssec}
\sigma^{res}(N,Q^{2},M)&=&F_{1}^{H_{1}}(N+iM/2,\mu^{2})F_{2}^{H_{2}}(N-iM/2,\mu^{2})\nonumber\\
&\times&C_{I}^{res}\left(N,\frac{Q^{2}}{\mu^{2}},\alpha_{s}(\mu^{2})\right).\label{rescrosssec}
\eea

This theorethical result is very important: it shows that, near threshold, the Mellin-Fourier transform of the coefficient function
does not depend on the Fourier moments and that this is valid to all orders of QCD perturbation theory.
Furthermore this result remains valid for all values of hadronic
center-of-mass rapidity, because we have introduced a suitable integral transform over rapidity.
The resummed rapidity-integrated Drell-Yan coefficient function to NLL order has been studied in Section \ref{nllresummation}. It is given
 by
\bea\label{explform}
C_{I}^{res}\left(N,\frac{Q^{2}}{\mu^{2}},\alpha_{s}(\mu^{2})\right)=\left[\exp\{\ln N g_{1}(\lambda,2)+g_{2}(\lambda,2)\}\right]_{\mu^{2}_{r}=\mu^{2}}
\eea
where $\lambda=b_{0}\alpha_{s}(\mu^{2}_{r})\ln N$ and where the resummation functions $g_{1}(\lambda,2)$ and $g_{2}(\lambda,2)$ are given in
Eqs.(\ref{g1},\ref{g2}) of Section \ref{nllresummation} with the resummation coefficients in the $\overline{MS}$ scheme given in
Eq.(\ref{bah}) of Section \ref{eiconalapp}.

Now, we want to arrive to a NLO and NLL expression of the rapidity-dependent dimensional cross section. This is achieved firstly taking the Mellin and
Fourier inverse transforms of $\sigma^{res}(N,Q^{2},M)$ Eq.(\ref{rescrosssec}) in order to turn back to the variables $x$ and $Y$:
\bea\label{inverse}
\sigma^{res}(x,Q^{2},Y)=\int_{-\infty}^{\infty}\frac{dM}{2\pi}e^{-iMY}\int_{C-i\infty}^{C+i\infty}\frac{dN}{2\pi i}\,x^{-N}\sigma^{res}(N,Q^{2},M).
\eea
In principle the contour in the complex $N$-space of the inverse Mellin transform in Eq.(\ref{inverse}) has to be chosen in such a way that
the intersection of $C$ with the real axis
lies to the right of all the singularities of the integrand. In practice, this is not possible, because
the resummed coefficient function Eqs.(\ref{g1},\ref{g2}) of section \ref{nllresummation} has a branch cut on the real positive axis for
\bea\label{lp}
N\geq N_{L}\equiv e^{\frac{1}{ab_{0}\alpha_{s}(Q^{2})}},
\eea
which corresponds to the Landau singularity of $\alpha_{s}(Q^{2}/N^{a})$ (see Eq.(\ref{tls}) in Appendix \ref{appA}).
This is due to the fact that if the $N$-space expression is expanded in powers of $\alpha_{s}$, and
the Mellin inversion is performed order by order, a divergent series is obtained.
The ``Minimal Prescription'' proposed in \cite{Catani:1996yz} gives a
well defined formula to obtain the resummed result in $x$-space to which the divergent series is asymptotic
and is simply obtained choosing $C=C_{MP}$ in such a way that all the poles
of the integrand are to the left, except the Landau pole Eq.(\ref{lp}).
Recently, another method has been prposed in Ref.\cite{Forte:2006mi}.
Here, we will adopt the ``Minimal Prescription '' formula, deforming the contour
in order to improve numerical convergence and to avoid the singularities of the parton densities of Eq.(\ref{rescrosssec})
which are transated out of the real axis by $\pm iM/2$.
Hence, we perform the $N$-integral in Eq.(\ref{inverse}) over a curve $\Gamma$ given by:
\begin{eqnarray}
\Gamma&=&\Gamma_{1}+\Gamma_{2}+\Gamma_{3}\\
\Gamma_{1}(t)&=&C_{MP}-i\frac{M}{2}+t(1+i),\quad t\in (-\infty,0)\\
\Gamma_{2}(s)&=&C_{MP}+is\frac{M}{2},\quad s\in (-1,1)\\
\Gamma_{3}(t)&=&C_{MP}+i\frac{M}{2}-t(1-i),\quad t\in (0,+\infty)
\end{eqnarray}
The double inverse tranform of Eq.(\ref{inverse}) over the curve $\Gamma$ then becomes:
\bea\label{trasff}
\sigma^{res}(x,Q^{2},Y)=\frac{1}{\pi}\int_{0}^{1}\frac{dm}{m}\cos(-Y\ln m)\sigma^{res}(x,Q^{2},-\ln m),
\eea
where we have done the change of variable$M=-\ln m$.
The factor $\sigma^{res}(x,Q^{2},M)$ of the integrand in Eq.(\ref{trasff}) is given by
\begin{eqnarray}\label{trasfm}
&&\qquad\qquad\qquad\qquad\qquad\qquad\sigma^{res}(x,Q^{2},M)=\\
&&\frac{1}{\pi}\int_{0}^{1}\frac{ds}{s}\Re\bigg[x^{-C_{MP}-\ln s+i(M/2+1)}\sigma^{res}(C_{MP}+\ln s-i(M/2+1),Q^{2},M)\nonumber\\
&&\times(1-i)+\frac{sM}{2}x^{-C_{MP}-isM/2}\sigma^{res}(C_{MP}+isM/2, Q^{2},M)\bigg],\nonumber
\end{eqnarray}
where we have done another change of variables ($t=-\ln s$). Eqs.(\ref{trasff},\ref{trasfm}) are the expressions that we use to
evaluate numerically the resummed adimensional cross section in the variables $x$ and $Y$ Eq.(\ref{inverse}).
Furthermore, we need to know the analytic continuations to all the complex plane of the parton densities
at the scale $\mu^{2}$ in Eq.(\ref{rescrosssec}). Here, we need to evolve up a partonic fit taken at a certain
scale solving the DGLAP evolution equations in Mellin space. The solution of the evolution equations
is given in Section \ref{appB}.

Finally, we want to obtain a NLO determination of the cross section improved
with NLL resummation. In order to do this, we must keep the resummed dimensionaless part of the cross
section Eq.(\ref{trasff}), multiply it by the correct dimensional
prefactors Eqs.(\ref{gamma},\ref{W},\ref{Z}) and parton densities, add to the resummed part the full NLO cross section and subtract the
double-counted logarithmic enhanced contributions. Thus, we have
\bea\label{matching}
\frac{d\sigma}{dQ^{2}dY}=\frac{d\sigma^{NLO}}{dQ^{2}dY}+\frac{d\sigma^{res}}{dQ^{2}dY}-\left[\frac{d\sigma^{res}}{dQ^{2}dY}\right]_{\alpha_{s}=0}-
\alpha_{s}\left[\frac{\partial}{\partial\alpha_{s}}\left(\frac{d\sigma^{res}}{dQ^{2}dY}\right)\right]_{\alpha_{s}=0}.
\eea
The first term is the full NLO cross section given in \cite{Mukherjee:2006uu,Gehrmann:1997ez,Sutton:1991ay,Altarelli:1979ub}.
We report the complete expression in Appendix \ref{appC}.

The third and the fourth terms in Eq.(\ref{matching}) are obtained in the same way as the second one, but with the substitutions
\begin{eqnarray}
C_{I}^{res}\left(N,\frac{Q^{2}}{\mu^{2}},\alpha_{s}(\mu^{2})\right)&\rightarrow& 1,\label{matchtrm}\\
C_{I}^{res}\left(N,\frac{Q^{2}}{\mu^{2}},\alpha_{s}(\mu^{2})\right)&\rightarrow& \alpha_{s}(\mu^{2})
2A_{1}\left\{\ln^{2}N+\ln N\left[2\gamma_{E}-\ln\left(\frac{Q^{2}}{\mu^{2}}\right)\right]\right\}\nonumber,
\end{eqnarray}
respectively.
The terms that appear in the second in the second line of  Eqs.(\ref{matchtrm}) are exactly the $\emph{O}(\alpha_{s})$
logarithmic enhanced contributions in the $\overline{MS}$ scheme.

We note that the resummed cross section Eq.(\ref{matching}) is relevant  even when the variable $x$ is not large.
In fact, the cross section can get the dominant
contributions from the integral in Eq.(\ref{universal}) for values of $z$ Eq.(\ref{z}) that are near the threshold even
when $x$ is not close to one, because
of the strong
suppression of parton densities $F_{i}(x_{i},\mu^{2})$ when $x_{i}$ are large.

\subsection{NLL impact of resummation at E866 experiment}

To show the importance of this resummation, we have calculated the Drell-Yan rapidity distribution for
proton-proton collisions at the Fermilab fixed-target experiment E866/NuSea \cite{Webb:2003ps}.
The center-of-mass energy has been fixed at $\sqrt{S}=38.76\, \textrm{GeV}$ and the
invariant mass of the virtual photon $\gamma^{*}$  has been chosen to be $Q^{2}=64\,
\textrm{GeV}^{2}$ in analogy with \cite{Anastasiou:2003yy}. Clearly the
contribution of the virtual $Z^{0}$ can be neglected, because its mass is
much bigger than $Q^{2}$. In this case $x=0.04260$ and the upper and lower bound of the
hadronic rapidity $Y$ Eq.(\ref{boundY}) are given by $\pm 1.57795$. We have evolved up the  MRST 2001 parton distributions
(taken at  $\mu^{2}=1\,\textrm{GeV}^{2}$)
in order to compare to Ref.\cite{Anastasiou:2003yy}, where the NNLO calculation is performed.
However, results obtained using more modern parton sets should not be very different.
The LO parton set is given in \cite{Martin:2002dr} with $\alpha_{s}^{LO}(m_{Z})=0.130$ and the NLO set is given in \cite{Martin:2001es} with
 $\alpha_{s}^{NLO}(m_{Z})=0.119$.
The evolution of parton densities at the scale $\mu^{2}$  has been performed in the variable
flavor number scheme. The quarks has been considered massless and, at the scale of the transition of the flavor number ($N_{f}\rightarrow N_{f}+1$),
the new flavor is generated
dynamically. The resummation formula Eq.(\ref{rescrosssec}) together with Eqs.(\ref{g1}-\ref{explform}) has been
used with the number of flavors $N_{f}=4$.

\begin{figure}
\begin{center}
\includegraphics[scale=0.56]{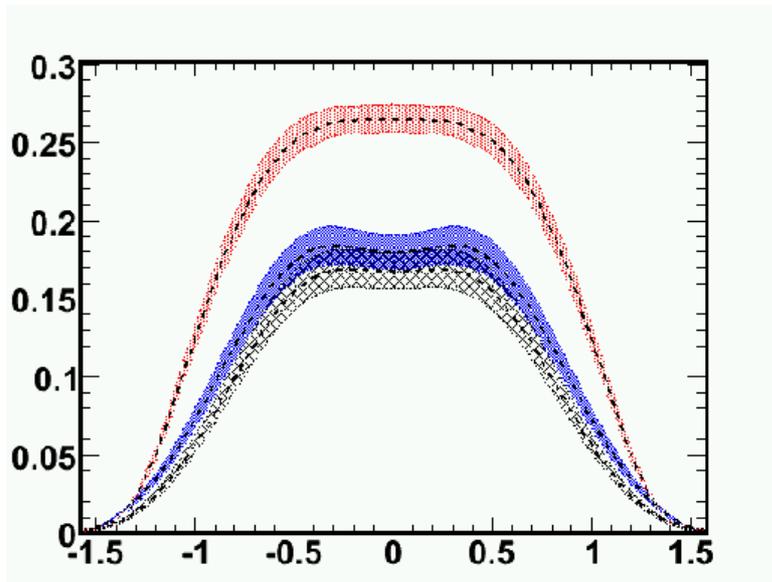}
\caption{\footnotesize{Y-dependence of $d^{2}\sigma/(dQ^{2}dY)$ in units of
$\textrm{pb}/\textrm{GeV}^2$. The curves are, from top to bottom, the NLO result (red band), the LO+LL
resummation (blue band) and the LO (black band). The bands are obtained varying
the factorization scale between $\mu^{2}= 2Q^{2}$ and $\mu^{2}=1/2 Q^{2}$.}}
\label{LOLLNLO}
\end{center}
\end{figure}

\begin{figure}
\begin{center}
\includegraphics[scale=0.6]{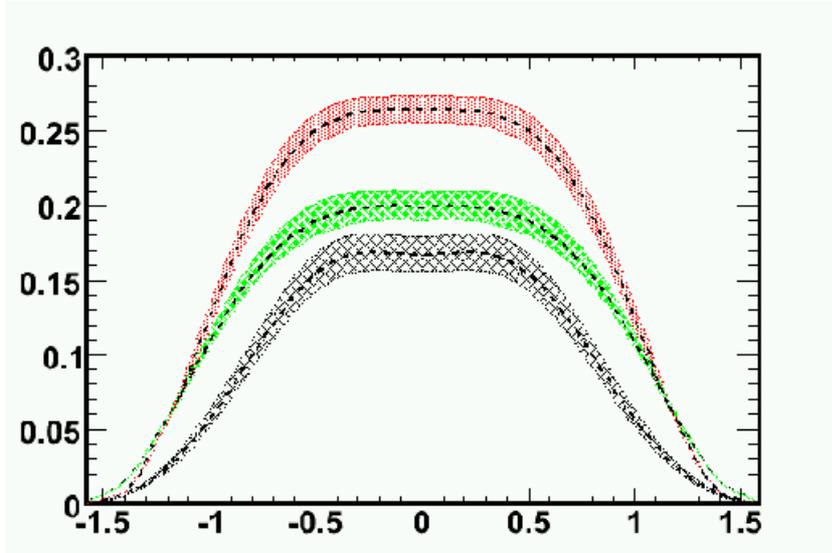}
\caption{\footnotesize{Y-dependence of $d^{2}\sigma/(dQ^{2}dY)$ in units of
$\textrm{pb}/\textrm{GeV}^2$. The curves are, from top to bottom, the NLO result (red band),
the NLO+NLL resummation (green band)
and the LO (black band). The bands are obtained as in figure \ref{LOLLNLO}.}}\label{LONLONLL}
\end{center}
\end{figure}

In figure \ref{LOLLNLO}, we plot  the rapidity-dependence of the cross section at LO, NLO and LO
improved with LL resummation. The effect of LL resummation
is small compared to the effect of the full NLO correction. We see that, at leading order, the impact
of the resummation is negligible in comparison to the NLO
fixed-order correction. This means that, at leading order, resummation is not necessary.

The LO, the NLO and its NLL improvement cross sections are shown in figure \ref{LONLONLL}.
The effect of the NLL resummation in the central rapidity region is almost as large as
the NLO correction, but it reduces the cross section instead of enhancing it for not large values of rapidity.
The origin of this suppression will be discussed in the next Section.
Going from the LO result to the NLO with NLL resummation,
we note a reduction of the dependence on the
factorization scale i.e. a reduction of the theoretical error.

\begin{figure}
\begin{center}
\includegraphics[scale=0.6]{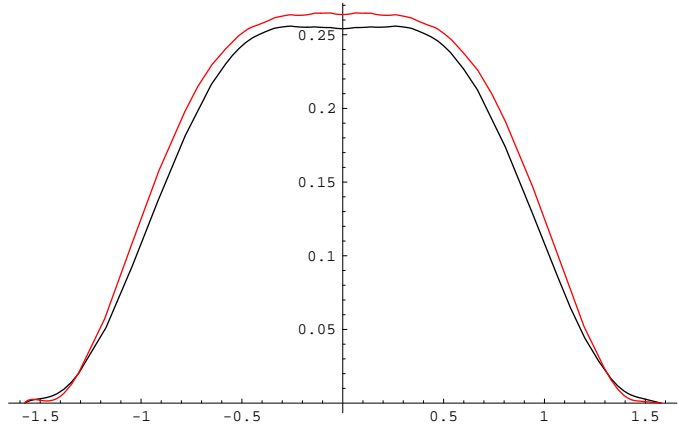}
\caption{\footnotesize{$d^{2}\sigma/(dQ^{2}dY)$ in units of
$\textrm{pb}/\textrm{GeV}^2$ for the full NLO computation (upper
red line) and for the leading terms of Eq.(\ref{lead}) (lower
black line). It has been calculated for one value of the
factorization scale $\mu^{2}=Q^{2}$.\label{fig}}}
\end{center}
\end{figure}

Now, we want to establish if the leading logarithmic terms that are included
in the resummed exptonent represent a good approximation to the exact fixed order computation.
Only if this is the case, we can believe that our resummation is reliable in
perturbative QCD.
In order to do this, we compare the full NLO DY rapidity cross section
with the one obtained including only the large-$N$ leading terms
of the coefficient function. For simplicity, we
choose the factorization scale $\mu^{2}$ to the scale of the process $Q^{2}$. The leading
large-$N$ coefficient function is given by:
\begin{equation}
C^{\rm
lead}(N,\alpha_{s}(Q^{2}))=1+\alpha_{s}(Q^{2})2A_{1}\left(\ln^{2}N+2\gamma_{E}\ln
N+\gamma_{E}^{2}-2+\frac{\pi^{2}}{3}\right)\label{lead},
\end{equation}
where we have added the constant terms at $\emph{O}(\alpha_{s})$ which are not resummed. For an
explicit derivation of these constant terms see for example Ref. \cite{Mukherjee:2006uu} Section 3.
We plot the result in Figure \ref{fig}. We see that the leading terms Eq.(\ref{lead}) represent
a good approximation to the exact NLO computation, because they account for more than 90\% of the
full NLO computation for all relevant rapidities.

In Figure \ref{fig2}, we plot only  the $\emph{O}(\alpha_{s})$
correction. Here we see that the $\emph{O}(\alpha_{s})$
contribution of Eq.(\ref{lead}) represent
a good approximation to the exact $\emph{O}(\alpha_{s})$ NLO contribution,
because it accounts for more than 80\% of the
full $\emph{O}(\alpha_{s})$ NLO correction for all relevant rapidities.

\begin{figure}
\begin{center}
\includegraphics[scale=0.7]{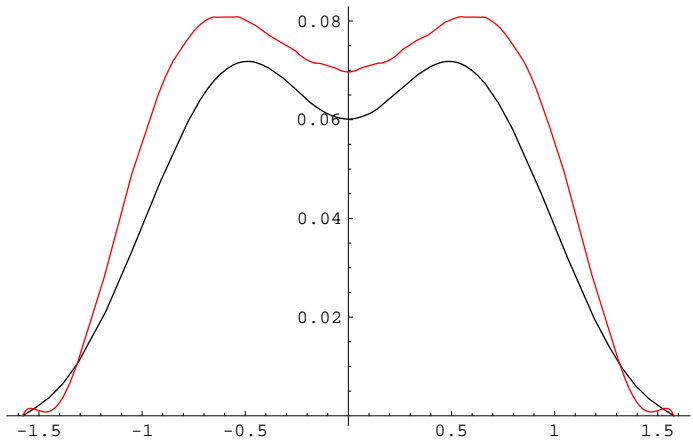}
\caption{\footnotesize{$d^{2}\sigma/(dQ^{2}dY)$ in units of
$\textrm{pb}/\textrm{GeV}^2$ for $\emph{O}(\alpha_{s})$ correction
of the full NLO computation (upper red line) and for the leading
$\emph{O}(\alpha_{s})$ term of Eq.(\ref{lead}) (lower black line).
\label{fig2}}}
\end{center}
\end{figure}

\begin{figure}
\begin{center}
\includegraphics[scale=0.6]{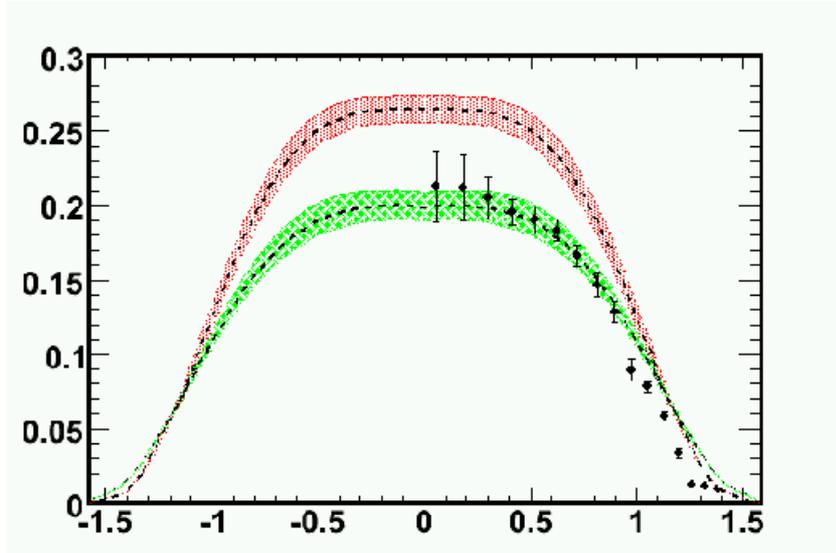}
\caption{\footnotesize{Y dependence of $d^{2}\sigma/(dQ^{2}dY)$ in units of
$\textrm{pb}/\textrm{GeV}^2$. The curves are, from top to bottom, the NLO result (red band) and the NLO+NLL resummation
(green band) together with the E866/NuSea data. The bands are obtained as in figure \ref{LOLLNLO}.}}
\label{NLONLLDT}
\end{center}
\end{figure}

In figure \ref{NLONLLDT}, we plot the experimental data of Ref.\cite{Webb:2003ps}
converted to the $Y$ variable
together with our NLO and NLL resummed predictions.

The data
in Ref.\cite{Webb:2003ps} are tabulated in invariant Drell-Yan pair mass $\sqrt{Q^{2}}$ and Feynman $x_{F}$
bins. To convert the data to the hadronic rapidity $Y$, we have used the definition of the Feynman $x_{F}$ which is
\begin{equation}\label{ciao}
x_{F}\equiv\frac{2Q_{z}}{\sqrt{S}}=\frac{2\sqrt{Q^{2}+Q_{\perp}^{2}}\sinh Y}{\sqrt{S}},
\end{equation}
where we have used Eq.(\ref{finalmente}).
Solving Eq.(\ref{ciao}) in $Y$ we have
\begin{equation}
Y=\ln\left[H+\sqrt{H^{2}+1}\right],\qquad\quad H=\frac{x_{F}\sqrt{S}}{2\sqrt{Q^{2}+Q_{\perp}^{2}}}.
\end{equation}
With this equation and with the aid of the $Q_{\perp}$ distribution, which is also given in Ref.\cite{Webb:2003ps},
we have converted the data from $x_{F}$ to $Y$. Furthermore, for each $x_{F}$ bin, we have done
the weighted average of three $\sqrt{Q^{2}}$ bins ($7.2\leq \sqrt{Q^{2}}\leq 7.7$;\, $7.7\leq\sqrt{Q^{2}}\leq 8.2$ and
$8.2\leq\sqrt{Q^{2}}\leq 8.7$ with the energies in GeV).

The agreement
with data is good and a great improvement  for not large rapidity is obtained with respect to the NLO calculation.
We note also that the NLL resummation gives better result than the NNLO calculation performed in \cite{Anastasiou:2003yy}.
The NNLO prediction has a worse agreement
with data than the NLO one for not large values of rapidity.
This result suggests that, for the case of rapidity distributions, NLL resummation is more important than high-fixed-order calculation
and that it can be so even at higher center-of-mass energies.

\section{The origin of suppression}

In this section, we shall show that the suppression of the cross section of the NLL correction with the parameter choices of
the experiment E866 is due to the shift in the complex plane of the dominant contribution of the resummed exponent. We shall do it
using a simplified toy-model.

Consider the collision of only two quarks with parton density
\begin{equation}
F(x)=(1-x)^{2}.
\end{equation}
Its Mellin transform is given by:
\begin{equation}
F(N)=\int_{0}^{1}dx\,x^{N-1}(1-x)^{2}=\frac{\Gamma(N)\Gamma(3)}{\Gamma(N+3)}=\frac{2}{N(N+1)(N+2)}.
\end{equation}
Furthermore, we take the double-log approximation (DLA) which is obtained performing the limit $\lambda\rightarrow 0$ in the resummed exponent
Eq.(\ref{explform}). Thus, in this simple model, the Mellin-Fourier transform of the NLL resummed cross section
Eq.(\ref{matching}) can be written in the following form:
\begin{equation}
\sigma(N,M)=\sigma^{FO}(N,M)+|F(N+iM/2)|^{2}\Delta\sigma^{\rm DLA}(N),
\end{equation}
where $\sigma^{FO}(N,M)$ are the exact NLO Mellin-Fourier moments and where
\begin{equation}
\Delta\sigma^{\rm DLA}(N)=\left[e^{\alpha_{s}2A_{1}\ln^{2}N}-1-\alpha_{s}2A_{1}\ln^{2}N\right].
\end{equation}
If there is a suppression, this means that the quantity
\begin{equation}
\sigma(N,M)-\sigma^{FO}(N,M)=\frac{4\Delta\sigma^{\rm DLA}(N)}{\left[N^{2}+\frac{M^{2}}{4}\right]\left[(N+1)^{2}+\frac{M^{2}}{4}
\right]\left[(N+2)^{2}+\frac{M^{2}}{4}\right]},
\end{equation}
should produce a negative contribution in performing the inverse Mellin and Fourier transform.
It is given by the integral
\begin{equation}
\int_{-\infty}^{\infty}\frac{dM}{2\pi}e^{-iMY}\int_{C-i\infty}^{C+i\infty}\frac{dN}{2\pi i}x^{-N}
\frac{4\Delta\sigma^{\rm DLA}(N)}{\left[N^{2}+\frac{M^{2}}{4}\right]\left[(N+1)^{2}+\frac{M^{2}}{4}
\right]\left[(N+2)^{2}+\frac{M^{2}}{4}\right]}.\label{invint}
\end{equation}
The integrand function of this expression has not only a cut on the negative real axis (as it happens in
the inclusive case), but also poles that are shifted in the complex plane:
\begin{equation}\label{poles}
-n\pm \frac{iM}{2};\qquad n=0,1,2.
\end{equation}
Because of the factor $x^{-N}$ in the inverse Mellin integral in Eq.(\ref{invint}), its dominant contribution comes from the poles with $n=0$
in Eq.(\ref{poles}). The contribution of the pole at $+iM/2$ is given by
\begin{equation}\label{stasecine}
\int_{-\infty}^{\infty}\frac{dM}{2\pi}e^{-iM(Y+\ln\sqrt{x})}\frac{4\Delta\sigma^{\rm DLA}(N=iM/2)}{iM(iM+1)(iM+2)},
\end{equation}
where
\begin{eqnarray}
\Delta\sigma^{\rm DLA}(N=iM/2)&=&\exp\left[\alpha_{s}2A_{1}\left(\ln^{2}\frac{|M|}{2}-\frac{\pi^{2}}{4}+i\pi\ln\frac{|M|}{2}\right)\right]
+\label{stasecine2}\\
&&-1-\alpha_{s}2A_{1}\left(\ln^{2}\frac{|M|}{2}-\frac{\pi^{2}}{4}+i\pi\ln\frac{|M|}{2}\right).\nonumber
\end{eqnarray}
The important thing to notice of this contribution is the fact that the imaginary pole has produced an oscillating prefactor
in front of the resummed exponent which, together with the oscillating factor of the Fourier inverse integral in Eq.(\ref{stasecine})
at zero hadronic rapidity $Y$, is given by
\begin{equation}\label{stasecine3}
\exp\left[i\left(\alpha_{s}2\pi A_{1}\ln\frac{|M|}{2}-iM\ln\sqrt{x}\right)\right].
\end{equation}
We note that, with the inclusion of the contribution of the other pole at $-iM/2$, the
real and immaginary part of Eq.(\ref{stasecine2}) contribute to the  integral of Eq.(\ref{invint}). The real and immaginary
part of Eq.(\ref{stasecine2}) are
given by:
\begin{eqnarray}
\Re[\Delta\sigma^{\rm DLA}(iM/2)]&=&e^{\alpha_{s}2A_{1}\left(\ln^{2}\frac{|M|}{2}-\frac{\pi^{2}}{4}\right)}
\cos\left(\alpha_{s}2\pi A_{1}\ln\frac{|M|}{2}\right)\label{festeggio}\\
&&-1-\alpha_{s}2A_{1}\left(\ln^{2}\frac{|M|}{2}-\frac{\pi^{2}}{4}\right)\nonumber\\
\Im[\Delta\sigma^{\rm DLA}(iM/2)]&=&e^{\alpha_{s}2A_{1}\left(\ln^{2}\frac{|M|}{2}-\frac{\pi^{2}}{4}\right)}
\sin\left(\alpha_{s}2\pi A_{1}\ln\frac{|M|}{2}\right)\label{festeggio1}\\
&&-\alpha_{s}2\pi A_{1}\ln\frac{|M|}{2}\nonumber
\end{eqnarray}
Now, to roughly estimate the effect of the oscillating factor in Eq.(\ref{stasecine}), we use the value of $M=M_{0}$ where the
phase of Eq.(\ref{stasecine3}) is stationary and is given by:
\begin{equation}
|M_{0}|=\frac{\alpha_{s}2\pi A_{1}}{|\ln\sqrt{x}|}.
\end{equation}
Substituting this in Eqs.(\ref{festeggio},\ref{festeggio1}), we find a suppression of about 35\% for the parameter choice
of E866 experiment and a suppression of about 10\% for the W boson production at RHIC with a center-of-mass energy of $\sqrt{S}=500\textrm{GeV}$
which has more or less the same value of $x$.
We should now recall that this is a rough estimation and that there is also the contribution of the cut on the negative real axis which
usually produces an enhancement. However,
comparing this estimation with the result for the W boson production at RHIC (see e.g. figure 1 and 2 in reference \cite{Mukherjee:2006uu}),
where resummation produces an enhancement of about 4\% we can believe that the ignored contribution in this section
produce an enhancement of about 15\%, thus giving a suppressio of about 20\% at E866 experiment.

\section{Resummation of prompt photon
rapidity distribution}

\subsection{General kinematics of prompt photon rapidity distribution}

Here, we consider the rapidity distribution of the prompt photon process discussed in chapter \ref{DP},
\begin{equation}
H_{1}(P_{1})+H_{2}(P_{2})\rightarrow \gamma(p_{\gamma})+X.
\end{equation}
Specifically, we are interested in the differential cross section $p_{\perp}^{3}\frac{d\sigma}{dp_{\perp}dY}(x_{\perp},p_{\perp}^{2},\eta_{\gamma})$,
 where as in section \ref{kinp}
of chapter \ref{DP} $p_{\perp}$
is the transverse momentum of the photon, $\eta_{\gamma}$ is its hadronic center-of-mass pseudorapidity
 and
\begin{equation}
x_{\perp}=\frac{4p_{\perp}^{2}}{S}.
\end{equation}
The pseudorapidity of the direct real photon
in the partonic center-of-mass frame $\hat{\eta}_{\gamma}$ is related to $\eta_{\gamma}$ through Eq.(\ref{etaetahat}) in section \ref{kinp}:
\begin{equation}\label{etaetahat2}
\hat{\eta}_{\gamma}=\eta_{\gamma}-\frac{1}{2}\ln\frac{x_{1}}{x_{2}}.
\end{equation}
Furthermore, as in chapter \ref{DP}, we use the following parametrizations of the photon and of the incoming partons' momenta
\begin{eqnarray}
p_{\gamma}&=&(p_{\perp}\cosh\hat{\eta}_{\gamma},\vec{p}_{\perp},p_{\perp}\sinh\hat{\eta}_{\gamma}),\\
p_{1}&=&x_{1}P_{1}=\frac{\sqrt{s}}{2}(1,\vec{0}_{\perp},1),\\
p_{2}&=&x_{2}P_{2}=\frac{\sqrt{s}}{2}(1,\vec{0}_{\perp},-1).
\end{eqnarray}
The transverse energy that the partons can transfer to the outgoing partons must be
less than the partonic center-of-mass energy $\sqrt{s}=\sqrt{x_{1}x_{2}S}$. This means that
\begin{equation}\label{encons}
z\cosh^{2}\hat{\eta}_{\gamma}\leq 1,
\end{equation}
where we have defined the parton scaling variable
\begin{equation}
z=\frac{Q^{2}}{s}=\frac{x_{\perp}}{x_{1}x_{2}},
\end{equation}
with, as in chapter \ref{DP}, $Q^{2}=4p_{\perp}^{2}$.
Eq.(\ref{encons}) implies that the upper and lower boundaries for the partonic
center-of-mass pseudorapidity are given by
\begin{equation}
\hat{\eta}_{-}\leq\hat{\eta}_{\gamma}\leq\hat{\eta}_{+},
\end{equation}
where
\begin{equation}
\hat{\eta}_{\pm}=\ln\left(\frac{1}{\sqrt{z}}\pm\sqrt{\frac{1}{z}-1}\right)=\pm\ln\left(\frac{1}{\sqrt{z}}+\sqrt{\frac{1}{z}-1}\right).
\end{equation}
Using Eq.(\ref{etaetahat2}), we can rewrite the transverse energy condition
Eq.(\ref{encons}) as a condition for the lower bound of
$x_{2}$:
\begin{equation}
x_{2}\geq\frac{x_{1}\sqrt{x_{\perp}}e^{-\eta_{\gamma}}}{2x_{1}-\sqrt{x_{\perp}}e^{\eta_{\gamma}}}\equiv x_{2}^{0}.
\end{equation}
Now, the requirement that $x_{2}\leq 1$ implies the lower bound for $x_{1}$:
\begin{equation}
x_{1}\geq \frac{\sqrt{x_{\perp}}e^{\eta_{\gamma}}}{2-\sqrt{x_{\perp}}e^{-\eta_{\gamma}}}\equiv x_{1}^{0}.
\end{equation}
The upper and lower bounds of the hadronic center-of-mass pseudorapidity can be found with the
obvious condition that $x_{1(2)}^{0}\leq 1$. In this way, we find,
\begin{equation}
\eta_{-}\leq\eta_{\gamma}\leq\eta_{+},
\end{equation}
where
\begin{equation}
\eta_{\pm}=\ln\left(\frac{1}{\sqrt{x_{\perp}}}\pm\sqrt{\frac{1}{x_{\perp}}-1}\right)=\pm\ln\left(\frac{1}{\sqrt{x_{\perp}}}+\sqrt{\frac{1}{x_{\perp}}-1}
\right).
\end{equation}

\subsection{Mellin-Fourier transform and all-order resummation}

The expression with the factorization of collinear singularities of this cross section in perturbative QCD is
\begin{eqnarray}
p_{\perp}^{3}\frac{d\sigma}{dp_{\perp}d\eta_{\gamma}}(x_{\perp},p_{\perp}^{2},\eta_{\gamma})&=&\sum_{a,b}\int_{x_{1}^{0}}^{1}dx_{1}\int_{x_{2}^{0}}^{1}
dx_{2}x_{1}F_{a}^{H_1}(x_1,\mu^2)
x_{2}F_{b}^{H_2}(x_2,\mu^2)\nonumber\\
&&\times C_{ab}\left(z,\frac{Q^{2}}{\mu^{2}},\alpha_{s}(\mu^{2}),\hat{\eta}_{\gamma}\right)\delta(x_{\perp}-zx_{1}x_{2}),\label{factdprap}
\end{eqnarray}
where $F_a^{H_1}(x_1,\mu^2)$, $F_b^{H_2}(x_2,\mu^2)$ are the
distribution functions of partons $a,b$ in the colliding hadrons and $\mu^{2}$ is the factorization
scale equal to the renormalization scale. $\hat{\eta}_{\gamma}$ is a function
of $\eta_{\gamma}$, $x_{1}$ and $x_{2}$ as defined by Eq.(\ref{etaetahat2}).
The coefficient function\phantom{a}
$C_{ab}(z,\frac{Q^2}{\mu^2},\as(\mu^2),\hat{\eta}_{\gamma})$ is defined in terms of
the partonic cross section for the process where partons $a$, $b$
are incoming as
\begin{equation}
C_{ab}\left(z,\frac{Q^2}{\mu^2},\as(\mu^2),\hat{\eta}_{\gamma}\right)
=p_\perp^3\frac{d\hat\sigma_{ab}}{dp_\perp d\eta_{\gamma}}.
\end{equation}
To allow the Mellin transform to deconvolute Eq.(\ref{factdprap}), we first perform the Fourier transform
with respect to $\eta_{\gamma}$, thus obtaining
\begin{eqnarray}
\sigma(N,Q^{2},M)&=&\int_{0}^{1}dx_{\perp}x_{\perp}^{N-1}\int_{\eta_{-}}^{\eta_{+}}d\eta_{\gamma}
p_{\perp}^{3}\frac{d\sigma}{dp_{\perp}d\eta_{\gamma}}(x_{\perp},p_{\perp}^{2},\eta_{\gamma})\\
&=&\sum_{a,b}F_{a}^{H_1}(N+1+iM/2,\mu^2)
F_{b}^{H_2}(N+1-iM/2,\mu^2)\nonumber\\
&&\times C_{ab}\left(N,\frac{Q^{2}}{\mu^{2}},\alpha_{s}(\mu^{2}),M\right)\label{factmfdp}.
\end{eqnarray}
where
\begin{eqnarray}
F_{c}^{H_{i}}(N+1\pm iM/2,\mu^{2})&=&\int_{0}^{1}dxx^{N\pm iM/2}F_{i}^{H_{i}}(x,\mu^{2}),\label{mfpd2}\\
C_{ab}\left(N,\frac{Q^{2}}{\mu^{2}},\alpha_{s}(\mu^{2}),M\right)&=&2\int_{0}^{1}dzz^{N-1}\int_{0}^{\hat{\eta}_{+}}d\hat{\eta}_{\gamma}
\cos(M\hat{\eta}_{\gamma})\nonumber\\
&\times&
C\left(z,\frac{Q^{2}}{\mu^{2}},\alpha_{s}(\mu^{2}),\hat{\eta}_{\gamma}\right).\label{mfcoefffunc3}
\end{eqnarray}
A resummed expression of Eq.(\ref{factmfdp}) in the threshold limit for the transverse energy ($z\rightarrow 1$ or equivalently $N\rightarrow\infty$)
is obtained in the same way as we have done at the beginning of section \ref{resproof}, since the upper
integration bound of $\hat{\eta}_{\gamma}$ in Eq.(\ref{mfcoefffunc3}) is
\begin{equation}
\hat{\eta}_{+}=\ln\left(\frac{1}{\sqrt{z}}+\sqrt{\frac{1}{z}-1}\right)=\sqrt{1-z}-(1-z)+\emph{O}((1-z)^{3/2}).
\end{equation}
Thus, up to terms suppressed by factors $1/N$, the resummed expression of Eq.(\ref{factmfdp}) is:
\begin{eqnarray}
\sigma^{res}(N,Q^{2},M)&=&\sum_{a,b}F_{a}^{H_1}(N+1+iM/2,\mu^2)
F_{b}^{H_2}(N+1-iM/2,\mu^2)\nonumber\\
&&\times C_{I\,ab}^{res}\left(N,\frac{Q^{2}}{\mu^{2}},\alpha_{s}(\mu^{2})\right)\label{factmfdp},
\end{eqnarray}
where $C_{I\,ab}^{res}$ is the resummed pseudorapidity-integrated coefficient function for the
prompt photon production for the subprocess which involves the initial partons $a,b$. These resummed
coefficient function has been studied in chapter \ref{DP} and in Refs. \cite{Catani:1998tm,Bolzoni:2006ky}.
This result is analogous to that of the Drell-Yan rapidity distributions case, in the sense that the resummed
formula is obtained through a translation of the parton densities' moments by $1\pm iM/2$ and the
pseudorapidity-integrated coefficient functions.

\chapter{Renormalization group resummation of  tranverse
distributions}\label{QT}

We prove the all-order exponentiation of soft logarithmic
corrections at small transverse momentum to the distribution of
Drell-Yan process. We apply the renormalization group approach
developed in the context of integrated cross sections. We show
that all large logs in the soft limit can be expressed in terms of
a single dimensional variable, and we use the renormalization
group to resum them. The resummed  result that we obtain is,
beyond the next-to-leading log accuracy, more general and less
predictive than those previously released. The origin of this
could be due to factorization properties of the cross section. The
understanding of this point is a work in progress.

\section{Drell-Yan distribution at small transverse momentum}

We consider the Drell-Yan process \bea
H_{1}(P_{1})+H_{2}(P_{2})\rightarrow \gamma^{*}(Q)+X(K), \eea and,
in particular, the differential cross section
$\frac{d\sigma}{dq_{\perp}^{2}dY}(Q^{2},q_{\perp}^{2},x_{1},x_{2})$,
where $q_{\perp}$ is the transverse momentum with respect to
colliding axis of the hadrons $H_{1}$ and $H_{2}$, $Q^{2}$ is the
virtuality of photon and $x_{1},x_{2}$ are useful dimensionless
variables, that, in terms of the hadronic center-of-mass squared
energy $S=(P_{1}+P_{2})^{2}$ and the photon center-of-mass
rapidity $Y$, are given by \bea\label{defv}
x_{1}=\sqrt{\frac{Q^{2}+q_{\perp}^{2}}{S}}e^{Y};\qquad
x_{2}=\sqrt{\frac{Q^{2}+q_{\perp}^{2}}{S}}e^{-Y}. \eea The
relation between these two variables and the fraction of energy
carried by the virtual photon is \bea
\frac{x_{1}+x_{2}}{2}=\frac{E_{\gamma^{*}}}{\sqrt{S}}. \eea
According to standard factorization of perturbative QCD, the
expression for the differential cross section is
\begin{eqnarray}\label{fatt0}
\frac{d\sigma}{dq_{\perp}^{2}dY}(Q^{2},q_{\perp}^{2},x_{1},x_{2})&=&\int_{z_{1}^{min}}^{1}dz_{1}\int_{z_{2}^{min}}^{1}
dz_{2}f_{1}(z_{1},\mu^{2})f_{2}(z_{2},\mu^{2})\nonumber\\
&&\times\frac{d\hat{\sigma}}{dq_{\perp}^{2}dy}(Q^{2},q_{\perp}^{2},s,y,\mu^{2},\alpha_{s}(\mu^{2})),
\end{eqnarray}
where $f_{1}(z_{1},\mu^{2}),f_{2}(z_{2},\mu^{2})$ are the parton
distribution functions of the colliding quark and anti-quark in
the hadrons $H_{1}$ and $H_{2}$ respectively. The arbitrary scale
$\mu^{2}$ is the factorization scale, which, for simplicity, is
chosen to be equal to the renormalization scale. The condition
that the invariant mass of the emitted particles $K^{2}$ cannot be
negative, imposes that $(z_{1}-x_{1})(z_{2}-x_{2})\geq
\frac{q_{\perp}^{2}}{S}$ and, taking the small $q_{\perp}^{2}$
limit, we obtain that \bea\label{cond} z_{1}^{min}=x_{1},\qquad
z_{2}^{min}=x_{2}. \eea The partonic center-of-mass squared energy
$s$ and  rapidity $y$ are related to the hadronic ones by a
scaling and a boost along the collision axis with respect to the
longitudinal momentum fraction $z_{1},z_{2}$ of the incoming
partons: \bea s=z_{1}z_{2}S; \qquad
y=Y-\frac{1}{2}\ln\frac{z_{1}}{z_{2}}. \eea We define analogous
variables to that of Eqs.(\ref{defv}) at the partonic level \bea
\xi_{1}\equiv\frac{x_{1}}{z_{1}}=\sqrt{\frac{Q^{2}+q_{\perp}^{2}}{s}}e^{y};\qquad
\xi_{2}\equiv\frac{x_{2}}{z_{2}}=\sqrt{\frac{Q^{2}+q_{\perp}^{2}}{s}}e^{-y},
\eea with inverse relations \bea
s=\frac{Q^{2}+q_{\perp}^{2}}{(x_{1}/z_{1})(x_{2}/z_{2})};\qquad
y=\frac{1}{2}\ln\frac{x_{1}/z_{1}}{x_{2}/z_{2}}. \eea Now, thanks
to these equations, we can define a dimensionless differential
cross section and coefficient function
\begin{eqnarray}
W(q_{\perp}^{2}/Q^{2},x_{1},x_{2})&=&\frac{Q^{4}}{x_{1}x_{2}}\frac{d\sigma}{dq_{\perp}^{2}dY}(Q^{2},q_{\perp}^{2},x_{1},x_{2}),\\
\hat{W}\left(\frac{Q^{2}}{\mu^{2}},\frac{q_{\perp}^{2}}{Q^{2}},\frac{x_{1}}{z_{1}},\frac{x_{2}}{z_{2}},\alpha_{s}(\mu^{2})\right)&=&
\frac{Q^{4}}{(x_{1}/z_{1})(x_{2}/z_{2})}
\frac{d\hat{\sigma}}{dq_{\perp}^{2}dy}(Q^{2},q_{\perp}^{2},s,y,\mu^{2},\alpha_{s}(\mu^{2})),\nonumber\\
\end{eqnarray} in such a way that Eq.(\ref{fatt0}), together with the
conditions Eqs.(\ref{cond}), takes the useful form of a
convolution product
\begin{eqnarray}\label{fatt}
W(q_{\perp}^{2}/Q^{2},x_{1},x_{2})&=&\int_{x_{1}}^{1}\frac{dz_{1}}{z_{1}}\int_{x_{2}}^{1}
\frac{dz_{2}}{z_{2}}f_{1}(z_{1},\mu^{2})f_{2}(z_{2},\mu^{2})\nonumber\\
&&\times\hat{W}(Q^{2}/\mu^{2},q_{\perp}^{2} /Q^{2},
x_{1}/z_{1},x_{2}/z_{2},\alpha_{s}(\mu^{2})),
\end{eqnarray}
which is valid only for small $q_{\perp}^{2}$.

\section{The role of standard factorization}

It is known that this expression (or equivalently
Eq.(\ref{fatt0})) is originated by the factorization of collinear
divergences in the impact parameter ($\vec{b}$) which is conjugate
upon Fourier transformation to the transverse momentum
($\vec{q}_{\perp}$): \bea W(Q^{2}b^{2},x_{1},x_{2})=\int
d^{2}q_{\perp}e^{i\vec{q}_{\perp}\vec{b}}W(q_{\perp}^{2}/Q^{2},x_{1},x_{2}).
\eea
In $d=4-2\epsilon$ dimensions this factorization has the form
\bea\label{fattcoll}
&&\qquad \qquad  \qquad \qquad \hat{W}(Q^{2}/\mu^{2},Q^{2}b^{2},x_{1},x_{2},\alpha_{s}(\mu^{2}))=\nonumber \\
&=&\int_{x_{1}}^{1}\frac{dz_{1}}{z_{1}}\int_{x_{2}}^{1}\frac{dz_{2}}{z_{2}}Z(z_{1},\alpha_{s}(\mu^{2}),\epsilon)
Z(z_{2},\alpha_{s}(\mu^{2}),\epsilon)\hat{W}^{(0)}(Q^{2},b^{2},\frac{x_{1}}{z_{1}},\frac{x_{2}}{z_{2}},\az,\epsilon).
\eea Note that the universal function $Z$ that extracts the
collinear divergences from the bare coefficient function doesn't
depend on the Fourier conjugate ($b$) of the transverse momentum
Ref.\cite{Collins:1989gx}. In Fourier space Eq.(\ref{fatt}),
becomes
\begin{eqnarray}
W(Q^{2}b^{2},x_{1},x_{2})&=&\int_{x_{2}}^{1}\frac{dz_{1}}{z_{1}}\int_{x_{2}}^{1}\frac{dz_{2}}{z_{2}}f_{1}(z_{1},\mu^{2})f_{2}(z_{2},\mu^{2})\nonumber\\
&&\times\hat{W}(Q^{2}/\mu^{2}, Q^{2}b^{2},
x_{1}/z_{1},x_{2}/z_{2},\alpha_{s}(\mu^{2}))
\end{eqnarray}
 Furthermore, Eq.(\ref{fatt}) tells us that the differential
cross section is a convolutional product which is diagonalized by
a double Mellin transform. Thus, performing the double Mellin and
Fourier transform, the coefficient function takes the simple
factorized form: \bea\label{fattmellin}
W(Q^{2}b^{2},N_{1},N_{2})=f_{1}(N_{1},\mu^{2})f_{2}(N_{2},\mu^{2})\hat{W}(Q^{2}/\mu^{2},Q^{2}b^{2},N_{1},N_{2},\alpha_{s}(\mu^{2})).
\eea Our goal is to resum the large logarithms $\ln Q^{2}b^{2}$ to
all logarithmic orders. These logs are present to all orders in
the contributions to this differential cross section. They come
from the kinematical region of soft and collinear emissions.
However, we know from Eq.(\ref{fattcoll}) that collinear
divergences that arises in the limit $q_{\perp}\rightarrow 0$  are
absorbed in parton distribution function evolution. Consequently,
we will resum only the large logarithms $\ln Q^{2}b^{2}$ that come
from soft contributions.
\\

We define the usual physical anomalous dimension: \bea
Q^{2}\frac{\partial}{\partial
Q^{2}}W(Q^{2}b^{2},N_{1},N_{2})=\gamma_{(W)}(Q^{2}b^{2},N_{1},N_{2},\alpha_{s}(Q^{2}))W(Q^{2}b^{2},N_{1},N_{2}).
\eea It is clear that
$\gamma_{(W)}(Q^{2}b^{2},N_{1},N_{2},\alpha_{s}(Q^{2}))$ is a
renormalization group invariant and we will show that it is also
independent of $N_{1}$ and $N_{2}$ when choosing the arbitrary
scale $\mu^{2}$ equal to $1/b^{2}$ and taking into account only
soft contributions. Thus, in the soft limit and with the
convenient choice $\mu^{2}=1/b^{2}$, we can write \bea
\gamma_{(W)}^{SOFT}(Q^{2}b^{2},N_{1},N_{2},\alpha_{s}(Q^{2}))=\gamma(1,Q^{2}b^{2},\alpha_{s}(Q^{2}))
\eea So, the resummed expression for the cross section
Eq.(\ref{fatt}) in Fourier space, in which the collinear
contributions to the large $\ln Q^{2}b^{2}$ are separated from the
soft ones, has the general form:
\begin{eqnarray}
W^{res}(Q^{2}b^{2},x_{1},x_{2})
&=&\int_{x_{2}}^{1}\frac{dz_{1}}{z_{1}}\int_{x_{2}}^{1}\frac{dz_{2}}{z_{2}}f_{1}(z_{1},1/b^{2})f_{2}(z_{2},1/b^{2})
K^{res}(b^{2},Q_{0}^{2},Q^{2})\nonumber\\
&&\times\hat{W}^{res}(Q_{0}^{2}b^{2},Q_{0}^{2}b^{2},
x_{1}/z_{1},x_{2}/z_{2},\alpha_{s}(1/b^{2})),\label{resummedexp}
\end{eqnarray}
where \bea\label{resummedexp3}
K^{res}(Q^{2}b^{2},Q_{0}^{2},Q^{2})=\exp\left\{\int_{Q_{0}^{2}}^{Q^{2}}\frac{d\bar{\mu}^{2}}{\bar{\mu}^{2}}
\Gamma^{res}(\frac{Q^{2}}{\bar{\mu}^{2}},\bar{\mu}^{2}b^{2},\alpha_{s}(\bar{\mu}^{2}))\right\},
\eea
 The scale $Q_{0}^{2}$, must be larger than the lower limit of the perturbative analysis ($Q_{0}^{2}>\Lambda_{QCD}^{2}$).
 Hence, in order to absorb the possible large correction
of the type $\ln Q_{0}^{2}b^{2}$ we will always choose
$Q_{0}^{2}=1/b^{2}$. Accordingly, the condition
$Q_{0}^{2}>\Lambda_{QCD}^{2}$ becomes $b^{2}<1/\Lambda_{QCD}^{2}$
and the resummed exponent of Eq.(\ref{resummedexp3}) is related to
the resummed physical anomalous dimension $\gamma^{res}$ through
the logarithmic derivative: \bea\label{rel}
\gamma^{res}(1,Q^{2}b^{2},\alpha_{s}(Q^{2}))=Q^{2}\frac{\partial}{\partial
Q^{2}}\int_{1/b^{2}}^{Q^{2}}\frac{d\bar{\mu}^{2}}{\bar{\mu}^{2}}\Gamma^{res}(\frac{Q^{2}}{\bar{\mu}^{2}},\bar{\mu}^{2}b^{2},\alpha_{s}(\bar{\mu}^{2})).
\eea

 It is now clear that resummation of collinear emissions is realized by the parton distribution
 evolution thanks to the fact that the factorization scale
$\mu^{2}$  is arbitrary. Resummation of soft gluon emissions can
be achieved by the resummation of the exponent that appears in
this expression. This is the subject of the next section.
\\

\section{The $q_{\perp}^{2}$ singularities of soft gluon contributions}

We now proceed through the calculation of the resummed exponent
using kinematics analysis and renormalization group improvement.
According to the Appendix \ref{phasespacedec}, the phase space
measure in $d=4-2\epsilon$ dimensions for $n$ extra emissions of
the partonic Drell-Yan subprocess can be written for $n=0$ and
$n\geq 1$ respectively as:
\begin{eqnarray}\label{n0}
\frac{d\phi_{1}(p_{1}+p_{2};q)}{dq_{\perp}^{2}dy}&=&\frac{1}{Q^{4}}\delta(1-\xi_{1})\delta(1-\xi_{2})\delta(\hat{q}_{\perp}^{2})\\
\label{n1}\frac{d\phi_{n+1}(p_{1}+p_{2};q,k_{1},\dots
,k_{n})}{dq_{\perp}^{2}dy}&=&N(\epsilon)(q_{\perp}^{2})^{-\epsilon}\int_{0}^{(\sqrt{s}-\sqrt{Q^{2}})^{2}}
\frac{dM^{2}}{2\pi}d\phi_{n}(k;k_{1},\dots ,k_{n})\nonumber\\
&&\times\delta(M^{2}-M_{0}^{2});\nonumber\\
&&\nonumber\\
k^{2}=M^{2};\qquad
M_{0}^{2}&=&\frac{Q^{2}}{\xi_{1}\xi_{2}}[(1-\xi_{1})(1-\xi_{2})+\hat{q}_{\perp}^{2}(1-\xi_{1}-\xi_{2})],
\end{eqnarray}
where $N(\epsilon)=1/(2(4\pi)^{2-2\epsilon})$,
$\xi_{i}=x_{i}/z_{i}$, $\hat{q}_{\perp}^{2}=q_{\perp}^{2}/Q^{2}$
and $d\Omega^{n-1}(\epsilon)$ stands for the integration of $n-1$
dimensionless variables ($z_{i}, i=1,\dots ,n-1$).  $\xi_{1}$ and
$\xi_{2}$ are related to the partonic center-of-mass rapidity (y)
and energy (s) by the relations: \bea
s=\frac{Q^{2}+q_{\perp}^{2}}{\xi_{1}\xi_{2}};\qquad \qquad
y=\frac{1}{2}\ln\frac{\xi_{1}}{\xi_{2}}. \eea

The phase space measure $d\phi_{n}(k;k_{1},\dots ,k_{n})$ is the
same as the phase space measure of the DIS process with an
incoming momentum with a nonzero invariant mass ($k^{2} =M^{2}$)
and $n$ outgoing massless particles. This phase space has been
analyzed in Section \ref{kindis} and is given by,
\begin{eqnarray}
d\phi_{1}(k;k_{1})&=&2\pi\delta(M^{2}), \qquad n=0\\
d\phi_{n}(k;k_{1},\dots
,k_{n})&=&2\pi\left[\frac{N(\epsilon)}{2\pi}\right]^{n-1}(M^{2})^{n-2-(n-1)\epsilon}d\Omega^{n-1}(\epsilon),\qquad
n\geq 1.
\end{eqnarray}
According to this, we can rewrite Eqs.(\ref{n0}),(\ref{n1}) in
this form:
\begin{eqnarray}
\frac{d\phi_{1}(p_{1}+p_{2};q)}{dq_{\perp}^{2}dy}&=&\frac{1}{Q^{4}}\delta(1-\xi_{1})\delta(1-\xi_{2})\delta(\hat{q}_{\perp}^{2})\\
\frac{d\phi_{2}(p_{1}+p_{2};q,k_{1})}{dq_{\perp}^{2}dy}&=&(q_{\perp}^{2})^{-\epsilon}N(\epsilon)\delta(M_{0}^{2})\\
M_{0}^{2}&=&\frac{Q^{2}}{\xi_{1}\xi_{2}}[(1-\xi_{1})(1-\xi_{2})+\hat{q}_{\perp}^{2}(1-\xi_{1}-\xi_{2})].\label{m0}
\end{eqnarray}
for $n=0,1$ respectively, and
\begin{equation}
\frac{d\phi_{n+1}(p_{1}+p_{2};q,k_{1},\dots
,k_{n})}{dq_{\perp}^{2}dy}=(q_{\perp}^{2})^{-\epsilon}2\pi\left[\frac{N(\epsilon)}{2\pi}\right]^{n}(M_{0}^{2})^{n-2-(n-1)\epsilon}
d\Omega^{n-1}(\epsilon)
\end{equation}
for $n\geq 2$. The dependence of the phase space on
$\hat{q}_{\perp}^{2}$ comes entirely from the factors: \bea
&&(q_{\perp}^{2})^{-\epsilon}\delta(M_{0}^{2}),\qquad n=1\label{pssingle}\\
&&(M_{0}^{2})^{n-1}(q_{\perp}^{2})^{-\epsilon}(M_{0}^{2})^{-(n-1)\epsilon-1},\qquad
n\geq 1. \label{psmultiple}\eea

The phase space measure must be combined with the square modulus
of the amplitude, in order to determine the logarithmic
singularities in $q_{\perp}=0$ which are regularized in
$d=4-2\epsilon$ dimensions. Studying the behavior of the
invariants that can be constructed with the external momenta, we
can  establish in which kinematical region the square modulus of
the amplitude can be singular in $\hat{q}_{\perp}^{2}\rightarrow
0$. From the study of the DIS-like emissions (see Section
\ref{kindy}) we know that \bea
k_{i}^{0}&=&\frac{\sqrt{M_{0}^{2}}}{2}(z_{n-1}\cdots z_{i+1})^{1/2}(1-z_{i}),\qquad  1\leq i \leq n-2\\
k_{n-1}^{0}&=&\frac{\sqrt{M_{0}^{2}}}{2}(1-z_{n-1})\\
k_{n}^{0}&=&k_{1}^{0}. \eea This means that all the invariants
that can appear in the function $D_{G}(\beta,P_{E})$ in
Eq.(\ref{amplitude}) of Section \ref{kindis} can be expressed in
terms of the following ones:  \bea
q^{2}&=&Q^{2},\qquad p_{1}^{2}=p_{2}^{2}=k_{i}^{2}=0,\qquad p_{1}\cdot p_{2}=\frac{s}{2}\\
k_{i}\cdot k_{j}&\sim& M_{0}^{2}= \frac{Q^{2}}{\xi_{1}\xi_{2}}[(1-\xi_{1})(1-\xi_{2})+\hat{q}_{\perp}^{2}(1-\xi_{1}-\xi_{2})]\label{integer}\\
p_{1}\cdot k_{i}&\sim&p_{2}\cdot k_{i}\sim \sqrt{sM_{0}^{2}}\label{halfinteger}\\
&=&
Q^{2}\left[\frac{(1+\hat{q}_{\perp}^{2})}{\xi_{1}\xi_{2}}[(1-\xi_{1})(1-\xi_{2})+\hat{q}_{\perp}^{2}(1-\xi_{1}-\xi_{2})]\right]^{1/2}.
\eea

In the case $n=1$, the single emission squared amplitude at tree
level has a $1/q_{\perp}^{2}$ singularity for
$q_{\perp}\rightarrow 0$. This can be easily seen by an explicit
$\emph{O}(\alpha_{s})$ computation (see for example
Refs.\cite{Altarelli:1984pt,Ellis:1981hk,Ellis:1981nt}). Hence, in
the general case, we expect that in the $q_{\perp}\rightarrow 0$
limit the squared amplitude has the following behavior
\begin{equation}\label{amplitudeqt}
|A_{n+1}|^{2}\sim
\frac{1}{\hat{q}_{\perp}^{2}}(M_{0}^{2})^{n_{1}}(M_{0}^{2})^{n_{2}\epsilon}g_{_{
n_{1}n_{2}}}(\xi_{1},\xi_{2}),
\end{equation}
where $N$ and $k$ are integer or half-integer numbers (see
Eqs.(\ref{integer},\ref{halfinteger})). However, as discussed in
Section \ref{kindy}, here we will assume that only the integer
powers of $M_{0}^{2}$ contribute. Then, we know that phase space
contributes the factors of Eqs.(\ref{pssingle},\ref{psmultiple})
and, hence, Eq.(\ref{amplitudeqt}) implies that a generic
contribution to the coefficient function $\hat{W}^{(0)}$ has the
folowing structure:
\begin{eqnarray}
(\hat{q}_{\perp}^{2})^{-1-\epsilon}\delta(M_{0}^{2})(M_{0}^{2})^{n_{1}}(M_{0}^{2})^{n_{2}\epsilon}g_{_{
nn_{2}}}(\xi_{1},\xi_{2})
;\,\,\,\qquad\qquad\quad n=1,\label{ciaomaddy}\\
(\hat{q}_{\perp}^{2})^{-1-\epsilon}(M_{0}^{2})^{-1-(n-n_{2}'-1)\epsilon}(M_{0}^{2})^{n-1+n_{1}}g_{_{nn_{2}'}}(\xi_{1},\xi_{2});\quad
\quad n>1\label{ciaomaddy2},
\end{eqnarray}
where for the moment we do not care about the overall dimensional
factor. Now, we are interested in taking into account only the
$1/\hat{q}_{\perp}^{2}$ singularity, because more singular terms
are forbidden and less singular ones are suppressed. As proven in
Appendix \ref{appG}, in the limit $q_{\perp}^{2}\rightarrow 0$,
\begin{equation}\label{appa}
\delta(M_{0}^{2})=\frac{\xi_{1}\xi_{2}}{Q^{2}}\left[\frac{\delta(1-\xi_{1})}{(1-\xi_{2})_{+}}+
\frac{\delta(1-\xi_{2})}{(1-\xi_{1})_{+}}-\ln\hat{q}_{\perp}^{2}\delta(1-\xi_{1})
\delta(1-\xi_{2})\right]+\emph{O}(\hat{q}_{\perp}^{2}),
\end{equation}
and, for $\eta=-(n-n_{2}'-1)\epsilon$ (with $\epsilon<0$
$n-n_{2}'-1>0$),
\begin{eqnarray}\label{appb}
&&\qquad\qquad\quad[(1-\xi_{1})(1-\xi_{2})+\hat{q}_{\perp}^{2}(1-\xi_{1}-\xi_{2})]^{\eta-1}=\\
&&=(1-\xi_{1})^{\eta-1}(1-\xi_{2})^{\eta-1}-\frac{(\hat{q}_{\perp}^{2})^{\eta}}{\eta^{2}}
\delta(1-\xi_{1})\delta(1-\xi_{2})+\emph{O}(\hat{q}_{\perp}^{2}).\nonumber
\end{eqnarray}
Therefore, $n_{1}=-n+1$,\, $n_{2}'<n-1$. Furthermore, we note that
the terms not proportional to $\delta(1-\xi_{1})\delta(1-\xi_{2})$
are divergent (in the $q_{\perp}\rightarrow 0$ limit) due to
collinear emissions. These divergences are absorbed and resummed
by the Altarelli-Parisi evolution of the parton distributions
$f_{1}(z_{1},\mu^{2})$ and $f_{2}(z_{2},\mu^{2})$ when they are
evaluated at the scale $\mu^{2}=1/b^{2}$ in Fourier space (see
Eq.(\ref{resummedexp})). As a first conclusion, we obtain that the
contributions that must be resummed in the limit
$\hat{q}_{\perp}^{2}\rightarrow 0$ are those which are
proportional to
  $\delta(1-\xi_{1})\delta(1-\xi_{2})$ and, thus, belong to the kinematical region of only the soft extra emissions.
  Hence, we obtain that the soft part of the coefficient function that must be
resummed has, after the inclusion of loops, the following general
form:
\begin{equation}\label{coefff}
\hat{W}(Q^{2},q_{\perp}^{2},\xi_{1},\xi_{2},\az,\epsilon)=\sum_{n=0}^{\infty}\az^{n}
\hat{{W}}_{n}(Q^{2},q_{\perp}^{2},\az,\epsilon)\delta(1-\xi_{1})\delta(1-\xi_{2}),
\end{equation}
with
\begin{eqnarray}
 \hat{{W}}_{n}(Q^{2},q_{\perp}^{2},\az,\epsilon)&=&(Q^{2})^{-n\epsilon}\big[C^{(0)}_{n0}(\epsilon)\delta(\hat{q}_{\perp}^{2})+
 \sum_{k=2}^{n}C^{(0)}_{nk}(\epsilon)(\hat{q}_{\perp}^{2})^{-1-k\epsilon}\nonumber\\
&&+\sum_{k=1}^{n}C'^{(0)}_{nk}(\epsilon)(\hat{q}_{\perp}^{2})^{-1-k\epsilon}\ln\hat{q}_{\perp}^{2}\big],
\end{eqnarray}
where the factor of  $(Q^{2})^{-n\epsilon}$ has been introduced
for dimensional reasons. We, now, perform the double Mellin
transform and the Fourier transform using the fact that:
\begin{eqnarray}
\int d^{2}\hat{q}_{\perp}e^{i\hat{b}\cdot\hat{q}_{\perp}}\delta(\hat{q}_{\perp}^{2})&=&\pi,\\
\int d^{2}\hat{q}_{\perp}e^{i\hat{b}\cdot\hat{q}_{\perp}}(\hat{q}_{\perp}^{2})^{-1-k\epsilon}&=&\pi F_{k}(\epsilon)(\hat{b}^{2})^{k\epsilon},\\
\int
d^{2}\hat{q}_{\perp}e^{i\hat{b}\cdot\hat{q}_{\perp}}(\hat{q}_{\perp}^{2})^{-1-k\epsilon}\ln\hat{q}_{\perp}^{2}&=&-\pi
F_{k}(\epsilon)(\hat{b}^{2})^{k\epsilon}\ln\hat{b}^{2}-\pi
\frac{F'_{k}(\epsilon)}{k}(\hat{b}^{2})^{k\epsilon},\\
\hat{b}^{2}\equiv Q^{2}b^{2};\qquad &&
F_{k}(\epsilon)=-\frac{4^{-k\epsilon}}{k\epsilon}\frac{\Gamma(1-k\epsilon)}{\Gamma(1+k\epsilon)}.
\end{eqnarray}
According to this, Eq.(\ref{coefff}) has, after Mellin and Fourier
transform, this structure: \bea
\hat{W}^{(0)}(Q^{2},b^{2},\az,\epsilon)&=&\sum_{n=0}^{\infty}\sum_{k=0}^{n}\tilde{C}^{(0)}_{nk}(\epsilon)
[(Q^{2})^{-\epsilon}\az]^{n-k}[(1/b^{2})^{-\epsilon}\az]^{k}+\\
&&+\ln
Q^{2}b^{2}\sum_{n=1}^{\infty}\sum_{k=1}^{n}\tilde{{C}}^{'(0)}_{nk}(\epsilon)[(Q^{2})^{-\epsilon}\az]^{n-k}[(1/b^{2})^{-\epsilon}\az]^{k}\nonumber
\eea

\section{The resummed exponent in renormalization group
approach}\label{bordello}

At this point, we calculate the resummed exponent that appears in
Eq.(\ref{resummedexp3}):
\begin{eqnarray}\label{bareexp}
\int_{1/b^{2}}^{Q^{2}}\frac{d\bar{\mu}^{2}}{\bar{\mu}^{2}}
\Gamma^{(0)}(Q^{2},\bar{\mu}^{2},b^{2},\az,\epsilon)&=&\ln\left(\frac{\hat{W}^{(0)}(Q^{2},b^{2},\az,\epsilon)}
{\hat{W}^{(0)}(1/b^{2},b^{2},\az,\epsilon)}\right)=\\
&=&\left[\sum_{n=1}^{\infty}\sum_{k=0}^{n-1}E^{(0)}_{nk}(\epsilon)[(\bar{\mu}^{2})^{-\epsilon}\az]^{n-k}
[(1/b^{2})^{-\epsilon}\az]^{k}\right]_{\bar{\mu}^{2}=1/b^{2}}
^{\bar{\mu}^{2}=Q^{2}}\nonumber\\
&&+\ln\bigg(1+\ln
Q^{2}b^{2}\sum_{n=1}^{\infty}\sum_{k=1}^{n}\tilde{E}^{(0)}_{nk}(\epsilon)[(Q^{2})^{-\epsilon}\az]^{n-k}\nonumber\\
&&\times[(1/b^{2})^{-\epsilon}\az]^{k}\bigg),\nonumber
\end{eqnarray}
where the last term has not been expanded because we must take
into account that in the bare coefficient function
Eq.(\ref{coefff}) there is only one explicit logarithm. From the
explicit calculation to order $\emph{O}(\az)$, we find that \bea
E^{(0)}_{10}(\epsilon)&=&\frac{2\pi}{3}\frac{(4\pi)^{\epsilon}}{\Gamma(1-\epsilon)}\left(-\frac{2}{\epsilon^{2}}-\frac{3}{\epsilon}+\pi^{2}-8\right),\\
\tilde{E}^{(0)}_{11}(\epsilon)&=&\frac{4\pi}{3}\frac{(4\pi)^{\epsilon}}{\Gamma
(1-\epsilon)}F_{1}(\epsilon). \eea Now, we want to rewrite
Eq.(\ref{bareexp}) in a renormalized form. To do this, we use, as
expalined in Chapter \ref{DISDY2}, the fact that
$(Q^{2})^{-\epsilon}\az$ and $(1/b^{2})^{-\epsilon}\az$ are
renormalization group invariant. Consequently, we may write: \bea
(Q^{2})^{-\epsilon}\az&=&\alpha_{s}(Q^{2})Z^{(\alpha_{s})}(\alpha_{s}(Q^{2}),\epsilon),\\
(1/b^{2})^{-\epsilon}\az&=&\alpha_{s}(1/b^{2})Z^{(\alpha_{s})}(\alpha_{s}(1/b^{2}),\epsilon),
\eea where $Z^{(\alpha_{s})}(\alpha_{s}(\mu^{2}),\epsilon)$ has
multiple poles at $\epsilon=0$ and $\mu^{2}$ is the
renormalization scale which for simplicity has been chosen equal
to the factorization scale. Furthermore we note that the universal
functions
$Z^{(\hat{W})}(N_{1},\alpha_{s}(\mu^{2}),\epsilon)Z^{(\hat{W})}(N_{2},\alpha_{s}(\mu^{2}),\epsilon)$
that extract the collinear poles from the coefficient function
simplify in the first line of Eq.(\ref{bareexp}). Thus the
renormalized expression of Eq.(\ref{bareexp}) has the form:
\begin{eqnarray}\label{renexp}
\int_{1/b^{2}}^{Q^{2}}\frac{d\bar{\mu}^{2}}{\bar{\mu}^{2}}\Gamma(\frac{Q^{2}}{\bar{\mu}^{2}},\bar{\mu}^{2}b^{2},\alpha_{s}(\bar{\mu}^{2}),\epsilon)
&=&\left[\sum_{m=1}^{\infty}\sum_{n=0}^{n-1}E^{R}_{mn}(\epsilon)\alpha_{s}^{m-n}(\bar{\mu}^{2})\alpha_{s}^{n}(1/b^{2})\right]_{\bar{\mu}^{2}=1/b^{2}}
^{\bar{\mu}^{2}=Q^{2}}\\
&&+\ln\bigg(1+\ln
Q^{2}b^{2}\sum_{m=1}^{\infty}\sum_{n=1}^{m}\tilde{E}^{R}_{mn}(\epsilon)\nonumber\\
&&\times\alpha_{s}(Q^{2})^{m-n}\alpha_{s}^{n}(1/b^{2})\bigg).\nonumber
\end{eqnarray}
The resummed exponent is clearly pole-free and so we can exploit
the cancellation of the poles that could be present in the
coefficients $E^{R}_{mn}(\epsilon)$ and
$\tilde{E}^{R}_{mn}(\epsilon)$. Furthermore we want to perform a
comparison with previously released resummation formulae given in
\cite{Collins:1984kg,Ellis:1981hk}. In order to do these two
things we rewrite Eq.(\ref{renexp}) in terms of the renormalized
physical anomalous dimension
$\Gamma(\frac{Q^{2}}{\bar{\mu}^{2}},\bar{\mu}^{2}b^{2},\alpha_{s}(\bar{\mu}^{2}),\epsilon)$
and therefore, we calculate, according to Eq.(\ref{rel}), the
logarithmic derivative of Eq.(\ref{renexp}):
\begin{eqnarray}\label{gamma}
\gamma(1,\bar{\mu}^{2}b^{2},\alpha_{s}(\bar{\mu}^{2}),\epsilon)=\qquad\qquad\qquad\qquad\qquad\qquad\qquad\qquad\qquad\quad&&\\
=\beta^{(d)}(\alpha_{s}(\bar{\mu}^{2}))\frac{\partial}{\partial\alpha_{s}(\bar{\mu}^{2})}
\sum_{m=1}^{\infty}\sum_{n=0}^{n-1}E^{R}_{mn}(\epsilon)\alpha_{s}^{m-n}(\bar{\mu}^{2})\alpha_{s}^{n}(1/b^{2})\qquad\,\,&&\nonumber\\
\nonumber\\
+\frac{\sum_{m=1}^{\infty}\sum_{n=1}^{m}\tilde{E}^{R}_{mn}(\epsilon)\alpha_{s}(\bar{\mu}^{2})^{m-n}\alpha_{s}^{n}(1/b^{2})}{1+\ln
\bar{\mu}^{2}b^{2}\sum_{m=1}^{\infty}\sum_{n=1}^{m}\tilde{E}^{R}_{mn}(\epsilon)\alpha_{s}(\bar{\mu}^{2})^{m-n}\alpha_{s}^{n}(1/b^{2})}
\qquad\qquad\,\,&&\nonumber\\
\nonumber\\+\frac{\ln
\bar{\mu}^{2}b^{2}\beta^{(d)}(\alpha_{s}(\bar{\mu}^{2}))\partial /
\partial\alpha_{s}\sum_{m=1}^{\infty}\sum_{n=1}^{m}\tilde{E}^{R}_{mn}(\epsilon)\alpha_{s}(\bar{\mu}^{2})^{m-n}\alpha_{s}^{n}(1/b^{2})}
{1+\ln
\bar{\mu}^{2}b^{2}\sum_{m=1}^{\infty}\sum_{n=1}^{m}\tilde{E}^{R}_{mn}(\epsilon)\alpha_{s}(\bar{\mu}^{2})^{m-n}\alpha_{s}^{n}(1/b^{2})},&&\nonumber
\end{eqnarray}
where
\begin{equation}
\beta^{(d)}(\alpha_{s}(\bar{\mu}^{2}))=-\epsilon\alpha_{s}(\bar{\mu}^{2})+\beta(\alpha_{s}(\bar{\mu}^{2})),
\end{equation}
and
$\beta(\alpha_{s}(\bar{\mu}^{2}))=-\beta_{0}^{2}\alpha_{s}(\bar{\mu}^{2})+\emph{O}(\alpha_{s}^{3})$
is the usual four-dimensional $\beta$-function. Now, in order to
isolate the terms which contain the explicit logs from the rest we
add and subtract the term
\begin{equation}
\sum_{m=1}^{\infty}\sum_{n=1}^{m}\tilde{E}^{R}_{mn}(\epsilon)\alpha_{s}(\bar{\mu}^{2})^{m-n}\alpha_{s}^{n}(1/b^{2}).
\end{equation}
After this, we re-expand the various terms in powers of
$\alpha_{s}(\bar{\mu}^{2})$ and $\alpha_{s}(1/b^{2})$, but not in
powers of $\ln\bar{\mu}^{2}b^{2}$. The result that we find in this
way has the following structure:
\begin{eqnarray}
\int_{1/b^{2}}^{Q^{2}}\frac{d\bar{\mu}^{2}}{\bar{\mu}^{2}}
\Gamma(\frac{Q^{2}}{\bar{\mu}^{2}},\bar{\mu}^{2}b^{2},\alpha_{s}(\bar{\mu}^{2}),\epsilon)=
\int_{1/b^{2}}^{Q^{2}}\frac{d\bar{\mu}^{2}}{\bar{\mu}^{2}}\left(\sum_{m=1}^{\infty}\sum_{n=0}^{m}
\Gamma^{R}_{mn}(\epsilon)\alpha_{s}^{m-n}(\bar{\mu}^{2})\alpha_{s}^{n}(1/b^{2})\right)&&\nonumber\\
+\int_{1/b^{2}}^{Q^{2}}\frac{d\bar{\mu}^{2}}{\bar{\mu}^{2}}\left(\frac{\ln
\bar{\mu}^{2}b^{2}\sum_{m=2}^{\infty}\sum_{n=1}^{m-1}\tilde{\Gamma}^{R}_{mn}(\epsilon)\alpha_{s}^{m-n}(\bar{\mu}^{2})\alpha_{s}^{n}(1/b^{2})}
{1+\ln
\bar{\mu}^{2}b^{2}\sum_{m=1}^{\infty}\sum_{n=1}^{m}\tilde{E}^{R}_{mn}(\epsilon)\alpha_{s}^{m-n}(\bar{\mu}^{2})
\alpha_{s}^{n}(1/b^{2})}\right)&&
\end{eqnarray}
To show the cancellation of divergences we rewrite the integrand
separating off the $b^{2}$-independent terms as in Chapter
\ref{DISDY2} and Chapter \ref{DP}:
\begin{eqnarray}
\gamma(1,\bar{\mu}^{2}b^{2},\alpha_{s}(\bar{\mu}^{2}),\epsilon)&=&\hat{\Gamma}^{(c)}(\alpha_{s}(\bar{\mu}^{2}),\epsilon)+
\hat{\Gamma}^{(l)}(\alpha_{s}(\bar{\mu}^{2}),\alpha_{s}(1/b^{2}),\epsilon)\nonumber\\
&&+\hat{\Gamma}^{'(l)}(\ln\bar{\mu}^{2}b^{2},\alpha_{s}(\bar{\mu}^{2}),\alpha_{s}(1/b^{2}),\epsilon)\nonumber,
\end{eqnarray}
where
\begin{eqnarray}
\hat{\Gamma}^{(c)}(\alpha_{s}(\bar{\mu}^{2}),\epsilon)&=&\sum_{m=1}^{\infty}\Gamma^{R}_{m}(\epsilon)\alpha_{s}^{m}(\bar{\mu}^{2})\\
\hat{\Gamma}^{(l)}(\alpha_{s}(\bar{\mu}^{2}),\alpha_{s}(1/b^{2}),\epsilon)&=&\sum_{m=0}^{\infty}\sum_{n=1}^{\infty}
\Gamma^{R}_{m+nn}(\epsilon)\alpha_{s}^{m}(\bar{\mu}^{2})\alpha_{s}^{n}(1/b^{2})
\end{eqnarray}
\begin{equation}
\label{logesp}\hat{\Gamma}^{'(l)}(\ln
\bar{\mu}^{2}b^{2},\alpha_{s}(\bar{\mu}^{2}),\alpha_{s}(1/b^{2}),\epsilon)=\frac{\ln
\bar{\mu}^{2}b^{2}\sum_{m=2}^{\infty}\sum_{n=1}^{m-1}\tilde{\Gamma}^{R}_{mn}(\epsilon)\alpha_{s}^{m-n}(\bar{\mu}^{2})\alpha_{s}^{n}(1/b^{2})}
{1+\ln
\bar{\mu}^{2}b^{2}\sum_{m=1}^{\infty}\sum_{n=1}^{m}\tilde{E}^{R}_{mn}(\epsilon)\alpha_{s}^{m-n}(\bar{\mu}^{2})\alpha_{s}^{n}(1/b^{2})}
\end{equation}
We know that
\begin{equation}
\alpha_{s}(1/b^{2})=f(\alpha_{s}(\bar{\mu}^{2}),\ln
\bar{\mu}^{2}b^{2}),
\end{equation}
so, in principle, it is possible to invert this relation in order
to obtain:
\begin{equation} \ln
\bar{\mu}^{2}b^{2}=g(\alpha_{s}(\bar{\mu}^{2}),\alpha_{s}(1/b^{2})).
\end{equation}
The function $g$ has the following perturbative expression:
\begin{equation}
g(\alpha_{s}(\bar{\mu}^{2}),\alpha_{s}(1/b^{2}))=\frac{1}{\beta_{0}}\left(\frac{1}{\alpha_{s}(\bar{\mu}^{2})}-
\frac{1}{\alpha_{s}(1/b^{2})}\right)\left(
1+\sum_{r=1}^{\infty}l_{r}\alpha_{s}^{r}(\bar{\mu}^{2})\right).
\end{equation}
Substituting this expression in Eq.(\ref{logesp}) and re-expanding
in powers of $\alpha_{s}(\bar{\mu}^{2})$ and
$\alpha_{s}(1/b^{2})$, we obtain that: \bea
\hat{\Gamma}^{'(l)}(\ln
\bar{\mu}^{2}b^{2},\alpha_{s}(\bar{\mu}^{2}),\alpha_{s}(1/b^{2}),\epsilon)=
\bar{\Gamma}^{{(l)}}(\alpha_{s}(\bar{\mu}^{2}),\alpha_{s}(1/b^{2}),\epsilon)-
\bar{\Gamma}^{{(l)}}(\alpha_{s}(\bar{\mu}^{2}),\alpha_{s}(\bar{\mu}^{2}),\epsilon),
\eea where \bea
\bar{\Gamma}^{{(l)}}(\alpha_{s}(\bar{\mu}^{2}),\alpha_{s}(\mu^{2}),\epsilon)=
\sum_{m=1}^{\infty}\sum_{n=1}^{\infty}\bar{\Gamma}^{R}_{m+nn}(\epsilon)\alpha_{s}^{m}(\bar{\mu}^{2})\alpha_{s}^{n}(\mu^{2}).
\eea We choose as counterterm, \bea
Z^{(\Gamma)}(\alpha_{s}(\bar{\mu}^{2}),\epsilon)=\hat{\Gamma}^{(l)}(\alpha_{s}(\bar{\mu}^{2}),\alpha_{s}(\bar{\mu}^{2}),\epsilon).
\eea With this choice we obtain: \bea\label{count}
\gamma(1,\bar{\mu}^{2}b^{2},\alpha_{s}(\bar{\mu}^{2}),\epsilon)&=&
\hat{\Gamma}^{(c)}(\alpha_{s}(\bar{\mu}^{2}),\epsilon)+
\hat{\Gamma}^{(l)}(\alpha_{s}(\bar{\mu}^{2}),\alpha_{s}(\bar{\mu}^{2}),\epsilon)+\nonumber\\
&&+\hat{\Gamma}^{(l)}(\alpha_{s}(\bar{\mu}^{2}),\alpha_{s}(1/b^{2}),\epsilon)-
\hat{\Gamma}^{(l)}(\alpha_{s}(\bar{\mu}^{2}),\alpha_{s}(\bar{\mu}^{2}),\epsilon)+\nonumber\\
&&+\bar{\Gamma}^{{(l)}}(\alpha_{s}(\bar{\mu}^{2}),\alpha_{s}(1/b^{2}),\epsilon)-
\bar{\Gamma}^{{(l)}}(\alpha_{s}(\bar{\mu}^{2}),\alpha_{s}(\bar{\mu}^{2}),\epsilon)\label{canc}.
\eea The first line is a power series with coefficients which are
pole-free for each $b^{2}$, because the second and the third lines
vanish when $b^{2}=1/\bar{\mu}^{2}$. Hence, the sum of the second
and the third line must be finite at $\epsilon =0$, but it is not
necessarily analytic in $\alpha_{s}(\mu^{2})$. To find its
perturbative expression in powers of $\alpha_{s}(\mu^{2})$ we
rewrite the last two lines of Eq.(\ref{count}) as
\begin{eqnarray}
\int_{1/b^{2}}^{\bar{\mu}^{2}}\frac{d\mu^{2}}{\mu^{2}}
\left(\frac{\partial}{\partial\ln\mu^{2}}
\hat{\Gamma}^{(l)}(\alpha_{s}(\bar{\mu}^{2}),\alpha_{s}(\mu^{2}),\epsilon)
+\frac{\partial}{\partial\ln\mu^{2}}
\bar{\Gamma}^{{(l)}}(\alpha_{s}(\bar{\mu}^{2}),\alpha_{s}(1/b^{2}),\epsilon)\right).
\end{eqnarray}
There could be other residual cancellations of $\epsilon =0$ poles
between these two terms, but  their sum must be finite at
$\epsilon=0$ and analytic in $\alpha_{s}(\mu^{2})$ and
$\alpha_{s}(1/b^{2})$. Thus, we get a perturbative expression of
the physical anomalous dimension with finite coefficient:
\begin{eqnarray}
\gamma(1,\bar{\mu}^{2}b^{2},\alpha_{s}(\bar{\mu}^{2}))&=&-\int_{1/b^{2}}^{\bar{\mu}^{2}}\frac{d\mu^{2}}{\mu^{2}}
A(\alpha_{s}(\mu^{2}))-B(\alpha_{s}(\bar{\mu}^{2}))-\nonumber\\
&&-\int_{1/b^{2}}^{\bar{\mu}^{2}}\frac{d\mu^{2}}{\mu^{2}}C(\alpha_{s}(\bar{\mu}^{2}),\alpha_{s}(\mu^{2})),\label{cssresform}
\end{eqnarray}
where
\begin{eqnarray}
\label{A}A(\alpha_{s}(\mu^{2}))&=&\sum_{n=1}^{\infty}A_{n}\alpha_{s}^{n}(\mu^{2})\\
\label{B}B(\alpha_{s}(\bar{\mu}^{2}))&=&\sum_{m=1}^{\infty}B_{m}\alpha_{s}^{m}(\bar{\mu}^{2})\\
\label{C}C(\alpha_{s}(\bar{\mu}^{2}),\alpha_{s}(\mu^{2}))&=&
\sum_{m=1}^{\infty}\sum_{n=1}^{\infty}C_{mn}\alpha_{s}^{m}(\bar{\mu}^{2})\alpha_{s}^{n}(\mu^{2})\\
\end{eqnarray}
After an integration by parts of the first term we obtain an
expression for the all-orders resummed exponent:
\bea\label{provres}
\int_{1/b^{2}}^{Q^{2}}\frac{d\bar{\mu}^{2}}{\bar{\mu}^{2}}
\Gamma^{res}(\frac{Q^{2}}{\bar{\mu}^{2}},\bar{\mu}^{2}b^{2},\alpha_{s}(\bar{\mu}^{2}))&=&
-\int_{1/b^{2}}^{Q^{2}}\frac{d\bar{\mu}^{2}}{\bar{\mu}^{2}}
\bigg[\ln\frac{Q^{2}}{\bar{\mu}^{2}}A(\alpha_{s}(\bar{\mu}^{2}))+B(\alpha_{s}(\bar{\mu}^{2}))+\\
&&+\int_{1/b^{2}}^{\bar{\mu}^{2}}\frac{d\mu^{2}}{\mu^{2}}C(\alpha_{s}(\bar{\mu}^{2}),\alpha_{s}(\mu^{2}))\bigg]\nonumber
\eea

To obtain a LL resummation we need only a coefficient ($A_{1}$)
and to obtain a NLL resummation we need four coefficients
$A_{1},A_{2},B_{1},C_{11}$. However it has been demonstrated by
explicit calculations \cite{Ellis:1981hk,Ellis:1981nt} that
\begin{equation}
C_{11}=0.
\end{equation}
Therefore for the resummation at the NLL level we need only three
coefficients $A_{1},A_{2},B_{1}$. Our general formula reduces to
that of Ref.\cite{Collins:1984kg} when
\begin{equation}\label{noncapcond}
C_{mn}=0.
\end{equation}
This restriction could be a consequence of the factorization of
soft emissions from the hard part of the coefficient function, but
this remains unproven.

The result for the anomalous dimension in Eq.(\ref{cssresform})
can be rewritten in our formalism performing the change of
variable
\begin{equation}
n'=\frac{\bar{\mu}^{2}}{\mu^{2}}.
\end{equation}
We get
\begin{equation}
\gamma(1,\bar{\mu}^{2}b^{2},\alpha_{s}(\bar{\mu}^{2}))=
-\int_{1}^{\bar{\mu}^{2}b^{2}}\frac{dn'}{n'}\,G(\alpha_{s}(\bar{\mu}^{2}),\alpha_{s}(\bar{\mu}^{2}/n'))
+\tilde{G}(\alpha_{s}(\bar{\mu}^{2})),\label{qtresform}
\end{equation}
where
\begin{eqnarray}
G(\alpha_{s}(\bar{\mu}^{2}),\alpha_{s}(\mu^{2}))&=&\sum_{m=0}^{\infty}G_{mn}\alpha_{s}^{m}(\bar{\mu}^{2})\alpha_{s}^{n}(\mu^{2})\\
\tilde{G}(\alpha_{s}(\bar{\mu}^{2}))&=&\sum_{m=1}^{\infty}\tilde{G}\alpha_{s}^{m}(\bar{\mu}^{2}).
\end{eqnarray}
The case of the resummation formula of Ref.\cite{Collins:1984kg}
is obtained when $G_{mn}$ is non-vanishing only when $m=0$. In
this case, we have
\begin{equation}
\gamma(1,\bar{\mu}^{2}b^{2},\alpha_{s}(\bar{\mu}^{2}))=
-\int_{1}^{\bar{\mu}^{2}b^{2}}\frac{dn'}{n'}\,G(\alpha_{s}(\bar{\mu}^{2}/n'))
+\tilde{G}(\alpha_{s}(\bar{\mu}^{2})).\label{qtresform2}
\end{equation}

\section{Logs of $q_{\perp}^{2}$ vs. logs of $b^{2}$ to all logarithmic orders}

Large logarithms of $q_{\perp}$ appear in the perturbative
coefficients in the form of plus distributions. We define \bea
\left[\frac{\ln^{p}(\hat{q}_{\perp}^{2})}{\hat{q}_{\perp}^{2}}\right]_{+}
\eea in such a way that \bea
\int_{0}^{1}d\hat{q}_{\perp}^{2}\left[\frac{\ln^{p}(\hat{q}_{\perp}^{2})}{\hat{q}_{\perp}^{2}}\right]_{+}=0.
\eea Let us consider the Fourier transforms \bea\label{defI}
I_{p}(Q^{2}b^{2})=\frac{1}{Q^{2}}\int
d^{2}q_{\perp}e^{i\vec{q}_{\perp}\cdot\vec{b}}\left[\frac{\ln^{p}(\hat{q}_{\perp}^{2})}{\hat{q}_{\perp}^{2}}\right]_{+}=
2\pi\int_{0}^{\infty}\hat{q}_{\perp}d\hat{q}_{\perp}J_{0}(\hat{q}_{\perp}\hat{b})
\left[\frac{\ln^{p}(\hat{q}_{\perp}^{2})}{\hat{q}_{\perp}^{2}}\right]_{+},
\eea where we have used the definition of the 0-order Bessel
function $J_{0}$: \bea
J_{0}(z)=\frac{1}{2\pi}\int_{0}^{2\pi}d\theta e^{iz\cos\theta}.
\eea We now exploit the definition of the plus distribution: \bea
I_{p}(Q^{2}b^{2})=2\pi\int_{0}^{1}d\hat{q}_{\perp}[J_{0}(\hat{q}_{\perp}\hat{b})-1]\frac{\ln^{p}\hat{q}_{\perp}^{2}}{\hat{q}_{\perp}}+
2\pi\int_{1}^{\infty}d\hat{q}_{\perp}J_{0}(\hat{q}_{\perp}\hat{b})\frac{\ln^{p}\hat{q}_{\perp}^{2}}{\hat{q}_{\perp}}.
\eea Writing $\ln^{p}\hat{q}_{\perp}^{2}$ as the $p^{th}$
$\alpha$-derivative of $(\hat{q}_{\perp}^{2})^{\alpha}$ at
$\alpha=0$, we get \bea
I_{p}(Q^{2}b^{2})&=&2\pi\frac{\partial^{p}}{\partial\alpha^{p}}\left\{\int_{0}^{1}d\hat{q}_{\perp}
[J_{0}(\hat{q}_{\perp}\hat{b})-1]\hat{q}_{\perp}^{2\alpha-1}+
\int_{1}^{\infty}d\hat{q}_{\perp}J_{0}(\hat{q}_{\perp}\hat{b})\hat{q}_{\perp}^{2\alpha-1}\right\}\nonumber\\
&=&2\pi\frac{\partial^{p}}{\partial\alpha^{p}}\left[\int_{0}^{\infty}d\hat{q}_{\perp}J_{0}(\hat{q}_{\perp}\hat{b})\hat{q}_{\perp}^{2\alpha-1}-
\int_{0}^{1}d\hat{q}_{\perp}\hat{q}_{\perp}^{2\alpha-1}\right]\nonumber\\
&=&\pi\frac{\partial^{p}}{\partial\alpha^{p}}\left[\left(\frac{Q^{2}b^{2}}{4}\right)^{-\alpha}\frac{\Gamma
(\alpha)}{\Gamma (1-\alpha)}-\frac{1}{\alpha}\right],
\label{genfun}\eea where the last equality follows from the
identity \bea
\int_{0}^{\infty}dxx^{\mu}J_{\nu}(ax)=2^{\mu}a^{-\mu-1}\frac{\Gamma(1/2+\nu/2+\mu/2)}{\Gamma(1/2+\nu/2-\mu/2)}\\
a>0;\qquad -\mathrm{Re}\nu-1<\mathrm{Re}\mu<1/2. \eea From
Eq.(\ref{genfun}), we read off the generating function $G(\alpha)$
of $I_{p}$ \bea\label{genfunI}
G(\alpha)=\frac{\pi}{\alpha}\left[\left(\frac{Q^{2}b^{2}}{4}\right)^{-\alpha}\frac{\Gamma(1+\alpha)}{\Gamma(1-\alpha)}-1\right],
\eea in the sense that \bea\label{dergen}
I_{p}(Q^{2}b^{2})=\left[\frac{d^{p}}{d\alpha^{p}}G(\alpha)\right]_{\alpha=0}.
\eea Now, the generating function of logarithms of $Q^{2}b^{2}$ is
$(Q^{2}b^{2})^{-\alpha}$ in the sense that \bea\label{dergenL}
L_{p}\equiv\ln^{p}(1/(Q^{2}b^{2}))=\left[\frac{d^{p}}{d\alpha^{p}}(Q^{2}b^{2})^{-\alpha}\right]_{\alpha=0}.
\eea Inverting Eq.(\ref{genfunI}), we find the relation between
the generating function of $L_{p}$ and the generating function of
$I_{p}$, which is \bea\label{relgen}
(Q^{2}b^{2})^{-\alpha}=\frac{1}{\pi}S(\alpha)[\alpha
G(\alpha)+\pi], \eea where \bea\label{salpha}
S(\alpha)=\frac{1}{4^{\alpha}}\frac{\Gamma(1-\alpha)}{\Gamma(1+\alpha)}.
\eea Performing the Taylor expansion of the r.h.s. of
Eq.(\ref{relgen}) around $\alpha=0$ and using Eq.(\ref{dergen}),
we obtain: \bea
(Q^{2}b^{2})^{-\alpha}=\frac{1}{\pi}\sum_{m=0}^{\infty}\frac{\alpha^{m}}{m!}\sum_{i=0}^{m}{m\choose
i}iI_{i-1}S^{(m-i)}(0), \eea where $S^{(j)}(0)$ is the j-th
derivative of $S(\alpha)$ evaluated at $\alpha=0$. Now, using
Eq.(\ref{dergenL}), we the relation between $L_{p}$ and $I_{p}$:
\bea L_{p}=\frac{1}{\pi}\sum_{i=1}^{p}{p\choose
i}iI_{i-1}S^{(p-1)}(0)=\frac{1}{\pi}\sum_{k=1}^{p}{p-1\choose
k-1}pS^{(k-1)}I_{p-k}. \eea Thanks to the first equality in
Eq.(\ref{defI}) and to the fact that \bea
\ln^{p-k}\hat{q}_{\perp}^{2}=\frac{1}{p(p-1)\cdots
(p-k+1)}\frac{d^{k}}{d\ln^{k}\hat{q}_{\perp}^{2}}\ln^{p}\hat{q}_{\perp}^{2},
\eea we arrive at a relation to all logarithmic orders between the
logs of $b^{2}$ and the logs of $q_{\perp}^{2}$: \bea\label{LI}
L_{p}=\frac{1}{\pi}\sum_{k=1}^{p}\frac{S^{(k-1)}(0)}{(k-1)!}
\int\frac{d^{2}q_{\perp}}{Q^{2}}e^{i\vec{q}_{\perp}\cdot\vec{b}}\left[\frac{1}{\hat{q}_{\perp}^{2}}\frac
{d^{k}}{d\ln^{k}\hat{q}_{\perp}^{2}}\ln^{p}\hat{q}_{\perp}^{2}\right]_{+}.
\eea This relation allows us to derive the relation between a
generic function of $\ln(1/Q^{2}b^{2})$ and a function of
$\ln\hat{q}_{\perp}^{2}$. Indeed, given a function \bea
h\left(\ln\frac{1}{Q^{2}b^{2}}\right)=\sum_{p=0}^{\infty}h_{p}\ln^{p}\frac{1}{Q^{2}b^{2}},
\eea Eq.(\ref{LI}) implies: \bea\label{generalfun}
h\left(\ln\frac{1}{Q^{2}b^{2}}\right)=\frac{1}{\pi}
\sum_{k=1}^{\infty}\frac{S^{(k-1)}(0)}{(k-1)!}\int\frac{d^{2}q_{\perp}}{Q^{2}}e^{i\vec{q}_{\perp}\cdot\vec{b}}
\left[\frac{1}{\hat{q}_{\perp}^{2}}\frac{d^{k}}{d\ln^{k}\hat{q}_{\perp}^{2}}h(\ln\hat{q}_{\perp}^{2})\right]_{+}.
\eea The r.h.s. of Eq.(\ref{generalfun}) can be viewed as the
Fourier transform of a function (more properly a distribution)
$\hat{h}(\ln\hat{q}_{\perp}^{2})$: \bea
h\left(\ln\frac{1}{Q^{2}b^{2}}\right)&=&
\int\frac{d^{2}q_{\perp}}{Q^{2}}e^{i\vec{q}_{\perp}\cdot\vec{b}}\hat{h}(\ln\hat{q}_{\perp}^{2}),\nonumber\\
\label{dabaq}\hat{h}(\ln\hat{q}_{\perp}^{2})&=&
\frac{1}{\pi}\sum_{k=1}^{\infty}\frac{S^{(k-1)}(0)}{(k-1)!}\left[\frac{1}{\hat{q}_{\perp}^{2}}\frac{d^{k}}{d\ln^{k}\hat{q}
_{\perp}^{2}}h(\ln\hat{q}_{\perp}^{2})\right]_{+}. \eea

\section{Resummation in $q_{\perp}$-space}

In this section, we investigate the consequences of our general
result Eq.(\ref{dabaq}) for the resummation at the NLL level of
logarithmic accuracy. According to eq(\ref{resummedexp3}) and the
discussion below and according to Eq.(\ref{provres}), we have that
our resummation factor formula in Fourier space is:
\begin{equation}
K^{res}(Q^{2}b^{2},1/b^{2},Q^{2})=\exp\left\{E^{res}(Q^{2}b^{2},1/b^{2},Q^{2})\right\},
\end{equation}
where
\begin{equation}\label{exp}
E^{res}(Q^{2}b^{2},1/b^{2},Q^{2})=
\int_{1/b^{2}}^{Q^{2}}\frac{d\bar{\mu}^{2}}{\bar{\mu}^{2}}
\Gamma^{res}(\frac{Q^{2}}{\bar{\mu}^{2}},\bar{\mu}^{2}b^{2},\alpha_{s}(\bar{\mu}^{2})),
\end{equation}
and where at NLL level
\begin{eqnarray}
\Gamma^{res}_{NLL}(\frac{Q^{2}}{\bar{\mu}^{2}},\bar{\mu}^{2}b^{2},\alpha_{s}(\bar{\mu}^{2}))&=&
-\ln\frac{Q^{2}}{\bar{\mu}^{2}}\big[A_{1}\alpha_{s}(\bar{\mu}^{2})+
A_{2}\alpha_{s}^{2}(\bar{\mu}^{2})\big]-B_{1}\alpha_{s}(\bar{\mu}^{2})\nonumber\\
&&-\frac{C_{11}}{\beta_{0}}\alpha_{s}(\bar{\mu}^{2})\ln\frac{\alpha_{s}(1/b^{2})}{\alpha_{s}(\bar{\mu}^{2})},
\label{coeffgamma}
\end{eqnarray}
where we have used the definition of the $\beta$-function: \bea
\mu^{2}\frac{d}{d\mu^{2}}\alpha_{s}(\mu^{2})=\beta{\alpha_{s}}=-\beta_{0}\alpha_{s}^{2}-\beta_{1}\alpha_{s}^{3}+\emph{O}(\alpha_{s}^{4})
\eea and where we have used the change of variable \bea
\frac{d\mu^{2}}{\mu^{2}}=\frac{d\alpha_{s}}{\beta(\alpha_{s})}
\eea to compute the integral that appears in the last term of
Eq.(\ref{provres}). Now, thanks to Eq.(\ref{dabaq}), we can
rewrite the resummed exponent in $b$-space Eq.(\ref{exp}) in terms
of a resummed exponent defined in $q_{\perp}$-space. Thus, up to
NNLL terms, we obtain: \bea
E^{res}_{NLL}(Q^{2}b^{2},1/b^{2},Q^{2})=\int
d^{2}q_{\perp}e^{i\vec{q}_{\perp}\cdot\vec{b}}
\left[\frac{\hat{\Gamma}_{NLL}^{res}(\hat{q}_{\perp}^{2},q_{\perp}^{2},Q^{2})}{q_{\perp}^{2}}\right]_{+},
\eea where \bea
\hat{\Gamma}_{NLL}^{res}(\hat{q}_{\perp}^{2},q_{\perp}^{2},Q^{2})=&=&-\ln\hat{q}_{\perp}^{2}\big[\hat{A}_{1}\alpha_{s}(q_{\perp}^{2})+
\hat{A}_{2}\alpha_{s}^{2}(q_{\perp}^{2})\big]-\hat{B}_{1}\alpha_{s}(q_{\perp}^{2})+\nonumber\\
&&-\frac{\hat{C}_{11}}{\beta_{0}}\alpha_{s}(q_{\perp}^{2})\ln\frac{\alpha_{s}(Q^{2})}{\alpha_{s}(q_{\perp}^{2})}
\eea and where the relation of the constant coefficients of this
last equation and the of Eq.(\ref{coeffgamma}) is \bea
\hat{A}_{1}&=&\frac{A_{1}}{\pi}\\
\hat{A}_{2}&=&-\left(\frac{A_{2}}{\pi}+\frac{A_{1}}{\pi}\beta_{0}\ln\frac{e^{2\gamma_{E}}}{4}\right)\\
\hat{B}_{1}&=&\frac{B_{1}}{\pi}-\frac{A_{1}}{\pi}\ln\frac{e^{2\gamma_{E}}}{4}\\
\hat{C}_{11}&=&\frac{C_{11}}{\pi}. \eea Here $\gamma_{E}$ is the
usual Euler gamma. Now, we want to define a resummation factor in
$q_{\perp}$-space. Looking at Eq.(\ref{resummedexp}), we note that
large $\ln Q^{2}b^{2}$ of collinear nature are resummed by the
parton distribution function. So, in order to define a resummation
in $q_{\perp}$-space, we must take them into account. For
simplicity, we consider the resummed part of non-singlet cross
section, because the non-singlet parton distribution functions,
which are defined as \bea
f'_{a'}(N,\mu^{2})=f_{a}(N,\mu^{2})-f_{b}(N,\mu^{2})\qquad a,b\neq
g, \eea evolve independently. In particular, in Mellin moments $N$
they satisfy the following evolution equations: \bea
\mu^{2}\frac{\partial}{\partial\mu^{2}}f'_{a'}(N,\mu^{2})=\gamma'(N,\alpha_{s}(\mu^{2}))f'_{a'}(N,\mu^{2}).
\eea Hence, the non-singlet parton distribution functions
evaluated at $\mu^{2}=1/b^{2}$ are related to the ones evaluated
at $\mu^{2}=Q^{2}$ by, \bea
f'_{a'}(N,1/b^{2})=\exp\left\{-\int_{1/b^{2}}^{Q^{2}}\frac{d\bar{\mu}^{2}}{\bar{\mu}^{2}}
\gamma'(N,\alpha_{s}(\bar{\mu}^{2}))\right\}f'_{a'}(N,Q^{2}). \eea
Thus, the resummed part of the cross section with the non-singlet
parton distribution functions evaluated at $\mu^{2}=Q^{2}$ becomes
\bea
\exp\left\{-\int_{1/b^{2}}^{Q^{2}}\frac{d\bar{\mu}^{2}}{\bar{\mu}^{2}}
\sum_{j=1}^{2}\gamma'(N,\alpha_{s}(\bar{\mu}^{2}))\right\}K^{res}(b^{2},1/b^{2},Q^{2})=\\
=\exp\left\{\int_{1/b^{2}}^{Q^{2}}\frac{d\bar{\mu}^{2}}{\bar{\mu}^{2}}
\left[\Gamma^{res}(\frac{Q^{2}}{\bar{\mu}^{2}},\bar{\mu}^{2}b^{2},\alpha_{s}(\bar{\mu}^{2}))-
\sum_{j=1}^{2}\gamma'(N_{j},\alpha_{s}(\bar{\mu}^{2}))\right]\right\}\nonumber
\eea The general relation between a function of $\ln Q^{2}b^{2}$
and its Fourier anti-transform Eq.(\ref{dabaq}), immediately
enables us to define a resummed exponent of the non-singlet part
of the cross section in $q_{\perp}$-space, which is:
\begin{eqnarray}
K^{res}(\hat{q}_{\perp}^{2},q_{\perp}^{2},Q^{2})=\frac{1}{\pi}
\sum_{k=1}^{\infty}\frac{S^{(k-1)}(0)}{(k-1)!}
\quad\qquad\qquad\qquad\,\,\qquad\qquad\qquad\qquad\qquad\qquad\qquad\qquad&&\nonumber\\
\times\left\{\frac{1}{\hat{q}_{\perp}^{2}}
\frac{d^{k}}{d\ln^{k}\hat{q}_{\perp}^{2}}\exp
\left[\int_{q_{\perp}^{2}}^{Q^{2}}\frac{d\bar{\mu}^{2}}{\bar{\mu}^{2}}
\left(\Gamma^{res}(\frac{Q^{2}}{\bar{\mu}^{2}},\bar{\mu}^{2}/q_{\perp}^{2},\alpha_{s}
(\bar{\mu}^{2}))-\sum_{j=1}^{2}
\gamma'(N_{j},\alpha_{s}(\bar{\mu}^{2}))\right)\right]\right\}_{+}\quad\qquad&&\nonumber\\
=\frac{1}{\pi}\sum_{k=1}^{\infty}\frac{S^{(k-1)}(0)}{(k-1)!}
\qquad\qquad\qquad\qquad\qquad\,\,\qquad\qquad\qquad\qquad\qquad\qquad\qquad\qquad\qquad\qquad\,\,&&\nonumber\\
\times\left\{\frac{d}{d\hat{q}_{\perp}^{2}}\frac{d^{k-1}}{d\ln^{k-1}
\hat{q}_{\perp}^{2}}\exp\left[\int_{q_{\perp}^{2}}^{Q^{2}}
\frac{d\bar{\mu}^{2}}{\bar{\mu}^{2}}
\left(\Gamma^{res}(\frac{Q^{2}}{\bar{\mu}^{2}},\bar{\mu}^{2}/q_{\perp}^{2},\alpha_{s}(\bar{\mu}^{2}))
-\sum_{j=1}^{2}\gamma'(N_{j},\alpha_{s}(\bar{\mu}^{2}))\right)\right]\right\}_{+}\quad\,\,&&\nonumber
\end{eqnarray}
\begin{eqnarray}
=\label{sigmaqt}\lim_{\eta\rightarrow 0^{+}}\frac{1}{\pi}
\sum_{k=0}^{\infty}\frac{S^{(k)}(0)}{k!}\frac{d^{k}}{d\ln^{k}\hat{q}_{\perp}^{2}}
\frac{d}{\hat{q}_{\perp}^{2}}\bigg\{\theta(\hat{q}_{\perp}^{2}-\eta)
\exp\bigg[\int_{q_{\perp}^{2}}^{Q^{2}}\frac{d\bar{\mu}^{2}}{\bar{\mu}^{2}}\qquad\qquad\qquad\qquad\,\,\,\quad&&\nonumber\\
\times\left(\Gamma^{res}(\frac{Q^{2}}{\bar{\mu}^{2}},\bar{\mu}^{2}/q_{\perp}^{2},\alpha_{s}(\bar{\mu}^{2}))-
\sum_{j=1}^{2}\gamma'(N_{j},\alpha_{s}(\bar{\mu}^{2}))\right)\bigg]\bigg\},\quad\quad\qquad\qquad\qquad\qquad\quad&&
\end{eqnarray} where the last equation defines implicity the
$q_{\perp}$-space resummation exponent:
\begin{equation}
K^{res}(\hat{q}_{\perp}^{2},q_{\perp}^{2},Q^{2})=\exp\left\{E^{res}(\hat{q}_{\perp}^{2},q_{\perp}^{2},Q^{2})\right\}.
\end{equation}
All the previously released expressions for this exponent given in
\cite{Ellis:1997ii,Frixione:1998dw,Kulesza:2001jc} are particular
cases of this general expression. They differ essentially in the
criteria according to which the subleading terms are kept.
\\

We want to calculate the NLL result in $q_{\perp}$-space. Thus,
keeping only the terms up to NNLL in Eq.(\ref{sigmaqt}) we obtain
\begin{eqnarray}
K^{res}_{NLL}(\hat{q}_{\perp}^{2},q_{\perp}^{2},Q^{2})=\frac{1}{\pi}\frac{d}{\hat{q}_{\perp}^{2}}\bigg\{
\theta(\hat{q}_{\perp}^{2}-\eta)\exp
\bigg[\int_{q_{\perp}^{2}}^{Q^{2}}\frac{d\bar{\mu}^{2}}{\bar{\mu}^{2}}
\bigg(\Gamma^{res}_{NLL}(\frac{Q^{2}}{\bar{\mu}^{2}},\bar{\mu}^{2}/q_{\perp}^{2},\alpha_{s}(\bar{\mu}^{2}))
\,&&\nonumber\\
-\sum_{j=1}^{2}\gamma'(N_{j},\alpha_{s}(\bar{\mu}^{2}))\bigg)\bigg]
\sum_{k=0}^{\infty}\frac{S^{(k)}(0)}{k!}[-\ln\hat{q}_{\perp}^{2}A_{1}\alpha_{s}(q_{\perp}^{2})]^{k}\bigg\}\label{sigmaqtnll}.\,&&
\end{eqnarray}
In order to compare this result at NLL to that of
\cite{Frixione:1998dw}, we define a new variable $h$: \bea h\equiv
2\ln\hat{q}_{\perp}^{2}A_{1}\alpha_{s}(q_{\perp}^{2}). \eea In
terms of this variable and using Eq.(\ref{salpha}) the series that
appears in Eq.(\ref{sigmaqtnll}) can be computed: \bea
\sum_{k=0}^{\infty}\frac{S^{(k)}(0)}{k!}[-\ln\hat{q}_{\perp}^{2}
A_{1}\alpha_{s}(q_{\perp}^{2})]^{k}=S(-h/2)=2^{h}\frac{\Gamma(1+h/2)}{\Gamma(1-h/2)}.
\eea In conclusion, we obtain that the NLL resummation factor
becomes
\begin{eqnarray}
K^{res}_{NLL}(\hat{q}_{\perp}^{2},q_{\perp}^{2},Q^{2})=\frac{1}{\pi}\frac{d}{\hat{q}_{\perp}^{2}}\bigg\{
\theta(\hat{q}_{\perp}^{2}-\eta)\exp\bigg[\int_{q_{\perp}^{2}}^{Q^{2}}
\frac{d\bar{\mu}^{2}}{\bar{\mu}^{2}}
\bigg(\Gamma^{res}_{NLL}(\frac{Q^{2}}{\bar{\mu}^{2}},\bar{\mu}^{2}/q_{\perp}^{2},\alpha_{s}(\bar{\mu}^{2}))&&\nonumber\\
-\sum_{j=1}^{2}\gamma'(N_{j},\alpha_{s}(\bar{\mu}^{2}))\bigg)\bigg]
2^{h}\frac{\Gamma(1+h/2)}{\Gamma(1-h/2)}\bigg\},&&
\end{eqnarray}
which gives the same result given in Ref.\cite{Frixione:1998dw} in
the case that the coefficient $C_{11}$ that appears in
Eq.(\ref{coeffgamma}) is equal to zero and that the arbitrary
constants $c_{1}$ and $c_{2}$ also defined in
\cite{Frixione:1998dw} are equal to one. It is clear that the last
two terms of the exponential of our result let the non-singlet
parton distribution densities, which enter the
$\hat{q}_{\perp}^{2}$ derivative, evolve from the scale $Q^{2}$ to
the scale $q_{\perp}^{2}$.

We conclude the chapter noting that also in this case the resummed
results using the renormalization group approach are less
predictive than results obtained with the approach of
Ref.\cite{Sterman:1986aj}, as it is shown in
Ref.\cite{Collins:1984kg}. Furthermore the conditions that reduce
our results to those of Ref.\cite{Collins:1984kg} in terms of
factorization properties is still an interesting open question.

\chapter{Predictive power of the resummation formulae}\label{predictivity}

We have shown that the renormalization group resummed expressions
are less predictive than those obtained with other approaches
discussed in Sec.\ref{alternative}. In this Chapter we shall
compare the various approaches quantitatively. We will show that
all the resummation coefficients can be determined by a fixed
order computation. This can be useful, because the determination
of the resummation coefficients from a fixed order computation
represents a possible way to check the correctness of the
resummation formulae with strong factorization properties.

In particular, in this chapter, we will show how the resummation
coefficients $g_{mnp}$ (for the prompt photon case), $g_{mn}$ (for
the DIS and DY cases) and $G_{mn},\tilde{G}_{m}$ (for the DY
transverse momentum distribution) can be determined. For the
rapidity distributions of DY and DIS they are the same of the
all-inclusive cases (see Chapter \ref{DYDPR}) The resummation
coefficients are determined by comparing the expansion of the
resummed anomalous dimension $\gamma$ in powers of $\as(Q^2)$ with
a fixed-order calculation, which in general has the form:
\begin{equation}\label{gammafo}
\gamma_{\rm FO}(N,\as)=\sum_{i=1}^{k_{\rm min}}\as^i\,\sum_{j=1}^i
\gamma^i_j\,\ln^j\frac{1}{N}+O(\as^{k_{\rm min}+1})+O(N^0),
\end{equation}
where $\gamma_{\rm FO}(N,\as)$ is the physical anomalous dimension
for each individual partonic subprocess for the promt photon case,
for the qq channel in the DY case and for the q channel in the DIS
case. For the case of the small transverse momentum DY
distribution it ha the form:
\begin{equation}\label{gammafo}
\gamma_{\rm FO}(\bar{\mu}^{2}b^{2},\as)=\sum_{i=1}^{k_{\rm
min}}\as^i\,\sum_{j=0}^i
\tilde{\gamma}^{i}_{j}\,\ln^j\frac{1}{\bar{\mu}b^{2}}+O(\as^{k_{\rm
min}+1})+O(\frac{1}{\bar{\mu}^{2}b^{2}}).
\end{equation}
The number $k_{\rm min}$ is the minimum order at which the
anomalous dimension must be calculated in order to determine its
$N^{k-1}LL$ resummation.

For prompt photon production, the number of coefficients $N_k$
that must be determined at each logarithmic order, and the minimum
fixed order which is necessary in order to determine them are
summarized in Table \ref{tab}, according to whether the
coefficient function is fully factorized
[Eq.(\ref{resformulanason}) sec.\ref{DPresform} ], or has
factorized $N$-dependent and $N$-independent terms
[Eq.(\ref{resformulaint}) sec.\ref{DPresform}], or not factorized
at all [Eq.~(\ref{resformula}) sec\ref{DPresform}]. In the
approach of Refs.\cite{Laenen:1998qw,Catani:1998tm} the
coefficient function is fully factorized, and furthermore some
resummation coefficients are related to universal coefficients of
Altarelli-Parisi splitting functions, so that $k_{\rm min}=k$. For
prompt-photon production, available results do not allow to test
factorization, and test relation of resummation coefficients to
Altarelli-Parisi coefficients only to lowest $O(\as)$.

\begin{table}
\begin{center}
\begin{tabular}{|c||c|c|c||}
\hline
     & \multicolumn{3}{|c||}{Prompt photon} \\
\hline
      & Eq.(\ref{resformulanason}) sec.\ref{DPresform}& Eq.(\ref{resformulaint}) sec.\ref{DPresform}&
      Eq.(\ref{resformula}) sec.\ref{DPresform}\\
\hline\hline
$N_k$ & $2k$ & $\frac{k(k+3)}{2}$  & $\frac{k(k+1)(k+5)}{6}$\\
\hline
$k_{\rm min}$ & $k+1$ & $2k$ & $3k-1$\\
\hline
\end{tabular}
\caption{\label{tab} Number of coefficients $N_k$ and minimum
order of the required perturbative calculation $k_{\rm min}$ for
inclusive prompt photon $N^{k-1}LL$ resummation.}
\end{center}
\end{table}

The results for DIS and Drell-Yan, according to whether the
coefficient function has factorized $N$-dependent and
$N$-independent terms as in
Refs.\cite{Contopanagos:1996nh,Catani:1989ne,Sterman:1986aj}
[Eq.(\ref{disdyresfac}) sec.\ref{DISDY3}] or no factorization
properties as in [(Eq.\ref{disdyres}) sec.\ref{DISDY3}], are
reported in table \ref{tab2}. Current fixed-order results support
factorization for Drell-Yan and DIS only to the lowest nontrivial
order $O(\as^2)$.

\begin{table}
\begin{center}
\begin{tabular}{|c||c|c||}
\hline &\multicolumn{2}{c||}{DIS/DY}\\
\hline & Eq.(\ref{disdyresfac}) sec.\ref{DISDY3} & Eq.(\ref{disdyres}) sec.\ref{DISDY3}\\
\hline\hline $N_{k}$ & $k$& $\frac{k(k+1)}{2}$   \\
\hline $k_{\rm min}$ & $k$ & $2k-1$\\
\hline
\end{tabular}
\caption{\label{tab2} Number of coefficients $N_k$ and minimum
order of the required perturbative calculation $k_{\rm min}$ for
inclusive DIS and DY $N^{k-1}LL$ resummation.}
\end{center}
\end{table}

In table \ref{tab3}, we report also the results for the small
transverse DY resummation. We list $N_{k}$ and $k_{\rm min}$ for
the approach of Ref.\cite{Collins:1984kg} [Eq.(\ref{qtresform2})
sec.\ref{bordello}] and for the renormalization group approach
[Eq.(\ref{qtresform}) sec.\ref{bordello}]. If the two cases are
related by factorization properties of the cross section is not
yet understood even if probable.

\begin{table}
\begin{center}
\begin{tabular}{|c||c|c||}
\hline &\multicolumn{2}{c||}{DY transverse distribution}\\
\hline & Eq.(\ref{qtresform2}) sec.\ref{bordello} & Eq.(\ref{qtresform}) sec.\ref{bordello}\\
\hline\hline $N_{k}$ & $2k-1$& $\frac{k^{2}+3k-2}{2}$   \\
\hline $k_{\rm min}$ & $k$ & $2k-1$\\
\hline
\end{tabular}
\caption{\label{tab3} Number of coefficients $N_k$ and minimum
order of the required perturbative calculation $k_{\rm min}$ for
small transverse momentum DY $N^{k-1}LL$ resummation.}
\end{center}
\end{table}

In the following, we present all the proofs of these results.

\section{Prompt photon production in the strongest factorization case}\label{sec2}

This is the case of Eq.(\ref{resformulanason}) in section
\ref{DPresform}. In this case there are $N_{k}=2k$ non-vanishing
coefficients $g_{00i}$ and $g_{0i0}$ $i=1,2,\dots,k$. The resummed
expression of the anomalous dimension at $N^{k-1}LL$ is given by:
\begin{equation}\label{inttot}
\gamma(N,\alpha_{s}(Q^{2}))=\int_{1}^{N}\frac{dn}{n}\left(\sum_{i=1}^{k}g_{0i0}\alpha_{s}^{i}(Q^{2}/n^{2})\right)
+\int_{1}^{N}\frac{dn}{n}\left(\sum_{i=1}^{k}g_{00i}\alpha_{s}^{i}(Q^{2}/n)\right).
\end{equation}
We consider first the second integral in Eq.(\ref{inttot}). Noting
that:
\begin{equation}
\frac{dn}{n}=-\frac{d\alpha_{s}(Q^{2}/n)}{\beta(\alpha_{s})},
\end{equation}
where
\begin{eqnarray}
\beta(\alpha_{s})&=&-b_{0}\alpha_{s}^{2}-b_{1}\alpha_{s}^{3}+\emph{O}(\alpha_{s}^{4})\\
b_{0}&\equiv&\frac{\beta_{0}}{4\pi}\qquad
b_{1}\equiv\frac{\beta_{1}}{(4\pi)^{2}},
\end{eqnarray}
with $\beta_{0}$ and $\beta_{1}$ given in Eq.(\ref{betapar}) in
section \ref{appA}, we can rewrite it in the form:
\begin{equation}\label{secint}
\int_{\alpha_{s}(Q^{2})}^{\alpha_{s}(Q^{2}/N)}d\alpha_{s}\frac{\sum_{i=1}^{k}g_{00i}\alpha_{s}^{i}}{\beta_{0}\alpha_{s}^{2}\left(
1+\frac{\beta_{1}}{\beta_{0}}\alpha_{s}+\frac{\beta_{2}}{\beta_{0}}\alpha_{s}^{2}+\cdots\right)}.
\end{equation}
Now, we expand up to order $\alpha_{s}^{k-2}$ each term that
compares in the integrand of this last expression and collect all
the coefficients that correspond to the same power of
$\alpha_{s}$. Doing this, we have that the integral (\ref{secint})
can be rewritten in the following form:
\begin{eqnarray}
\frac{1}{b_{0}}\Bigg\{\int_{\alpha_{s}(Q^{2})}^{\alpha_{s}(Q^{2}/N)}d\alpha_{s}
\bigg[\frac{g_{001}}{\alpha_{s}}+(b_{1}^{1}g_{001}+g_{002})+(b_{2}^{1}g_{001}+b_{2}^{2}g_{002}+g_{003})\alpha_{s}+\cdots+\nonumber\\
+(b_{k-1}^{1}g_{001}+\cdots+b_{k-1}^{k-1}g_{00k-1}+g_{00k})\alpha_{s}^{k-2}\bigg]\Bigg\},
\end{eqnarray}
where $k>1$ and the numbers  $b_{i}^{j}$ are build up with the
coefficients of the $\beta$ function. Now, we perform the integral
over $\alpha_{s}$. We get:
\begin{eqnarray}\label{primo}
&&\int_{1}^{N}\frac{dn}{n}\left(\sum_{i=1}^{k}g_{00i}\alpha_{s}^{i}(Q^{2}/n)\right)=\nonumber\\
&=&\frac{1}{b_{0}}\Bigg\{g_{001}\ln\left(\frac{\alpha_{s}(Q^{2}/N)}{\alpha_{s}(Q^{2})}\right)+(b_{1}^{1}g_{001}+g_{002})
[\alpha_{s}(Q^{2}/N)-\alpha_{s}(Q^{2})]+\nonumber\\
&&+\frac{1}{2}(b_{2}^{1}g_{001}+b_{2}^{2}g_{002}+g_{003})[\alpha_{s}^{2}(Q^{2}/N)-\alpha_{s}^{2}(Q^{2})]+\cdots+\nonumber\\
&&+\frac{1}{k-1}(b_{k-1}^{1}g_{001}+\cdots+b_{k-1}^{k-1}g_{00k-1}+g_{00k})[\alpha_{s}^{k-1}(Q^{2}/N)-\alpha_{s}^{k-1}(Q^{2})]\Bigg\}
\end{eqnarray}
To perform the first integral of Eq.(\ref{inttot}), we it is
sufficient to note that in this case
\begin{equation}
\frac{dn}{n}=-\frac{d\alpha_{s}(Q^{2}/n^{2})}{2\beta(\alpha_{s})}.
\end{equation}
and proceed as before. The result that we obtain is:
\begin{eqnarray}\label{secondo}
&&\int_{1}^{N}\frac{dn}{n}\left(\sum_{i=1}^{k}g_{0i0}\alpha_{s}^{i}(Q^{2}/n^{2})\right)=\nonumber\\
&=&\frac{1}{2b_{0}}\Bigg\{g_{010}\ln\left(\frac{\alpha_{s}(Q^{2}/N^{2})}{\alpha_{s}(Q^{2})}\right)+(b_{1}^{1}g_{010}+g_{020})
[\alpha_{s}(Q^{2}/N^{2})-\alpha_{s}(Q^{2})]+\nonumber\\
&&+\frac{1}{2}(b_{2}^{1}g_{010}+b_{2}^{2}g_{020}+g_{030})[\alpha_{s}^{2}(Q^{2}/N^{2})-\alpha_{s}^{2}(Q^{2})]+\cdots+\nonumber\\
&&+\frac{1}{k-1}(b_{k-1}^{1}g_{010}+\cdots+b_{k-1}^{k-1}g_{0k-10}+g_{0k0})[\alpha_{s}^{k-1}(Q^{2}/N^{2})-\alpha_{s}^{k-1}(Q^{2})]\Bigg\}
\end{eqnarray}
At this point, we take the first term of Eq.(\ref{primo}) together
with the first term of Eq.(\ref{secondo}) in order to isolate the
first contributions of Eq.(\ref{inttot}). We have:
\begin{equation}
\frac{1}{2b_{0}}\left[g_{010}\ln
\left(\frac{\alpha_{s}(Q^{2}/N^{2})}{\alpha_{s}(Q^{2})}\right)+2g_{001}\ln\left(\frac{\alpha_{s}(Q^{2}/N)}{\alpha_{s}(Q^{2})}\right)\right]
\end{equation}
From this contribution, we want to extract the firs two LL terms.
Hence, using the one loop running of $\alpha_{s}(Q^{2}/N^{a}),
a=1,2$ (Eq.\ref{run} of section \ref{appA}), we obtain for this
contribution:
\begin{eqnarray}\label{LL}
\alpha_{s}(Q^{2})\ln\frac{1}{N}[-(g_{001}+g_{010})]&+&\alpha_{s}^{2}(Q^{2})\ln^{2}\frac{1}{N}[b_{0}/2(g_{001}+2g_{010})]\nonumber\\
&&+\emph{O}(\alpha_{s}^{2}\ln(N))+\emph{O}(\alpha_{s}^{3+i}\ln^{j}(N)),\label{LL}
\end{eqnarray}
where $i\geq 0,1\leq j\leq i+3$. Now, we take the second term of
Eq.(\ref{primo}) together with the second term of
Eq.(\ref{secondo}) in order to keep the second the second
contributions of Eq.(\ref{inttot}) and we have:
\begin{equation}\label{PLL}
\frac{1}{2b_{0}}\left[(b_{1}^{1}g_{010}+g_{020})(\alpha_{s}(Q^{2}/N^{2})-\alpha_{s}(Q^{2}))+
2(b_{1}^{1}g_{001}+g_{002})(\alpha_{s}(Q^{2}/N)-\alpha_{s}(Q^{2}))\right].
\end{equation}
From this contribution, we want to extract the first two NLL
terms. In order to do this, we use the one loop running of
$\alpha_{s}(Q^{2}/N^{a}), a=1,2$ and observe that the coefficients
of $g_{001}$ and $g_{010}$ are modified by the last two terms of
Eq.(\ref{LL}). We get:
\begin{eqnarray}\label{NLL}
&&\alpha_{s}^{2}(Q^{2})\ln(1/N)[-(c_{1}^{1}g_{010}+d_{1}^{1}g_{001}+g_{020}+g_{002})]+\nonumber\\
&&\alpha_{s}^{3}(Q^{2})\ln^{2}(1/N)[b_{0}(2\tilde{c}_{1}^{1}g_{010}+\tilde{d}_{1}^{1}g_{001}+2g_{020}+g_{002})]\nonumber\\
&&+\emph{O}(\alpha_{s}^{3}\ln(N))+\emph{O}(\alpha_{s}^{4+i}\ln^{j}(N)),\label{NLL}
\end{eqnarray}
where $c_{1}^{1},d_{1}^{1},\tilde{c}_{1}^{1},\tilde{d}_{1}^{1}$
are coefficients (that are of no concern to us) and $i\geq 0,1\leq
j\leq i+3$. This procedure can be repeated for all the other
contributions. So, we take the k-th term of Eq.(\ref{primo})
together with the k-th term of Eq.(\ref{secondo}) in order to keep
the k-th contributions of Eq.(\ref{inttot}). For the general k-th
contribution, we get:
\begin{eqnarray}
&&\frac{1}{2b_{0}}\bigg[\frac{1}{k-1}(b_{k-1}^{1}g_{010}+\cdots+b_{k-1}^{k-1}g_{0k-10}+g_{0k0})(\alpha_{s}^{k-1}(Q^{2}/N^{2})-
\alpha_{s}^{k-1}(Q^{2}))+\nonumber\\
&&+\frac{2}{k-1}(b_{k-1}^{1}g_{001}+\cdots+b_{k-1}^{k-1}g_{00k-1}+g_{00k})(\alpha_{s}^{k-1}(Q^{2}/N)-\alpha_{s}^{k-1}(Q^{2}))\bigg]\label{PNk-1LL}.
\end{eqnarray}
From this contribution, we want to extract the first two
$N^{k-1}LL$ terms. In order to do this, again, we use the one loop
running of $\alpha_{s}(Q^{2}/N^{a}), a=1,2$ and note that the
coefficients $g_{00i}$ and $g_{0i0}$ with $i=1,2,\dots,k-1$ are
modified by the previous terms. For this generic term, we get:
\begin{eqnarray}\label{Nk-1LL}
&&\alpha_{s}^{k}\ln\frac{1}{N}[-(c_{k-1}^{1}g_{010}+\cdots+c_{k-1}^{k-1}g_{0k-10}+d_{k-1}^{1}g_{001}+\cdots+d_{k-1}^{k-1}g_{00k-1}+g_{0k0}+g_{00k})]+
\nonumber\\
&&+\alpha_{s}^{k+1}\ln^{2}\frac{1}{N}[b_{0}k/2(2\tilde{c}_{k-1}^{1}g_{010}+\cdots+2\tilde{c}_{k-1}^{k-1}g_{0k-10}+
\tilde{d}_{k-1}^{1}g_{001}+\cdots+\tilde{d}_{k-1}^{k-1}g_{00k-1}\nonumber\\
&&+2g_{0k0}+g_{00k})]+\emph{O}(\alpha_{s}^{k+1}\ln(N))+\emph{O}(\alpha_{s}^{k+2+i}\ln^{j}(N)),
\end{eqnarray}
where $i\geq 0,1\leq j\leq i+3$.

To summarize, Eqs. (\ref{LL},\ref{NLL},\ref{Nk-1LL}) tell us that
from the expression of the physical anomalous dimension
Eq.(\ref{inttot}), we can extract the following linear
combinations of the coefficients $g_{010},\dots ,
g_{0k0},g_{001},\dots , g_{00k}$:
\begin{eqnarray}
l_{1}&=&-(g_{001}+g_{010})\nonumber\\
l_{2}&=&\frac{b_{0}}{2}(g_{001}+2g_{010})\nonumber\\
l_{3}&=&-(c_{1}^{1}g_{010}+d_{1}^{1}g_{001}+g_{020}+g_{002})\nonumber\\
l_{4}&=&b_{0}(2\tilde{c}_{1}^{1}g_{010}+\tilde{d}_{1}^{1}g_{001}+2g_{020}+g_{002})\nonumber\\
&&\cdots\nonumber\\
l_{2k-1}&=&-(c_{k-1}^{1}g_{010}+\cdots +c_{k-1}^{k-1}g_{0k-10}+d_{k-1}^{1}g_{001}+\cdots +d_{k-1}^{k-1}g_{00k-1}+g_{0k0}+g_{00k})\nonumber\\
l_{2k}&=&\frac{kb_{0}}{2}(2\tilde{c}_{k-1}^{1}g_{010}+\cdots
+2\tilde{c}_{k-1}^{k-1}g_{0k-10}+\tilde{d}_{k-1}^{1}g_{001}+\cdots
+\tilde{d}_{k-1}^{k-1}g_{00k-1}+2g_{0k0}+g_{00k})\nonumber,
\end{eqnarray}
with $k\geq 2$ and $l_{i}$, $i=1,\dots,2k$ the known terms. These
are $2k$ independent linear combinations that determine the $2k$
coefficients $g_{010},\dots , g_{0k0},g_{001},\dots , g_{00k}$ of
a $N^{k-1}LL$ resummation comparing them  with the correspondent
terms of Eq.(\ref{gammafo}) up to order $\alpha_{s}^{k+1}$. This
is a direct consequence of the fact that the two vectors $(1,1)$
and $(1,2)$ are independent. This shows that, in order to obtain a
$N^{k-1}LL$ resummation in the case of the strongest factorization
(Eq.(\ref{resformulanason}) in section \ref{DPresform}) , we need
to know a $N^{k+1}LO$ fixed order calculation of the physical
anomalous dimension. Hence, in this case $k_{\rm min}=k+1$.

\section{Prompt photon production in the weaker factorization case}\label{dpcasoint}

This is the case of Eq.(\ref{resformulaint}) in section
\ref{DPresform}. In this case in order to perform a LL
resummation, we need two coefficients ($g_{010},g_{001}$); to
perform a NLL resummation three more coefficients are added
($g_{020},g_{002},g_{011}$); in general to perform a $N^{k-1}LL$
resummation $k+1$ coefficients are added ($g_{0ij}\,,i+j=k$) to
those of the $N^{k-2}LL$ resummation. Hence, in order to perform a
$N^{k-1}LL$ resummation, we need to determine
\begin{equation}
N_{k}=\sum_{p=1}^{k}(p+1)=\frac{k(k+3)}{2},
\end{equation}
coefficients. We want to determine $k_{\rm min}$ in
Eq.(\ref{gammafo}) so that all the $k(k+3)/2$ are fixed by the
same number of independent conditions obtained from the fixed
order expansion of the resummed physical anomalous dimension. We
note, first of all, that this happens if we can extract $2$
independent conditions from the LL contributions, $3$ from the NLL
one and $k+1$ from the $N^{k-1}LL$. The $N^{k-1}LL$ expression of
the physical anomalous dimension in this case is given by:
\begin{eqnarray}\label{resformfattfort}
\gamma(N,\alpha_{s}(Q^{2}))&=&\int_{1}^{N}\frac{dn}{n}\left(\sum_{i=1}^{k}g_{0i0}\alpha_{s}^{i}(Q^{2}/n^{2})\right)
+\int_{1}^{N}\frac{dn}{n}\left(\sum_{i=1}^{k}g_{00i}\alpha_{s}^{i}(Q^{2}/n)\right)+\nonumber\\
&&+\int_{1}^{N}\frac{dn}{n}\sum_{s=2}^{k}\sum_{i=1}^{s-1}g_{0is-i}\alpha_{s}^{i}(Q^{2}/n^{2})\alpha_{s}^{s-i}(Q^{2}/n).
\end{eqnarray}
The first two LL contributions have been already computed and are
given in Eq.(\ref{LL}).

Now, we extract the first $3$ NLL contributions of
Eq.(\ref{resformfattfort}). Recalling the derivation of
Eq.(\ref{PLL}), we obtain that these contributions are contained
in the following expression:
\begin{eqnarray}\label{27}
&&\frac{1}{2b_{0}}\left[(b_{1}^{1}g_{010}+g_{020})(\alpha_{s}(Q^{2}/N^{2})-
\alpha_{s}(Q^{2}))+2(b_{1}^{1}g_{001}+g_{002})(\alpha_{s}(Q^{2}/N)-\alpha_{s}(Q^{2}))\right]\nonumber\\
&&+\int_{1}^{N}\frac{dn}{n}g_{011}\alpha_{s}(Q^{2}/n^{2})\alpha_{s}(Q^{2}/n).
\end{eqnarray}
Since
\begin{eqnarray}
&&\alpha_{s}(Q^{2}/p^{a})=\frac{\alpha_{s}(Q^{2})}{1+a\beta_{0}\alpha_{s}(Q^{2})\ln\frac{1}{p}}+\emph{O}(\alpha_{s}^{2+i}\ln^{j}\frac{1}{p}),
\quad i\geq 0,\quad1\leq j\leq i\nonumber\\
&&\frac{1}{1+ab_{0}\alpha_{s}(Q^{2})\ln\frac{1}{p}}=
\sum_{j=0}^{\infty}(-)^{j}a^{j}b_{0}^{j}\alpha_{s}^{j}(Q^{2})\ln^{j}\frac{1}{p},
\end{eqnarray}
and keeping in mind that corrections to the NLL come from
Eq.(\ref{LL}), we have that Eq.(\ref{27}) become:
\begin{eqnarray}\label{29}
&&\frac{1}{2b_{0}}\sum_{j=1}^{\infty}(-)^{j}[2^{j}(c_{1j}^{1}g_{010}+g_{020})+
2(d_{1j}^{1}g_{001}+g_{002})]b_{0}^{j}\alpha_{s}^{j+1}(Q^{2})\ln^{j}\frac{1}{N}+\\
&&+\int_{1}^{N}\frac{dn}{n}g_{011}\alpha_{s}^{2}(Q^{2})
\left(\sum_{j=0}^{\infty}(-)^{j}b_{0}^{j}\alpha_{s}^{j}(Q^{2})\ln^{j}\frac{1}{n}\right)\left(
\sum_{i=0}^{\infty}(-)^{i}2^{i}b_{0}^{i}\alpha_{s}^{i}(Q^{2})\ln^{i}\frac{1}{n}\right).\nonumber
\end{eqnarray}
The Cauchy product of the two series in Eq.(\ref{29}) is given by
\begin{eqnarray}
&&(\sum_{j=0}^{\infty}(-)^{j}b_{0}^{j}\alpha_{s}^{j}(Q^{2})\ln^{j}\frac{1}{n})(
\sum_{i=0}^{\infty}(-)^{i}2^{i}b_{0}^{i}\alpha_{s}^{i}(Q^{2})\ln^{i}\frac{1}{n})=\nonumber\\
&&=\frac{1}{2b_{0}}\sum_{j=1}^{\infty}(-)^{j-1}
\left(\sum_{i=1}^{j}2^{i}\right)b_{0}^{j}\alpha_{s}^{j-1}(Q^{2})\ln^{j-1}\frac{1}{n}
\end{eqnarray}
and
\begin{equation}
\sum_{i=1}^{j}2^{i}=2(2^{j}-1).
\end{equation}
Now, because
\begin{equation}\label{varchange}
\frac{dn}{n}=-d\ln\frac{1}{n},
\end{equation}
we can perform the integration. We get
\begin{equation}\label{32}
\frac{1}{2b_{0}}\sum_{j=1}^{3}(-)^{j}\left[2^{j}(c_{1j}^{1}g_{010}+g_{020})+2(d_{1j}^{1}g_{001}+g_{002})+
\frac{2(2^{j}-1)}{j}g_{011}\right]
b_{0}^{j}\alpha_{s}^{j+1}(Q^{2})\ln^{j}\frac{1}{n},
\end{equation}\\
where $c_{1}^{1j},d_{1}^{1j}$ are certain coefficients we don not
need to worry about. The first three NLL contributions are given
by $j=1,2,3$.

The last step is to extract the first $k+1$ $N^{k-1}LL$
contributions that come from Eq.(\ref{resformfattfort}). Recalling
how Eq.((\ref{PNk-1LL})) was computed, we have that the desired
contributions are contained in the following expression:
\begin{eqnarray}\label{33}
&&\frac{1}{2b_{0}}\bigg[\frac{1}{k-1}(b_{k-1}^{1}g_{010}+
\cdots+b_{k-1}^{k-1}g_{0k-10}+g_{0k0})(\alpha_{s}^{k-1}(Q^{2}/N^{2})-\alpha_{s}^{k-1}(Q^{2}))+\nonumber\\
&&+\frac{2}{k-1}(b_{k-1}^{1}g_{001}+\cdots+b_{k-1}^{k-1}g_{00k-1}+g_{00k})(\alpha_{s}^{k-1}(Q^{2}/N)-\alpha_{s}^{k-1}(Q^{2}))\bigg]+\nonumber\\
&&+\int_{1}^{N}\frac{dn}{n}\sum_{i=1}^{k-1}g_{0ik-i}\alpha_{s}^{i}(Q^{2}/n^{2})\alpha_{s}^{k-i}(Q^{2}/n)\label{intfond}
\end{eqnarray}
We use the following relations
\begin{eqnarray}\label{sviluppodialfa}
&&\alpha_{s}^{r}(Q^{2}/p^{a})=\frac{\alpha_{s}^{r}(Q^{2})}{(1+ab_{0}\alpha_{s}(Q^{2})\ln\frac{1}{p})^{r}}+
\emph{O}(\alpha_{s}^{2+i}\ln^{j}\frac{1}{p}),\quad i\geq 0,1\leq j\leq i+1,\nonumber\\
&&\frac{1}{(1+ab_{0}\alpha_{s}(Q^{2})\ln\frac{1}{p})^{r}}=
\sum_{m=0}^{\infty}(-)^{m}{r+m-1 \choose
m}a^{m}b_{0}^{m}\alpha_{s}^{m}(Q^{2})\ln^{m}\frac{1}{p},
\end{eqnarray}
where  ${r \choose m}$ are the usual binomial coefficients. With
this, we can compute the integral in Eq.(\ref{intfond}) performing
the Cauchy product of the two series expansion of
$\alpha_{s}(Q^{2}/n^{2})$ and of $\alpha_{s}(Q^{2}/n)$ and
performing explicitly the integral using the change of variable
Eq.(\ref{varchange}). We get:
\begin{equation}
\int_{1}^{N}\frac{dn}{n}\sum_{i=1}^{k-1}g_{0ik-i}\alpha_{s}^{i}(Q^{2}/n^{2})\alpha_{s}^{k-i}(Q^{2}/n)=
\sum_{i=1}^{k-1}g_{0ik-i}\sum_{m=0}^{\infty}C_{m}^{(i,k-i)}b_{0}^{m}\alpha_{s}^{k+m}(Q^{2})\ln^{m+1}\frac{1}{N},
\end{equation}
where
\begin{equation}
C_{m}^{(i,j)}=\frac{(-1)^{m+1}}{m+1}\sum_{l=0}^m 2^l
\left(\begin{array}{c}l+i-1\\i-1\end{array}\right)
\left(\begin{array}{c}m-l+j-1\\j-1\end{array}\right) \label{Cijm},
\end{equation}
and where
\begin{equation}
{n\choose -1}=\frac{\Gamma(n+1)}{\Gamma(0)\Gamma(n+2)}
\end{equation}
is equal to $1$ for $n=-1$ and $0$ otherwise. Thefore, keeping in
mind the calculation of Eq.(\ref{PNk-1LL}) and taking the first
$k+1$ $N^{k-1}LL$ contributions of Eq.(\ref{33}), we have
\begin{eqnarray}
&&\sum_{m=0}^{k}\bigg\{\frac{(-)^{m+1}}{k-1}{k+m-1 \choose m+1}
[2^{m}(c_{k-1m}^{1}g_{010}+\cdots +c_{k-1m}^{k-1}g_{0k-10}+g_{0k0})+\nonumber\\
&&+(d_{k-1m}^{1}g_{001}+\cdots +d_{k-1j}^{k-1}g_{00k-1}+g_{00k})+\sum_{t=2}^{k-1}\sum_{i=1}^{t-1}g_{0it-i}f^{(k-1)}_{itm}]
+\nonumber\\
&&+\sum_{i=1}^{k-1}g_{0ik-i}C_{m}^{(i,k-i)}\bigg\}b_{0}^{m}\alpha_{s}^{k+m}(Q^{2})\ln^{m+1}\frac{1}{N},\label{noncistodentro}
\end{eqnarray}
where  $c^{i}_{k-1m},d^{i}_{k-1m},f^{(k-1)}_{itm}$ are certain
coefficients we do not have to worry about. At this point, we can
make some simplifications. In fact, since
\begin{equation}
C_{m}^{(0,k)}=\frac{(-1)^{m+1}}{m+1}{m+k-1\choose
k-1}=\frac{(-1)^{m+1}}{k-1}{m+k-1\choose m+1}
\end{equation}
and (see Appendix \ref{combinatorics})
\begin{equation}
C_{m}^{(k,0)}=2^{m}C_{m}^{(0,k)},
\end{equation}
we can write Eq.(\ref{noncistodentro}) in the following form:
\begin{eqnarray}
&&\sum_{m=0}^{k}\bigg\{
[C_{m}^{(k,0)}(c_{k-1m}^{1}g_{010}+\cdots +c_{k-1m}^{k-1}g_{0k-10})+\nonumber\\
&&+C_{m}^{(0,k)}(d_{k-1m}^{1}g_{001}+\cdots
+d_{k-1j}^{k-1}g_{00k-1})+C_{m}^{(0,k)}\sum_{t=2}^{k-1}\sum_{i=1}^{t-1}g_{0it-i}f^{(k-1)}_{itm}]
+\nonumber\\
&&+\sum_{i=0}^{k}g_{0ik-i}C_{m}^{(i,k-i)}\bigg\}b_{0}^{m}\alpha_{s}^{k+m}(Q^{2})\ln^{m+1}\frac{1}{N},\label{noncistodentro2}
\end{eqnarray}
What this result tells us, is that passing from the $N^{k-2}LL$ to
the $N^{k-1}LL$ resummation, $k+1$ new resummation coefficients
are added. In Eq.(\ref{noncistodentro2}), we have $k+1$ conditions
for this coefficients (one for each $m$) to be set equal to the
corresponding fixed order contribution of Eq.(\ref{gammafo}) . We
shall now show that this conditions are independent. This is
equivalent to showing that the $k+1$ linear combinations
\begin{equation}
\sum_{i=0}^{k}g_{0ik-i}\tilde{C}_{m}^{(i,k-i)}\qquad m=0,1,\dots,k
\end{equation}
with
\begin{equation}\label{relaz1bis}
\tilde{C}_{m}^{(i,j)}\equiv
(-)^{m+1}(m+1)C_{m}^{(i,j)}=\sum_{l=0}^{m}2^{l}{l+i-1 \choose
i-1}{m-l+j-1 \choose j-1}
\end{equation}
are independent. Moreover, this is equivalent to showing that for
each $k$ the columns of the $(k+1)\times(k+1)$ matrix
$A_{mi}^{k}\equiv \tilde{C}_{m}^{(i,k-1)}$ are independent
vectors. To show this, we need to use two identities proved in
Appendix \ref{combinatorics}:
\begin{eqnarray}
\tilde{C}_{m}^{(i,0)}&=&2^{m}\tilde{C}_{m}^{(0,i)}\label{relaz2}\\
\tilde{C}_{m}^{(i,j)}&=&2\tilde{C}_{m}^{(i,j-1)}-\tilde{C}_{m}^{(i-1,j)};\qquad\quad
i,j\geq 1,\label{relaz1}\\
\tilde{C}_{m}^{(0,i)}&=&{m+i-1\choose i-1}\label{relaz3}
\end{eqnarray}
which allows us to to compute the columns of the matrix
$A^{(k)}_{mj}$ explicitly for all $k$. To show their independence,
we use induction on $k$, i.e. we demonstrate the independence of
the columns for $k=1$ and then we assume that the property is
valid for $k-1$ to prove that it remains valid for $k$. In the
case $k=1$, we have a $2\times 2$ matrix:
\begin{equation}
(\tilde{C}_{m}^{(0,1)}\quad \tilde{C}_{m}^{(1,0)})=(1\quad 2^{m}),
\end{equation}
and now it is clear the two columns are independent, because $1$
and $2^{m}$ are independent functions of $m$. Now, using the
induction hypothesis
\begin{equation}\label{ipind}
\sum_{i=0}^{k-1}\alpha_{i}\tilde{C}_{m}^{(i,k-1-i)}=0
\Longleftrightarrow \alpha_{i}=0,\quad i=0,\dots ,k-1,
\end{equation}
we want to show that it is sufficient to prove that:
\begin{equation}\label{fin}
\sum_{i=0}^{k}\beta_{i}\tilde{C}_{m}^{(i,k-i)}=0
\Longleftrightarrow \beta_{i}=0,\quad i=0,\dots ,k.
\end{equation}
Using the relations (\ref{relaz2},\ref{relaz1}), we have:
\begin{eqnarray}
\sum_{j=0}^{k}\beta_{j}\tilde{C}_{m}^{(j,k-j)}&=&
(\beta_{0}+2^{m}\beta_{k})\tilde{C}_{m}^{(0,k)}+\sum_{j=1}^{k-1}\beta_{j}\tilde{C}_{m}^{(j,k-j)}\nonumber\\
&=&(\beta_{0}+2^{m}\beta_{k})\tilde{C}_{m}^{(0,k)}+
2\sum_{j=1}^{k-1}\beta_{j}\tilde{C}_{m}^{(j,k-1-j)}-
\sum_{j=1}^{k-1}\beta_{j}\tilde{C}_{m}^{(j-1,k-j)}\nonumber\\
&=&(\beta_{0}+2^{m}\beta_{k})\tilde{C}_{m}^{(0,k)}+
2\sum_{j=1}^{k-1}\beta_{j}\tilde{C}_{m}^{(j,k-1-j)}-
\sum_{j'=0}^{k-2}\beta_{j'+1}\tilde{C}_{m}^{(j',k-1-j')}\nonumber\\
&=&(\beta_{0}+2^{m}\beta_{k})\tilde{C}_{m}^{(0,k)}-
\beta_{1}\tilde{C}_{m}^{(0,k-1)}+\sum_{j=1}^{k-2}(2\beta_{j}-\beta_{j+1})\tilde{C}_{m}^{(j,k-1-j)}\nonumber\\
&&+2\beta_{k-1}\tilde{C}_{m}^{(k-1,0)}=0\label{indfin}
\end{eqnarray}
Now, from Eq.(\ref{relaz3}), we know that $\tilde{C}_{m}^{(0,k)}$
is a degree-($k-1$) polynomial in $m$. Furthermore, from
Eq.(\ref{relaz1}), we know that the vectors
$\tilde{C}_{m}^{(j,k-1-j)}$ with $j=0,\dots ,k-1$ are at most
polynomials of degree $k-2$ in $m$. Consequently,
Eq.(\ref{indfin}) can be satisfied if and only if
\begin{eqnarray}
(\beta_{0}+2^{m}\beta_{k})\tilde{C}_{m}^{(0,k)}=0,\\
-\beta_{1}\tilde{C}_{m}^{(0,k-1)}+\sum_{j=1}^{k-2}(2\beta_{j}-\beta_{j+1})\tilde{C}_{m}^{(j,k-1-j)}+2\beta_{k-1}\tilde{C}_{m}^{(k-1,0)}=0.
\end{eqnarray}
From the first, it follows that:
\begin{equation}\label{fin1}
\beta_{0}=\beta_{k}=0,
\end{equation}
while from the second, thanks to the induction hypothesis
Eq.(\ref{ipind}), we have that
\begin{equation}\label{fin2}
\beta_{1}=\beta_{k-1}=0,
\end{equation}
and that
\begin{equation}\label{fin3}
\beta_{1}=\frac{1}{2}\beta_{2}=\frac{1}{4}\beta_{3}=\dots
=\frac{1}{2^{k-2}}\beta_{k-1}.
\end{equation}
In conclusion from Eqs.(\ref{fin1},\ref{fin2},\ref{fin3}) it
follows that
\begin{equation}
\beta_{j}=0,\qquad\qquad j=0,\dots ,k.
\end{equation}
This completes the proof that the columns of the squared
$(k+1)\times (k+1)$ matrices $A^{(k)}_{mj}\equiv
\tilde{C}_{m}^{(j,k-j)}$ are independent for all $k$. This shows
that, in order to obtain a $N^{k-1}LL$ resummation in the case of
the weaker factorization (Eq.(\ref{resformulaint}) in section
\ref{DPresform}), we need to know a $N^{2k}LO$ fixed order
calculation of the physical anomalous dimension. Hence, in this
case $k_{\rm min}=2k$.

\section{Prompt photon in the general case}

Let us now consider the most general case, in which the
coefficient function does not satisfy  any factorization property.
This is the case of Eq.(\ref{resformula}) in section
\ref{DPresform}. In this case, in order to perform a LL
resummation we need $2$ coefficients $(g_{001},g_{010})$; to
perform a NLL resummation $5$ coefficients are added
$(g_{002},g_{101},g_{020},g_{110},g_{011})$ and to perform a
$N^{k-1}LL$ resummation $k(k+3)/2$ coefficients are added
($g_{mnp}$ with $m+n+p=k$ but without $n=p=0$). Thus, in order to
perform a $N^{k-1}LL$ resummation, we need to determine
\begin{equation}
\label{nofnum}
N_k=\sum_{p=1}^k\frac{p(p+3)}{2}=\frac{k(k+1)(k+5)}{6}
\end{equation}
coefficients. The $N^{k-1}LL$ expression of the physical anomalous
dimension is given, in this case, by
\begin{equation}
\gamma(N,\alpha_{s}(Q^{2}))=\int_{1}^{N}\frac{dn}{n}\sum_{s=1}^{k}\sum_{i=0}^{s-1}\sum_{j=0}^{s-i}g_{ijs-i-j}\alpha_{s}^{i}(Q^{2})
\alpha_{s}^{j}(Q^{2}/n^{2})\alpha_{s}^{s-i-j}(Q^{2}/n).
\end{equation}
Now, we proceed in the same way as we have done in section
\ref{dpcasoint} and we find that the $k(k+3)/2$ new coefficients
that are added passing from the $N^{k-2}LL$ to the $N^{k-1}LL$
contributions appear only in the following combinations:
\begin{equation}
\sum_{i=0}^{k-1}\sum_{j=0}^{k-i} g_{ijk-i-j}\, \sum_{m=0}^\infty
C^{(j,k-i-j)}_m \,b_0^m\as(Q^2)^{k+m}\,\ln^{m+1}\frac{1}{N}.
\label{gammak2}
\end{equation}
Each term with fixed $m$ in the expansion  Eq.~(\ref{gammak2})
provides a new condition on these coefficients. However, these
conditions are not linearly independent for all choices of $m$.
Indeed, let us define  the matrix $C_{m}^{(j,k-i-j)}\equiv
D_{m(i,j)}^{(k)}$, where the lines are labelled by the index $m$
and the columns by the multi-index $(i,j)$. This matrix  gives the
linear combination of the coefficients $g_{mnp}$ in
Eq.(\ref{gammak2}) to be determined and it turns out to be of rank
\begin{equation}\label{statement}
rg(D_{m(i,j)}^{(k)})=2k\leq \frac{k(k+3)}{2}.
\end{equation}
We shall now prove this statement:

$D^{(k)}_{m(i,j)}$ is a $M\times\frac{k(k+3)}{2}$ matrix , whose
columns are the $M$-component vectors
\begin{equation}
D^{(k)}_m=C^{(j,k-i-j)}_m;\qquad 0\leq i\leq k-1;\quad 0\leq j\leq
k-i;\qquad 0\leq m\leq M.
\end{equation}
We use induction on $k$. For $k=1$,
$D^{(1)}$ is a $2\times 2$ matrix with columns
\begin{equation}
D^{(1)}_m=\left(C_m^{(0,1)},\quad C_m^{(1,0)}\right)
=\frac{(-1)^{m+1}}{m+1}\,(1,\quad 2^m),
\end{equation}
that are linearly independent; the rank of $D^{(1)}$ is 2. Let us
check explicitly also the case $k=2$. In this case
\begin{equation}
D^{(2)}_m= \left(C_m^{(0,1)},\quad C_m^{(1,0)}, \quad
C_m^{(0,2)},\quad C_m^{(1,1)},\quad C_m^{(2,0)}\right).
\end{equation}
The first two columns are the same as in the case $k=1$: they span
a 2-dimensional subspace. The last three columns are independent
as a consequence of Eq.(\ref{fin}) with $k=1$. Furthermore,
$C_m^{(0,2)}$ and $C_m^{(2,0)}=2^m C_m^{(0,2)}$ are independent of
all other columns, because they are the only ones that are
proportional to a degree-1 polynomial in $m$. Finally,
$C_m^{(1,1)}$ is a linear combination of the first two columns, as
a consequence of Eqs.(\ref{relaz1bis},\ref{relaz1}) with $i=j=1$.
Thus, the rank of $D^{(2)}$ is $2+2=4$.

We now assume that $D^{(k-1)}$ has rank $2(k-1)$, and we write the
columns of $D^{(k)}$ as \bea &&D^{(k)}_m=
\left(C_m^{(j,k-1-i-j)},\quad C_m^{(l,k-l)}\right)
\\
&&0\leq i\leq k-2,\qquad 0\leq j\leq k-1-i\qquad \qquad 0\leq
l\leq k. \eea By the induction hypothesis, only $2(k-1)$ of the
columns $C_m^{(j,k-1-i-j)}$ are independent. The columns
$C_m^{(l,k-l)}$ are all independent as a consequence of Eq.(
\ref{fin}); among them, those with $1\leq l\leq k-1$ can be
expressed as linear combinations of $C_m^{(j,k-1-i-j)}$ by
Eq.~(\ref{relaz1}). Only $C_m^{(0,k)}$ and $C_m^{(k,0)}$ are
independent of all other columns because they are proportional to
a degree-$(k-1)$ polynomial in $m$, while all others are at most
of degree $(k-2)$. Hence, only two independent vectors are added
to the $2(k-1)$-dimensional subspace spanned by
$C_m^{(j,k-1-i-j)}$, and the rank of $D^{(k)}$ is \beq
2(k-1)+2=2k. \eeq It follows that each individual terms in the sum
over $m$ in Eq.~(\ref{gammak2}) depends only on $2k$ independent
linear combinations of the coefficients $g_{ijk-i-j},\> 0,\le i
\le k-1,\>0\le j\le k-i$.

This means that the N$^{k-1}$LL order resummed result depends only
on $2k$ independent linear combinations of the $k(k+3)/2$ new
coefficients that are added passing from the $N^{k-2}LL$
resummation to the $N^{k-1}LL$ one and that the remaining
coefficients are arbitrary. Because a term with fixed $m$ in
Eq.(\ref{gammak2}) is of order $\as^{k+m}$, this implies that a
computation of the anomalous dimension up to fixed order $k_{\rm
min}=3k-1$ is sufficient for the N$^{k-1}$LL resummation, because
$m=0,1,\dots,k-1$. Note that when going from N$^{k-1}$LL to
N$^{k}$LL, at this higher order, in general some new linear
combinations of the $k(k+3)/2$ coefficients, that we have added
from the $N^{k-2}LL$ to the $N^{k-1}LL$, will appear through terms
depending on $\beta_1$. Hence, some of the combinations of
coefficients that were left undetermined in the N$^{k-1}$LL
resummation will now become determined. However, this does not
affect the value $k_{\rm min}$ of the fixed-order accuracy needed
to push the resummed accuracy at one extra order. In conclusion,
even in the absence of any factorization, despite the fact that
now the number of coefficients which must be determined grows
cubically according to Eq.~(\ref{nofnum}), the required order in
$\alpha_s$ of the computation which determines them grows only
linearly.

\section{DIS and DY in all cases}

This is the case of Eq.(\ref{disdyres}) in section \ref{DISDY3}.
Here we discuss the case without assuming any factorization
property, because the factorized case will be recovered as a
particular case. So, in the general case, at LL we need to
determine $1$ coefficient ($g_{01}$); at NLL $2$ coefficients are
added ($g_{02},g_{11}$) and at $N^{k-1}LL$, $k$ coefficients
($g_{ik-i}$ with $i=0,1,\dots,k-1$) are added to the $N^{k-2}LL$
ones . Thus, in order to perform a $N^{k-1}LL$ resummation, we
need to determine
\begin{equation}
N_{k}=\sum_{p=1}^{k}p=\frac{k(k+1)}{2}
\end{equation}
coefficients. The $N^{k-1}LL$ expression of the physical anomalous
dimension is given by
\begin{equation}
\gamma(N,\alpha_{s}(Q^{2}))=a\int_{1}^{N}\frac{dn}{n}\sum_{s=1}^{k}\sum_{i=0}^{s-1}g_{is-i}\alpha_{s}^{i}(Q^{2})
\alpha_{s}^{s-i}(Q^{2}/n^{a}),
\end{equation}
where $a=1$ for DIS and $a=2$ for DY. Now, we proceed in the same
way as we have done in the previous sections and we find that the
$k$ new coefficients that are added passing from the $N^{k-2}LL$
to the $N^{k-1}LL$ contributions appear only in the following
combinations:
\begin{equation}
\sum_{i=0}^{k-1}g_{ik-i}\, \sum_{m=0}^\infty a^{m+1} C^{(0,k-i)}_m
\,b_0^m\as(Q^2)^{k+m}\,\ln^{m+1}\frac{1}{N}. \label{gammak2ciao}
\end{equation}
Again, each term with fixed $m$ in this expansion provides a new
condition on these coefficients. These conditions are all linearly
independent for any choice of $m$, because the $k\times k$ matrix
$A^{(k)}_{mi}\equiv C_{m}^{0,k-i}$ with $m,i=0,1,\dots,k-1$ has
all independent columns. This is a direct consequence of the fact
that each column is a polynomial with different degree in $m$.
This implies that, in order to determine a $N^{k-1}LL$
resummation, a computation of the physical anomalous dimension up
to order $k_{\rm min}=2k-1$ is sufficient. In the more restrictive
case Eq.(\ref{disdyresfac}) in Section \ref{DISDY3}, the only
non-vanishing coefficients are $g_{0,k}$. This means that going
from the $N^{k-2}LL$ to the $N^{k-1}LL$ we need to take only one
combination of the expansion Eq.(\ref{gammak2ciao}). Hence, in
this more restrictive case, $N_{k}=k$ and $k_{\rm min}=k$.

\section{DY small transverse momentum distribution}

This is the case of Eq.(\ref{qtresform}) in section \ref{bordello}
obtained with the renormalization group approach. The result of
the approach of Ref.\cite{Collins:1984kg} (reported in
Eq.(\ref{qtresform}) of section \ref{bordello}) will be recovered
as a particular case. In the most general case, at LL we need to
determine $1$ coefficient ($G_{01}$); at NLL $3$ more coefficients
are added ($G_{11},G_{02},\tilde{G}_{1}$) and at $N^{k-1}LL$,
$k+1$ coefficients ($G_{ik-i}$ with $i=0,1,\dots,k-1$ and
$\tilde{G}_{k-1}$) are added. Therefore, in order to perform a
$N^{k-1}LL$ resummation, we need to determine
\begin{equation}
N_{k}=\sum_{p=1}^{k}(p+1)-1=\frac{k^{2}+3k-2}{2}
\end{equation}
coefficients. The $N^{k-1}LL$ expression of the physical anomalous
dimension is given by
\begin{equation}
\gamma(1,\bar{\mu}^{2}b^{2},\alpha_{s}(\bar{\mu}^{2}))=
-\int_{0}^{\bar{\mu}^{2}b^{2}}\frac{dn'}{n'}\sum_{s=1}^{k}\sum_{i=0}^{s-1}
G_{is-i}\alpha_{s}^{i}(\bar{\mu}^{2})\alpha_{s}^{s-i}(\bar{\mu}^{2}/n')
+\sum_{i=1}^{k-1}\tilde{G}_{i}\alpha_{s}^{i}(\bar{\mu}^{2}).
\end{equation}
Now, we proceed as before and we find that the $k$ new
coefficients that are added passing from the $N^{k-2}LL$ to the
$N^{k-1}LL$ contributions appear only in the following
combinations:
\begin{equation}
-\sum_{i=0}^{k-1}G_{ik-i}\, \sum_{m=0}^\infty C^{(0,k-i)}_m
\,b_0^m\as(\bar{\mu}^{2})^{k+m}\,\ln^{m+1}\left(\frac{1}{\bar{\mu}^{2}b^{2}}\right)
+\tilde{G}_{k-1}\alpha_{s}^{k-1}(\bar{\mu}^{2}). \label{gammak2}
\end{equation}
As before, each term with fixed $m$ in this expansion provides a
new independent condition on this coefficients. This implies that,
in order to determine a $N^{k-1}LL$ resummation, a computation of
the physical anomalous dimension up to order $k_{\rm min}=2k-1$ is
sufficient. In the more restrictive case of Eq.(\ref{qtresform2}
sec.\ref{bordello}), the only non-vanishing coefficients are
$G_{0,k}$ and $\tilde{G}_{k-1}$. This means that going from the
$N^{k-2}LL$ to the $N^{k-1}LL$ only $2$ coefficients are added but
the LL where we have only one coefficient. Hence, in this more
restrictive case, $N_{k}=2k-1$ and $k_{\rm min}=k$.

\chapter{Conclusions}

In this thesis, we have studied the renormalization group approach
to resummation for all inclusive deep-inelastic and Drell-Yan
processes. The advantage of this approach is that it does not rely
on factorization of the physical cross section, and in fact it
simply follows from general kinematic properties of the phase
space. Then we have analyzed some of its generalizations.

In particular, we have presented a generalization to prompt photon
production of the approach to Sudakov resummation which has been
described in Chapter \ref{DISDY2} for deep-inelastic scattering
and Drell-Yan production.  It is interesting to see that also with
the more intricate two-scale kinematics that characterizes prompt
photon production in the soft limit, it remains true that
resummation simply follows from general kinematic properties of
the phase space. Also, this approach does not require a separate
treatment of individual colour structures when more than one
colour structure contributes to fixed order results.

The resummation formulae obtained here turn out to be less
predictive than previous results: a higher fixed-order computation
is required in order to determine the resummed result. This
depends on the fact that here neither specific factorization
properties of the cross section in the soft limit is assumed, nor
that soft emission satisfies eikonal-like relations which allow
one to determine some of the resummation coefficients in terms of
universal properties of collinear radiation. Currently,
fixed-order results are only available up to $O(\alpha_s^2)$ for
prompt photon production. An order $\as^3$ computation is required
to check nontrivial properties of the structure of resummation:
for example, factorization, whose effects only appear at the
next-to-leading log level, can only be tested at $O(\alpha_s^3)$.
The greater flexibility of the approach presented here would turn
out to be necessary if the prediction obtained using the more
restrictive resummation  were to fail at order $\alpha_s^3$.

We have also proved a resumation formula for the Drell-Yan
rapidity distributions to all logarithmic accuracy and valid for
all values of rapidity. Isolating a universal dimensionless
coefficient function, which is exactly that ones of the Drell-Yan
rapidity-integrated, we have shown a general procedure to obtain
resummed results to NLL for the rapidity distributions of a
virtual photon $\gamma^{*}$ or of a real vector boson
$W^{\pm},Z^{0}$. Furthermore, we have outlined a general method to
calculate numerical predictions and analyzed the impact of
resummation for the fixed-target experiment E866/NuSea. This shows
that NLL resummation has an important effects on predictions of
differential rapidity cross sections giving an agreement with data
that is better than NNLO full calculations. We have found a
suppression of the cross section for not large values of hadronic
rapidity instead of enhancing it. This suppression arises due to
the shift in the complex plane of the dominant contribution of
resummed exponent. These leaves open questions for future studies
about possible suppression of the rapidity integrated cross
sections at small $x$.

The study of the renormalization group resummation applied to the
case of small transverse momentum distribution of Drell-Yan pairs
has opened further interestingly aspects about the relation
between factorization properties of the cross section and the
final structure of the resummed results which has been not yet
well understood.

Furthermore, because of its generality, renormalization group
resummation lends itself naturally to some important future
applications. They are the factorization of resummation of rare
meson decay processes (like $B\rightarrow X_{s}\gamma$) and the
resummation of generalized parton densities in deeply virtual
proton Compton scattering.

\appendix
\chapter{Relations between logarithms of $N$ and logarithms of
$(1-z)$}\label{lognlogx}

In this Appendix we want to find the general relations between the
logarithms of $N$ and the logarithms of $(1-z)$. Let's consider a
generic logarithmical enhanced term in $z$ space
\begin{equation}\label{logz}
\left[\frac{\ln^{p}(1-z)}{1-z}\right]_{+}
\end{equation}
and take its Mellin transform:
\begin{equation}\label{ip}
I_{p}\equiv \int_{0}^{1}dz\frac{z^{N-1}-1}{1-z}\ln^{p}(1-z).
\end{equation}
To find a general relation between the terms of the type of
Eq.(\ref{logz}) and the logs of $N$, we first notice that all
integrals $I_{p}$ Eq.(\ref{ip}) can be obtained from a generating
function $G(\eta)$:
\begin{eqnarray}
I_{p}&=&\frac{d^{p}}{d\eta^{p}}G(\eta)\vert_{\eta=0}, \label{ip2}\\
G(\eta)&=&\int_{0}^{1}dz(z^{N-1}-1)e^{(\eta-1)\ln(1-z)}=\frac{\Gamma(N)\Gamma(\eta)}{\Gamma(N+\eta)}-\frac{1}{\eta}.
\end{eqnarray}
Now, using the Stirging expansion of the $\Gamma$ function at
large $N$
\begin{equation}\label{stirling}
\Gamma(N+1)=\sqrt{2\pi N}e^{N\ln
N-N}+\emph{O}\left(\frac{1}{N}\right),
\end{equation}
we get
\begin{equation}\label{genfuncrel}
G(\eta)=\frac{1}{\eta}\left[\frac{\Gamma(1+\eta)}{N^{\eta}}-1\right]+\emph{O}\left(\frac{1}{N}\right).
\end{equation}
At this point, we notice that $N^{-\eta}$ is just the generating
function of the log of $N$:
\begin{equation}
L_{p}\equiv \ln^{p}\frac{1}{N}=\frac{d^{p}}{d\eta^{p}}e^{\eta\ln
1/N}\vert_{\eta=0}.
\end{equation}
Hence, Eq.(\ref{genfuncrel}) can be viewed as a relation between
the generating function for $I_{p}$ and for $L_{p}$. In
particular, Taylor-expanding $\Gamma(1+\eta)$ in
Eq.(\ref{genfuncrel}) leads to leading, next-to-leading,\dots $\ln
N$ relations:
\begin{eqnarray}
G(\eta)&=&\frac{1}{\eta}\left[\frac{1}{N^{\eta}}\sum_{k=0}^{\infty}\frac{\Gamma^{(k)}(1)}{k!}\eta^{k}-1\right]\nonumber\\
&=&\sum_{k=0}^{\infty}\frac{\Gamma^{(k)}(1)}{k!}\frac{1}{\eta}\frac{d^{k}}{d\ln^{k}1/N}[e^{\eta\ln
1/N}-1]\nonumber\\
&=&-\sum_{k=0}^{\infty}\frac{\Gamma^{(k)}(1)}{k!}\frac{d^{k}}{d\ln^{k}1/N}\int_{0}^{1-\frac{1}{N}}dze^{(\eta-1)\ln(1-z)}.
\end{eqnarray}
Now, if we put this last equation in Eq.(\ref{ip2}), we obtain:
\begin{eqnarray}
I_{p}&=&-\sum_{k=0}^{\infty}\frac{\Gamma^{(k)}(1)}{k!}\frac{d^{k}}{d\ln^{k}1/N}\int_{0}^{1-\frac{1}{N}}dz\frac{\ln^{p}(1-z)}{1-z}+
\emph{O}\left(\frac{1}{N}\right)\label{uffa}\\
&=&\frac{1}{p+1}\sum_{k=0}^{p+1} {p+1\choose
k}\Gamma^{(k)}(1)\left(\ln\frac{1}{N}\right)^{p+1-k}+\emph{O}\left(\frac{1}{N}\right),
\end{eqnarray}
where in the last equality we have used the identity
\begin{equation}
\frac{d^{k}}{d\ln^{k}1/N}\int_{0}^{1-\frac{1}{N}}dz\frac{\ln^{p}(1-z)}{1-z}=
-\frac{1}{p+1}k!{p+1\choose k}\left(\ln\frac{1}{N}\right)^{p+1-k}.
\end{equation}
This last expression is equal to zero when $k>p+1$. This result
can be expressed in terms of derivatives with respect to
$\ln(1-z)$ with the identity
\begin{equation}
\frac{d^{k}}{d\ln^{k}1/N}\int_{0}^{1-\frac{1}{N}}dz\frac{\ln^{p}(1-z)}{1-z}=
\int_{0}^{1-\frac{1}{N}}\frac{dz}{1-z}\frac{d^{k}\ln^{p}(1-z)}{d\ln^{k}(1-z)}
\end{equation}
thus obtaining for $I_{p}$ Eq.(\ref{uffa}) the following
all-logarithmic-order relation:
\begin{eqnarray}
I_{p}&=&-\sum_{k=0}^{p}\frac{\Gamma^{(k)}(1)}{k!}\int_{0}^{1-\frac{1}{N}}\frac{dz}{1-z}\frac{d^{k}\ln^{p}(1-z)}{d\ln^{k}(1-z)}\nonumber\\
&&+\frac{\Gamma^{(p+1)}(1)}{p+1}+\emph{O}\left(\frac{1}{N}\right).\label{pindep}
\end{eqnarray}

The inverse result, expressing $L_{p}$ in terms of $I_{p}$, can be
analogously found inverting the relation between the generating
functions Eq.(\ref{genfuncrel}):
\begin{equation}
N^{-\eta}=\frac{\eta G(\eta)+1}{\Gamma(1+\eta)}.
\end{equation}
Proceeding as before, we get
\begin{eqnarray}
\ln^{n}\frac{1}{N}&=&\sum_{i=1}^{n}{n\choose
i}\Delta^{(n-i)}(1)iI_{i-1}+\Delta^{(n)}(1)+\emph{O}\left(\frac{1}{N}\right)\\
&=&\sum_{k=1}^{n}\frac{\Delta^{(k-1)}(1)}{(k-1)!}\int_{0}^{1}dz\,\frac{z^{N-1}-1}{1-z}\frac{d^{k}\ln^{n}(1-z)}{d\ln^{k}(1-z)}\nonumber\\
&&+\Delta^{(n)}(1)+\emph{O}\left(\frac{1}{N}\right),
\end{eqnarray}
where $\Delta^{(k)}(\eta)$ is the $k$th derivative of
\begin{equation}
\Delta(\eta)\equiv\frac{1}{\Gamma(\eta)}.
\end{equation}

Because of the $p$-independence of the coefficients of the
expansion Eq.(\ref{pindep}), we can determine explicitly the
Mellin transform of a generic logarithmical enhanced function
\begin{equation}
\left[\frac{\hat{g}(\ln(1-z)^{a})}{1-z}\right]_{+}=\sum_{p=0}^{\infty}\hat{g}_{p}\left[\frac{\ln^{p}(1-z)^{a}}{1-z}\right]_{+}.
\end{equation}
Indeed, its Mellin transform up to non-logarithmical terms is
given by
\begin{eqnarray}
\int_{0}^{1}dz\,\frac{z^{N-1}-1}{1-z}\hat{g}(\ln(1-z)^{a})=\sum_{p=0}^{\infty}\hat{g}_{p}a^{p}I_{p}\nonumber\\
=\sum_{p=0}^{\infty}\frac{\hat{g}_{p}a^{p}}{p+1}\sum_{k=0}^{p+1}{p+1\choose
k}\Gamma^{(k)}(1)\left(\ln\frac{1}{N}\right)^{p+1-k}\\
=-\sum_{k=0}^{\infty}\frac{\Gamma^{(k)}(1)}{k!}\int_{0}^{1-\frac{1}{N}}\frac{dz}{1-z}\frac{d^{k}}{d\ln^{k}(1-z)^{a}}
\hat{g}(\ln(1-z)^{a})\nonumber\\
=\int_{0}^{1-\frac{1}{N}}\frac{dz}{1-z}g(\ln(1-z)^{a})=\frac{1}{a}\int_{1}^{N^{a}}\frac{dn}{n}g(\ln\frac{1}{n})\label{calcespl},
\end{eqnarray}
where in the last equality we have done the change of variable
$n=(1-z)^{-a}$ and where
\begin{equation}
g(\ln \mathcal{K})\equiv
-\sum_{k=0}^{\infty}\frac{\Gamma^{(k)}(1)a^{k}}{k!}\frac{d^{k}}{d\ln^{k}\mathcal{K}}\hat{g}(\ln
\mathcal{K}).
\end{equation}
The inverse relation can be analogously derive. Namely, we can
cast the integral of any function
\begin{equation}
g(\ln(1-z)^{a})=\sum_{p=0}^{\infty}g_{p}\ln^{p}(1-z)^{a}
\end{equation}
as a Mellin transform, up to non-logarithmic terms:
\begin{eqnarray}\label{fiorentina}
\frac{1}{a}\int_{1}^{N^{a}}\frac{dn}{n}g(\ln\frac{1}{n})=\int_{0}^{1}dz\frac{z^{N-1}-1}{1-z}\hat{g}(\ln(1-z)^{a}),
\end{eqnarray}
where
\begin{equation}\label{fiorentina2}
\hat{g}(\ln
\mathcal{K})\equiv-\sum_{k=0}^{\infty}\frac{\Delta^{(k)}(1)a^{k}}{k!}\frac{d^{k}}{d\ln^{k}\mathcal{K}}g(\ln
\mathcal{K}).
\end{equation}
For completeness, we recall that $\Gamma'(1)=-\gamma_{E}=-0.5772$
and that in some cases it can be useful to perform the change of
variable $n=n^{'a}$ so that we have
\begin{equation}
\frac{1}{a}\int_{1}^{N^{a}}\frac{dn}{n}g(\ln\frac{1}{n})=\int_{1}^{N}\frac{dn'}{n'}g(\ln\frac{1}{n^{'a}})
\end{equation}

\chapter{Phase space}\label{phasespacedec}

The $n$-body phase space can be expressed in terms of $(n-m)$-body
and $(m+1)$-body phase spaces. To prove this statement, let us
consider the definition of the phase space in $d=4-2\epsilon$
dimensions for a generic process with incoming momentum $P$ and
$n$ on-shell particles in the final state with outgoing momenta
$p_{1},p_{2},\dots,p_{n}$:
\begin{equation}\label{ps}
d\phi_{n}(P;p_{1},\dots,p_{n})=\frac{d^{d-1}p_{1}}{(2\pi)^{d-1}2p_{1}^{0}}
\dots\frac{d^{d-1}p_{n}}{(2\pi)^{d-1}2p_{n}^{0}}(2\pi)^{d}\delta_{d}(P-p_{1}-\dots-p_{n}).
\end{equation}
The momentum-conservation delta function can be rewritten as
\begin{eqnarray}
\delta_{d}(P-p_{1}-\dots-p_{n})&=&\int
d^{d}Q\delta_{d}(P-Q-p_{1}-\dots-p_{m})\nonumber\\
&&\times\delta_{d}(Q-p_{m+1}-\dots-p_{n}).\label{deltaprop}
\end{eqnarray}
The integration measure of this equation can be rewritten in this
way:
\begin{equation}\label{varchange}
d^{d}Q=(2\pi)^{d}\frac{d^{d-1}Q}{(2\pi)^{d-1}2Q^{0}}\frac{d(Q^{0})^{2}}{2\pi}
=(2\pi)^{d}\frac{d^{d-1}Q}{(2\pi)^{d-1}2Q^{0}}\frac{dQ^{2}}{2\pi},
\end{equation}
where
\begin{eqnarray}
Q^{2}&=&(Q^{0})^{2}-|\vec{Q}|^{2}\label{qsq1}\\
&=&(p_{m+1}+\dots+p_{n})^{2}\label{qsq2}\\
&=&(P-p_{1}-\dots-p_{m})^{2}.\label{qsq3}
\end{eqnarray}
We shall now find the minimum and the maximum value of the
variable $Q^{2}$. If we use Eq.(\ref{qsq2}) in the center-of-mass
frame of the momentum $p_{m+1}+\dots+p_{n}$, we have
\begin{equation}
Q^{2}=(p_{m+1}^{0}+\dots+p_{n}^{0})^{2}.
\end{equation}
The minimum value of $Q^{2}$ is achieved when all the energies
$p_{m+1}^{0},\dots,p_{n}^{0}$ are equal to their invariant masses
($m_{i}=\sqrt{p_{i}^{2}}$). Hence
\begin{equation}\label{qsqmin}
Q^{2}_{min}=(\sqrt{p_{m+1}^{2}}+\dots+\sqrt{p_{n}^{2}})^{2},
\end{equation}
because $Q^{2}$ is a Lorentz ivariant. From Eq.(\ref{qsq3}), we
have that in the center-of-mass frame of the momentum
$P-p_{1}-\dots-p_{m}$,
\begin{equation}
Q^{2}=(P^{0}-p_{1}^{0}-\dots-p_{m}^{0})^{2}.
\end{equation}
Now, the minimum value of $p_{1}^{0}+\dots+p_{m}^{0}$ is achieved
when all these terms reduces to their invariant masses
($m_{i}=\sqrt{p_{i}^{2}}$). In this case
\begin{equation}
\vec{p}_{1}=\dots=\vec{p}_{m}=0
\end{equation}
and this implies that $\vec{P}=0$ and that $P^{0}=\sqrt{P^{2}}$.
Therefore, we obtain
\begin{equation}\label{qsqmax}
Q^{2}_{max}=(\sqrt{P^{2}}-\sqrt{p_{1}^{2}}-\dots-\sqrt{p_{m}^{2}})^{2},
\end{equation}
again as a consequence of the Lorentz invariance of $Q^{2}$. We
obtain immediately the general phase space decomposition formula
using Eqs.(\ref{ps},\ref{deltaprop},\ref{varchange}) together:
\begin{eqnarray}
d\phi_{n}(P;p_{1},\dots,p_{n})&=&\int_{Q^{2}_{min}}^{Q^{2}_{max}}\frac{dQ^{2}}{2\pi}d\phi_{m+1}(P;Q,p_{1},\dots,p_{m})\nonumber\\
&&\times d\phi_{n-m}(Q;p_{m+1},\dots,p_{n}),\label{gendecps}
\end{eqnarray}
where $Q^{2}_{min}$ and $Q^{2}_{max}$ are given by
Eq.(\ref{qsqmin}) and Eq.(\ref{qsqmax}) respectively.  Using
recursively Eq.(\ref{gendecps}) with $m=1$, it is possible to
rewrite a $n$-body phase space in terms of $n$ two-body phase
spaces. We shall now compute the two-body phase space in the
center-of-mass frame. We have
\begin{eqnarray}
d\phi_{2}(P;Q,p)&=&\frac{d^{d-1}Q}{(2\pi)^{d-1}2Q^{0}}\frac{d^{d-1}p}{(2\pi)^{d-1}2p^{0}}
(2\pi)^{d}\delta_{d}(P-Q-p)\nonumber\\
&=&\frac{(2\pi)^{2\epsilon-2}}{4}\frac{d^{d-1}p}{Q^{0}p^{0}}\delta(P^{0}-Q^{0}-p^{0}).\label{pezzo1}
\end{eqnarray}
In the center-of-mass frame $\vec{P}=0$ and thus
$P^{0}=\sqrt{P^{2}}$. If we can neglect the invariant mass of $p$,
we have that in this frame $|\vec{Q}|=|\vec{p}|=p^{0}$ and that
\begin{eqnarray}
\delta(P^{0}-Q^{0}-p^{0})&=&\frac{\sqrt{(p^{0})^{2}+Q^{2}}}{\sqrt{P^{2}}}\delta\left(p^{0}-\frac{P^{2}-Q^{2}}{2\sqrt{P^{2}}}\right)\nonumber\\
&=&\frac{Q^{0}}{\sqrt{P^{2}}}\delta\left(p^{0}-\frac{P^{2}-Q^{2}}{2\sqrt{P^{2}}}\right).\label{pezzo2}
\end{eqnarray}
Now,
\begin{equation}
d^{d-1}p=|\vec{p}|^{d-2}d|\vec{p}|d\Omega_{d-1}=(p^{0})^{d-2}dp^{0}d\Omega_{d-1}.\label{pezzo3},
\end{equation}
where $d\Omega_{d-1}$ is the solid angle in $d-1$ dimensions.
Substituting Eqs.(\ref{pezzo2},\ref{pezzo3}) in Eq.(\ref{pezzo1})
and integrating over $p^{0}$, we obtain
\begin{eqnarray}
d\phi_{2}(P;Q,p)&=&\frac{(2\pi)^{2\epsilon-2}}{4}\frac{(p^{0})^{1-2\epsilon}}{\sqrt{P^{2}}}d\Omega_{d-1}\nonumber\\
&=&N(\epsilon)(P^{2})^{-\epsilon}\left(1-\frac{Q^{2}}{P^{2}}\right)^{1-2\epsilon}d\Omega_{d-1},\label{twobody}
\end{eqnarray}
where
\begin{equation}
N(\epsilon)=\frac{1}{2(4\pi)^{2-2\epsilon}}.
\end{equation}
 Finally, we want to calculate
$d\Omega_{d-1}$. To do this, we use its definition
\begin{eqnarray}
d\Omega_{d-1}&=&d\theta_{d-1}\sin^{d-3}\theta_{d-1}d\Omega_{d-2},\label{solidangred}
\end{eqnarray}
which can be applied recursively $i$ times till $d-2-i>0$ with
$0\leq\theta_{d-i}\leq\pi$. The normalization of the solid angles
can be obtained performing the gaussian integral in spherical
coordinates thus giving
\begin{equation}\label{omeganorm}
\Omega_{d-1}=\int
d\Omega_{d-1}=\frac{2\pi^{(d-1)/2}}{\Gamma((d-1)/2)},
\end{equation}
where $\Gamma$ is the usual gamma function
\begin{equation}
\Gamma(\alpha)=\int_{0}^{\infty}dt\,e^{-t}t^{\alpha-1}
\end{equation}
  In our
case ($d=4-2\epsilon$) we can use Eq.(\ref{solidangred}) two times
and we odtain
\begin{equation}
d\Omega_{d-1}=d\Omega_{3-2\epsilon}=\sin^{1-2\epsilon}\theta\,d\theta\sin^{-2\epsilon}\phi
\,d\phi\,d\Omega_{1-2\epsilon}.
\end{equation}
In many cases it is useful to rewrite this equation in terms of
other variables variables
\begin{equation}
y_{1}=\frac{1+\cos\theta}{2},\qquad
y_{2}=\frac{1+\cos\phi}{2},\qquad 0\leq y_{i}\leq 1.
\end{equation}
Doing this change of variables and recalling that the two-body
squared amplitude cannot depend on more than two angles, we obtain
\begin{equation}
d\Omega_{3-2\epsilon}=\frac{4^{1-2\epsilon}\pi^{1/2-\epsilon}}{\Gamma(1/2-\epsilon)}\int_{0}^{1}dy_{1}[y_{1}(1-y_{1})]^{-\epsilon}
\int_{0}^{1}dy_{2}[y_{2}(1-y_{2})]^{-1/2-\epsilon}.\label{domega}
\end{equation}
We recall for completeness that the integrals in
Eqs.(\ref{gendecps},\ref{domega}) are indicated only to remind the
integration range. They can be performed without the matching with
the square amplitude of the corrispondig process only for the
determination of their normalization. We can check explicitly that
Eq.(\ref{domega}) is correct performing its integral. This is
easily done using the definition of the $B$ function
\begin{equation}
B(z,w)\equiv\int_{0}^{1}dt\,t^{z-1}(1-t)^{w-1}=\frac{\Gamma(z)\Gamma(w)}{\Gamma(z+w)},
\end{equation}
and the Legendre duplication formula
\begin{equation}
\Gamma(2z)=(2\pi)^{-1/2}2^{2z-1/2}\Gamma(z)\Gamma(z+1/2).
\end{equation}
We find
\begin{equation}
\Omega_{3-2\epsilon}=\frac{2\pi^{3/2-\epsilon}}{\Gamma(3/2-\epsilon)},
\end{equation}
which is in agreement with Eq.(\ref{omeganorm}).

\chapter{Full NLO expression for DY rapidity
distributions}\label{appC}

We report here the complete expression of the NLO DY distributions given in \cite{Mukherjee:2006uu,Gehrmann:1997ez,Sutton:1991ay,Altarelli:1979ub} with
the factorization scale equal to the renormalization scale:
\bea
&&\frac{d\sigma^{NLO}}{dQ^{2}dY}=\emph{N}(Q^{2})\sum_{q,q'}c_{qq'}\int_{x_{1}^{0}}^{1}\frac{dx_{1}}{x_{1}}\int_{x_{2}^{0}}^{1}\frac{dx_{2}}{x_{2}}
\nonumber\\
&&\times\bigg\{\left[C^{(0)}_{q\bar{q}}(x_{1},x_{2},Y)+\frac{\alpha_{s}(\mu^{2})}{2\pi}C^{(1)}_{q\bar{q}}\left(x_{1},x_{2},Y,\frac{Q^{2}}{\mu^{2}}\right)\right]\nonumber\\
&&\times\left\{q(x_{1},\mu^{2})\bar{q'}(x_{2},\mu^2)+\bar{q}(x_{1},\mu^{2})q'(x_{2},\mu^{2})\right\}\nonumber\\
&&+\frac{\alpha_{s}(\mu^{2})}{2\pi}C^{(1)}_{gq}(x_{1},x_{2},Y)g(x_{1},\mu^{2})\left\{q'(x_{2},\mu^{2})+\bar{q'}(x_{2},\mu^{2})\right\}\nonumber\\
&&+\frac{\alpha_{s}(\mu^{2})}{2\pi}C^{(1)}_{qg}(x_{1},x_{2},Y)\left\{q(x_{1},\mu^{2})+\bar{q}(x_{1},\mu^{2})\right\}g(x_{2},\mu^{2})\bigg\},
\eea where \bea N(Q^{2})&=&\frac{4\pi\alpha^{2}}{9Q^{2}S}\quad
\textrm{for
$\gamma^{*}$},\\
N(Q^{2})&=&\frac{\pi G_{F}Q^{2}\sqrt{2}}{3S}\delta(Q^{2}-M^{2}_{V}) \quad \textrm{for $Z^{0}$ and $W^{\pm}$},
\eea
and where
\begin{eqnarray}
&&C^{(0)}_{q\bar{q}}(x_{1},x_{2},Y)=x_{1}x_{2}\delta(x_{1}-x_{1}^{0})\delta(x_{2}-x_{2}^{0}),
\quad x_{1(2)}^{0}=\sqrt{x}e^{\pm
Y},\\
&&C^{(1)}_{q\bar{q}}\left(x_{1},x_{2},Y,\frac{Q^{2}}{\mu^{2}}\right)=x_{1}x_{2}C_{F}\bigg\{\delta(x_{1}-x_{1}^{0})\delta(x_{2}-x_{2}^{0})\bigg[
\frac{\pi^2}{3}-8\nonumber\\
&&+2\textrm{Li}_{2}(x_{1}^{0})+2\textrm{Li}_{2}(x_{2}^{0})+\ln^{2}(1-x_{1}^{0})+\ln^{2}(1-x_{2}^{0})+2\ln\frac{x_{1}^{0}}{1-x_{1}^{0}}\nonumber\\
&&\times2\ln\frac{x_{2}^{0}}{1-x_{2}^{0}}\bigg]+\bigg(\delta(x_{1}-x_{1}^{0})\bigg[\frac{1}{x_{2}}-\frac{x_{2}^{0}}{x_{2}^{2}}-
\frac{x_{2}^{0\,2}+x_{2}^{2}}{x_{2}^{2}(x_{2}-x_{2}^{0})}\nonumber\\
&&\times\ln\frac{x_{2}^{0}}{x_{2}}+\frac{x_{2}^{0\,2}+x_{2}^{2}}{x_{2}^{2}}\left(\frac{\ln(1-x_{2}^{0}/x_{2})}{x_{2}-x_{2}^{0}}\right)_{+}
+\frac{x_{2}^{0\,2}+x_{2}^{2}}{x_{2}^{2}}\nonumber\\
&&\times\frac{1}{(x_{2}-x_{2}^{0})_{+}}\ln\frac{2x_{2}^{0}(1-x_{1}^{0})}{x_{1}^{0}(x_{2}+x_{2}^{0})}\bigg]+(1\leftrightarrow
2)\bigg)\nonumber\\
&&+\frac{G^{A}(x_{1},x_{2},x_{1}^{0},x_{2}^{0})}{[(x_{1}-x_{1}^{0})(x_{2}-x_{2}^{0})]_{+}}+H^{A}(x_{1},x_{2},x_{1}^{0},x_{2}^{0})+\ln\frac{Q^{2}}{\mu^{2}}\nonumber\\
&&\times\bigg\{\delta(x_{1}-x_{1}^{0})\delta(x_{2}-x_{2}^{0})\big[3+2\ln\frac{1-x_{1}^{0}}{x_{1}^{0}}+2\ln\frac{1-x_{2}^{0}}{x_{2}^{0}}\big]\nonumber\\
&&+\left(\delta(x_{1}-x_{1}^{0})\frac{x_{2}^{0\,2}+x_{2}^{2}}{x_{2}^{2}}\frac{1}{(x_{2}-x_{2}^{0})_{+}}+(1\leftrightarrow
2)\right)\bigg\}\bigg\},\\
&&C^{(1)}_{gq}(x_{1},x_{2},Y)=x_{1}x_{2}T_{F}\bigg\{\frac{\delta(x_{2}-x_{2}^{0})}{x_{1}^{3}}\bigg[(x_{1}^{0\,2}+(x_{1}-x_{1}^{0})^{2})
\nonumber\\
&&\times\ln\frac{2(x_{1}-x_{1}^{0})(1-x_{2}^{0})}{(x_{1}+x_{1}^{0})x_{2}^{0}}
+2x_{1}^{0}(x_{1}-x_{1}^{0})\bigg]+\frac{G^{C}(x_{1},x_{2},Y)}{(x_{2}-x_{2}^{0})_{+}}\nonumber\\
&&+H^{C}(x_{1},x_{2},Y)
+\ln\frac{Q^{2}}{\mu^{2}}\bigg\{\frac{\delta(x_{2}-x_{2}^{0})}{x_{1}^{3}}(x_{1}^{0\,2}+(x_{1}-x_{1}^{0})^{2})\bigg\}\bigg\},\\
&&C^{(1)}_{qg}(x_{1},x_{2},Y)=C^{(1)}_{gq}(x_{2},x_{1},-Y),
\end{eqnarray}
with
\bea
\textrm{Li}_{2}(x)&=&-\int_{0}^{x}dt\frac{\ln(1-t)}{t},\\
G^{A}(x_{1},x_{2},Y)&=&\frac{2(x_{1}x_{2}+x_{1}^{0}x_{2}^{0})(x_{1}^{2}x_{2}^{2}+x_{1}^{0\,2}x_{2}^{0\,2})}{x_{1}^{2}x_{2}^{2}(x_{1}+x_{1}^{0})(x_{2}+x_{2}^{0})},\\
H^{A}(x_{1},x_{2},Y)&=&-\frac{4x_{1}^{0}x_{2}^{0}(x_{1}x_{2}+x_{1}^{0}x_{2}^{0})}{x_{1}x_{2}(x_{1}x_{2}^{0}+x_{2}x_{1}^{0})^{2}},\\
G^{C}(x_{1},x_{2},Y)&=&\frac{2x_{2}^{0}(x_{1}^{0\,2}x_{2}^{0\,2}+(x_{1}x_{2}-x_{1}^{0}x_{2}^{0})^{2})(x_{1}x_{2}+x_{1}^{0}x_{2}^{0})}
{x_{1}^{3}x_{2}^{2}(x_{1}x_{2}^{0}+x_{2}x_{1}^{0})(x_{2}+x_{2}^{0})},
\eea
and
\beq
H^{C}(x_{1},x_{2},Y)=\frac{2x_{1}^{0}x_{2}^{0}(x_{1}^{0}x_{2}^{0}+x_{1}x_{2})(x_{1}x_{1}^{0}x_{2}^{2}+x_{1}^{0}x_{2}^{0}(x_{1}x_{2}^{0}+2x_{1}^{0}x_{2}))}
{x_{1}^{2}x_{2}^{2}(x_{1}x_{2}^{0}+x_{2}x_{1}^{0})^{3}}.
\eeq

\chapter{Proof of some identities of chapter \ref{QT}}\label{appG}

\section{Proof of Eq.(\ref{appa})}\label{appE}

We compute in the limit $\hat{q}_{\perp}^{2}\rightarrow 0$ the
distribution $\delta (M_{0}^{2})$ where $M_{0}^{2}$ is given by
Eq.(\ref{m0}). Firstly we have \bea\label{propdelta} \delta
(M_{0}^{2})=\frac{\xi_{1}\xi_{2}}{Q^{2}}\delta
((1-\xi_{1})(1-\xi_{2})+\hat{q}_{\perp}^{2}(1-\xi_{1}-\xi_{2})).
\eea Then taking a generic test function $f(\xi_{1},\xi_{2})$,
\bea
&&\int d\xi_{1}\int d\xi_{2}f(\xi_{1},\xi_{2})\delta ((1-\xi_{1})(1-\xi_{2})+\hat{q}_{\perp}^{2}(1-\xi_{1}-\xi_{2}))\nonumber\\
&=&\int d\xi_{1}\int
d\xi_{2}\frac{f(\xi_{1},\xi_{2})}{1-\xi_{2}+\hat{q}_{\perp}^{2}}\delta\left(1-\xi_{1}-\frac{\xi_{2}\hat{q}_{\perp}^{2}}{1-\xi_{2}+\hat{q}_{\perp}^{2}}\right)\nonumber\\
&=&\int d\xi_{1}\int
d\xi_{2}\frac{f(\xi_{1},\xi_{2})-f(\xi_{1},1)}{1-\xi_{2}+\hat{q}_{\perp}^{2}}\delta\left(1-\xi_{1}-\frac{\xi_{2}\hat{q}_{\perp}^{2}}{1-\xi_{2}+\hat{q}_{\perp}^{2}}\right)\nonumber\\
&&+\int d\xi_{1}f(\xi_{1},1)\int
d\xi_{2}\frac{1}{1-\xi_{2}+\hat{q}_{\perp}^{2}}\delta\left(1-\xi_{1}-\frac{\xi_{2}\hat{q}_{\perp}^{2}}{1-\xi_{2}+\hat{q}_{\perp}^{2}}\right)\nonumber\\
&=&\int d\xi_{1}\int
d\xi_{2}\frac{f(\xi_{1},\xi_{2})-f(\xi_{1},1)}{1-\xi_{2}+\hat{q}_{\perp}^{2}}\delta\left(1-\xi_{1}-\frac{\xi_{2}\hat{q}_{\perp}^{2}}{1-\xi_{2}+\hat{q}_{\perp}^{2}}\right)\nonumber\\
&&+\int
d\xi_{1}\frac{f(\xi_{1},1)}{1-\xi_{1}+\hat{q}_{\perp}^{2}}\int
d\xi_{2}\delta\left(1-\xi_{2}-\frac{\xi_{1}\hat{q}_{\perp}^{2}}{1-\xi_{1}+\hat{q}_{\perp}^{2}}\right)\nonumber\\
&=&\int d\xi_{1}\int
d\xi_{2}\frac{f(\xi_{1},\xi_{2})-f(\xi_{1},1)}{1-\xi_{2}+\hat{q}_{\perp}^{2}}\delta\left(1-\xi_{1}-\frac{\xi_{2}\hat{q}_{\perp}^{2}}{1-\xi_{2}+\hat{q}_{\perp}^{2}}\right)\nonumber\\
&&+\int d\xi_{1}\frac{f(\xi_{1},1)}{1-\xi_{1}+\hat{q}_{\perp}^{2}}\nonumber\\
&=&\int d\xi_{1}\int
d\xi_{2}\frac{f(\xi_{1},\xi_{2})-f(\xi_{1},1)}{1-\xi_{2}+\hat{q}_{\perp}^{2}}\delta\left(1-\xi_{1}-\frac{\xi_{2}\hat{q}_{\perp}^{2}}{1-\xi_{2}+\hat{q}_{\perp}^{2}}\right)\nonumber\\
&&+\int
d\xi_{1}\frac{f(\xi_{1},1)-f(1,1)}{1-\xi_{1}+\hat{q}_{\perp}^{2}}+f(1,1)\ln\frac{1+\hat{q}_{\perp}^{2}}{\hat{q}_{\perp}^{2}}.
\eea Now, taking the limit $\hat{q}_{\perp}^{2}\rightarrow 0$ and
using Eq.(\ref{propdelta}), we obtain the following identity in
the distribution sense \bea \delta
(M_{0}^{2})=\frac{\xi_{1}\xi_{2}}{Q^{2}}\left[\frac{\delta(1-\xi_{1})}{(1-\xi_{2})_{+}}+\frac{\delta(1-\xi_{2})}{(1-\xi_{1})_{+}}-\ln\hat{q}_{\perp}^{2}\delta(1-\xi_{1})
\delta(1-\xi_{2})\right]+\emph{O}(\hat{q}_{\perp}^{2}), \eea which
is exactly Eq.(\ref{appa}).

\section{Proof of Eq.(\ref{appb})}\label{appF}

Let us compute for a generic test function $f(\xi_{1},\xi_{2})$
the distribution \bea T_{\hat{q}_{\perp}^{2}}=\int d\xi_{1}\int
d\xi_{2}[(1-\xi_{1})(1-\xi_{2})+\hat{q}_{\perp}^{2}(1-\xi_{1}-\xi_{2})]^{\eta-1}f(\xi_{1},\xi_{2})
\eea in the small-$\hat{q}_{\perp}^{2}$ limit with
$\eta=a|\epsilon|$, $a>0$ and $\epsilon \neq 0$. The integration
range is fixed by the requirement \bea
M_{0}^{2}&=&\frac{Q^{2}}{\xi_{1}\xi_{2}}[(1-\xi_{1})(1-\xi_{2})+\hat{q}_{\perp}^{2}(1-\xi_{1}-\xi_{2})]\geq
0, \eea which gives \bea 0\leq\xi_{1}\leq 1\qquad
0\leq\xi_{2}\leq\frac{(1-\xi_{1})(1+\hat{q}_{\perp}^{2})}{1-\xi_{1}+\hat{q}_{\perp}^{2}}=\bar{\xi}_{2}
\eea or equivalently \bea 0\leq\xi_{2}\leq 1\qquad
0\leq\xi_{1}\leq\frac{(1-\xi_{2})(1+\hat{q}_{\perp}^{2})}{1-\xi_{2}+\hat{q}_{\perp}^{2}}=\bar{\xi}_{1}.
\eea We observe also that \bea\label{0}
(1-\xi_{1})(1-\xi_{2})+\hat{q}_{\perp}^{2}(1-\xi_{1}-\xi_{2})&=&(1-\xi_{2}+\hat{q}_{\perp}^{2})\left(1-\xi_{1}-\frac{\xi_{2}\hat{q}_{\perp}^{2}}{1-\xi_{2}+\hat{q}_{\perp}^{2}}\right)\nonumber\\
&=&(1-\xi_{1}+\hat{q}_{\perp}^{2})\left(1-\xi_{2}-\frac{\xi_{1}\hat{q}_{\perp}^{2}}{1-\xi_{1}+\hat{q}_{\perp}^{2}}\right).
\eea Next, we decompose \bea
T_{\hat{q}_{\perp}^{2}}=T^{1}_{\hat{q}_{\perp}^{2}}+T^{2}_{\hat{q}_{\perp}^{2}}
\eea where \bea
T^{1}_{\hat{q}_{\perp}^{2}}&=&\int_{0}^{1}d\xi_{1}(1-\xi_{1}+\hat{q}_{\perp}^{2})^{\eta-1}\int_{0}^{\bar{\xi_{2}}}d\xi_{2}\left(1-\xi_{2}-\frac{\xi_{1}\hat{q}_{\perp}^{2}}{1-\xi_{1}+\hat{q}_{\perp}^{2}}\right)^{\eta-1}[f(\xi_{1},\xi_{2})-f(\xi_{1},1)]\nonumber\\
T^{2}_{\hat{q}_{\perp}^{2}}&=&\int_{0}^{1}d\xi_{1}(1-\xi_{1}+\hat{q}_{\perp}^{2})^{\eta-1}\int_{0}^{\bar{\xi_{2}}}d\xi_{2}\left(1-\xi_{2}-\frac{\xi_{1}\hat{q}_{\perp}^{2}}{1-\xi_{1}+\hat{q}_{\perp}^{2}}\right)^{\eta-1}f(\xi_{1},1).
\eea The $\xi_{2}$ integral in $T^{2}_{\hat{q}_{\perp}^{2}}$ is
immediately performed: \bea\label{int}
\int_{0}^{\bar{\xi_{2}}}d\xi_{2}\left(1-\xi_{2}-\frac{\xi_{1}\hat{q}_{\perp}^{2}}{1-\xi_{1}+\hat{q}_{\perp}^{2}}\right)^{\eta-1}=
\frac{1}{\eta}\left[\frac{(1-\xi_{1})(1+\hat{q}_{\perp}^{2})}{1-\xi_{1}+\hat{q}_{\perp}^{2}}\right]^{\eta}.
\eea Therefore, \bea
T^{2}_{\hat{q}_{\perp}^{2}}=\frac{(1+\hat{q}_{\perp}^{2})^{\eta}}{\eta}\int_{0}^{1}d\xi_{1}f(\xi_{1},1)\frac{(1-\xi_{1})^{\eta}}{1-\xi_{1}+\hat{q}_{\perp}^{2}}.
\eea Proceeding as above, we regularize the $\xi_{1}$ integral:
\bea
T^{2}_{\hat{q}_{\perp}^{2}}=T^{21}_{\hat{q}_{\perp}^{2}}+T^{22}_{\hat{q}_{\perp}^{2}},
\eea where \bea
\label{1}T^{21}_{\hat{q}_{\perp}^{2}}&=&\frac{(1+\hat{q}_{\perp}^{2})^{\eta}}{\eta}\int_{0}^{1}d\xi_{1}\frac{(1-\xi_{1})^{\eta}}{1-\xi_{1}+\hat{q}_{\perp}^{2}}[f(\xi_{1},1)-f(1,1)]\\
\label{2}T^{22}_{\hat{q}_{\perp}^{2}}&=&\frac{(1+\hat{q}_{\perp}^{2})^{\eta}}{\eta}f(1,1)\int_{0}^{1}d\xi_{1}\frac{(1-\xi_{1})^{\eta}}{1-\xi_{1}+\hat{q}_{\perp}^{2}}\nonumber\\
&=&f(1,1)\frac{(1+\hat{q}_{\perp}^{2})^{\eta}}{\eta}\frac{(1+\hat{q}_{\perp}^{2})^{\eta}-(\hat{q}_{\perp}^{2})^{\eta}}{\eta}.
\eea The two distributions $T^{21}_{\hat{q}_{\perp}^{2}}$ and
$T^{22}_{\hat{q}_{\perp}^{2}}$ are now well defined as
$\hat{q}_{\perp}^{2}\rightarrow 0$. Similarly, \bea
T^{1}_{\hat{q}_{\perp}^{2}}=T^{11}_{\hat{q}_{\perp}^{2}}+T^{12}_{\hat{q}_{\perp}^{2}},
\eea where \bea
\label{3}T^{11}_{\hat{q}_{\perp}^{2}}&=&\int_{0}^{1}d\xi_{1}(1-\xi_{1}+\hat{q}_{\perp}^{2})^{\eta-1}\int_{0}^{\bar{\xi_{2}}}d\xi_{2}\left(1-\xi_{2}-\frac{\xi_{1}\hat{q}_{\perp}^{2}}{1-\xi_{1}+\hat{q}_{\perp}^{2}}\right)^{\eta-1}\\
&&\{[f(\xi_{1},\xi_{2})-f(\xi_{1},1)]-[f(1,\xi_{2})-f(11)]\}\nonumber\\
T^{12}_{\hat{q}_{\perp}^{2}}&=&\int_{0}^{1}d\xi_{1}(1-\xi_{1}+\hat{q}_{\perp}^{2})^{\eta-1}\int_{0}^{\bar{\xi_{2}}}d\xi_{2}\left(1-\xi_{2}-\frac{\xi_{1}\hat{q}_{\perp}^{2}}{1-\xi_{1}+\hat{q}_{\perp}^{2}}\right)^{\eta-1}\nonumber\\
&&[f(1,\xi_{2})-f(1,1)] \eea The term
$T^{12}_{\hat{q}_{\perp}^{2}}$ can be computed by changing the
order of integration and exploiting Eq.(\ref{0}) and using
eq(\ref{int}), thus obtaining \bea
\label{4}T^{12}_{\hat{q}_{\perp}^{2}}=\frac{(1+\hat{q}_{\perp}^{2})^{\eta}}{\eta}\int_{0}^{1}d\xi_{2}\frac{(1-\xi_{2})^{\eta}}{1-\xi_{2}+\hat{q}_{\perp}^{2}}[f(1,\xi_{2})-f(1,1)].
\eea Collecting Eqs.(\ref{1},\ref{2},\ref{3},\ref{4}), taking the
limit $\hat{q}_{\perp}^{2}\rightarrow 0$ and using the identity
\bea
(1-\xi)^{\eta-1}=(1-\xi)^{\eta-1}_{+}+\frac{1}{\eta}\delta(1-\xi),
\eea we obtain the following identity in sense of distributions:
\bea
&&\qquad\qquad\quad[(1-\xi_{1})(1-\xi_{2})+\hat{q}_{\perp}^{2}(1-\xi_{1}-\xi_{2})]^{\eta-1}=\\
&&=(1-\xi_{1})^{\eta-1}(1-\xi_{2})^{\eta-1}-\frac{(\hat{q}_{\perp}^{2})^{\eta}}{\eta^{2}}\delta(1-\xi_{1})\delta(1-\xi_{2})+\emph{O}(\hat{q}_{\perp}^{2})\nonumber
\eea which is exactly Eq.(\ref{appb}).

\chapter{Proof of combinatoric properties of chapter
\ref{predictivity}}\label{combinatorics}

Let us consider the coefficients defined by
\begin{equation}
C_{m}^{(i,j)}=\frac{(-1)^{m+1}}{m+1}\sum_{l=0}^{m}2^{l}{l+i-1
\choose i-1}{m-l+j-1 \choose j-1}.
\end{equation}
We shall now prove that
\begin{eqnarray}
C_{m}^{(i,0)}&=&2^{m}C_{m}^{(0,i)}\\
C_{m}^{(i,j)}&=&2C_{m}^{(i,j-1)}-C_{m}^{(i-1,j)};\qquad\quad
i,j\geq 1,
\end{eqnarray}
or, equivalently, that
\begin{eqnarray}
\tilde{C}_{m}^{(i,0)}&=&2^{m}\tilde{C}_{m}^{(0,i)}\label{rel2}\\
\tilde{C}_{m}^{(i,j)}&=&2\tilde{C}_{m}^{(i,j-1)}-\tilde{C}_{m}^{(i-1,j)};\qquad\quad
i,j\geq 1,\label{rel1},
\end{eqnarray}
where
\begin{equation}
\tilde{C}_{m}^{(i,j)}\equiv
(-)^{m+1}(m+1)C_{m}^{(i,j)}=\sum_{l=0}^{m}2^{l}{l+i-1 \choose
i-1}{m-l+j-1 \choose j-1}.
\end{equation}

Eq.(\ref{rel2}) follows immediately. Indeed:
\begin{eqnarray}
\tilde{C}_{m}^{(i,0)}=2^{m}{m+i-1 \choose i-1}\\
\tilde{C}_{m}^{(0,i)}={m+i-1 \choose i-1}\label{rel3}.
\end{eqnarray}

We turn now to the proof of  Eq.(\ref{rel1}). In the case $i=j=1$
it follows again immediately. In fact we have:
\begin{eqnarray}
\tilde{C}_{m}^{(0,1)}=1,\tilde{C}_{m}^{(1,0)}=2^{m},\tilde{C}_{m}^{(1,1)}=\sum_{l=0}^{m}2^{l}=2^{m+1}-1.
\end{eqnarray}
In the cases $i=1,j=2$ e $i=2,j=1$ it is quite easy:
\begin{eqnarray}
\tilde{C}_{m}^{(1,1)}&=&2^{m+1}-1,\quad\tilde{C}_{m}^{(0,2)}=m+1,\quad\tilde{C}_{m}^{(2,0)}=2^{m}(m+1)\\
\tilde{C}_{m}^{(1,2)}&=&2(2^{m+1}-1)-(m+1),\quad\tilde{C}_{m}^{(2,1)}=2^{m+1}(m+1)-(2^{m+1}-1),
\end{eqnarray}
where we have use the equality:
\begin{equation}
\sum_{l=0}^{m}l2^{l}=\lim_{\alpha\rightarrow
1}\frac{1}{\ln2}\frac{d}{d\alpha}\sum_{l=0}^{m}(e^{\alpha\ln2})^{l}=2^{m+1}(m+1)-2(2^{m+1}-1).
\end{equation}
In the other cases, i.e. for $i,j\geq 2$, it follows in a
straightforward way:
\begin{eqnarray}
2\tilde{C}_{m}^{(i,j-1)}-\tilde{C}_{m}^{(i-1,j)}=\qquad\qquad\qquad\qquad\qquad\qquad&&\nonumber\\
=2\sum_{l=0}^{m}2^{l}{l+i-1 \choose i-1}{m-l+j-2 \choose j-2}-\sum_{l=0}^{m}2^{l}{l+i-2 \choose i-2}{m-l+j-1 \choose j-1}&&\nonumber\\
=\tilde{C}_{m}^{(i,j)}-\sum_{l=0}^{m}2^{l+1}{l+i-1 \choose i-1}{m-l+j-2 \choose j-1}+\sum_{l=0}^{m}2^{l}{l+i-2 \choose i-1}{m-l+j-1 \choose j-1}&&\nonumber\\
=\tilde{C}_{m}^{(i,j)}-\sum_{l'=1}^{m+1}2^{l'}{l'+i-2 \choose i-1}{m-l'+j-1 \choose j-1}+\sum_{l=0}^{m}2^{l}{l+i-2 \choose i-1}{m-l+j-1 \choose j-1}&&\nonumber\\
=\tilde{C}_{m}^{(i,j)}-2^{m+1}{m+i-1 \choose i-1}{j-2 \choose
j-1}+{i-2 \choose i-1}{m+j-1 \choose j-1}\nonumber\\
=\tilde{C}_{m}^{(i,j)},\qquad\qquad&&
\end{eqnarray}
where $l'=l+1$ and where in the second, the third and the last
line we have used the following identities:
\begin{eqnarray}
{n+r-2 \choose r-2}&=&{n+r-1 \choose r-1}-{n+r-2 \choose r-1},\\
{r-2 \choose r-1}&=&\frac{\Gamma(r-1)}{\Gamma(r)\Gamma(0)}=0.
\end{eqnarray}
We note that these last two properties are valid only for $r\geq
2$ and this is the reason why the cases $i=j=1$ e $i=1,j=2$ and
$i=2,j=1$ have been treated separately.

\bibliographystyle{hunsrt}

\backmatter
\chapter{Acknowledgements}

I want to thank S. Forte and G. Ridolfi for giving me the
opportunity to work with them, for all the things that I have
learnt from them and for their friendship.

I thank my parents, because they wanted me to watch how much
beautiful is the world. Furthermore, I want to thank them for
their supporting and for all the times that they had reason.

I want to thank all the students of the Modern Physics, Quantum
Mechanics, General Physics and Mathematical Analysis courses,
because with their questions they have helped me not to forget
that I study physics because I love reality.

Finally, I want to thank all my friends, because I am not yet able
to give them back what they gave to me: the reason for not to quit
in any circumstance of life.

\end{document}